\documentclass[aps,rmp,twocolumn,superscriptaddress,floatfix,amsmath,amssymb]{revtex4}

\usepackage{hyperref}
\usepackage{epsfig}
\usepackage{bm}
\usepackage{graphicx}
\usepackage{dcolumn}
\usepackage{bm}
\usepackage{longtable}

\begin{document}


\title{Spinodal nanodecomposition in semiconductors doped with transition metals}

\author{T. Dietl}
\affiliation{Institute of Physics, Polish Academy of Sciences, PL-02-668 Warszawa, Poland}
\affiliation{Institute of Theoretical Physics, Faculty of Physics, University of Warsaw, PL-02-093 Warszawa, Poland}
\affiliation{WPI-Advanced Institute for Materials Research (WPI-AIMR), Tohoku University, Sendai 980-8577, Japan}

\author{K. Sato}
\email{ksato@mat.eng.osaka-u.ac.jp}
\affiliation{Graduate School of Engineering, Osaka University, Suita, Osaka 565-0871, Japan}
\affiliation{Institute for NanoScience Design, Osaka University, Toyonaka, Osaka 560-8531, Japan}

\author{T. Fukushima}
\affiliation{Graduate School of Engineering Science, Osaka University,  Toyonaka, Osaka 560-8531, Japan}

\author{A. Bonanni}
\email{alberta.bonanni@jku.at}
\affiliation{Institut f\"ur Halbleiter-und-Festk\"orperphysik, Johannes Kepler University, A-4040 Linz, Austria}

\author{M. Jamet}
\affiliation{Commissariat \`a l'Energie Atomique, INAC/SP2M-UJF, F-38054 Grenoble, France}

\author{A. Barski}
\affiliation{Commissariat \`a l'Energie Atomique, INAC/SP2M-UJF, F-38054 Grenoble, France}

\author{S. Kuroda}
\affiliation{Institute of Materials Science, University of Tsukuba, Tsukuba, Ibaraki 305-8573, Japan}

\author{M. Tanaka}
\affiliation{Department of Electrical Engineering and Information Systems, The University of Tokyo, Tokyo 113-8656, Japan}

\author{Pham~Nam~Hai}
\affiliation{Department of Physical Electronics, Tokyo Institute of Technology, Tokyo 152-8552, Japan}

\author{H.~Katayama-Yoshida}
\affiliation{Graduate School of Engineering Science, Osaka University,  Toyonaka, Osaka 560-8531, Japan}
\affiliation{Institute for NanoScience Design, Osaka University, Toyonaka, Osaka 560-8531, Japan}

\date{\today}

\begin{abstract}
This review presents the recent progress in computational materials
design, experimental realization, and control methods of spinodal
nanodecomposition under three- and two-dimensional crystal-growth
conditions in spintronic materials, such as magnetically doped semiconductors.
The computational description of nanodecomposition, performed
by combining first-principles calculations with kinetic Monte Carlo simulations, is discussed together
with extensive electron microscopy, synchrotron radiation, scanning probe, and ion beam methods that have
been employed to visualize binodal and spinodal nanodecomposition (chemical phase separation) as well as nanoprecipitation
(crystallographic phase separation) in a range of semiconductor compounds with a concentration of
transition metal (TM) impurities beyond the solubility limit.
The role of growth conditions, codoping by shallow impurities, kinetic barriers, and surface reactions in controlling
the aggregation of magnetic cations is highlighted. According to theoretical simulations
 and experimental results the TM-rich regions appear in the form of either nanodots (the {\em dairiseki} phase) or nanocolumns
(the {\em konbu} phase) buried in the host semiconductor. Particular attention is paid to Mn-doped group
III arsenides and antimonides,
TM-doped group III nitrides, Mn- and Fe-doped Ge, and Cr-doped group II chalcogenides, in which ferromagnetic features
persisting up to above room temperature correlate with the presence of nanodecomposition and account for
the application-relevant magnetooptical and magnetotransport properties of these compounds.
Finally, it is pointed out that spinodal nanodecomposition can be
viewed as a new class of bottom-up approach to nanofabrication.

\end{abstract}



\maketitle
\tableofcontents

\section{Introduction}
\label{sec:introduction}
The detection of ferromagnetic features persisting up to above room temperature in a variety of magnetically doped semiconductors and oxides has been one of the most surprising developments in materials science over recent years
\cite{Bonanni:2007_SST,Dietl:2003_NM,Liu:2005_JMSME,Fukumura:2013_B,Pearton:2003_JAP,Coey:2010_NJP,Roever:2011_PRB,Kuzemsky:2013_JMPB,Nealon:2012_NS, Sawicki:2013_PRB,Kobayashi:2008_NJP,Makarova:2010_B,Yao:2012_APL}. In particular, the presence of robust ferromagnetism in technology-relevant semiconductors  (e.g., GaN and Si) and oxides  (e.g., ZnO and TiO$_2$) has promised to open the door to a wide exploitation in devices of remarkable spintronic functionalities found in dilute ferromagnetic semiconductors [such as (Ga,Mn)As]  below their Curie temperature $T_{\text{C}}$, so far not exceeding 200~K \cite{Dietl:2014_RMP,Jungwirth:2014_RMP,Sato:2010_RMP}. However, it is increasingly clear that high-$T_{\text{C}}$ nonmetal ferromagnets represent a distinct class of systems in which the strength of ferromagnetic (FM) features correlates neither with the density of band holes necessary to mediate long-range interactions between diluted spins nor with an average concentration of magnetic ions.

In this review we focus on the latest progress in the investigations of structural and magnetic properties of dilute magnetic semiconductors (DMSs), such as group III pnictides, e.g., GaAs and GaN, group II chalcogenides, e.g., ZnTe, and group IV elemental semiconductors, e.g., Ge, doped with transition metal (TM) ions, either Cr, Mn, or Fe. These studies have pointed out that, except for Mn in II-VI compounds \cite{Pajaczkowska:1978_PCGC,Giebultowicz:1993_PRB}, the fundamental premise of DMSs on the random distribution of magnetic ions over cation sites is often not valid, and that the aggregation of magnetic cations is strictly correlated with the presence of high-$T_{\text{C}}$ ferromagnetism in these systems. As described here, this striking conclusion has been drawn from state-of-the-art theoretical modeling and experimental investigations.  More specifically, the present understanding of semiconductors doped with transition metals has been built by combining:
\begin{itemize}
\item {\em Ab initio} computational studies providing the magnitude of chemical interactions between magnetic cations depending on their distance, charge state, and location (surface versus bulk); the interaction energies constitute the input parameters for kinetic simulations of spinodal nanodecomposition within the Monte Carlo method or by solving the relevant Cahn-Hilliard equation.
\item Extensive epitaxy and post-growth processing protocols, including codoping and annealing, allowing to establish the relationship between fabrication conditions and physical properties.
\item Comprehensive nanocharacterization investigations employing ever improving methods of electron microscopy, synchrotron radiation, ion beams, and scanning probes that provide element-specific information on atom distributions, charge states, magnetic properties, and defects with spatial resolution down to the nm range.
\item Scrupulous magnetization and magnetic resonance studies involving also the extensive examination of reference samples containing nominally no magnetic ions but otherwise grown and processed in the same way as the films under investigation.
\end{itemize}

According to the results reviewed in this paper, this methodology has made it possible to bring to light theoretically and experimentally a number of unanticipated properties of magnetically doped semiconductors. The new findings can be summarized as follows:
 \begin{enumerate}
\item In DMSs, open orbitals of TM impurities not only provide localized spins but also, via $p$--$d$ hybridization, contribute to the bonding, which usually results in attractive forces between magnetic cations and in a miscibility gap in the thermodynamic phase diagram of the alloy.
 \item The actual TM distribution, for given fabrication conditions, post-growth processing, and epitaxial strain, is determined by a competition between attractive forces, entropy terms, and kinetic barriers at the growth surface or in the sample volume. This results in the coexitance of TM ions in random cation-substitutional positions and in regions with high concentrations of magnetic constituents, either commensurate with the host lattice (chemical phase separation) or in the form of precipitates (crystallographic phase separation). The process of TM aggregation is referred to as spinodal nanodecomposition although nucleation mechanisms (binodal decomposition) are involved in many cases.
\item The buried regions with high concentrations of magnetic constituents [called condensed magnetic semiconductors (CMSs), even if an insulator-to-metal transition occurred locally] appear as TM-rich nanodots (distributed randomly over the sample volume or clustered in defined planes) or in the form of TM-rich nanocolumns extending along specific crystal directions.
\item The incorporation and distribution of magnetic ions depend on codoping by shallow impurities or electrically active defects, which rather than changing the carrier concentration alter the valence of the TM ions and, thus, the chemical forces and kinetic barriers controlling the aggregation of magnetic cations either at the growth surface or in the film volume.
\item Because of the bonding to host atoms and the effects of strains, the structural and magnetic properties of particular CMS nanocrystals (NCs) may not yet be listed in chemical compendia and have to be assessed experimentally.
\item Depending on the predominant character of the exchange interactions, the spins within the {\em individual} CMS NCs exhibit  FM, ferrimagnetic (FR), or antiferromagnetic (AF) spin ordering that owing to the high concentrations of TM ions persists typically to above room temperature (RT). At the same time, local strains, the character of interfaces, internal spin-orbit interactions, and the shapes of TM-rich regions control the magnetic anisotropy of particular CMS NCs.
\item Samples of nominally the same DMS with a given average TM concentration, can show diverse structural and macroscopic magnetic properties, as the relative  abundance of randomly distributed TM ions and NCs formed by various phase separation processes depends on fabrication conditions, codoping, and defect content. Except for DMSs with a large concentration of holes, randomly distributed TM spins can order only at low temperatures, whereas the NCs can lead a FM, AF, and/or superparamagnetic  behavior persisting up to high temperatures.
\item As known, noninteracting ferromagnetic nanoparticles exhibit a superparamagnetic behavior. However, decomposed magnetic alloys show typically temperature-independent narrow and leaning magnetic hysteresis loops up to the ordering temperature of the relevant CMS, even if the diameters of the TM-rich regions are in the sub 10\,nm range. These superferromagnetic characteristics point to the important role of dipole interactions or strain-mediated spin-dependent coupling between the NCs. Ferromagnetic like features can result also from uncompensated spins at the surface of AF NCs.
\item The outstanding microscopic structure of decomposed alloys points to novel functionalities beyond those foreseen for hole-mediated uniform dilute FM semiconductors. Nevertheless, similar to the case of uniform systems, the macroscopic magnetotransport and magnetooptical properties of decomposed magnetic alloys are usually correlated with macroscopic magnetic characteristics.  However, decomposed alloys additionally show a mixing between diagonal and nondiagonal components of the conductivity tensors and the persistence of magnetooptical effects in the spectral region below the band gap.
\item Magnetic studies of thin epitaxial layers containing a small quantity of magnetic impurities are challenged by the uncontrolled contamination of samples by FM nanoparticles and microparticles.
\end{enumerate}

Altogether, one can now obtain various semiconductors with spatial distributions of TM cations on demand. In particular, by selecting appropriate growth conditions, layers' layout, codoping, and/or processing protocols it becomes possible to fabricate either a uniform DMS or a decomposed alloy with a predefined (i) character of the phase separation (chemical versus crystallographic), (ii)  type of TM aggregation (nanodots versus nanocolumns), (iii) plane at which nanodots assemble, and (iv) chemical and magnetic properties of the precipitates. This command over the structural and magnetic properties offers a spectrum of opportunities for the design and realization of modulated systems with properties and functionalities encountered neither in uniform DMSs nor in the case of self-assembled quantum dots or nanowires of nonmagnetic semiconductors.

As seen in the Table of Contents, the main body of the present review consists of three major parts.

First, we discuss theoretical aspects of nanodecomposition in magnetically doped semiconductors (Sec.\,\ref{sec:theory}). We start (Sec.\,\ref{sec:ab-initio}) by presenting first-principles {\em ab initio} approaches and their applications (Sec.\,\ref{sec:pairing}) for determining the chemical forces between magnetic ions, either in bulk or at the growth surface, as well as their dependence on codoping with shallow impurities. In Secs.\,\ref{sec:dairiseki} and \ref{sec:konbu} we describe Monte Carlo simulations of three-dimensional (3D) and two-dimensional (2D) spinodal decomposition, corresponding respectively to the spontaneous formation of TM-rich nanodots in the film volume and to the self-assembly of TM-rich nanocolumns during the epitaxy. Taking into account shape magnetic anisotropy, samples containing nanodots and nanocolums exhibit quite different superparamagnetic properties, as discussed in Sec.\,II.D. Next, in Sec.\,\ref{sec:mixing} the computed phase diagrams of some binary alloy are presented, allowing us to recall the notions of miscibility gap, nucleation, and spinodal decomposition. We also emphasize the importance of kinetic barriers in the process of TM aggregation. These concepts make it possible to discuss, in Sec.\,\ref{sec:CH}, the Cahn-Hilliard equation quantifying the dynamics of spinodal decomposition.  As a whole, theoretical modeling elucidates the nature of nanodecomposition for different combinations of hosts, TM ions, shallow impurities, and fabrication procedures.

Second, in Secs.\,\ref{sec:GaAs}--\ref{sec:ZnTe} experimental studies of nanodecomposition in specific families of magnetically doped semiconductors are presented. Whenever possible the findings are discussed in the context of the theoretical predictions outlined in Sec.\,\ref{sec:theory}. We start (Sec.\,\ref{sec:GaAs}) with the model FM semiconductor (Ga,Mn)As. For this system pioneering works aiming at the fabrication of GaAs containing MnAs nanoprecipitates were carried out \cite{De_Boeck:1996_APL,Shi:1996_S}. Furthermore, early {\em ab initio} studies of (Ga,Mn)As and related systems revealed the presence of strong attractive forces between TM cation pairs in DMSs \cite{Schilfgaarde:2001_PRB}. We describe the chemical and crystallographic phase separations occurring in (Ga,Mn)As under annealing or epitaxy at appropriately high temperatures. Structural information on Mn-rich (Mn,Ga)As NCs is linked to FM-like features as well as to device-relevant magnetic circular dichroism (MCD) and magnetotransport characteristics of this nonocomposite system. The understanding of ferromagnetism in decomposed (Ga,Mn)As allows us to assess the origin of high-$T_{\text{C}}$ in (Ga,Mn)P, (In,Mn)As, (Ga,Mn)Sb, and (In,Mn)Sb.

In Secs.\,\ref{sec:GaN-Mn} and \ref{sec:GaN-Fe} we describe investigations of group III nitrides, doped with either Mn or Fe and, in some studies, codoped with Si or Mg. In the case of (Ga,Mn)N (Sec.\,\ref{sec:GaN-Mn}) a rich collection of magnetic properties at similar average Mn content is observed. The nanodecomposition scenario is confirmed by structural nanocharacterization demonstrating the correlation of high-$T_{\text{C}}$ and low-$T_{\text{C}}$ with respectively the presence and absence of a phase separation. Similar experiments point to spinodal decomposition, in the form of nanocolumns, in (Al,Cr)N \cite{Gu:2005_JMMM}. Section \ref{sec:GaN-Fe} is devoted to (Ga,Fe)N, the subject of particularly comprehensive structural and magnetic studies. The collected data provide evidences for spinodal decomposition as well as for the precipitation of various Fe-rich (Fe,Ga)$_x$N NCs whose composition $x$ (and thus magnetic properties), abundance, and location in predefine planes can be controlled by growth conditions, codoping by Si and Mg, and architecture.

Sections \ref{sec:Ge-Mn} and \ref{sec:Ge-Fe} present the outcome of nanodecomposition studies on Mn- and Fe-doped Ge, respectively, in which the TM distribution is uniform or either chemical or crystallographic phase separation is observed depending on the growth temperature. In addition to detailed structural and magnetic investigations, comprehensive MCD and magnetotransport data are available for these systems. Particularly relevant are works on the formation of Mn-rich nanocolumns along the [001] growth direction in (Ge,Mn) deposited onto Ge(001) substrates.

As described in Sec.\,\ref{sec:ZnTe}, in the case of (Zn,Cr)Te spinodal decomposition results in either Cr-rich nanodots or nanocolumns assuming a $\langle 111\rangle$ orientation, even for epitaxy on (001) substrates. Despite chemical heterogeneity there is a strict correlation between magnetization, MCD, and anomalous Hall effect (AHE) in these alloys. Furthermore, nanodecomposition and, thus, other associated properties are strongly affected by changing the concentrations of donors or acceptors, either by manipulating with stoichiometry via altering the intensities of the molecular beams or by codoping with I or N, respectively. Pioneering work drawing attention to the possibility of spinodal nanodecomposition in (Zn,Cr)Se \cite{Karczewski:2003_JSNM} is also discussed.

In the third part of our review, Sec.\,\ref{sec:prospects}, we discuss application prospects of the remarkable properties revealed for decomposed magnetically doped semiconductors over the recent years. We emphasize the possibility of bottom-up nanotechnology based on the control of spinodal nanodecomposition ("spinodal nanotechnology"). It is pointed out that decomposed magnetic alloys consisting, for instance, of a semiconductor with embedded  NCs of a FM metal can exhibit functionalities that cannot be realized employing either uniform FM metal films or semiconductor quantum dot layers. Finally, in Sec.\,\ref{sec:summary}, we summarize the main conclusions stemming from studies of heterogeneous DMSs and present an outlook on open questions and challenges ahead. Furthermore, a list of abbreviations is provided.

As seen, the physics of high-$T_{\text{C}}$ ferromagnetism in magnetically doped oxides \cite{Fukumura:2013_B,Coey:2010_NJP,Sawicki:2013_PRB,Li:2012_PRB}, carbon derivatives \cite{Kuzemsky:2013_JMPB,Makarova:2010_B,Wang:2014_PRB}, and some other systems \cite{Roever:2011_PRB,Nealon:2012_NS,Yao:2012_APL,Rylkov:2012_JETPL} is beyond the scope of this review. In particular, the question of ferromagnetism originating from spins residing on defects \cite{Coey:2010_NJP,Zhou:2014_NIMPRB} or on open $p$ shells \cite{Volnianska:2010_JPCM}, or mediated by defects or by residual impurities like hydrogen \cite{Li:2012_PRB} is not addressed here. We  note that not only several authors have considered theoretically whether ferromagnetism would be possible without magnetic elements \cite{Elfimov:2002_PRL,Kenmochi:2004_JJAP,Kenmochi:2004_JPSJ,Kenmochi:2005_JJAP,Bouzerar:2006_PRL,Ivanovskii:2007_PU,Mavropoulos:2009_PRB,Volnianska:2010_JPCM,Droghetti:2009_PRB,Du:2012_PRL} but also the role of spinodal nanodecomposition was examined in this context \cite{Seike:2012_JJAP,Seike:2013_JKPS,Seike:2013_SSC}.

As emphasized in our survey, the recent progress in understanding high-$T_{\text{C}}$ DMSs results, to a large extent, from the application of various powerful nanocharacterization tools. Nevertheless, we do not discuss in detail relevant experimental techniques, as many of them were recently reviewed in a collection \cite{Bonanni:2011_SST} that contains also useful information about the methodology of magnetic measurements on thin films employing superconducting quantum interference device (SQUID) magnetometry \cite{Sawicki:2011_SST}. A related experimental challenge is the contamination by FM nanoparticles and microparticles that already reside in the substrate or can be incorporated during the growth, annealing, etching, handling, or storing of particular samples. Some instructive examples were disclosed \cite{Abraham:2005_APL,Grace:2009_AM,Matsubayashi:2002_N,Makarova:2006_N} but presumably much more cases, even if spotted, have not been published.

\section{Theory of spinodal nanodecomposition}
\label{sec:theory}

\subsection{{\em Ab initio} materials design}
\label{sec:ab-initio}

Ideally theoretical studies should predict the microscopic TM distribution, including the presence of chemical and crystallographic phase separations, and the corresponding electronic and magnetic properties at given growth conditions and codoping by shallow impurities. As already reviewed elsewhere \cite{Sato:2010_RMP}, first-principles methods for electronic structure calculations have played the important role in predicting various properties of homogeneous and heterogenous DMSs without referring to any experimental parameters. This is mainly due to the success of the local density approximation (LDA) in the density functional theory (DFT)  \cite{Hohenberg:1964_PR,Kohn:1965_PR,Dreizler:1995_B}.

In the DFT, a many electron system is described by using the electron density
$n(\mathbf{r})$, i.e., it is proven that the many electron wave function $\Phi$ and
the expectation value of a physical quantity $\langle A \rangle$ can be written
as a functional of $n(\mathbf{r})$, such as $\Phi [n(\mathbf{r})]$ and
$A [n(\mathbf{r})]$, respectively. In particular, one can search for the
ground state by minimizing the total energy functional $E [n(\mathbf{r})]$ with
respect to  $n(\mathbf{r})$. In the Kohn-Sham scheme, the total energy functional
is formulated by referring to a one-electron system and given as,
\begin{eqnarray}
E[n(\mathbf{r})] = T_{\text{op}}[n(\mathbf{r})]+\int n(\mathbf{r})v(\mathbf{r})d^{3}r + \nonumber \\
+ \frac{1}{2}\int\int e^{2}\frac{n(\mathbf{r})\cdot n(\mathbf{r}^{\prime})}
{\mid\mathbf{r}-\mathbf{r}^{\prime}\mid}d^{3}r d^{3}r^{\prime}
+E_{\text{xc}}[n(\mathbf{r})].
\end{eqnarray}
Here $T_{\text{op}}[n(\mathbf{r})]$ is the electrons' kinetic energy,
the second term is the energy of the electrons in an external potential $v(\mathbf{r})$,
the third term is the classical Hartree interaction energy, and
the last term is the exchange-correlation energy.
For the exchange-correlation energy, a rigorous expression is not known and an efficient
approximation is needed  \cite{Dreizler:1995_B}.

The standard approximation is the LDA within which $E_{\text{xc}}$ is taken as
the exchange-correlation energy of homogeneous electron gas \cite{Dreizler:1995_B,Ceperly:1980_PRL,Perdew:1986_PRB}.
Thus, within the LDA, $E_{\text{xc}}$ is calculated as a function of the
local electron density, i.e., $E_{\text{xc}}[n(\mathbf{r})] \rightarrow E_{\text{xc}}(n(\mathbf{r}))$.
This approximation sounds inaccurate since the electron density distribution in a
real material varies strongly depending on the position $\mathbf{r}$.
Furthermore, the computations are carried out for supercells containing typically less than 100 atoms and assuming periodic boundary conditions.
Actually, despite these approximations, the LDA, or its spin-polarized version, the local spin density approximation (LSDA), is accurate enough for explaining
many physical properties of various materials \cite{Martin:2004_B}. Recently, LDA, LSDA, and their modifications are used not only
for understanding experimental results but also for designing new functional materials.

It is well known that the phase diagrams of a number of alloys exhibit a solubility gap in a certain concentration range.  Particularly low is the solubility of TM impurities in semiconductors. An exception here is a large solubility of Mn in II-VI compounds, in which Mn atoms remain distributed randomly over the substitutional cation sites up to concentrations often exceeding 50\% \cite{Pajaczkowska:1978_PCGC,Furdyna:1988_B}. The large solubility of Mn in II-VI compounds can be assigned to the truly divalent character of Mn whose $d$ states little perturb the $sp^3$ tetrahedral $sp$ bonds as both the lower $d^5$ (donor) and the upper $d^6$ (acceptor) Hubbard levels are respectively well below and above the band edges \cite{Dietl:1981_B,Dietl:2002_SST,Zunger:1986_B}.

We describe now two {\em ab initio} approaches that have provided quantitative information on the {\em thermodynamic} stability of particular DMS alloys A$_{1-x}$TM$_x$B as well as input parameters [such as the pairing energy $E_d$ and mixing energy $\Delta E(x)$] for studies of decomposition dynamics by the Monte Carlo simulations and by the Cahn-Hilliard equation, respectively. Within the first method one evaluates a change of the system energy associated with bringing two or more TM ions to neighboring cation positions. If this pairing energy $E_d$ is negative, i.e., there is an attractive chemical force between TM ions, their distribution may not be random. The second approach involves the evaluation of the alloy energy in comparison to weighted energies of the end compounds. A positive value of this mixing energy $\Delta E(x)$ points to the instability of the alloy.

\subsection{Pairing energy}
\label{sec:pairing}
The pairing energy $E_d$ (known also as heat of reaction or pair interaction energy) for a DMS alloy (A,TM)B is evaluated from total energies $E$ corresponding to three different contents of TM cations in the supercell \cite{Schilfgaarde:2001_PRB},
\begin{eqnarray}
E_d &=& E[(\mbox{A}_{N-2},\mbox{TM}_2)\mbox{B}_N] + E[\mbox{A}_N\mbox{B}_N]-\nonumber \\
 &-& 2E[(\mbox{A}_{N-1},\mbox{TM})\mbox{B}_N],
\end{eqnarray}
where $N$ is the total number of cations in the supercell. The magnitudes of $E_d$ for TMs at the nearest-neighbor cation positions were determined employing various DFT implementations for (Ga,Mn)As \cite{Schilfgaarde:2001_PRB,Mahadevan:2005_APL,Sato:2005_JJAP,Birowska:2012_PRL}; (Ga,V)As, (Ga,Cr)As, and (Ga,Fe)As \cite{Mahadevan:2005_APL}; (Ga,Mn)N \cite{Schilfgaarde:2001_PRB,Das:2003_PRB,Sato:2005_JJAP,Boguslawski:2006_APL,Chan:2008_PRB,Gonzalez:2011_PRB,Raebiger:2014_JAP}; (Ga,Cr)N \cite{Schilfgaarde:2001_PRB,Mahadevan:2005_APL,Cui:2005_PRL,Gonzalez:2011_PRB}; (Ga,Fe)N \cite{Gonzalez:2011_PRB,Navarro:2011_PRB}; (Ge,Mn,Cr,Co) \cite{Continenza:2006_APL}; (Zn,Mn)Te \cite{Kuroda:2007_NM}, and (Zn,Cr)Te \cite{Fukushima:2006_JJAP,Kuroda:2007_NM,Da_Silva:2008_NJP,Raebiger:2014_JAP}.

\begin{figure}[hb]
\includegraphics[angle=0,width=3.1in]{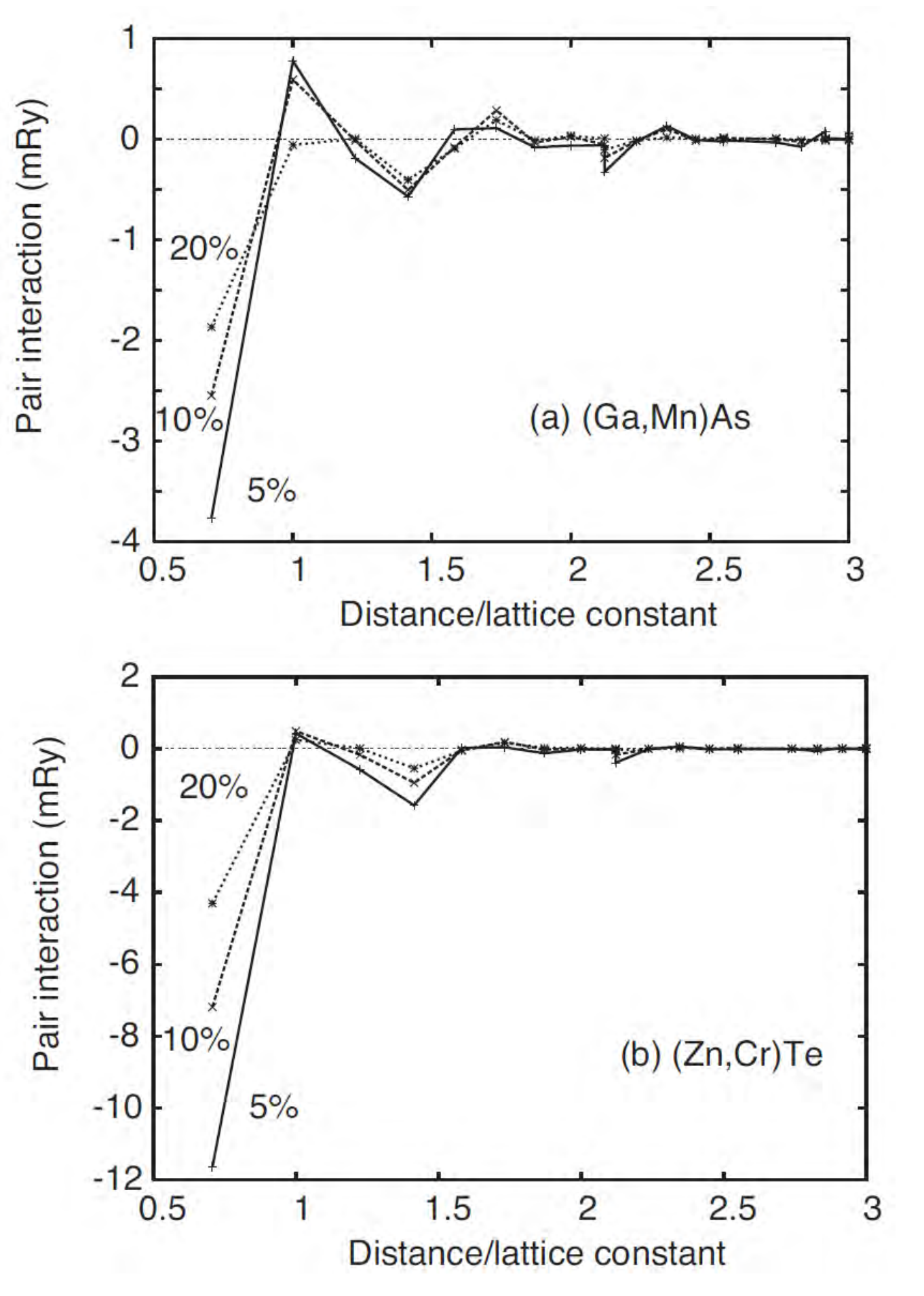}
\caption{\label{fig:ksato-pair} Calculated pair interaction energies $E_d$ as a function of the distance between TM pairs
in (a) Ga$_{1-x}$Mn$_{x}$As (from \onlinecite{Sato:2005_JJAP}) and (b) Zn$_{1-x}$Cr$_x$Te (from \onlinecite{Fukushima:2006_JJAP}) with various average concentrations $x$ of TM ions.}
\end{figure}

In the case of the nearest-neighbor Mn cations in ZnTe, $E_d = 21$~meV \cite{Kuroda:2007_NM}, the result consistent with a large solubility of Mn in II-VI compounds, as mentioned previously. In contrast, $E_d = -160$~meV for Cr cations in Zn$_{0.95}$Cr$_{0.05}$Te \cite{Fukushima:2006_JJAP}, which indicates that Cr cations should aggregate provided that diffusion barriers can be overcome at the growth or annealing temperature. Similarly, the pairing energy of Mn cation dimers was computed to be $E_d = -120$~meV in GaAs and $-300$~meV in GaN \cite{Schilfgaarde:2001_PRB}.  Being determined by $p$--$d$ hybridization, the magnitude of $|E_d|$ decays quickly with the distance between TM cations, as shown in Fig.~\ref{fig:ksato-pair}.

A number of experimentally relevant extensions of this approach has been proposed.

\begin{figure}
\includegraphics[width=3.3in]{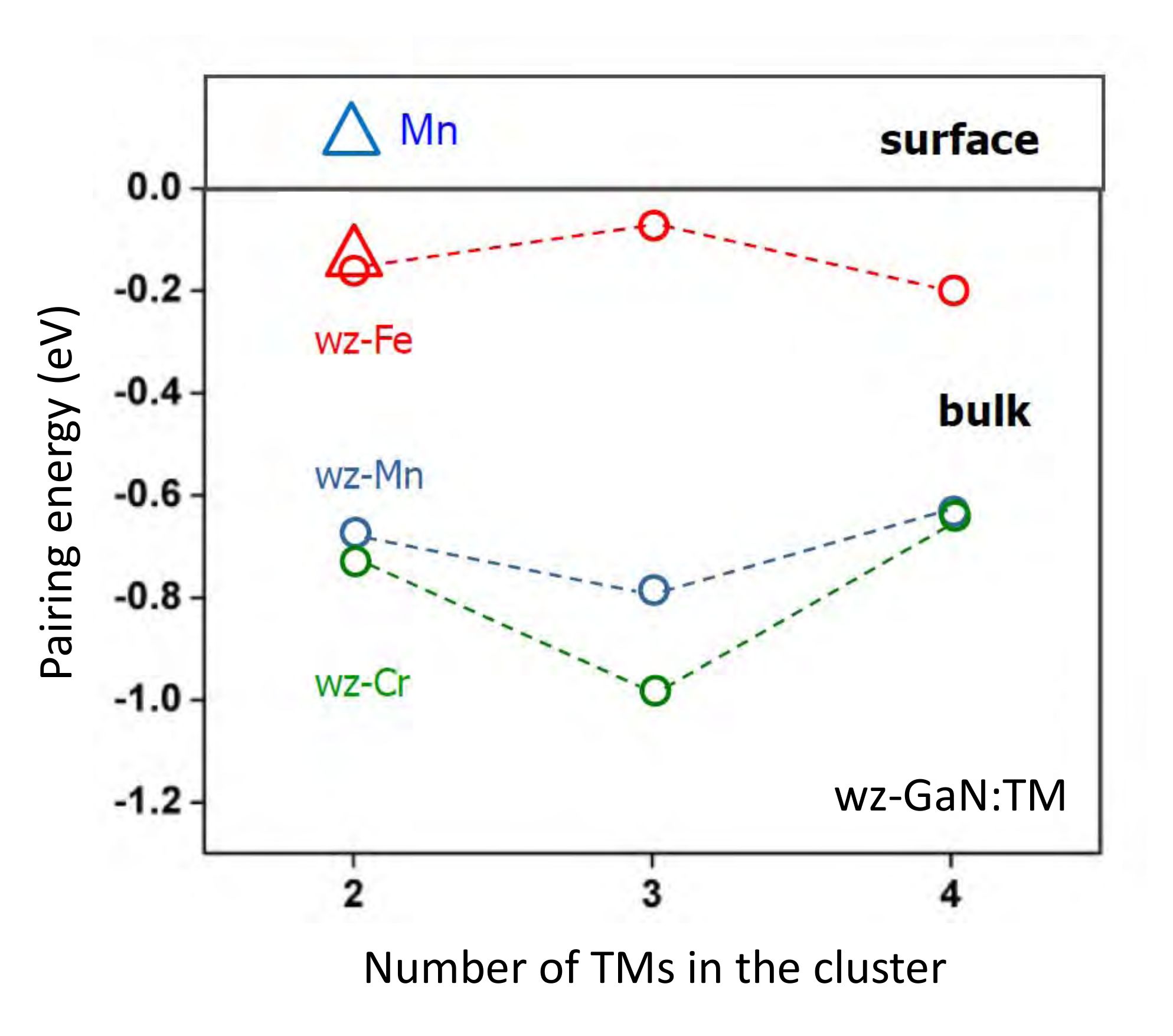}
\caption{(Color online) Pairing energies $E_d^{(n)}$ as a function of the number $n$ of TM cations in Cr, Mn, and Fe clusters in bulk wz-GaN (circles) and on Ga(0001) surface (triangles). Adapted from \onlinecite{Gonzalez:2011_PRB}.}
\label{fig:FIG_2_NGS}
\end{figure}

{\em Beyond dimers}: The formations of larger TM cation clusters than dimers were considered and their magnetic properties assessed \cite{Schilfgaarde:2001_PRB,Das:2003_PRB,Mahadevan:2005_APL,Gonzalez:2011_PRB,Cui:2007_PRB,Hynninen:2006_JPC,Navarro:2011_PRB}. In this case \cite{Schilfgaarde:2001_PRB},
 \begin{eqnarray}
E_d^{(n)} = E[(\mbox{A}_{N-n},\mbox{TM}_n)\mbox{B}_N] + E[\mbox{A}_N\mbox{B}_N]-\nonumber \\
 - E[(\mbox{A}_{N -n+1},\mbox{TM})\mbox{B}_N] - E[(\mbox{A}_{N-1},\mbox{TM})\mbox{B}_N].
\end{eqnarray}
This represents the energy change associated with the trapping of one more TM cation by a cluster consisting of $n-1$ TM cations. Early studies \cite{Schilfgaarde:2001_PRB} suggested that $E_d^{(n)} \simeq 0$ for $n=3$ in the case of (Ga,Mn)As, (Ga,Mn)N, and (Ga,Cr)N, i.e., that decomposition of these alloys into systems containing clusters with three TM cations is favored energetically. This conclusion is not supported by a more recent investigation indicating that $E_d^{(n)}$ remains strongly negative up to at least $n = 4$ \cite{Gonzalez:2011_PRB}, as shown in Fig.\,\ref{fig:FIG_2_NGS}. This work also pointed out that coupling between pairs of Mn and of Cr ions is FM in wurtzite (wz) GaN, whereas the interaction between Fe ions is AF. These expectations are discussed in comparison to experimental results in Secs.~\ref{sec:GaN-Mn} and \ref{sec:GaN-Fe}.

The structure of Mn clusters in (Ga,Mn)N and the interactions of
the magnetic Mn ions and clusters have been
studied through first-principles calculations and $T_{\mathrm{C}}$ was
evaluated using mean-field and Monte Carlo methods \cite{Hynninen:2007_PRB}. It was found that joining
substitutional Mn ions to clusters is energetically favorable and especially the
structures of two to four Mn ions formed around a single N ion are most stable.
These clusters are always found to have a FM ground state,
and FM intercluster interactions are also present even at relatively
long distances. For randomly distributed Mn clusters, high $T_{\mathrm{C}}$s
are obtained at high Mn concentrations \cite{Hynninen:2007_PRB}. The calculated
$T_{\mathrm{C}}$ is expected to depend dramatically on the microscopic cluster
distribution \cite{Hynninen:2006_APL}, giving a possible explanation for the
remarkable spread in experimental $T_{\mathrm{C}}$ values for (Ga,Mn)N (see Sec.\,\ref{sec:GaN-Mn_magnetic}).
The complexity of the Mn-clustering has been highlighted by further DFT calculations
where the coexistence of AF  and FM coupling has been found
for various configurations, different charge states and spatial distribution of
the clusters \cite{Cui:2007_PRB}.

{\em Nanocolumns and nanodots}: It was noted that in the presence of attractive forces between TM cations, the layer-by-layer epitaxy can result in the growth of TM-rich nanocolumns \cite{Fukushima:2006_JJAP}. The resulting structure was named the "konbu phase", where konbu means seaweed in Japanese. The presence of the konbu phase was experimentally demonstrated in the case of (Al,Cr)N \cite{Gu:2005_JMMM}, (Ge,Mn) (Sec.\,\ref{sec:Ge-Mn}), and (Zn,Cr)Te (Sec.\,\ref{sec:ZnTe}). Often, however, the TM-rich regions appear in the form of nanodots, the case of, annealed (Ga,Mn)As (see Sec.\,\ref{sec:GaAs}). Since the cross section looks then like a marble (dairiseki in Japanese), the decomposed state is referred to as the "dairiseki phase" \cite{Sato:2005_JJAP}. Results of Monte-Carlo simulations designed to show the formation of either konbu or dairiseki phases are presented in Secs.\,\ref{sec:dairiseki} and \ref{sec:konbu}.

{\em Surface aggregation}:  Since the initial step of the TM aggregation is expected to occur on the epitaxy plane, pairing energies of TM impurities residing on surfaces relevant to epitaxial processes were evaluated \cite{Gonzalez:2011_PRB,Navarro:2011_PRB,Birowska:2012_PRL}. As presented in Fig.\,\ref{fig:FIG_2_NGS}, $E_d$ values for Mn and Cr cation dimers on the (0001) wz-GaN surface become positive ($E_{\text{d}} \simeq 170$\,meV and 280\,meV, respectively), whereas $E_d$ remains negative for Fe pairs ($E_{\text{d}} = -120$\,meV), favoring the formation of Fe-rich clusters during the epitaxy \cite{Gonzalez:2011_PRB}.  This explains a much lower solubility limit of Fe comparing to Mn in epitaxial films of wz-GaN but does not elucidate why it is difficult to grow epitaxially (Ga,Cr)N with a large concentration of randomly distributed Cr ions (see, Secs.~\ref{sec:GaN-Mn} and \ref{sec:GaN-Fe}). Furthermore, a question addressed was how the spatial distribution of TM ions in epitaxial films could be affected by nonequivalence of certain crystal directions on the surface. It was found for (Ga,Mn)As deposited onto an unreconstructed surface of (001)GaAs that $E_d$ is by 1\,eV smaller for Mn pairs residing along the $[\bar{1}10]$ axis compared to the [110] case \cite{Birowska:2012_PRL}. Puzzling uniaxial anisotropies found in (Ga,Mn)As were explained by the lowering of crystal symmetry associated with the nonrandom distribution of Mn dimer orientations,  setting in during the epitaxy \cite{Birowska:2012_PRL}.

\begin{figure}
\includegraphics[width=3.3in]{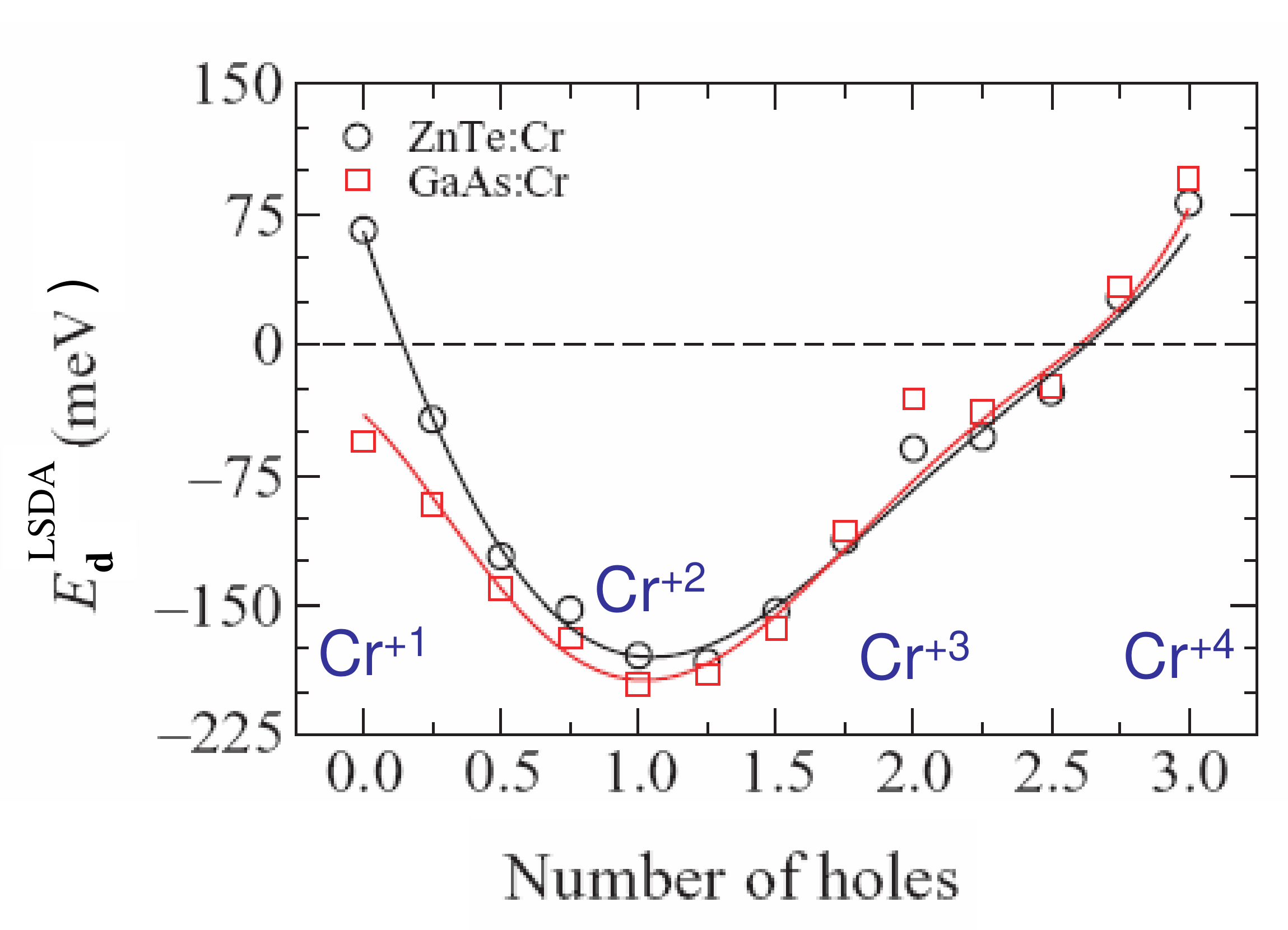}
\caption{Computed energy change resulting from bringing two Cr impurities to the nearest-neighbor cation positions in ZnTe and GaAs depending on the charge state, i.e., the number of holes in the Cr $d^5$ shell.  Adapted from \onlinecite{Da_Silva:2008_NJP}.}
\label{fig:ZnCrTe_Da_Silva}
\end{figure}

{\em Role of codoping}: It was suggested that codoping of DMSs with shallow donors or acceptors constitutes an efficient method of controlling the TM aggregation  \cite{Dietl:2006_NM}. This way of affecting the process of nanodecomposition operates if there are band-gap states derived from TM $d$ orbitals that can trap carriers supplied by shallow impurities (for a compilation of TM-related levels in various hosts see, e.g., \onlinecite{Dietl:2003_MRS}). The corresponding  change in the charge state and valency of the magnetic ions modifies chemical and spin-dependent interactions between TM impurities.  As an example, the energy of the {\em screened} Coulomb repulsion between two elementary charges residing on the nearest-neighbor cation sites in the GaAs lattice is 280\,meV.  Accordingly, a surplus of charge on TM ions brought by codoping with shallow dopants or by electrically active defects can outweigh the gain of energy stemming from $p$--$d$ hybridization and impede the NC assembling \cite{Dietl:2006_NM,Ye:2006_PRB,Boguslawski:2006_APL}. This intuitive picture was checked by \emph{ab initio} computations for (Ga,Mn)N \cite{Boguslawski:2006_APL}, (Ga,Fe)N \cite{Navarro:2011_PRB},  (Zn,Cr)Te \cite{Kuroda:2007_NM,Da_Silva:2008_NJP}, and (Ga,Cr)As \cite{Da_Silva:2008_NJP}. As shown in Fig.~\ref{fig:ZnCrTe_Da_Silva}, the value of $E_{\mathrm{d}}$ attains a minimum in ZnTe when the two Cr cations are in the $2+$ charge state \cite{Da_Silva:2008_NJP}. However, the computation results shown in the same plot also indicate that in GaAs $E_{\mathrm{d}}$ goes through a minimum for the Cr$^{2+}$ case, rather than for the Cr$^{3+}$ pairs, as might be expected for III-V compounds. Experimental verifications of these ideas have been found for (Ga,Fe)N codoped with Si (Sec.\,\ref{sec:GaN-Fe}) and for (Zn,Cr)Te codoped with I and N or deposited under growth conditions allowing to control the concentration of electrically active point defects (Sec.\,\ref{sec:ZnTe}). The effect of codoping is also discussed theoretically in Sec.\,\ref{sec:mixing} exploiting the concept of the mixing energy.

{\em Impurity and defect complexes}: Electrically active impurities and defects not only alter the position of the Fermi level and, thus, the TM charge and spin state but can form with TM ions complexes characterized by specific structural and magnetic properties. Prompted by experimental results summarized in Secs.\,\ref{sec:GaN-Mn} and \ref{sec:GaN-Fe}, magnitudes of $E_d^{(n,k)}$, where $k$ denotes the number of shallow dopants in the cluster were evaluated for various complexes in wz-GaN, including Mn-Mg$_k$ \cite{Devillers:2012_SR} as well as Fe$_n$-Mg$_k$ and Fe$_n$-Si$_k$ in the bulk and at the surface \cite{Navarro:2011_PRB}. It was also established theoretically that Mn-D cation dimers, where D denotes either P, As, or Sb donor, allow to increase the incorporation of Mn in substitutional positions in both Si and Ge \cite{Zhu:2008_PRL}.

\begin{figure}[htb]
    \centering
        \includegraphics[width=0.9\columnwidth]{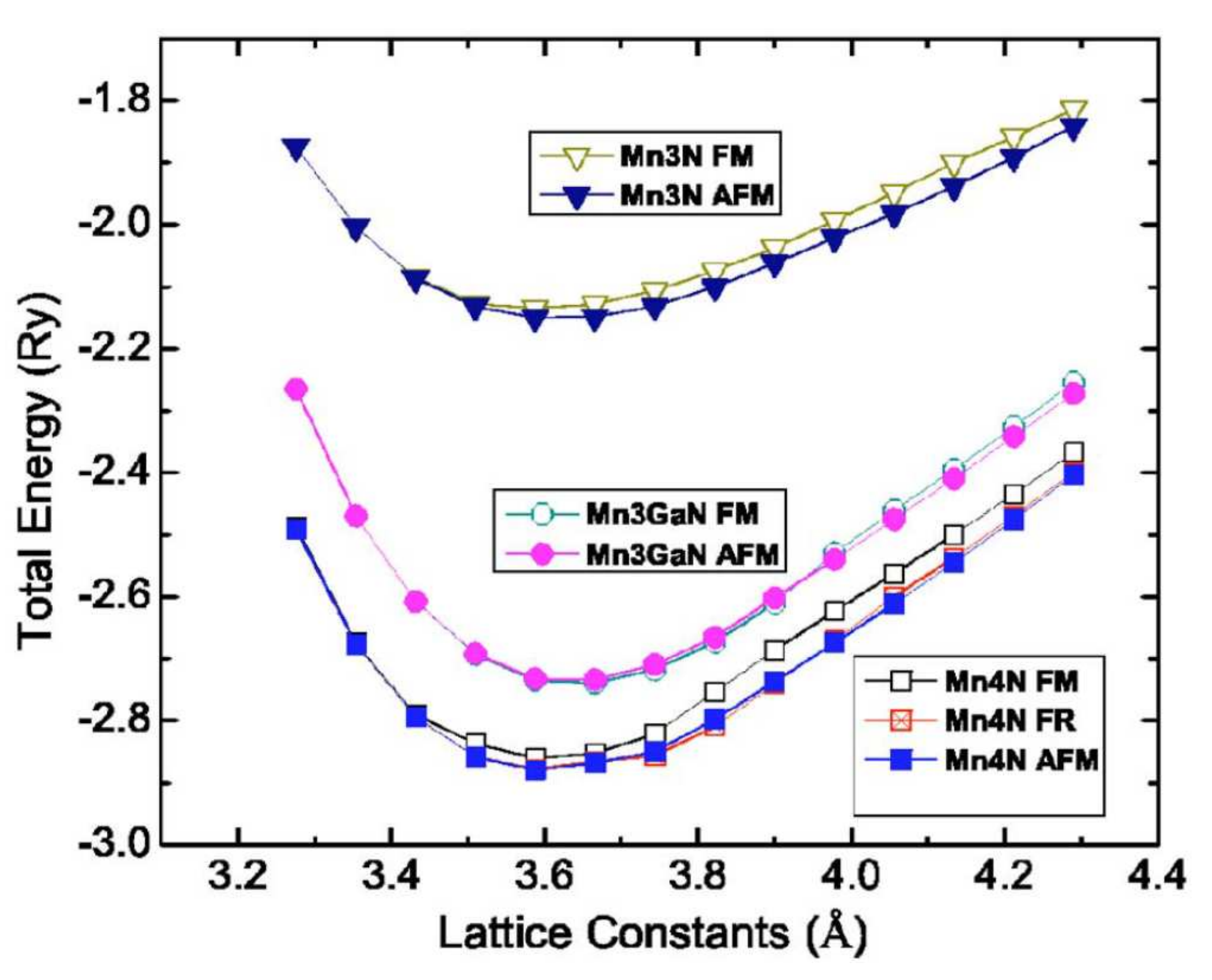}
    \caption{(Color online) Total energy for Mn$_{4}$N, GaMn$_{3}$N, and Mn$_{3}$N in different magnetic ordering configurations. From \onlinecite{Miao:2005_PRB}.}
    \label{fig:Miao}
\end{figure}

{\em Crystallographic phase separation}:
The electronic structure and magnetic properties of antiperovskite
GaMn$_{3}$N---frequently identified as a precipitated phase in (Ga,Mn)N beyond
the solubility limit of Mn into GaN (see Sec.\,\ref{sec:GaN-Mn_decomposition})---have
been studied by means of full potential linear muffin tin orbital (FP-LMTO)
methods and compared to the results for cubic Mn$_{4}$N and Mn$_{3}$N.
The total energy of these phases in different magnetic states, respectively FM,
FR, and AF is reported in Fig.\,\ref{fig:Miao}.

\subsection{Monte Carlo simulation of nanodecomposition}
\label{sec:Monte-Carlo}
From pairing energy considerations given in Sec.\,\ref{sec:pairing},
nanodecomposition is expected to occur in a number of DMSs.
In order to predict the resulting TM distribution, it is relevant
to simulate crystal growth. For this purpose, a hybrid method, combining {\em ab initio} and Monte Carlo simulations, was developed \cite{Sato:2005_JJAP}.

Within this method,  the alloy is described by an Ising model with the Hamiltonian of the system written as,
\begin{equation}
H =  \frac{1}{2} \sum_{i\neq j} V_{ij} \sigma_{i} \sigma_{j},
\end{equation}
where $V_{ij}$ is the effective pair interaction energy between two TM cations at sites $i$ and $j$,
$\sigma_{i}$ is the occupation index of the site $i$ by the TM ion, i.e.,
$\sigma_{i} = 1$ if site $i$ is occupied by a TM atom while
$\sigma_{i} = 0$ if site $i$ is occupied by a host atom.
The effective interaction is calculated as $V_{ij} = V_{AA}+V_{BB}-2V_{AB}$ for
a two component alloy AB, where $V_{AA}$, $V_{BB}$ and $V_{AB}$ are the potential energies
for the sites $ij$  occupied by AA, BB, and AB atoms, respectively.
The effective pair interactions $V_{ij}$ are evaluated for the medium obtained within the coherent potential approximation (CPA) by using
the generalized perturbation method proposed by  \onlinecite{Ducastelle:1976_JPF} in
the formulation of \onlinecite{Turchi:1988_PRB}.


The calculated effective pair interactions in (Ga,Mn)As and (Zn,Cr)Te
have been plotted in Fig.\,\ref{fig:ksato-pair} as a function of the pair distance.
By definition of $V_{ij}$, a negative $V_{ij}$ indicates effective attractive interactions
between the same kind of atoms and repulsive interactions between
different kind of atoms. As shown in Fig.\,\ref{fig:ksato-pair},
 pair interactions between magnetic impurities are effectively attractive.

Once the pair interactions in the Ising model are obtained, one can perform the
Monte Carlo simulation of the Ising model to obtain the distribution of
TM cations in a given semiconductor host at nonzero temperature.
The Kawasaki dynamics is used to relax the system  \cite{Binder:2000_B}. First, a large supercell
[typically $14\times 14 \times 14$ face-centered cubic (fcc) conventional cubic cell] is considered and
the magnetic cations are randomly distributed in the cell. Then, one picks up one host-TM pair 
and their position is exchanged. When the change in the energy due to
the exchange is negative, this process is allowed. When the energy is positive,
it is decided whether this process is allowed or not by using the Metropolis
criterion. This Monte Carlo step is repeated many times until the system reaches
the thermal equilibrium. In the present study, a nonequilibrium
state, namely the system is quenched within a very short time interval which is not
sufficient for the complete relaxation of the impurity distribution and then frozen.
This situation is simulated by interrupting
the iteration after a certain number of Monte Carlo steps \cite{Sato:2005_JJAP}.
This procedure assumes that $V_{ij}$ is the only relevant energy scale and, in particular,
that there are no kinetic barriers for cation exchange and diffusion over the lattice sites. The role of such
barriers is discussed in Secs.\,\ref{sec:mixing} and \ref{sec:CH}.

\subsubsection{Dairiseki phase}
\label{sec:dairiseki}
\begin{figure}
\includegraphics[angle=0,width=0.9\columnwidth]{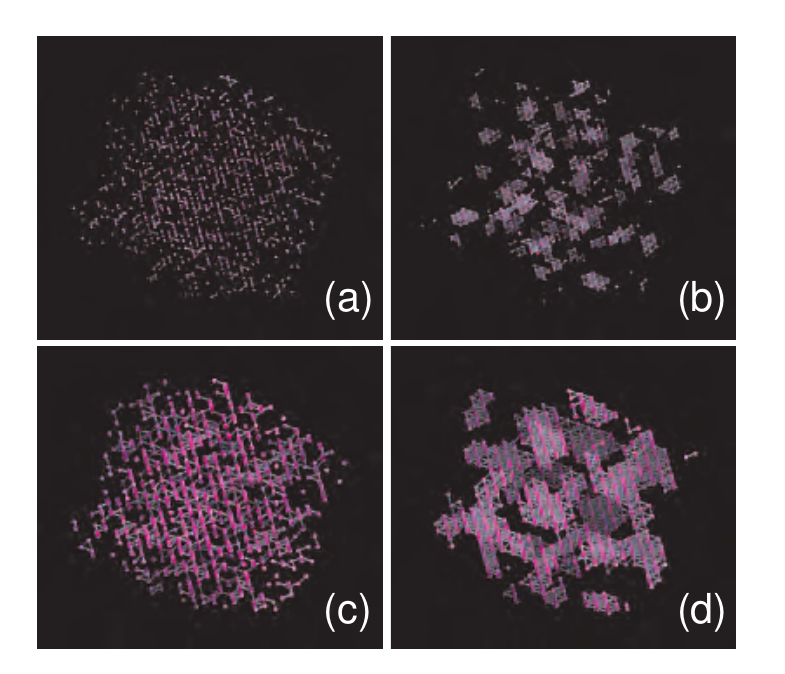}
\caption{(Color online) Mn configuration in (Ga,Mn)N. (a) and (c) refer to completely
random configurations, (b) and (d)
after 100 Monte Carlo steps (decomposed phase). Mn
concentrations are 5\% for (a) and (b), and 20\% for (c) and (d). Nearest
neighbor Mn–Mn pairs are connected by bars. From \onlinecite{Sato:2005_JJAP}.}.
\label{fig:ksato-phase}
\end{figure}

Figure\,\ref{fig:ksato-phase} visualizes decomposition in (Ga,Mn)N, as revealed by Monte Carlo simulations.
In the studied example, the average concentration of Mn in the simulation cell is 5\% or 20\%,
and 100 Monte Carlo steps per Mn site are performed at
temperature $k_BT_{\rm b}/V_{01}=0.5$, where $V_{01}$ is the chemical pair interaction energy between
the nearest neighbors. Because of the chemical pair interactions,
the Mn atoms attract each other, which results in the formation of clusters. In each cluster, the Mn atoms occupy
nearest-neighbor sites in order to decrease the energy as much as possible.
As seen in Fig.\,\ref{fig:ksato-phase}(b), the Mn concentration in the cluster is almost 100\% and the shape
of the cluster is nearly spherical. If the Mn average concentration is low,
the clusters are separated. For higher concentrations, the average size of
the clusters becomes larger and at 20\% the clusters are connected and
percolate through the cell [Fig.\,\ref{fig:ksato-phase}(d)]. Since the cross section of the
decomposed system looks like a marble, the decomposed state of
magnetically doped semiconductors was named the  "dairiseki phase" \cite{Sato:2005_JJAP}, where dairiseki means marble in Japanese.

\begin{figure}
\includegraphics[angle=0,width=3.1in]{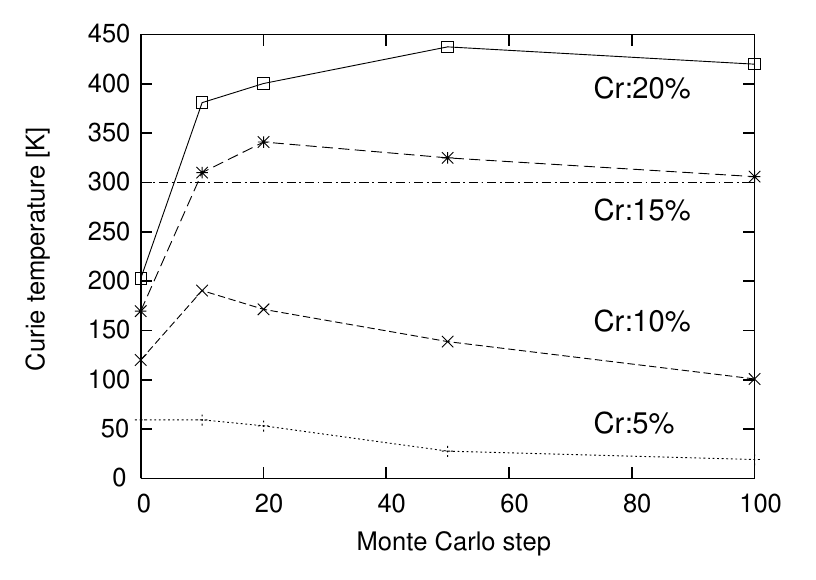}
\caption{Calculated Curie temperature of (Zn,Cr)Te as a function of Monte Carlo steps in the
simulation of decomposition. Curie temperatures are calculated by using the random
phase approximation. From \onlinecite{Fukushima:2006_PSSA}.}
\label{fig:TC_fukushima}
\end{figure}

As shown in the previous studies on ferromagnetism in DMSs with short-range spin-spin interactions,
at low concentrations $T_{\text{C}}$ is suppressed \cite{Sato:2010_RMP,Stefanowicz:2013_PRB}.
For example, below the percolation threshold [20\% for the fcc structure  \cite{Stauffer:1994_B}]
$T_{\text{C}}$ is 0 for systems with interactions only between the nearest neighbors.
As shown in Fig.\,\ref{fig:ksato-phase}(d), owing to decomposition there are
many percolating paths already at 20\% and the decomposition considerably affects the magnetic
properties of the system. In order to see the effects of
decomposition, $T_{\text{C}}$ values were calculated
as a function of Monte Carlo steps \cite{Fukushima:2006_PSSA}.
In order to include in the $T_{\text{C}}$ calculations the distribution of magnetic impurities, the random phase
approximation (RPA) proposed by \onlinecite{Bouzerar:2005_EL} and
\onlinecite{Hilbert:2004_PRB} was employed \cite{Fukushima:2006_PSSA}. It is known that this method correctly reproduces the magnetic
percolation effects and predicts reasonable $T_{\text{C}}$ magnitudes,  close to the exact Monte Carlo values.
Figure \ref{fig:TC_fukushima} shows the calculated $T_{\text{C}}$ of (Zn,Cr)Te within the RPA as a function of
Monte Carlo steps.

As shown in Fig.\,\ref{fig:TC_fukushima}, at low concentrations the decomposition works
to lower $T_{\text{C}}$ values. Due to decomposition many spatially isolated clusters are
formed, as shown in Fig.\,\ref{fig:ksato-phase}(b).
Since the exchange interactions between TM atoms are short-range in this case,
there is strong FM coupling
between TM moments in the cluster, but FM correlations between clusters
are weak. This is why $T_{\text{C}}$ of the whole system is suppressed by
 decomposition. In the case of (Zn,Cr)Te after 100 MC steps,
nonzero values of $T_{\text{C}}$ are predicted, and the system is  FM. However when
the exchange interactions are more short range, the system becomes
superparamagnetic after the decomposition. For higher concentrations,
because of the increase in the number of the nearest-neighbor pairs,
$T_{\text{C}}$ rises as the decomposition proceeds.
The increase of $T_{\text{C}}$ is observed already at 15\%, which is below
the percolation threshold; thus it is found that the decomposition
effectively lowers the magnetic percolation threshold.

\subsubsection{Konbu phase}
\label{sec:konbu}
In addition to growth temperature and TM concentration, one can control
the dimensionality of the growth process.
In the simulations presented in Sec.\,\ref{sec:dairiseki}, the magnetic impurities can
diffuse in any direction, therefore TM aggregation occurs in three dimensions.
In actual experiments, molecular beam epitaxy (MBE) or metalorganic vapor phase epitaxy (MOVPE) are standard
methods for crystal growth. On a surface with low coverage, the migration of atoms on the surface
occurs relatively freely, but it is difficult to diffuse into
the layers beneath, i.e., the diffusion is limited only to the uppermost layer.
Thus under such crystal-growth conditions,
the dimensionality of the decomposition is automatically controlled.
Since the decomposition occurs layer by layer and
the exchange of atoms between layers is prohibited under this condition,
one can expect an anisotropic shape for the clusters formed by the decomposition.
On the first layer, small clusters are formed due to the decomposition, and then
the impurity atoms deposited on the second layer gather around the nuclei in the
first layer. This process is repeated many times, and finally clusters with
elongated shape along the crystal-growth direction are produced \cite{Fukushima:2006_JJAP}.

\begin{figure}
\includegraphics[angle=0,width=3.3in]{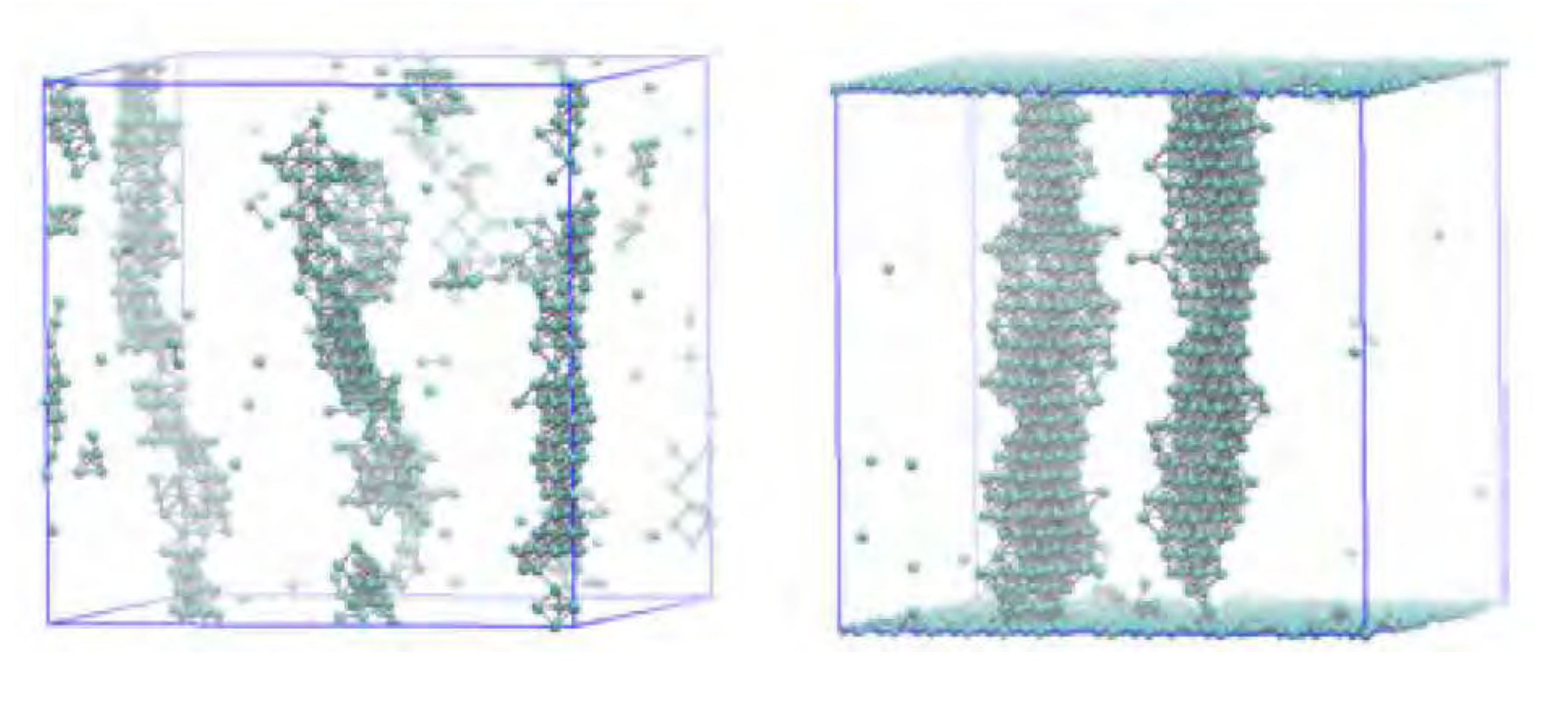}
\caption{(Color online) Monte Carlo simulations of phase separation in (Zn,Cr)Te under
the layer-by-layer crystal growth condition (konbu phase).
Average concentration of Cr is 5\% for both (a) and (b), except for the first and
the last layer in (b). In (b) delta-doping is simulated by depositing 80\% of Cr
on the first and the last layers. Cr positions in
the crystal are indicated by small spheres and they are connected by bars
if they occupy the nearest-neighbor sites. Adapted from \onlinecite{Fukushima:2006_JJAP}.}
\label{fig:ksato-konbu}
\end{figure}

In Fig.\,\ref{fig:ksato-konbu} results of Monte Carlo simulations are shown
for the layer-by-layer growth of (Zn,Cr)Te with the average Cr concentration of 5\%.
The growth direction is from the bottom to the top of the cube. As expected,
the Cr cations form clusters with columnar shape.
This state was named the konbu phase \cite{Fukushima:2006_JJAP}, where  konbu  means seaweed in Japanese.
Apparently the size of the clusters is much larger than the clusters found in
the dairiseki phase. In the konbu phase, the clusters are
spatially well separated.  The $T_{\text{C}}$ value of each cluster is expected
as high as the one of CrTe in a zinc-blende (zb) structure, presumably close to RT.
The $T_{\text{C}}$ magnitude of the
konbu phase shown in Fig.\,\ref{fig:ksato-konbu}(a) by using the RPA method is only 15\,K.
This low value is due to
the absence of magnetic interactions between separated columnar structures.
However, once the magnetic percolation paths are
introduced between the structures, $T_{\text{C}}$ should increase considerably.
To visualize this effect,
delta doping of Cr for the first and the last layers was assumed in the simulations [Fig.\,\ref{fig:ksato-konbu}(b)], leading to
$T_{\text{C}}$ of 346\,K \cite{Fukushima:2006_JJAP}. In reality,
magnetization blocking phenomena brought about by
magnetic anisotropy (see Sec.\,\ref{sec:sp}) and long-range dipole interactions between clusters' magnetizations
can lead to a large magnitude of apparent $T_{\text{C}}$.

\subsection{\label{sec:sp}Superparamagnetic blocking phenomena}

\begin{figure}[t]
\includegraphics[angle=0,width=9cm]{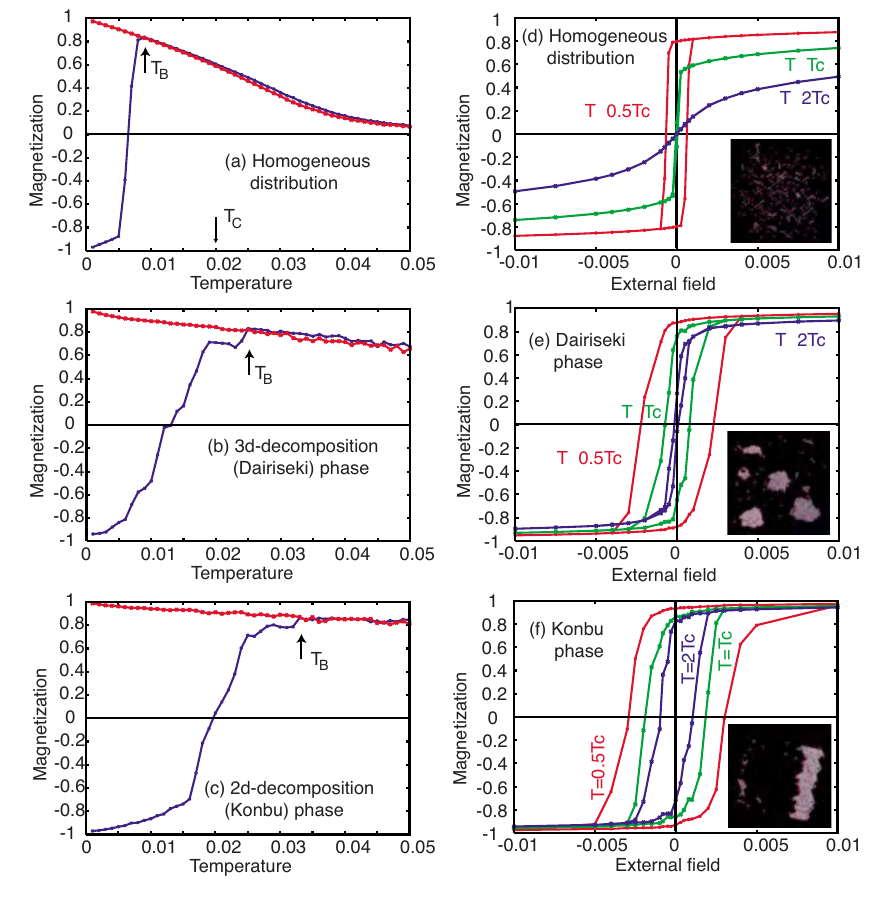}
\caption{(Color online) Superparamagnetic
blocking phenomenon in (Ga,Mn)N according to Monte Carlo simulations. Left-hand panels:
Magnetization as function of temperature starting from parallel
or antiparallel configurations of initial magnetization to the
external field. Right-hand panels: Magnetization as a function
of an external field. The simulations were performed for the uniform
Mn distribution (a) and (d), the dairiseki phase (b) and
(e), and the konbu phase (c) and (f). The insets in the lower
panels show snapshots of the Mn distribution in (Ga,Mn)N for the
respective phases. Temperature $(k_BT)$ and the external magnetic field are
scaled by the strength of the nearest-neighbor exchange energy $J_{01}$.
From \onlinecite{Sato:2007_JJAP}.}
\label{fig:ksato-super}
\end{figure}

If we focus only on a thermodynamic quantity such as a $T_{\text{C}}$ value,
there is not much difference
between the dairiseki phase and the konbu phase for low average TM concentrations.
For both phases as a result of decomposition, $T_{\text{C}}$ goes to 0 and
the system becomes superparamagnetic, if the interactions are short-range.
On the other hand, if we look into the magnetization process,
the dairiseki phase and the konbu phase behave very differently due to
superparamagnetic blocking phenomena associated with shape anisotropy.

To reverse the magnetization direction of nanomagnets within a single magnetic domain state,
magnetic anisotropy energy should be overcome. For a system with uniaxial
magnetic anisotropy, the magnetic anisotropy energy is $KV\sin^2 \theta$, where $K, V$
and $\theta$ are the anisotropy energy density, the volume of the system, and the angle between
the magnetization direction and the magnetic easy axis, respectively. In general, shape and crystalline
anisotropies contribute to the magnitude of $K$.  The existence of the energy barrier between
the easy directions results in a finite relaxation rate of the magnetization direction 
 $1/\tau \propto \exp(-KV/k_BT)$. Accordingly,
for larger systems (but small enough to keep single domain nature),
the relaxation time $\tau$ becomes longer and the magnetization direction is effectively
fixed along the initial direction at low temperatures.
This is the so-called superparamagnetic blocking phenomenon \cite{Aharoni:1996_B}.

In the present cases of decomposed systems,
the columnar structure contains high concentration of magnetic impurities.
Because of the FM exchange interactions between magnetic impurities,
the self-assembled structure with a high TM concentration behaves as a nanomagnet.
The radius of the columnar structure is a few nanometers and
within this length scale, we can expect that the structure contains a single magnetic domain.

As shown in Secs.\,\ref{sec:dairiseki} and \ref{sec:konbu},
it is possible to control the volume of the clusters formed
by decomposition and, hence, sample magnetization via the superparamagnetic blocking
phenomenon.
In order to demonstrate this idea, the magnetization process was simulated by the
Monte Carlo technique  \cite{Katayama-Yoshida:2007_JMMM} employing the method
elaborated by  \onlinecite{Dimitrov:1996_PRB}. In Fig.\,\ref{fig:ksato-super} the calculated temperature
and field dependence of magnetization
are shown taking the shape part of magnetic anisotropy into account.
The simulations were performed for a uniform TM distribution,
the dairiseki phase, and
the konbu phase of (Ga,Mn)N. As shown in the magnetization curves,
for the system with larger clusters,
the cohesive field becomes greater and hysteresis loop opens wider, i.e.,
one obtains a harder magnet.
Furthermore, magnetization as a function of temperature was calculate for increasing
temperature
starting from anti-parallel and parallel configurations of magnetization
with respect to the external magnetic field. Above the blocking temperature $T_{\rm b}$, the direction of the magnetization
flips and the two lines coincide.
It is demonstrated in Figs.\,\ref{fig:ksato-super}(a-c) that $T_{\rm b}$ becomes higher for
the dairiseki and konbu phases.

Until now many experimental reports on the magnetism of TM-doped semiconductors have been
published, but often they are not consistent, particularly for
wide
band-gap hosts. As shown in Sec.\,\ref{sec:pairing}, energies of chemical pair interactions between magnetic impurities are rather large, especially in wide band-gap DMSs,
so that these systems have a strong tendency toward decomposition. Since typically epitaxy
is a nonequilibrium
process, its outcome, particularly the degree of decomposition,
depends sensitively on growth and processing conditions. As a result, the size, shape, and content of TM-rich NCs resulting from
decomposition
vary strongly from experiment to experiment.
These considerations explain, at least partly, inconsistency in magnetic properties
reported by various groups for the same material.

\subsection{\label{sec:mixing}Mixing energy; nucleation and spinodal regions}
\label{sec:mixing}
In order to determine the temperature and TM concentration range corresponding to nanodecomposition,  a lattice model, originally developed by Flory and Huggins for the polymer-mixing problem \cite{Flory:1942_JCP,Huggins:1941_JCP,Rubinstein:2003_B}, was adapted for DMSs \cite{Sato:2007_JJAP,Chan:2008_PRB,Hai:2011_JAP}. Within this approach, the free energy  $F(x) = \Delta E(x) - TS(x)$
of a random alloy A$_{1-x}$TM$_x$B is described by the mixing energy $\Delta E(x)$ and the entropy $S(x)$ per one cation of the form
\begin{eqnarray}
\Delta E(x) &=& E(\mbox{A}_{1-x}\mbox{TM}_x\mbox{B}) -(1-x)E(\mbox{AB})  \nonumber \\
 &-& xE(\mbox{TMB});
\label{eq:mixing} \\
S(x) &=& -k_B[x\ln x + (1-x)\ln(1-x)].
\label{eq:entropy}
\end{eqnarray}

\begin{figure}
\includegraphics[angle=0,width=0.9\columnwidth]{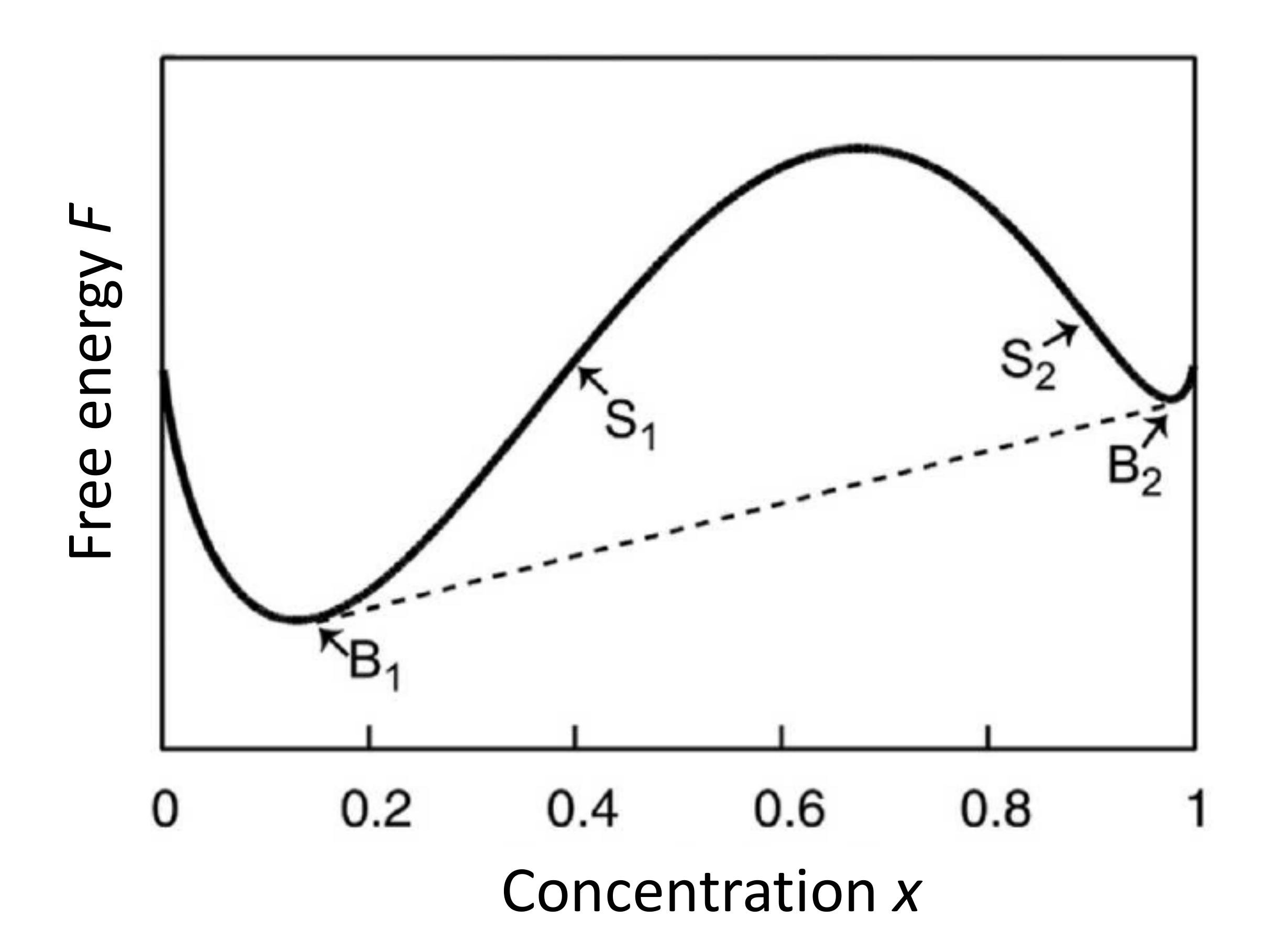}
\caption{\label{fig:tanaka_fig1}
Typical dependence of DMS free energy $F$ on TM concentration $x$. Binodal decomposition can occur between points B$_{1}$ and B$_{2}$, the ends of the common tangent segment of $F(x)$ (dashed line). Spinodal decomposition occurs between points S$_{1}$ and S$_{2}$ corresponding to $\partial^{2}F/\partial x^{2} = 0$. From \onlinecite{Hai:2011_JAP}.}
\end{figure}

If $\Delta E(x) > 0$, the alloy is unstable against decomposition into the end compounds \cite{Swalin:1970_B} but at $T > 0$ a tendency to randomization described by the entropy term becomes important.  Figure \ref{fig:tanaka_fig1} shows schematically the behavior of $F(x)$ for a convex function $\Delta E(x) > 0$ and $T >0$. It is seen that for such a case $F(x)$ there are two points $B_{1}$ and $B_{2}$ at compositions $x_1$ and $x_2$, between which $F(x) > F(x_1) + F(x_2)$.  Hence, for $x_1 < x < x_2$ the thermal equilibrium state corresponds to the {\em binodal decomposition}, i.e., in the case of DMSs, to the nucleation of TM-rich NCs of content $x_2$ in the TM-poor matrix of the concentration $x_1$. The region $x_1 < x < x_2$ corresponds to the miscibility gap at given temperature $T$.

Often, however,  the alloy in some range between $x_1$ and $x_2$ is stable against {\em small} fluctuations $\pm \Delta x$ of local composition. Since changes in the local free energy density integrated over the volume encompassing many fluctuations is proportional to  $(\Delta x)^2\partial^{2}F/\partial x^{2}$, the alloy can remain in a metastable state as long as $\partial^2F/\partial x^{2} > 0$, i.e., in the composition range, in which $F(x)$ is a concave function. In contrast, in the region between the points $S_{1}$ and $S_{2}$ in Fig.\,\ref{fig:tanaka_fig1}, where $\partial^2F/\partial x^{2} < 0$, there is no thermodynamic barrier separating the homogenous and non-uniform alloy.

The resulting spontaneous up hill diffusion process, called {\em spinodal decomposition}, leads to a system that is {\em structurally} homogenous but shows a modulation of the TM concentration. This mechanism of chemical phase separation is suitable for fabricating a NC system with rather uniform NC size and inter-NC distance \cite{Jones:2002_B}. An ordering in NC positions can be enhanced further on by strain, similar to the case of self-assembling quantum dots in semiconductors \cite{Stangl:2004_RMP}. As discussed in Secs.~\ref{sec:GaAs} and \ref{sec:Ge-Mn}, a periodic-like arrangement of TM-rich NCs has been found in Mn-doped GaAs and Mn-doped Ge, respectively.

The CPA approach \cite{Sato:2007_JJAP} and a cluster expansion method \cite{Chan:2008_PRB} were employed to evaluate from first principles the energy $E(\mbox{A}_{1-x}\mbox{TM}_x\mbox{B})$ and, thus, the mixing energy $\Delta E(x)$ of DMSs containing randomly distributed substitutional TM cations.

The CPA is particularly suitable for electronic structure calculations in the case of substitutional alloys \cite{Shiba:1971_PTP,Akai:1989_JPCM,Akai:1993_PRB}, such as DMSs. It is
most efficiently combined with the Korringa-Kohn-Rostoker (KKR) method for
the band structure calculation, particularly employing the MACHIKANEYAMA package \cite{Sato:2010_RMP} developed by Akai  \cite{machikaneyama}. Within the CPA, a configuration average of the alloy electronic structure  is calculated by using a mean-field like procedure. For the first step one considers a hypothetical atom that describes the averaged system. The crystal of hypothetical atoms constitutes an effective CPA medium. By using the multiple scattering theory the scattering path operator of the host and of the impurity atoms in the CPA medium is determined.
The weighted average of these scattering path operators with respect to the concentration should be equal to the scattering path operator of the hypothetical atom itself. This is the self-consistent equation to be solved by an iterative method for determining the CPA medium, providing electronic density of states and the total energy $E$ of the system. A more detailed explanation and practical formulation
is provided by \onlinecite{Gonis:2000_B}.

\begin{figure}
\includegraphics[angle=0,width=3.1in]{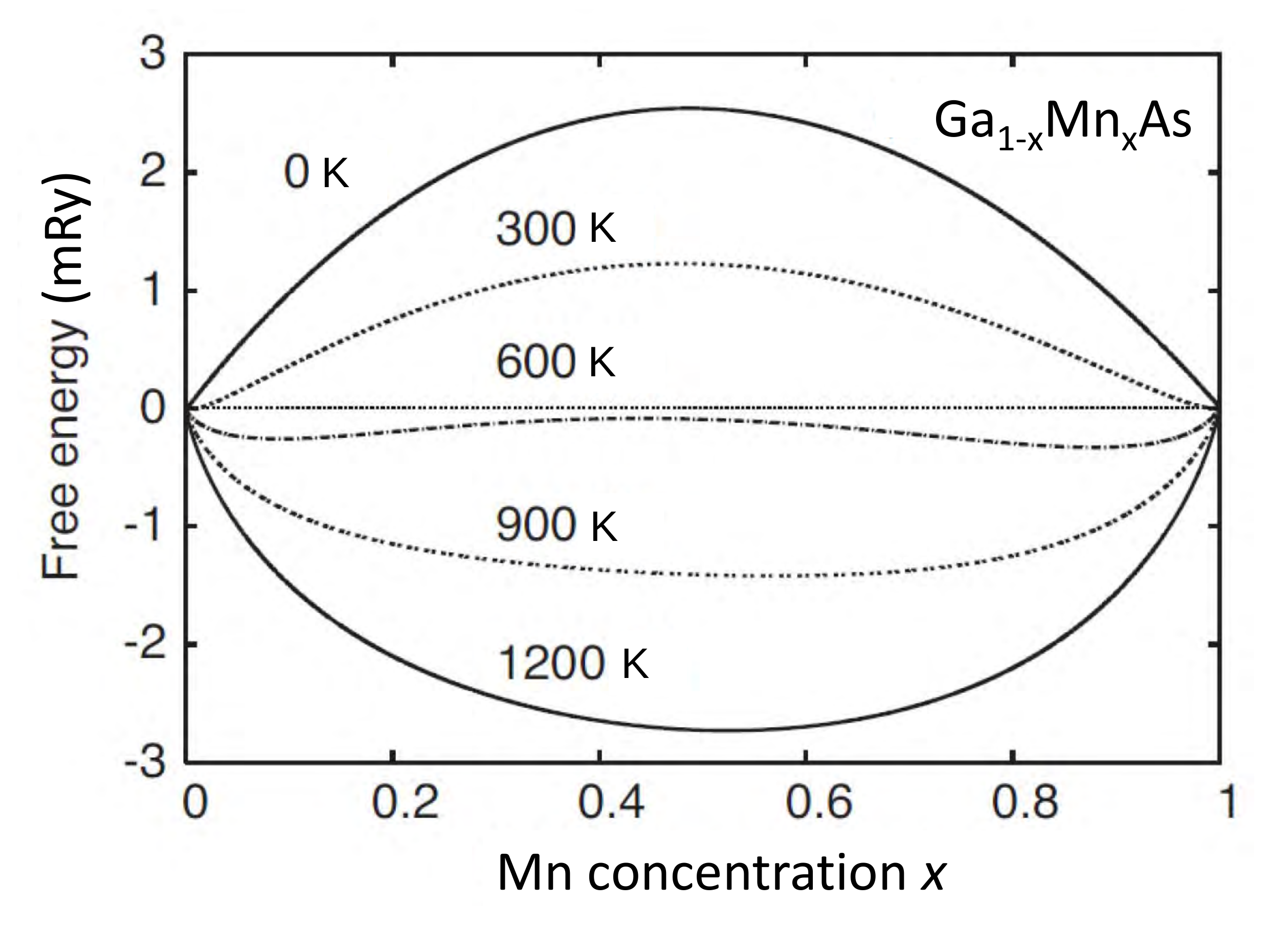}
\caption{\label{fig:ksato-free} Computed free energy $F(x)$ for Ga$_{1-x}$Mn$_x$As at various temperatures. As shown schematically in Fig.\,\ref{fig:tanaka_fig1} this plot allows one to determine composition ranges corresponding to binodal and spinodal decompositions at particular temperatures.
From \onlinecite{Sato:2007_JJAP}.}
\end{figure}

\begin{figure}
\includegraphics[angle=0,width=3.1in]{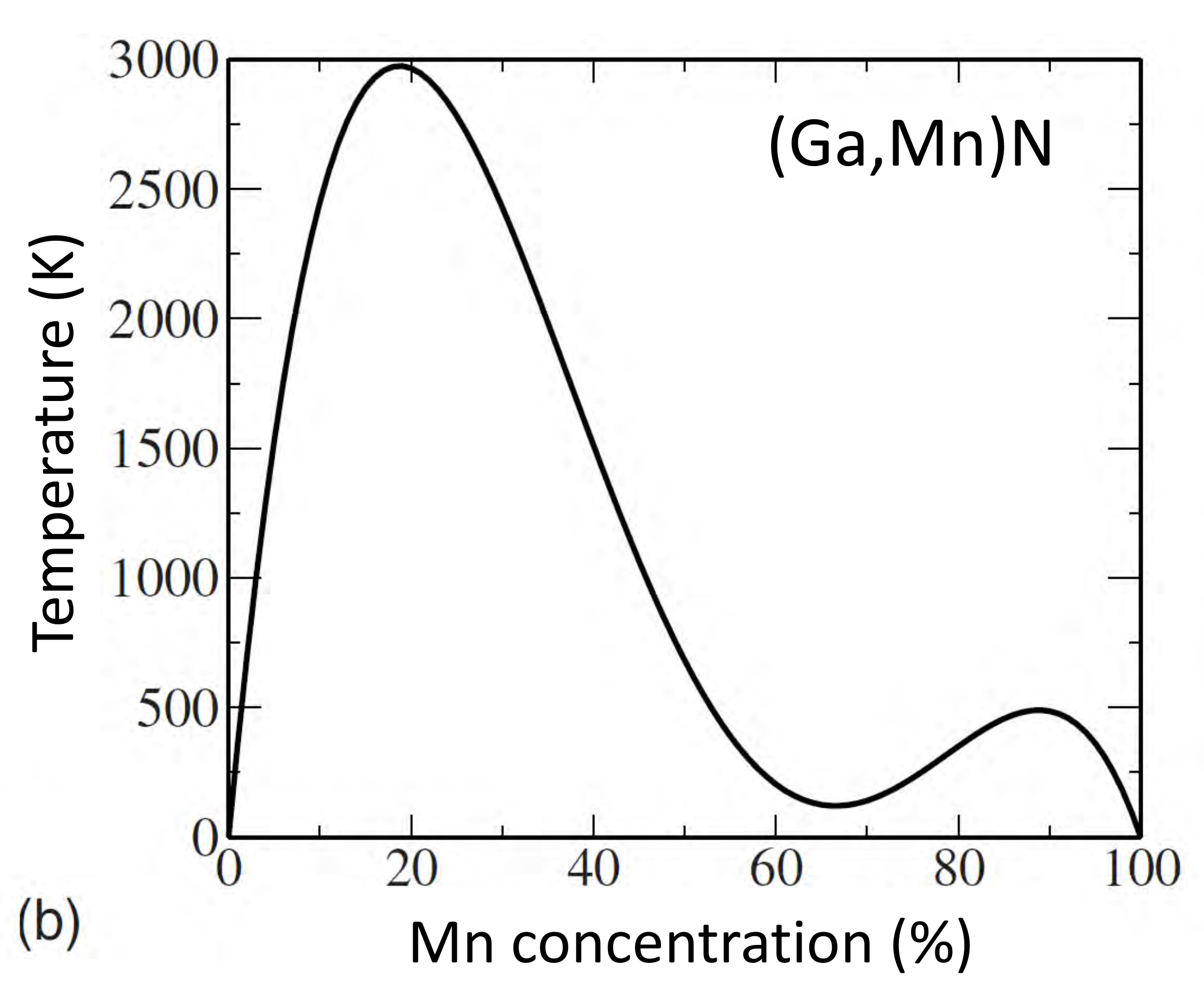}
\caption{\label{fig:GaMnN_Chan} Computed spinodal line for zinc-blende Ga$_{1-x}$Mn$_x$N. In the range of compositions $x$ between the lines at given temperature $T$ the alloy can undergo spinodal decomposition, so that at, e.g., 1000\,K Ga$_{1-x}$Mn$_x$N decomposes into compounds with $x = 0.03$ and 0.45. From \onlinecite{Chan:2008_PRB}.}
\end{figure}

The CPA method was employed to evaluate $\Delta E(x)$ for Zn$_{1-x}$Cr$_x$Te,  Ga$_{1-x}$Mn$_x$As, Ga$_{1-x}$Cr$_x$N, and Ga$_{1-x}$Mn$_x$N assuming zb structure and the experimental values of the lattice constants \cite{Sato:2007_JJAP}. The convex form of $\Delta E(x)$ demonstrated that spinodal decomposition can appear in these systems. A similar conclusion was derived from cluster expansion studies of zb-Ga$_{1-x}$Mn$_x$N \cite{Chan:2008_PRB}. Figure~\ref{fig:ksato-free} shows, as an example, $F(x)$ for Ga$_{1-x}$Mn$_x$As at various temperatures \cite{Sato:2007_JJAP}, which allows one to determine the spinodal line, i.e., the position of $S_{1}$ and $S_{2}$ points as a function of temperature. Such a spinodal line is presented in Fig.\,\ref{fig:GaMnN_Chan} for in zb-Ga$_{1-x}$Mn$_x$N \cite{Chan:2008_PRB}. It is seen that the upper critical solution temperature, above which the alloy is miscible for any $x$, is as high as 3\,000\,K in this case.  Hence, within the temperature range relevant for epitaxial growth and post-growth annealing, the DMSs in question can undergo spinodal decomposition.

According to Figs.\,\ref{fig:tanaka_fig1}--\ref{fig:GaMnN_Chan}, the range of compositions, at which alloy decomposition could appear, diminishes with increasing temperature. This is in qualitative {\em disagreement} with the long staying experimental observations which imply that {\em lowering} of epitaxy temperature is necessary for obtaining a uniform DMS \cite{Munekata:1989_PRL}, whereas annealing at {\em high} temperatures allows to generate phase separation \cite{De_Boeck:1996_APL}. This apparent contradiction, i.e., the existence of lower critical solution temperature, points to the importance of {\em diffusion} (kinetic) barriers precluding aggregation of TM cations if growth or annealing temperature is too low even if there is no thermodynamic barrier for spinodal decomposition. To our knowledge, in the case of DMSs, only the barrier height for diffusion of interstitial Mn and Li ions in GaAs has so far been theoretically evaluated \cite{Edmonds:2004_PRL,Bergqvist:2011_PRB}. At the same time, a phenomenological Cahn-Hillard theory of spinodal decomposition, presented in Sec.\,\ref{sec:CH}, was employed to describe properties of phase separation in DMSs \cite{Hai:2011_JAP}.  The influence of growth and processing conditions on nanodecomposition is discussed from the experimental perspective in Secs.\,\ref{sec:GaAs}--\ref{sec:ZnTe}.

Another important ingredient that has not been taken into account by the approaches leading to the results presented in Figs.\,\ref{fig:ksato-free} and \ref{fig:GaMnN_Chan} is the omission of interfacial energies. In particular, the comparison of free energies corresponding to the random alloy and the decomposed case referred to the thermodynamic limit, i.e., energies of TM ions at interfaces between TM-rich and TM-poor phases have been disregarded. In general, energy lowering associated with the TM aggregation is smaller for interfacial ions compared to the ions inside the cluster. This promotes the Oswald ripening, i.e., the formation of larger clusters at the expense of smaller ones. In extreme cases, small clusters, up to the so-call nucleation radius, are energetically unstable, which imposes a barrier for the nucleation process. Such a barrier can be lowered at crystal defects acting as nucleation centers or diffusion channels. According to the discussion of the pairing energies in Sec.\,\ref{sec:pairing}, even small clusters of TM cations are stable in a number of semiconductor compounds.

\begin{figure}
\includegraphics[angle=0,width=0.4\textwidth]{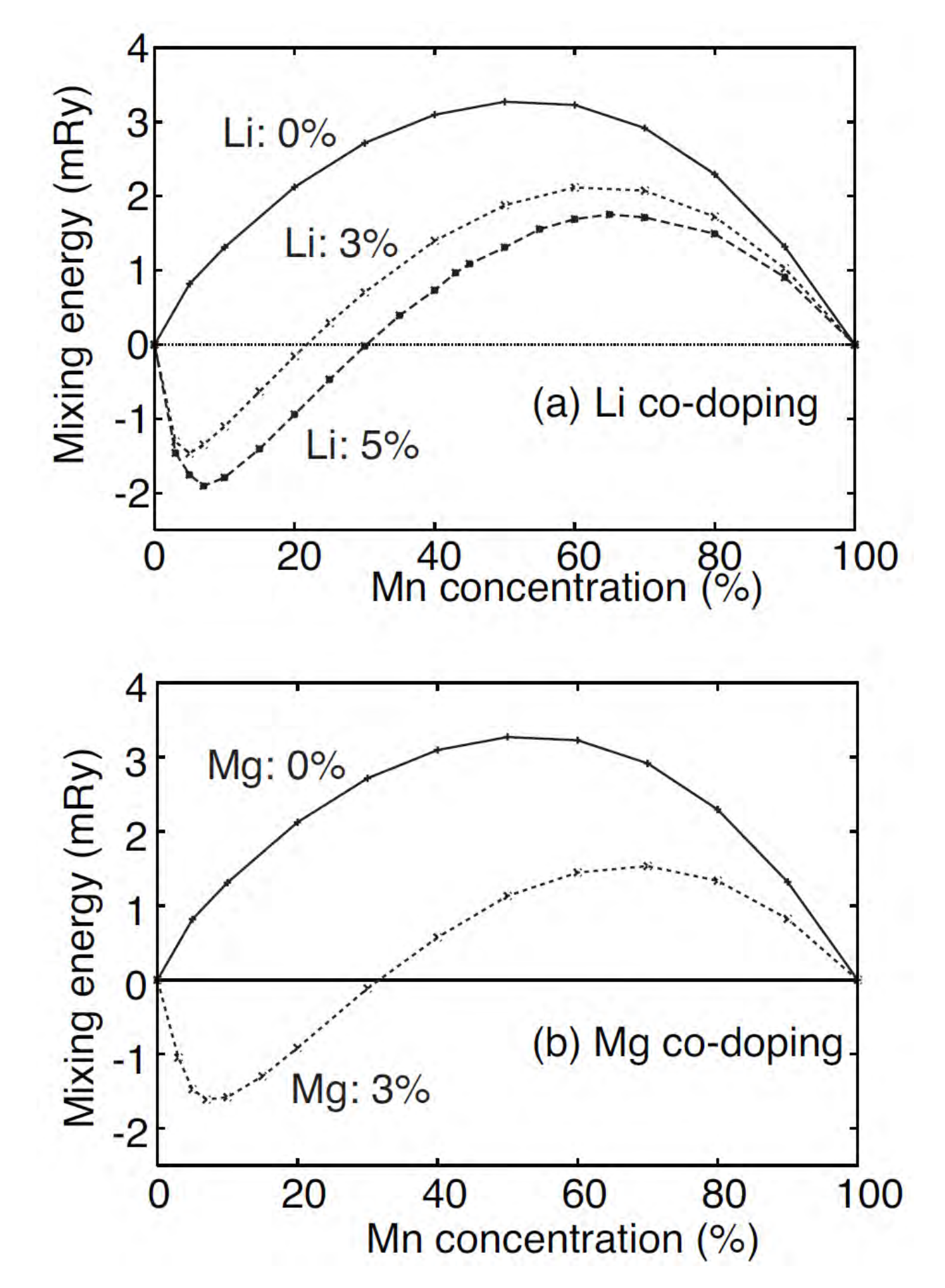}
\caption{\label{fig:ksato-codoping} Calculated mixing energy of (Ga,Mn)As
as a function of Mn concentration with (a) interstitial Li codoping and
(b) interstitial Mg codoping. From \onlinecite{Bergqvist:2011_PRB}.}
\end{figure}

As already mentioned in Sec.\,\ref{sec:pairing}, according to {\em ab initio} studies, codoping of DMSs by electrically active impurities or defects affects considerably the TM pairing energies. This conclusion is supported by studies of the mixing energy $\Delta E(x)$ of (Ga,Mn)As \cite{Sato:2007_JJAP_c,Bergqvist:2011_PRB}, (Ga,Mn)N, (Ga,Mn)Cr, and (Zn,Cr)Te \cite{Sato:2007_JJAP_c} codoped with donors.
In particular, the mixing energies
of (Ga,Mn)As with interstitial impurities acting as donors, such as
Li (Li$_{\rm int}$) and Mg (Mg$_{\rm int}$), were computed \cite{Bergqvist:2011_PRB}
As shown in Fig.\,\ref{fig:ksato-codoping}, the effect of codoping is dramatic and the
the mixing energy is significantly modified.
Particularly at low concentrations of Mn, a negative and concave region
is found in the calculated values of $\Delta E(x)$.
This means that
even at thermal equilibrium homogeneous doping of Mn is possible over a wide $x$ range.
Similar effects were also found for other combinations of
the host semiconductors and different substitutional donor
impurities \cite{Sato:2007_JJAP_c}.

The advantage of interstitial impurities (particularly Li$_{\rm int}$)
over the substitutional ones is that
interstitials can easily be removed by postgrowth annealing.
To check the effectiveness of this process,
Monte Carlo simulations of diffusion of interstitial impurities in (Ga,Mn)As
were performed  \cite{Bergqvist:2011_PRB}.
The results imply that since the binding energy between Mn$_{\rm int}$ and Mn$_{\rm Ga}$ is much greater
than in the case of Li$_{\rm int}$,
the low-temperature annealing more effectively wipes
away Li$_{\rm int}$ than Mn$_{\rm int}$ in (Ga,Mn)As.

\subsection{{Cahn-Hilliard} theory of spinodal decomposition}
\label{sec:CH}
One of the goal of decomposition theory is to describe spatial and temporal evolution of the local composition $x(\vec{r},t)$ at a surface or in a volume of the alloy containing initially randomly distributed TM cations of the concentration $\bar{x}$. While it is rather difficult to determine $x(\vec{r},t)$ in the nucleation regime, there exists a time-honored \onlinecite{Cahn:1958_JChP} theory of spinodal decomposition, discussed also in the context of phase separation in DMSs \cite{Hai:2011_JAP}. This approach exploits the Ginzburg-Landau theory of the continuous phase transitions, taking $u(\vec{r},t) = x(\vec{r},t) - \bar{x}$ as an order parameter. Furthermore, the interfacial effects are included to the lowest order in the gradient in the TM concentration $x(\vec{r})$. Under these assumptions the functional of the free energy assumes the form,
\begin{equation}
F_{\text{tot}}[u(\vec{r})] = N_0\int d \vec{r}{F[u(\vec{r})] + \frac{1}{2}\kappa|\nabla u(\vec{r})|^2}.
\label{eq:F_tot}
\end{equation}
where $N_0$ is the cation density and $F(x)$, in addition to the contributions given in Eqs.\,\ref{eq:mixing} and \ref{eq:entropy}, can contain a term $ku^2(\vec{r},t)/2$ describing an increase of the system energy associated with strain that builds in a decomposed system if the lattice constant depends on $x$. Along with the mixing energy $\Delta E(x)$, $\kappa$ and $k$ are parameters of the model, which in principle can be determined experimentally or by {\em ab initio} methods.

The distribution $u(\vec{r})$ under conditions of thermal equilibrium is provided by the variational minimization of $F_{\text{tot}}[u(\vec{r})]$,
$\delta F_{\text{tot}}[u(\vec{r})]/\delta u(\vec{r}) = 0$ with the constrain, imposed by the mass conservation, that  $u(\vec{r})$ integrated over the relevant crystal surface or volume vanishes. This procedure implies that (i) the decomposition occurs at temperatures below the critical temperature $T_c$, where $\partial^2 F(u,T)/\partial u^2)|_{0} < 0$; (ii) the compositions of TM-rich NCs and TM-poor matrix are determined from $\partial [F(u) - u(\partial F/\partial u)|_{0}] /\partial u =0$ and (iii) the spatial width of the transition region between these two domains scales with $\kappa^{1/2}$.

However, from the experimental viewpoint particularly relevant is the evaluation of $u(\vec{r},t)$ during the early stage of decomposition, i.e., far from thermal equilibrium. The starting point is the continuity equation,
\begin{equation}
\partial N_0u(\vec{r},t)/\partial t = -\nabla\cdot \vec{j}(\vec{r},t),
\label{eq:continuity}
\end{equation}
where the current of TM cations $\vec{j}(\vec{r},t)$ is driven by the gradient in the chemical potentials $\mu$,
\begin{equation}
\vec{j}(\vec{r},t) = M\nabla\mu(\vec{r},t).
\label{eq:current}
\end{equation}
Here, $M$ is the TM mobility, one more material parameter. Since the chemical potential corresponds to a change of the free energy by adding or removing one atom, its local nonequilibrium value is given by a variational derivative,
\begin{equation}
\mu(\vec{r},t)= -\delta F_{\text{tot}}[u(\vec{r},t)]/\delta u(\vec{r},t),
\label{eq:chemical}
\end{equation}
with the constrain $\int d \vec{r} u(\vec{r},t) = 0$. This variational derivative and, thus, the chemical potential  is not zero as $u(\vec{r},t)$ does not yet correspond to the thermal equilibrium distribution. Inserting Eq.\,\ref{eq:chemical} into Eq.\,\ref{eq:current} and then Eq.\,\ref{eq:current} into \ref{eq:continuity}, one obtains the Cahn-Hilliard equation in the form
\begin{equation}
\partial u(\vec{r},t)/\partial t = M\nabla^2{\partial [F(u) - u(\partial F(u)/\partial u)|_0]/\partial u - \kappa\nabla^2 u}.
\label{eq:C-H}
\end{equation}

There is a comprehensive literature devoted to mathematical aspects of this differential equation \cite{Novick-Cohen:2008_B}. It is convenient to  write its solution $u(\vec{r},t)$ as a Fourier transform of a function $A(\vec{q},t)$. If only the two lowest order terms are retained in the Taylor expansion of $F(u)$, a justified approximation at the early stage of decomposition, then
\begin{equation}
A(\vec{q},t) = A_0(\vec{q})\exp[R(\vec{q})t],
\end{equation}
where the aggregation rate
\begin{equation}
R(q) = -M[\partial^2 F(u)/\partial u^2|_0 + \kappa q^2]q^2.
\label{eq:R}
\end{equation}
This formula shows that $R$ is positive, at least for sufficiently small $q$ values, in the spinodal decomposition range $\partial^2 F(u)/\partial u^2|_0 <0$. In such a case, $R$ is governed by the cation mobility $M$ which, because of diffusion barriers, decreases strongly on lowering temperature. This explains why low-temperature epitaxy can result in a metastable state corresponding to a DMS with a uniform distribution of magnetic ions.  However, growth or annealing at appropriately high temperatures, or the presence of lattice defects, can enhance $M$ and promote TM aggregation.

Because of the interfacial energy, on approaching thermal equilibrium conditions, the size of TM-rich NCs should increase whereas their concentration decrease. However, the formation of a periodic structure by spinodal decomposition at early times may render the Oswald ripening a prohibitively slow process, particularly in DMSs with $\bar{x} \ll 1/2$. In contrast, both NCs and the surrounding matrix can attain uniform TM concentrations $x_2$ and $x_1$, respectively, corresponding to the binodal points, $B_2$ and $B_1$ in Fig.\,\ref{fig:tanaka_fig1} or even a transformation of TM-rich NCs to another crystallographic phase can take place. In either case, by determining $q_c$ that maximizes $R(q)$ one can evaluate the expected distance $\lambda_c = 2\pi/q_c$ between TM-rich NCs,
\begin{equation}
\lambda_c = 4\pi[\kappa/|2\partial^2 F(u)/\partial u^2|_0|]^{1/2},
\end{equation}
where $\kappa$, representing the interfacial energy, is of the order of $|E_d|a^2$ and $a$ is the lattice parameter. Since the magnitudes of pairing energy $E_d$ and $\partial^2 F(u)/\partial u^2|_0$ are similar (cf.\,Figs.\,\ref{fig:ksato-pair} and \ref{fig:ksato-free}), the distance between TM-rich NCs is expected to set at about ten lattice constants, the value in reasonable agreement with experimental results collected in Secs.\,\ref{sec:GaAs}--\ref{sec:ZnTe}. Simple geometrical considerations provide an average radius of NCs at given $\bar{x}$, $x_1$ and $x_2$, and $\lambda_c$.

Another interesting aspect is the dependence of $R(\vec{q})$ on the crystallographic direction of $\vec{q}$ brought about by the elastic term $\frac{1}{2}ku^2$ in $F(u)$ \cite{Hai:2011_JAP}. According to the theory developed for cubic crystals \cite{Cahn:1962_AM}, if elastic moduli fulfil the relation $2C_{44} - C_{11} + C_{12} > 0$, the spinodal decomposition is predicted to proceed along $\langle 100\rangle$ cubic directions. This inequality is obeyed in zb compounds of interest, GaAs, Ge, and ZnTe. Similar to the case of self-assembling semiconductor quantum dots obtained in the Stranski–Krastanov heteroepitaxy regime \cite{Stangl:2004_RMP}, strain minimization can govern the NC arrangement.

Typically, the TM concentration of the host containing TM-rich NCs is nonzero, $x_1 > 0$. Accordingly,  the magnetic response of decomposed alloy shows characteristics of the uniform DMS with TM content $x_1$ superimposed on magnetism of a system of NCs with the TM concentration $x_2$. Since the crystallographic and chemical structure of the NCs is usually imposed by the host,  magnetic properties of the individual NCs may not be listed in the existing materials compendia. Furthermore, magnetism of the NC ensemble is strongly affected by their distribution and coupling, either dipole-dipole type or mediated by strain \cite{Korenev:2015_NP} and/or spins in the host.

\section{Spinodal nanodecomposition in $\mbox{(Ga,Mn)As}$}
\label{sec:GaAs}
It is well known that (Ga,Mn)As containing uniformly distributed magnetic atoms has become a model system for the entire class of dilute FM semiconductors \cite{Dietl:2014_RMP,Jungwirth:2014_RMP,Tanaka:2014_APR}.  Similarly, the decomposed (Ga,Mn)As system, consisting of MnAs or Mn-rich (Mn,Ga)As NCs embedded in Mn-poor (Ga,Mn)As, since its first fabrication  \cite{De_Boeck:1996_APL,Shi:1996_S}, has revealed  properties relevant to the whole family of high $T_{\text{C}}$ semiconductors. For instance, both crystallographic \cite{De_Boeck:1996_APL} and chemical \cite{Moreno:2002_JAP} phase separations were put into the evidence, depending on fabrication and processing conditions. Furthermore, despite a small diameter down to 2\,nm, the ensemble of zb Mn-rich NCs shows FM-like features persisting up to 360\,K \cite{Moreno:2002_JAP,Yokoyama:2005_JAP}, to be compared to $T_{\text{C}}$ of 318\,K specific to freestanding samples of MnAs that crystallizes in a hexagonal NiAs-type structure. An enhanced magnetooptical response was found in decomposed films of (Ga,Mn)As \cite{Akinaga:2000_APL,Shimizu:2000_JVSTB}. The functionalities and device implications of such decomposed films (MnAs NCs in GaAs) are also described elsewhere \cite{Tanaka:2008_B}.

\subsection{Fabrication methods and nanocomposite structure}

Various methods were found to provide decomposed (Ga,Mn)As. It can be obtained by MOVPE \cite{Lampalzer:2004_JCG,Krug_von_Nida:2006_JPCM} or MBE at sufficiently high substrate temperatures \cite{Hai:2011_JAP} or by post-growth annealing of either GaAs implanted with Mn \cite{Ando:1998_APL,Shi:1996_S, Chen:2000_JAP,Wellmann:1997_APL} or fabricated by low-temperature (LT) MBE \cite{De_Boeck:1996_APL,Shimizu:2001_APL,Moreno:2002_JAP,Yokoyama:2005_JAP,Kwiatkowski:2007_JAP,Rench:2011_PRB,Sadowski:2011_PRB,DiPietro:2010_APL}.

At the growth temperature specific to MOVPE, 500--600$^{\circ}$C, one observes assembling of hexagonal MnAs NCs and their segregation toward the layer surface \cite{Krug_von_Nida:2006_JPCM}. They have typically elongated shape in the growth direction and their length reaches 100\,nm.

\begin{figure}
\includegraphics[angle=0,width=0.8\columnwidth]{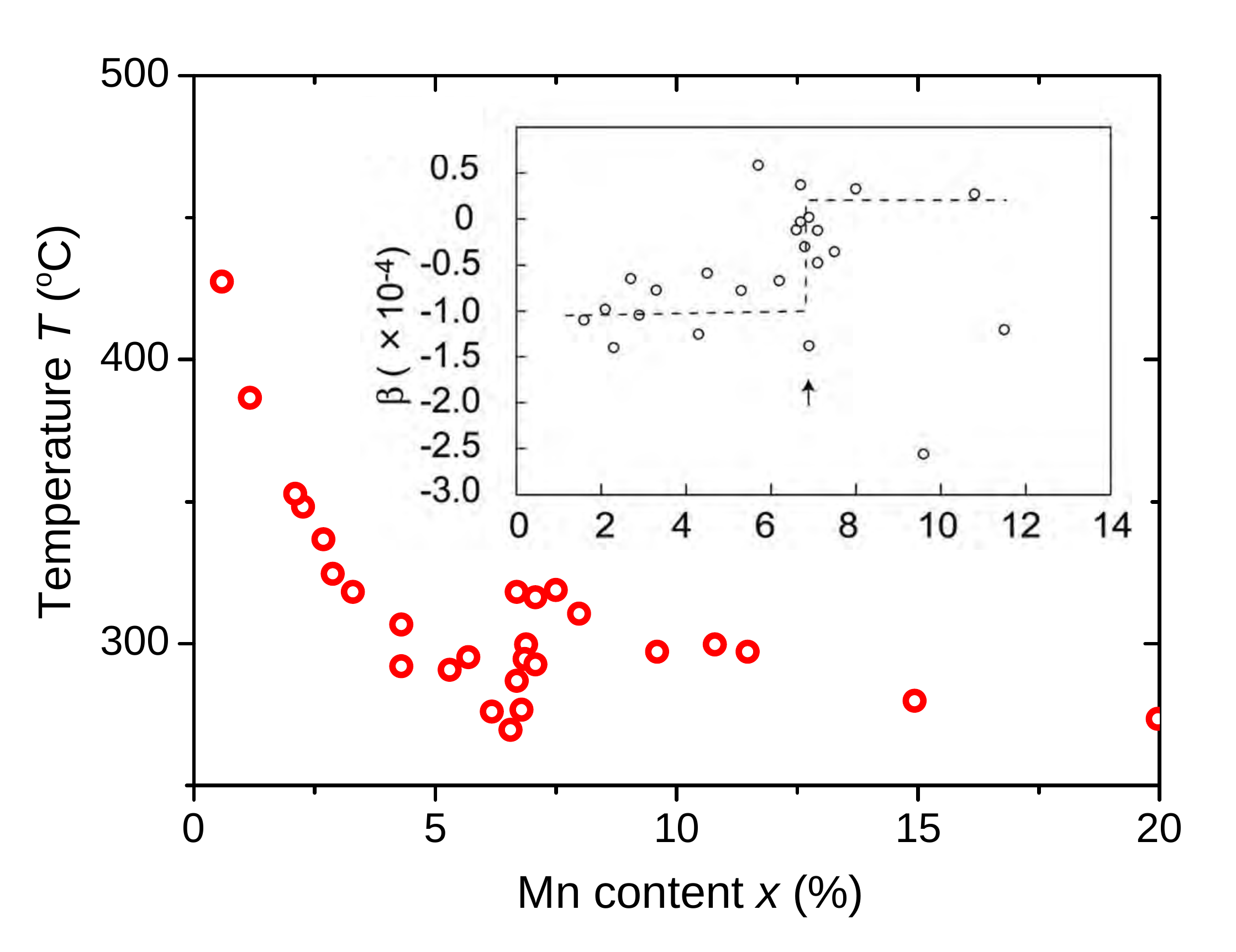}
\caption{(Color online)\label{fig:tanaka_fig2}
Phase decomposition diagram of (Ga,Mn)As alloy. Open circles are experimental
values of temperature $T_{\rm B}$ at which RHEED streaks, witnessing 2D growth, change into dots.
Inset shows the behavior of the in-plane lattice parameter. Adapted from \onlinecite{Hai:2011_JAP}.}
\end{figure}

According to the MBE growth diagram of Ga$_{1-x}$Mn$_x$As \cite{Matsukura:2002_B,Dietl:2014_RMP,Ohno:1996_APL,Hayashi:1997_JCG}, at temperatures above 350$^{\circ}$C at $x = 1$\% and above 200$^{\circ}$C at $x =10$\% an onset of TM aggregation is observed. Growth temperatures below 200$^{\circ}$C are required for maintaining   2D growth of (Ga,Mn)As incorporating more than 10\% of uniformly distributed Mn cations \cite{Mack:2008_APL,Chiba:2007_APL,Ohya:2007_APL,Wang:2008_APL}. Results of detailed investigation by reflection high-energy electron diffraction (RHEED) of temperature $T_{\rm B}(x)$ at which 2D growth disappears entirely, i.e., RHEED stripes transform into dots, are shown in Fig.\,\ref{fig:tanaka_fig2} \cite{Hai:2011_JAP}. As seen, $T_{\rm B}(x)$ decreases monotonously with $x$ reaching 270$^{\circ}$C for $x = 6.7$\% at which $T_{\rm B}$ abruptly rises to 320$^{\circ}$C, the effect accompanied by a steplike increase in the in-plane lattice constant. For $x > 6.7$\% a gradual decrease of $T_{\rm B}$ continues down to about 250$^{\circ}$C at $x = 20$\%. This behavior, may indicate that zb-MnAs or Mn-rich (Mn,Ga)As NCs prevail for $x < 6.7$\%, whereas at higher $x$  hexagonal MnAs NCs nucleate at $T \geq T_{\rm B}$ on the growth surface.

High annealing temperatures, typically 500--700$^{\circ}$C, are required to promote (Ga,Mn)As decomposition in films grown by low-temperature MBE \cite{De_Boeck:1996_APL,Shimizu:2001_APL,Moreno:2002_JAP,Yokoyama:2005_JAP,Kwiatkowski:2007_JAP,Rench:2011_PRB,Sadowski:2011_PRB,DiPietro:2010_APL}, although at onset of Mn aggregation was already noted in films annealed at 400$^{\circ}$C \cite{Sadowski:2011_PRB}.  Similar or higher temperatures were employed to fabricate decomposed (Ga,Mn)As out of Mn-implanted GaAs \cite{Ando:1998_APL,Shi:1996_S,Chen:2000_JAP,Wellmann:1997_APL}. These temperatures, significantly surpassing $T_{\rm B}$ of Fig.\,\ref{fig:tanaka_fig2}, collaborate the fact that aggregation of Mn cations is kinetic limited (see, Sec.\,\ref{sec:CH}), as at given temperature atom diffusion is slower in the bulk than on the growth surface. Furthermore, there is an interfacial energy barrier for forming a nucleus of NCs in the bulk, which requires higher annealing temperatures compared with those formed on the surface \cite{Hai:2011_JAP}.

\begin{figure}
\includegraphics[width=3.1in]{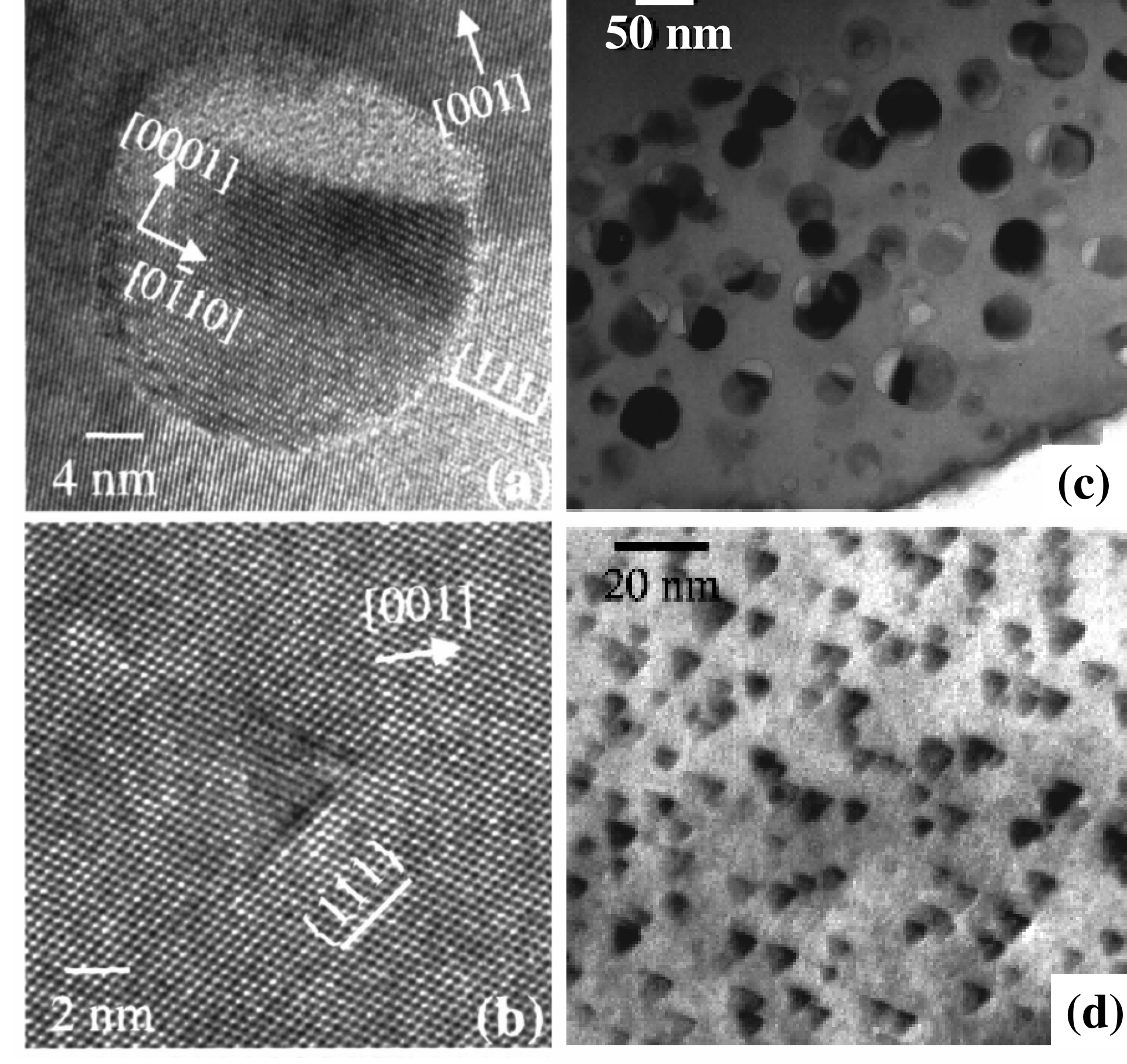}\vspace{-0mm}
\caption{Evidence for crystallographic and chemical phase separation in (Ga,Mn)As. High-resolution TEM (a,b) and cross-sectional bright-field TEM (c,d) images of Mn(Ga)As clusters embedded in a Ga(Mn)As matrix for: (a,c) sample I, which was subjected to rapid thermal annealing at 700$^{\circ}$C and (b,d) sample II, which was annealed {\em in situ} at 600$^{\circ}$C under MBE conditions. The formation of hexagonal MnAs NCs in sample I is the crystallographic phase separation, whereas the zb NCs are the evidence for the chemical separation. From \onlinecite{Moreno:2002_JAP}.}
\label{fig:MnAs_Moreno_ab}
\end{figure}

Depending on the magnitudes of the initial Mn concentration and annealing temperature,  either hexagonal MnAs NCs with an NiAs structure and diameter ranging from 5 to 500\,nm  precipitate usually with no dislocations and with orientation MnAs(0001)$\parallel$ GaAs(111)B (crystallographic phase separation) \cite{Ando:1998_APL,De_Boeck:1996_APL,Moreno:2002_JAP,Yokoyama:2005_JAP,Shi:1996_S,Chen:2000_JAP,Wellmann:1997_APL,Rench:2011_PRB,Krug_von_Nida:2006_JPCM} or  zb Mn-rich (Ga,Mn)As NCs aggregate (chemical phase separation) \cite{Moreno:2002_JAP,Yokoyama:2005_JAP,Kwiatkowski:2007_JAP}, according to high-resolution transmission electron microscopy (HRTEM), as shown in Fig.\,\ref{fig:MnAs_Moreno_ab}. A comprehensive x-ray diffraction (XRD) \cite{Moreno:2003_PRB,Moreno:2005_PRB} and extended x-ray absorption fine structure (EXAFS) studies  \cite{Demchenko:2007_JPCM} of hexagonal MnAs buried in the GaAs lattice provided detailed information on the magnitude of strain in reference to free-standing MnAs. The zb NCs can assume either tetrahedral  \cite{Moreno:2002_JAP} or spherical form \cite{Yokoyama:2005_JAP,Kwiatkowski:2007_JAP,Sadowski:2011_PRB} with diameter typically from 2 to 6\,nm; they are not stable against a transformation to the hexagonal phase for diameters exceeding 15\,nm \cite{Sadowski:2011_PRB}. Hexagonal and cubic zb NCs coexist in annealed (Ga,Mn)As with Mn content of 2\% and smaller \cite{Sadowski:2011_PRB}. It appears that the main features of decomposed (Ga,Mn)As and (Ga,Mn)As:Be \cite{Rench:2011_PRB} are similar.

The exact relative concentrations of Mn and Ga in the cubic NCs is barely known and may depend on details of the fabrication protocols;  the Mn concentration as low as 20\% was suggested \cite{Lawniczak-Jablonska:2011_PSS}.  Under some processing conditions the obtained nanoparticles were found to be structurally disordered  \cite{Moreno:2002_JAP,Kwiatkowski:2007_JAP}. In certain cases, presumably if growth of (Ga,Mn)As results in a substantial concentration of As antisites, hexagonal MnAs NCs obtained by annealing are accompanied by As precipitation and voids \cite{Kovacs:2011_JPCS}.

\begin{figure}
\includegraphics[angle=0,width=0.8\columnwidth]{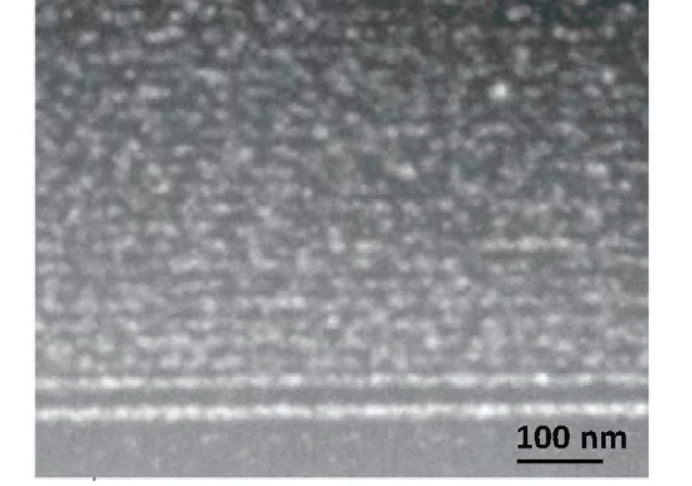}
\caption{\label{fig:sadowski}
Cross-sectional STEM (scanning transmittance electron microscope) image of the region close to the
GaAs(001) substrate of a GaAs/(Ga,Mn)As superlattice with Mn content of 3.7\%
annealed at 560$^{\circ}$C. Adapted from \onlinecite{Sadowski:2013_JPCM}.}
\end{figure}

From an application viewpoint particularly relevant is the preparation of magnetically active
NCs that reside in a predefine plane. This can be achieved by growing appropriately thin (Ga,Mn)As layers, in which Mn-rich NCs gather after annealing below 600$^{\circ}$C \cite{Shimizu:2001_APL,Sadowski:2013_JPCM}, as shown in Fig.\,\ref{fig:sadowski}.

\begin{figure}
\includegraphics[angle=0,width=1.0\columnwidth]{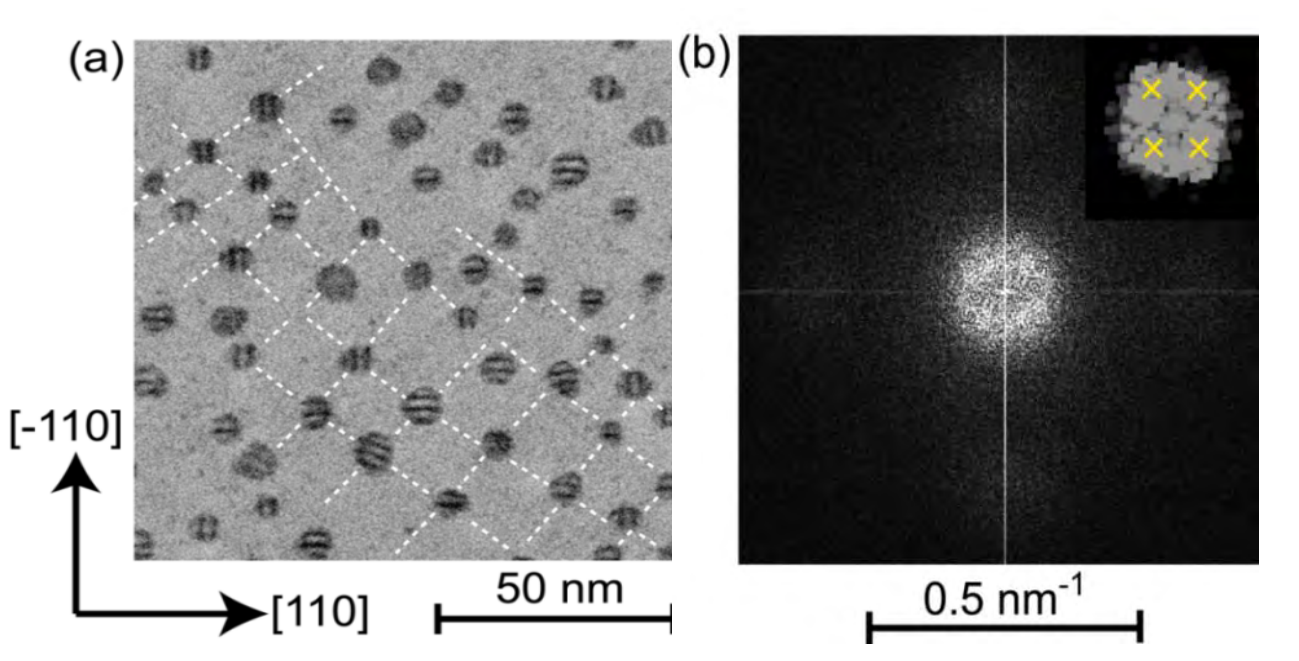}
\caption{\label{fig:tanaka_fig3}
(Color online) Periodic arrangement of MnAs NCs in GaAs. (a) Plain-view TEM image of hexagonal MnAs NCs fabricated using spinodal decomposition of a 5 nm-thick Ga$_{0.8}$Mn$_{0.2}$As layer. (b) Fourier transform of the contrast of a $0.36\times0.36$\,$\mu$m$^2$ TEM image. A ring structure with a radius of 0.063\,nm$^{-1}$ corresponding to a period of 16\,nm is observed. The inset shows the pattern of the ring structure after applying a steplike high pass filter, revealing four fold symmetry, that is, rectangular lattice structure. Adapted from \onlinecite{Hai:2011_JAP}.}
\end{figure}

Furthermore, according to results presented in Fig.\,\ref{fig:tanaka_fig3}, rectangularlike distribution of MnAs NCs with a narrow dispersion of diameters was obtained by using the phase decomposition diagram of the (Ga,Mn)As alloy and spinodal decomposition induced by annealing.

More recently, the influence of laser irradiation on the self-assembly of MnAs NCs was investigated \cite{Hai:2012_APL}. It was found that laser irradiation suppresses the temperature-induced transformation of small into larger hexagonal MnAs NCs,  and that the median diameter $D_1$ in the size distribution of small NCs depends on the incident photon energy $\hbar \omega$ following $D_1 \sim \omega^{-1/5}$. This behavior was explained by the desorption of Mn atoms from small NCs due to energy gain from optical transitions between their quantized energy levels.

\subsection{Magnetic properties}
\begin{figure}
\includegraphics[width=3.1in]{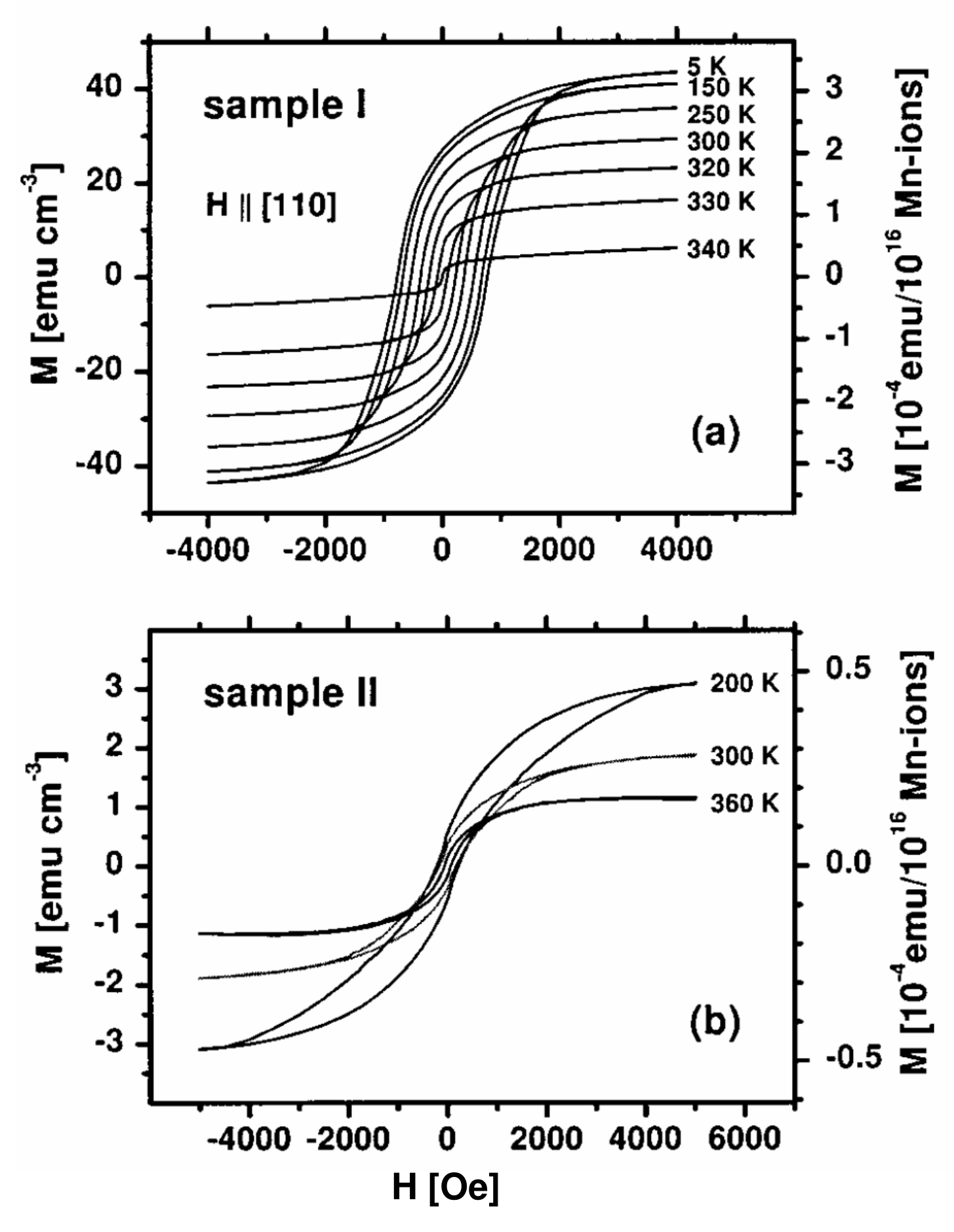}\vspace{-0mm}
\caption{Magnetization loops at different temperatures of annealed
(Ga,Mn)As containing hexagonal (a) and zinc-blende (b)
NCs, visualized in Fig.\,\ref{fig:MnAs_Moreno_ab}. From \onlinecite{Moreno:2002_JAP}.}
\label{fig:MnAs_Moreno_mag}
\end{figure}

Figures \ref{fig:MnAs_Moreno_mag}(a,b) show magnetization loops at various temperatures for (Ga,Mn)As samples with hexagonal and cubic NCs, whose micrographs have been displayed in Figs.\,\ref{fig:MnAs_Moreno_ab}(a,b), respectively. As seen in \ref{fig:MnAs_Moreno_mag}(a), square hystereses pointing to the $T_{\text{C}}$ value specific to bulk MnAs, $T_{\text{C}}\simeq 318$\,K are observed for the film with hexagonal NCs \cite{Moreno:2002_JAP}, the finding also reported by others \cite{Ando:1998_APL,Yokoyama:2005_JAP,Rench:2011_PRB,Wellmann:1997_APL}.  The data indicate that for those large NCs the magnitude of  superparamagnetic blocking temperature  $T_{\text{b}} = K(T)V/25k_B$ becomes smaller than the temperature $T$ just below $T_{\text{C}}$. It was also found that Si codoping enhanced the formation of bigger MnAs nanoclusters \cite{Shimizu:2002_PE}.

In contrast, the behavior of magnetization in the case of the sample with cubic NCs [Fig.\,\ref{fig:MnAs_Moreno_mag}(b)] is intriguing and reveals surprising but generic aspects of decomposed magnetically doped systems as follows:
\begin{itemize}
\item
Despite a rather small size of zb NCs, even at high temperatures magnetization loops are opened and weakly temperature dependent. This indicates that $M(H,T)$ cannot be modeled by the Langevin function describing superparamagnetic systems above $T_{\rm b}$, the conclusion emerging also from other studies of decomposed (Ga,Mn)As \cite{Yokoyama:2005_JAP,Sadowski:2011_PRB}. Given the NC volume $V$, this FM-like behavior points to a rather large magnitude of $K$ or to a significant role played by spin-dependent interactions between the NCs \cite{Hai:2007_JMMM}. For specific spatial distributions and densities of FM nanoparticles, long-range dipole interactions and/or exchange coupling lead to the so-called superferromagnetic phase \cite{Bolcal:2012_APPA,Panov:2012_APL,Morup:2010_BJN}, whose characteristics are consistent with the data displayed in Fig.\,\ref{fig:MnAs_Moreno_mag}(b). In particular, a smaller magnitude of low temperature magnetization, as compared to the value expected from spin counting, is consistent with the superferromagnetic scenario.
\item
This set of data points to $T_{\text{C}} \gtrsim 360$\,K, for zb Mn-rich (Mn,Ga)As stabilized by the GaAs host \cite{Moreno:2002_JAP,Yokoyama:2005_JAP,Sadowski:2011_PRB}, the value substantially higher than $T_{\text{C}} \simeq 318$\,K of free standing MnAs that crystallizes in the hexagonal structure. This $T_{\text{C}}$ enhancement can be assigned to higher crystal symmetry, which typically leads to greater density of states at the Fermi level and, hence, to elevated $T_{\text{C}}$.
\item
While decomposed films discussed above show superferromagnetic features, in some other samples, containing similar NCs according to TEM studies, a standard superparamagnetic behavior was found \cite{Rench:2011_PRB,DiPietro:2010_APL,Sadowski:2011_PRB}.
Differing properties of nominally similar samples imply that magnetic characteristics of decomposed systems are rather sensitive to pertinent details of the system fabrication and the resulting morphology, including the following:
\begin{itemize}
\item
the chemical structure and composition of the individual NCs (i.e., TM concentration), which determine the electronic structure (e.g., the itinerant versus localized character of $d$ electrons) and, thus,  magnetic characteristics such as the magnitude of $T_{\text{C}}$, magnetic moment, and energy of bulk and interfacial magnetic anisotropy
\item
the spatial distribution and density of the NCs' ensemble, which underlines the magnetic ground state originating from long-range dipole interactions
\item
the nature of host-mediated spin coupling among the NCs, which can depend on strain \cite{Korenev:2015_NP} and interface between NCs and the host as well as on the TM, carrier, and defect concentrations in the host.
\end{itemize}
\end{itemize}

In the case of superparamagnetic samples, the temperature dependence of zero-field cooled magnetization demonstrates that the distribution of NCs' size is log-normal \cite{Rench:2011_PRB,DiPietro:2010_APL}. At the same time, characteristics of magnetization relaxation are consistent with the superparamagnetic scenario and allow excluding spin-glass freezing as an origin of the magnetization maximum as a function of temperature \cite{Sadowski:2011_PRB}.

In general, magnetic response of decomposed (Ga,Mn)As contains contributions from both NCs with a high Mn concentration and (Ga,Mn)As with lower Mn content. In the cases discussed previously, the NCs dominate magnetic response in the entire temperature range. However, for (Ga,Mn)As grown by MBE under conditions corresponding to the onset of phase separation, the FM response of the (Ga,Mn)As host can prevail at low temperatures. Here  the NCs lead to the pinning of domain walls and to enlargement of the coercive force \cite{Wang:2006_APL}.  Only at high temperatures, above $T_{\text{C}}$ of the host,  is magnetization dominated by the NCs \cite{Wang:2006_APL}. A similar situation takes place in magnetic resonance studies: FM resonance of MnAs NCs dominates at high temperature, whereas a signal of dilute Mn ions takes over at low temperatures \cite{Krug_von_Nida:2006_JPCM}.

\subsection{Magneto-optical phenomena}
\label{sec:MCD_GaAs}
One of the attractive aspects of decomposed semiconductor alloys is that they show FM characteristics
to above RT and at the same time they can be easily integrated with existing semiconductor devices.
For instance, one can envisage magnetooptical insulators and modulators that would exploit a large magnitude
of the Kerr effect characterizing FM metals and weak optical losses specific to semiconductors.
Indeed, the giant Faraday effect and magnetic circular dichroism was found in GaAs with hexagonal metallic MnAs NCs at RT \cite{Akinaga:2000_APL}.
The Faraday rotation angle of the GaAs:MnAs layer reached about 0.2$^{\circ}$/$\mu$m at 0.98\,$\mu$m.

Optical reflectivity of GaAs with hexagonal MnAs and zb-(Mn,Ga)As NCs is similar to that of the host GaAs matrix
\cite{Yokoyama:2005_JAP}. However, magnetooptical spectra and intensities are different in these two kinds of decomposed (Ga,Mn)As;
the intensity is weaker in the case zb NCs.

While the magnitude of the Verdet constant (i.e., the normalized Faraday rotation) is attractive and comparable or larger than the one characterizing magnetic materials employed in commercial optical insulators, it is hard to obtain a GaAs:MnAs layer thick enough to insure the required angle of polarization rotation, 45$^{\circ}$. However, it is
possible to enhance the rotation angle by extending the effective optical path
length through a magnetic layer by using multiple
reflections.  By inserting a 140\,nm-thick GaAs:MnAs film in between distributed Bragg reflectors (DBR) of GaAs/AlAs, the Faraday rotation angle was enhanced sevenfold at a wavelength of 970\,nm at
RT \cite{Shimizu:2001_APL}. The  MCD signal at RT as a function of the magnetic field showed  saturation at about 1\,kOe, the dependence corresponding to magnetization of GaAs:MnAs system. Furthermore, a similar structure showed the Kerr rotation of 1.54$^{\circ}$ at the designed wavelength of ~980 nm under a relatively low magnetic field at RT \cite{Ueda:2003_JJAP}.
Theoretical modeling of optical and magnetooptical properties of these magneto-photonic structures properly describes the experimental data.

As a further step, in order to reduce transmission losses associated with large MnAs NCs, the central GaAs:MnAs layer was replaced by a  superlattice (SL) consisting of 2.8\,nm-thick AlAs and 5\,nm-thick GaAs:MnAs formed by annealing of (Ga,Mn)As with Mn content of $x =0.047$ \cite{Shimizu:2001_JAP}. In this way,  the transmission coefficient at a local
maximum at 990\,nm was 30\%, greatly improved from  2\% in the previous multilayer without a SL. However, in this case, the enhancement in the Faraday rotation per unit magnetic layer thickness, compared to the GaAs:MnAs/AlAs SL structure without DBR, is only 3.3, presumably because of lower structural quality.

\begin{figure}
\includegraphics[width=3.1in]{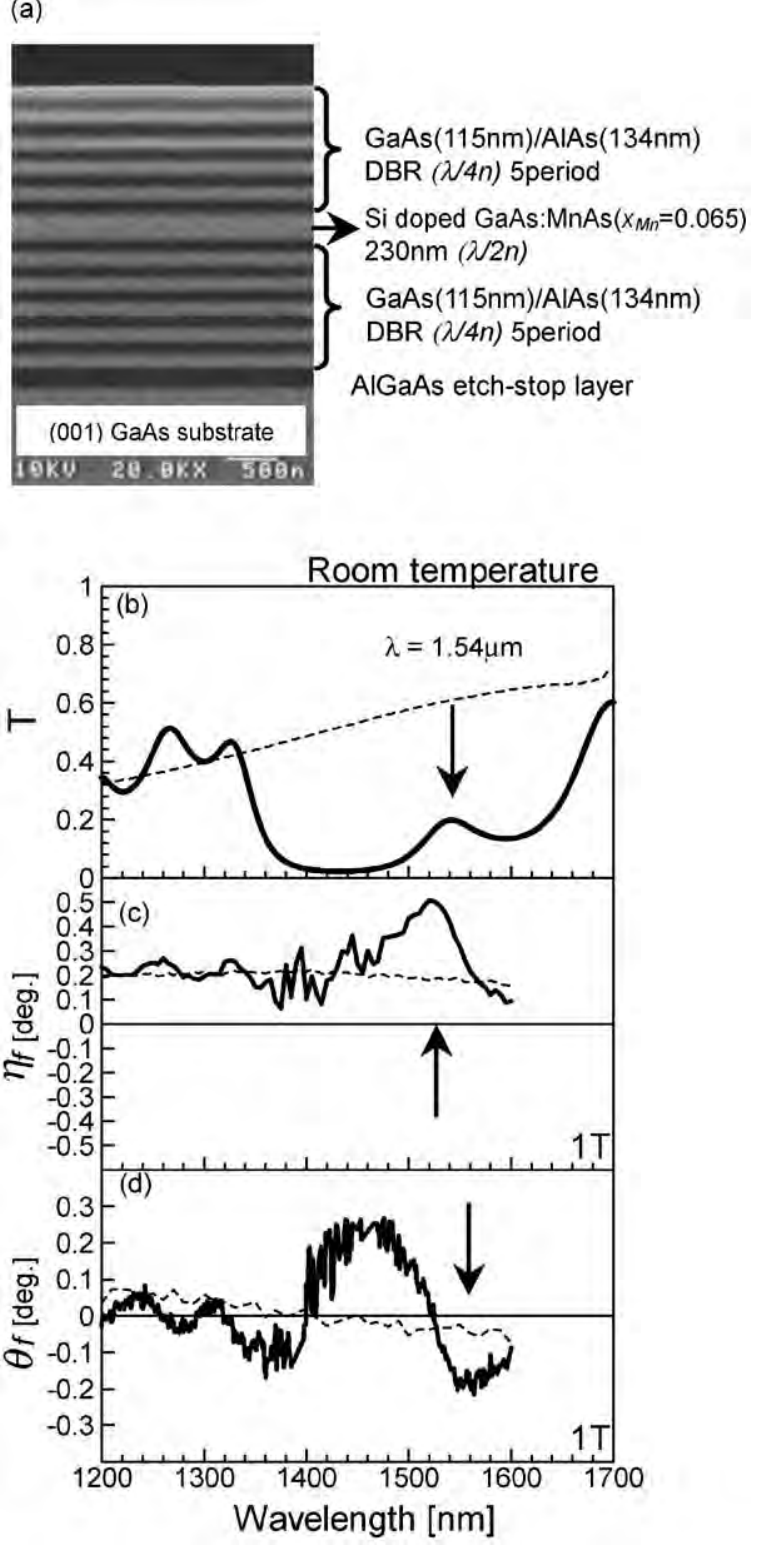}\vspace{-0mm}
\caption{An optimized structure with MnAs NCs for optical isolation. (a) A cross-sectional micrograph taken by
scanning electron microscope of a Si-doped GaAs:MnAs layer
placed between DBRs whose operational wavelength was set at 1.55\,$\mu$m.
The solid lines in (a,b,c) show respectively optical transmission $T$;
Faraday ellipticity $\eta_f$ (transmission MCD) and Faraday rotation $\theta_f$
at room temperature in the magnetic field of 1\,T perpendicular to the
plane of the sample shown in (a). Spectra of a
Si-doped GaAs:MnAs single layer (200\,nm) having the same Mn
and Si concentration are also plotted as references
(dotted lines).  From \onlinecite{Shimizu:2002_PE}.}
\label{fig:MnAs_Shimizu}
\end{figure}

Another interesting structure, presented in Fig.\,\ref{fig:MnAs_Shimizu}, was optimized for the photon wavelength of 1.55\,$\mu$m \cite{Shimizu:2002_PE}. Here, the central 230\,nm thick layer consisted of GaAs:MnAs decomposed film codoped with Si donors in order to reduce optical absorption associated with holes delivered by residual Mn ions in the GaAs host. The figure of merit (FOM), which is defined by the ratio of Faraday ellipticity (shown in Fig.\ref{fig:MnAs_Shimizu}) to optical losses, is 0.074$^{\circ}$/dB at the wavelength
of 1.54\,$\mu$m, which is twice as large as the FOM of
0.037$^{\circ}$/dB at 0.98\,$\mu$m obtained in the studies \cite{Shimizu:2001_APL} referred to previously.

At the same time, properties of {\em waveguide-type} optical insulators based on
higher losses of backward propagating light due to a nonreciprocal change of the refractive index
were theoretically analyzed for an (In,Al)As waveguide with embedded  MnAs nanoclusters \cite{Shimizu:2002_APL}.
Since in this case the proposed optical isolator structure is also composed of all semiconductor-based
materials, it can be integrated with III-V based optoelectronic devices such as
edge-emitting laser diodes. The device was assumed to be grown on an InP substrate and the operation
wavelength was 1.55\,$\mu$m. Two designs were considered corresponding to the propagation of TM and TE modes,
for which 119\,dB/cm and 36\,dB/cm of isolation were predicted, respectively.

\subsection{Magnetoresistance}
\subsubsection{Films}

Transport phenomena in granular metals exhibit a number of outstanding properties \cite{Beloborodov:2007_RMP}
that can be enriched by spin phenomena in decomposed magnetically doped semiconductors \cite{Michel:2008_APL,Binns:2005_JPD}.

Magnetoresistance (MR) properties of GaAs/MnAs granular hybrid structures consisting of
FM MnAs clusters within a paramagnetic GaAs:Mn host differ considerably from those
of paramagnetic and FM (Ga,Mn)As alloys. According to experimental studies of decomposed samples obtained
by various methods, giant positive MR dominates at high temperatures (typically above $\sim 30$\,K) whereas
negative MR takes over at low temperatures \cite{Akinaga:2000_APLc,Wellmann:1998_APL,Ye:2003_APL,Michel:2005_SM}.
Large positive MR was also observed at RT in the impact ionization regime \cite{Yokoyama:2006_JAP}.

The positive MR effect is similar to the giant MR found in
other metal-semiconductor hybrid systems \cite{Akinaga:2000_APLc,Solin:2000_S,Sun:2004_JJAP}.
One of relevant mechanisms could be an admixture to the longitudinal voltage of the field-dependent Hall voltage, possibly enhanced by the
AHE, a well known phenomenon in other heterogeneous systems \cite{Solin:2000_S}. A strong decrease
of the positive MR for the in-plane magnetic field \cite{Ye:2003_APL} supports this scenario.

Two quantum mechanisms, considered widely in the case of homogeneous DMSs \cite{Dietl:2008_JPSJ}, can contribute to negative MR:
(i) an orbital effect originating from the influence of the magnetic field upon interferences of self-crossing trajectories in
either diffusing  or hopping regime; (ii) a spin effect brought about by a destructive effect of the magnetic field upon spin-disorder scattering that controls interference of carrier-carrier interaction amplitudes
in disordered systems. At the same time, giant MR (GMR) or tunneling MR (TMR) effects associated with the field-induced ordering of NCs' magnetic moments or
spin-splitting of host's bands produced by stray fields can contribute to the magnitude of MR \cite{Michel:2005_SM}.
Furthermore, independently of the dominating MR mechanism, the spatial distribution of magnetic NCs is a factor substantially affecting the magnitude of MR and allowing its control \cite{Michel:2008_APL}.

\subsubsection{Spin-valve structures}
The decomposed (Ga,Mn)As layers containing MnAs NCs with diameter of 10\,nm were also formed in the fully epitaxial magnetic tunnel junctions (MTJs), GaAs:MnAs/Al(Ga)As/MnAs, in which a 20\,nm MnAs layer constituted the second FM electrode and the barrier was either of AlAs \cite{Hai:2006_APL} or of GaAs \cite{Hai:2008_PRB}.  The magnitude of TMR was examined as a function of temperature and the barrier thickness $d$ attending, at 7\,K, 17\% for $d_{\text{AlAs}} = 2.9$\,nm \cite{Hai:2006_APL} and 8\% for $d_{\text{GaAs}} = 10$\,nm \cite{Hai:2008_PRB}.   The observed oscillatory behavior of the TMR ratio with the increasing AlAs barrier thickness was explained by  quantum interference of two X-valley related wave functions in the AlAs barrier. The ensemble of the results demonstrates that GaAs:MnAs layers can act as an efficient spin injector and a spin detector at low temperatures.

\begin{figure}
\includegraphics[width=3.1in]{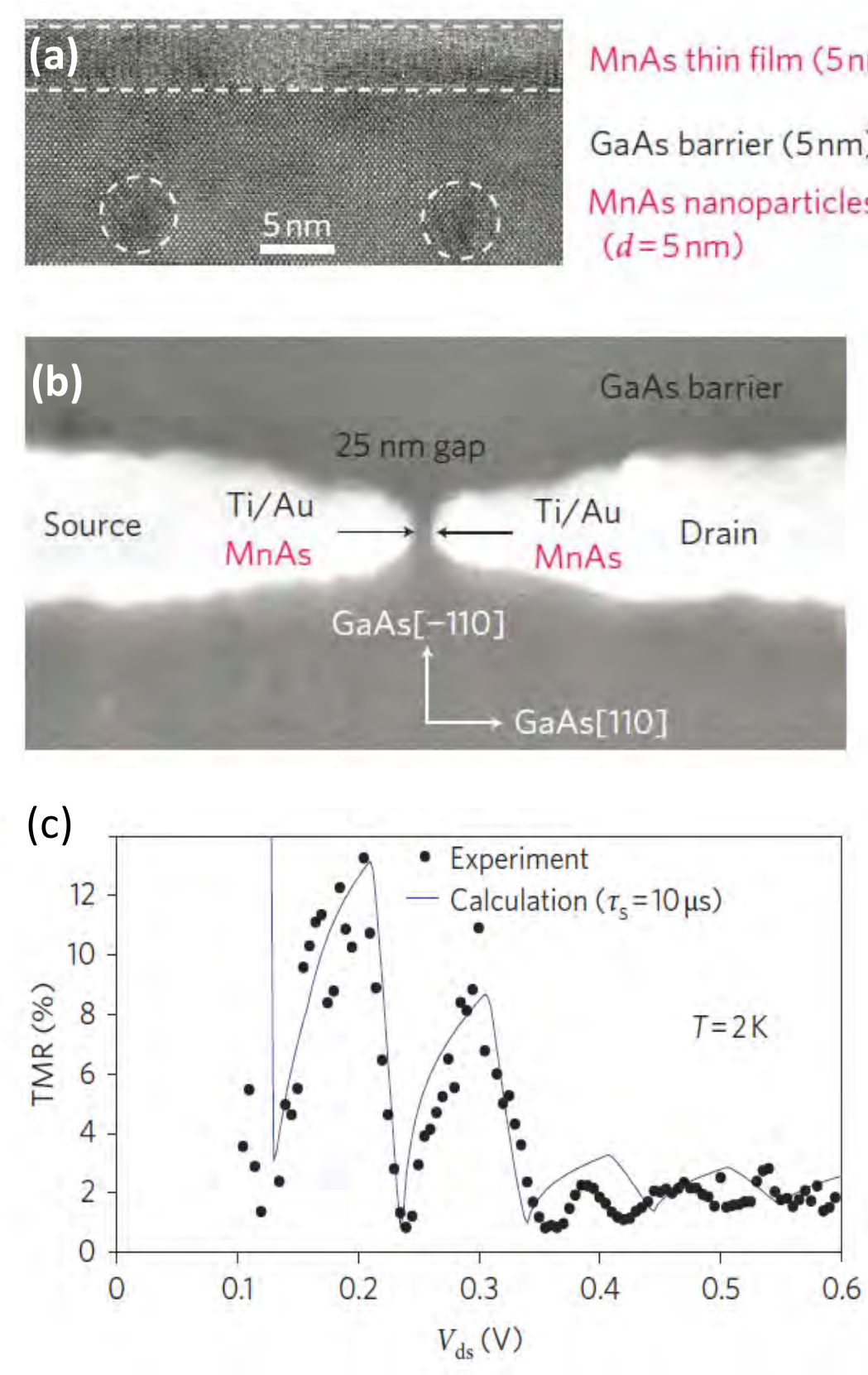}\vspace{-0mm}
\caption{(Color online) Magnetic tunnel junction involving a single MnAs NC. (a) Cross-sectional TEM image prior to patterning of electrodes 
and the top-view SEM image of the device (b). The values of TMR = $[R(\mu_0H=0) - R(\mu_0H = 1\,\mbox{T})]/R(\mu_0H = 1\,\mbox{T})$ as a function of $V_{\text{ds}}$ (c).  The line represents the theoretical calculation for carrier spin relaxation time within the MnAs NC $\tau_s = 10$\,$\mu$s. Adapted from \onlinecite{Hai:2010_NN}.}
\label{fig:SET}
\end{figure}

Figures \ref{fig:SET}(a,b) present micrographs of a lateral nanodevice patterned of a MnAs/GaAs/GaAs:MnAs MTJ, which allowed one to study TMR in the limit of tunneling across a {\em single} MnAs NC \cite{Hai:2010_NN}. Sizable oscillations in differential conductance $dI_{\text{ds}}/dV_{\text{ds}}$ demonstrated that the Coulomb blockade regime was achieved at 2\,K. As shown in Fig.\,\ref{fig:SET}(c), clear oscillations were also detected in TMR, defined as a relative difference in the resistance values without an external magnetic field (antiparallel arrangement of magnetization in the MnAs contacts) and in the magnetic field of 1\,T (parallel magnetizations across the device). Modeling of the TMR results by the theory of \onlinecite{Barnas:1998_EPL} led to the spin relaxation time $\tau_s$ of carriers in MnAs NCs as long as $10\,\mu$s. The enhancement of $\tau_s$ was assigned to dimensional quantization of electronic states specific to nanoparticles \cite{Hai:2010_NN}.

On the other hand, the magnetizing process of zb-MnAs NCs incorporated into tunnel junctions  was found to generate a voltage and a large magnetoresistance effect, which lasted for a time scale of $10^2$--$10^3$ s \cite{Hai:2009_N}. This spin-motive force was explained by the transformation of magnetic energies of the NCs into the electric energy of the electron system in the Coulomb blockade regime. A similar long-living spin-motive force was observed by \onlinecite{Miao:2014_NC} in tunnel junctions with EuS barriers and Al nanoparticles. However, an external energy, brought about by, e.g., thermal or electromagnetic noise, should be provided to these systems to maintain such a long-living spin-motive force \cite{Miao:2014_NC,Ralph:2011_N}. The problem of spin-motive force and magnetotransport in nanoscale magnetic systems remains an outstanding experimental and theoretical topic in spinodal-decomposed magnetic semiconductors.

\subsection{Spinodal nanodecomposition in other magnetic III-V compounds}

Several properties outlined previously for decomposed (Ga,Mn)As have been found in other III-V  semiconductors doped with TM ions.  These systems, if fabricated under specific growth conditions, show FM-like features with $T_{\text{C}}$ magnitudes independent of an average Mn concentration but virtually identical to values characterizing relevant Mn pnictides, i.e., $T_{\text{C}}$ of 291, 318, and 573\,K, for MnP, MnAs, and MnSb, respectively. To this category belong (Ga,Mn)P \cite{deAndres:2011_JAP}, (In,Ga,Mn)As/InP \cite{Hara:2005_NT,Hara:2006_APL}, (In,Mn)As \cite{Khodaparast:2013_PRB}, and (In,Mn)Sb \cite{Lari:2012_JAP} grown by MOVPE as well as (Ga,Sb)Mn \cite{Abe:2000_PE} obtained by MBE and GaP:C implanted with Mn \cite{Theodoropoulou:2002_PRL}. A similar situation takes place in Mn-doped II-IV-V$_2$ compounds crystallizing in chalcopyrite structures \cite{Kilanski:2010_JAP,Uchitomi:2015_JAP}. Particularly extensive studies have been performed for magnetically-doped GaN, as discussed in Secs.\,\ref{sec:GaN-Mn} and \ref{sec:GaN-Fe}.

\section{\label{sec:GaN-Mn}Spinodal nanodecomposition in $\mbox{(Ga,Mn)N}$}
\subsection{A controversial system}
\label{sec:GaN-Mn_magnetic}
Various contradicting information about the magnetism of (Ga,Mn)N has been reported in the literature.

Fabrication of single crystalline Ga$_{1-x}$Mn$_{x}$N epitaxial films with an experimentally documented
random distribution of Mn cations and a small concentration of donorlike defects or residual impurities
 \cite{Sarigiannidou:2006_PRB,Stefanowicz:2010_PRB,Bonanni:2011_PRB,Kunert:2012_APL} pointed to the presence of low-temperature
 ferromagnetism \cite{Sarigiannidou:2006_PRB,Sawicki:2012_PRB,Stefanowicz:2013_PRB}. In such samples
the Fermi level is pinned by the midgap Mn$^{2+}$/Mn$^{3+}$ impurity band \cite{Graf:2003_PSSB,Wolos:2009_PSSC,Dietl:2014_RMP} but Mn$^{3+}$ ions prevail. A
semi-insulating  character of this material, exploited as a buffer in recent magnetotransport experiments on
GaN:Si layers \cite{Stefanowicz:2014_PRB}, indicates that owing to strong $p$--$d$ coupling holes
are tightly bound to parent Mn acceptors \cite{Dietl:2008_PRB,Wolos:2009_PSSC}.
Accordingly, the impurity-band carriers remain localized by, presumably,
a combine effect of the Mott-Hubbard and Anderson-Mott mechanism up to at least $x = 10$\%.
For the resulting high spin $d$-shell configuration of Mn$^{3+}$ ions
FM superexchange was predicted for tetrahedrally bound DMSs \cite{Blinowski:1996_PRB}, in agreement with the corresponding experimental results for Ga$_{1-x}$Mn$_{x}$N \cite{Kondo:2002_JCG,Sarigiannidou:2006_PRB,Bonanni:2011_PRB,Sawicki:2012_PRB,Stefanowicz:2013_PRB}.
Tight-binding and Monte-Carlo computations carried out
within the superexchange scenario and taking into the Jahn-Teller distortion described quantitatively the experimental phase diagram
$T_{\text{C}}(x)$ \cite{Sawicki:2012_PRB,Stefanowicz:2013_PRB}.
The determined dependence $T_{\text{C}}(x) \propto x^{\alpha}$,
where $\alpha = 2.2\pm 0.2$ \cite{Sawicki:2012_PRB,Stefanowicz:2013_PRB}, characteristic for a short-range superexchange of either sign \cite{Twardowski:1987_PRB,Swagten:1992_PRB,Sawicki:2013_PRB},
leads to a rather low magnitude of $T_{\text{C}} \lesssim 13$\,K at small $x \lesssim 10$\% in Ga$_{1-x}$Mn$_{x}$N \cite{Stefanowicz:2013_PRB}. Within this scenario,  ferromagnetism at RT is expected for $x \gtrsim 0.5$, if impurity-band carriers will still remain localized at such high Mn concentrations. In contrast, within the $p$--$d$ Zener model, long-range FM interactions mediated by band holes would result in $T_{\text{C}} > 300 $\,K at $x$ as small as 5\% in Ga$_{1-x}$Mn$_{x}$N  \cite{Dietl:2000_S}.

However, in other series of Ga$_{1-x}$Mn$_{x}$N samples, grown by bulk-like methods with $x \leqslant 0.1$ \cite{Zajac:2001_APLb} or  by sputtering that allows $x$ to rise up 36\% \cite{Granville:2010_PRB}, the interaction between Mn ions was established to be AF. This is consistent with the dominance of Mn in the 2+ state in those cases \cite{Granville:2010_PRB,Zajac:2001_APLa,Graf:2003_PSSB} for which the tight-binding theory mentioned previously predicts AF coupling \cite{Kacman:2001_SST}. A negative sign of the Curie-Weiss temperature, found for MBE-grown films with $x$ between 8\% and 12\% \cite{Dhar:2003_PRB}, pointed also to the AF character of the spin interactions. The presence of spin-glass freezing at $T_{\text{f}}$ between 3 and 4.4\,K was revealed for these samples \cite{Dhar:2003_PRB}. The nature of compensating donors providing electrons to Mn ions has not yet been firmly established though N vacancies are one of the possibilities \cite{Piskorska:2015_JAP}.

Surprisingly, the appearance of FM-like features in (Ga,Mn)N, persisting up to temperatures well above 300\,K, was reported by quite a few groups \cite{Liu:2005_JMSME,Pearton:2003_JAP}, and assigned to double exchange that may appear when Mn$^{3+}$ and Mn$^{2+}$ ions coexist \cite{Reed:2005_APL,Sonoda:2007_APL}. In view of the technological importance of group III nitrides in today's photonics and electronics, the fabrication of a GaN-based functional FM semiconductor would constitute a major breakthrough. Since, however, neither spintronic devices nor phase diagram $T_{\text{C}}(x)$ have so far been reported in this case, it is tempting to assume that the high-temperature ferromagnetism in question is not under control. It is likely that spinodal nanodecomposition was involved in the samples reported to show RT ferromagnetism. Theoretical aspects of nanodecomposition, for which (Ga,Mn)N has often served as a model system, are addressed in Sec.\,\ref{sec:theory}. No magnetooptical effects associated with high-$T_{\text{C}}$ ferromagnetism have been detected in (Ga,Mn)N \cite{Ando:2006_S}.

\subsection{Experimental evidences for spinodal nanodecomposition in {(Ga,Mn)N}}
\label{sec:GaN-Mn_decomposition}
The FM  response persisting at elevated temperature and in the absence of carriers
mediating the interaction between magnetic ions is presently widely recognized
to originate in TM-doped semiconductors from regions of spinodal nanodecomposition
in the form of either coherent chemical separation or crystallographic phase precipitation.
In the case of (Ga,Mn)N the inclusion of FM NCs with different transition
temperatures, such as Mn$_{4}$N ($T_{\mathrm{C}}$=784~K) \cite{Pop:1994_MCP},
Mn$_{3-\delta}$Ga ($T_{\mathrm{C}}=765$~K) \cite{Niida:1996_JAP}, or
Ga$_{0.8}$Mn$_{3.2}$N ($T_{\mathrm{C}}$=235~K) \cite{Garcia:1985_JMMM} has been identified.

As the appearance and the form of spinodal nanodecomposition
depends sensitively on fabrication conditions, codoping by shallow impurities, defect concentration, and
post-growth processing, the detection of the subtle presence of phase separation
requires a thorough characterization of individual samples with a combination
of both local [high-resolution TEM (HRTEM),
3D atom probe (3DAP),...]  and averaging [SQUID magnetometry, synchrotron-based
XRD (SXRD), x-ray XMCD, magnetotransport,... ] techniques.


\begin{figure}[htb]
    \centering
        \includegraphics[width=0.9\columnwidth]{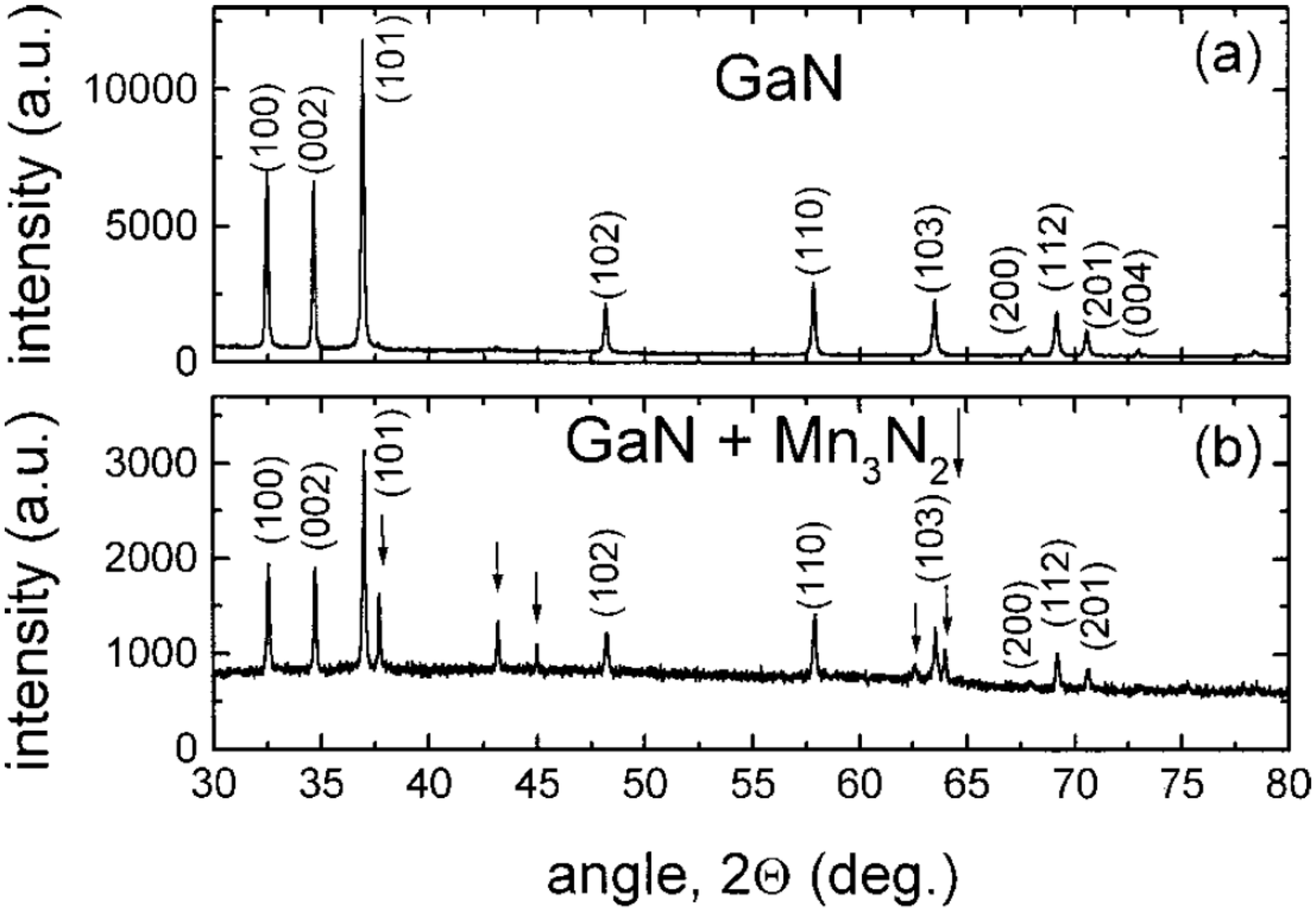}
    \caption{XRD pattern of (Ga,Mn)N grown by ammonothermal
method with (a) low and (b) high  Mn content. The Miller indices of the GaN
crystal lattice planes are given for each diffraction line and arrows indicate
diffraction lines originating from the Mn$_{3}$N$_{2}$ phase. From \onlinecite{Zajac:2001_APLa}.}
    \label{fig:Zajac2}
\end{figure}
\begin{figure}[htb]
    \centering
        \includegraphics[width=\columnwidth]{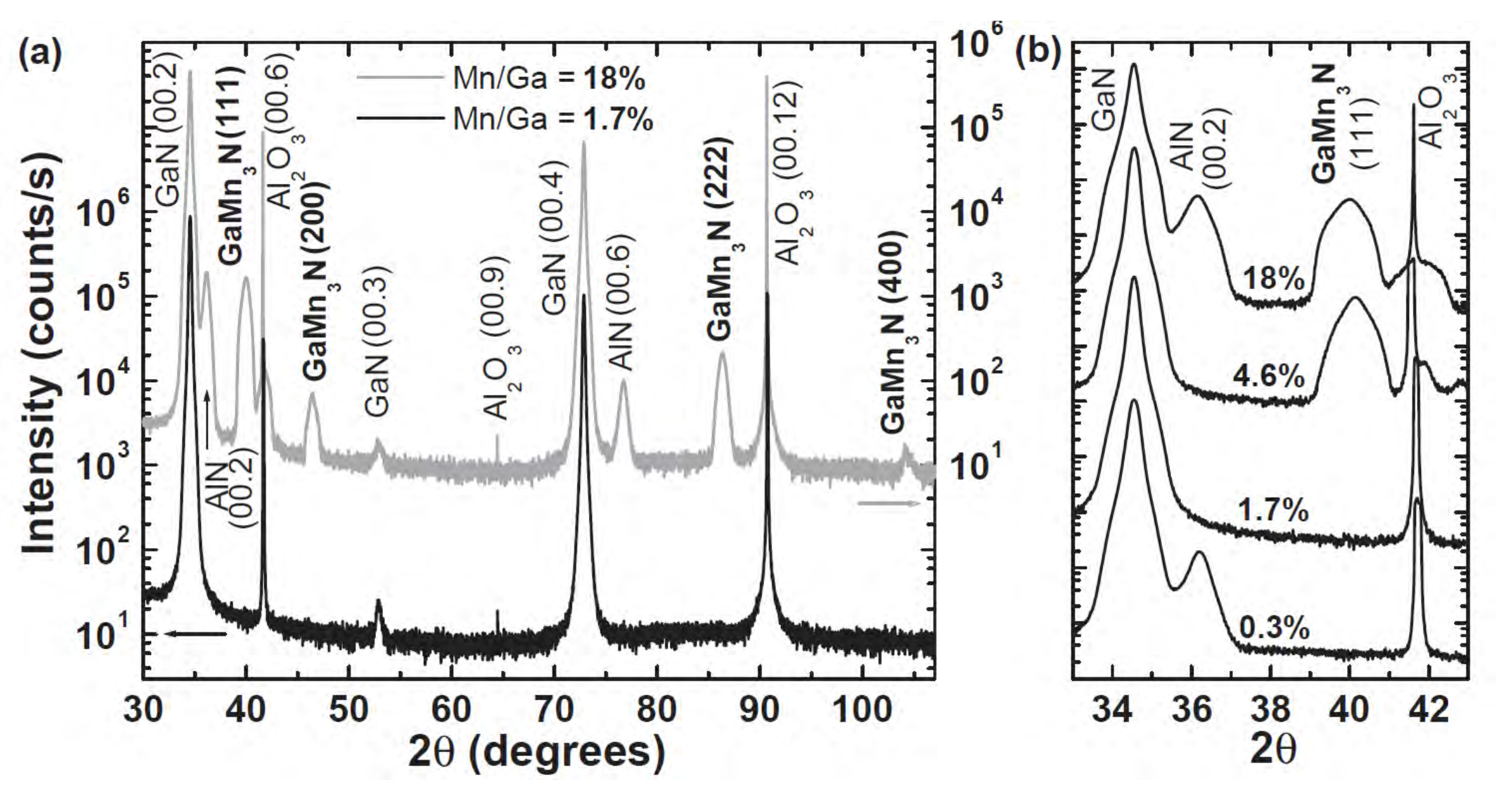}
    \caption{Determination of NCs' structure in (Ga,Mn)N. (a) XRD patterns of (Ga,Mn)N epilayers. For a 18\% Mn-doped GaN epilayer:
presence of oriented clusters with perovskite structure. (b) X-ray absorption: the perovskite clusters at Mn
contents larger than 2\% are identified as GaMn$_{3}$N NCs. From \onlinecite{Giraud:2004_EL}.}
    \label{fig:Giraud}
\end{figure}

Early examples of extended characterization of (Ga,Mn)N and detection of crystallographic phase separation
 are found in works reporting on samples fabricated by means of
ammonothermal technique with 0.25\% of Mn \cite{Zajac:2001_APLa}. Here,
the formation of a Mn$_{3}$N$_{2}$ was detected through XRD analysis, as reported
in Fig.\,\ref{fig:Zajac2} and diffraction measurements as in Fig.\,\ref{fig:Giraud} allowed also one
to identify GaMn$_{3}$N phases in MBE layers containing more than 2\% Mn
 \cite{Giraud:2004_EL,Yoon:2006_MSEB}.

\begin{figure}[htb]
    \centering
        \includegraphics[width=0.8\columnwidth]{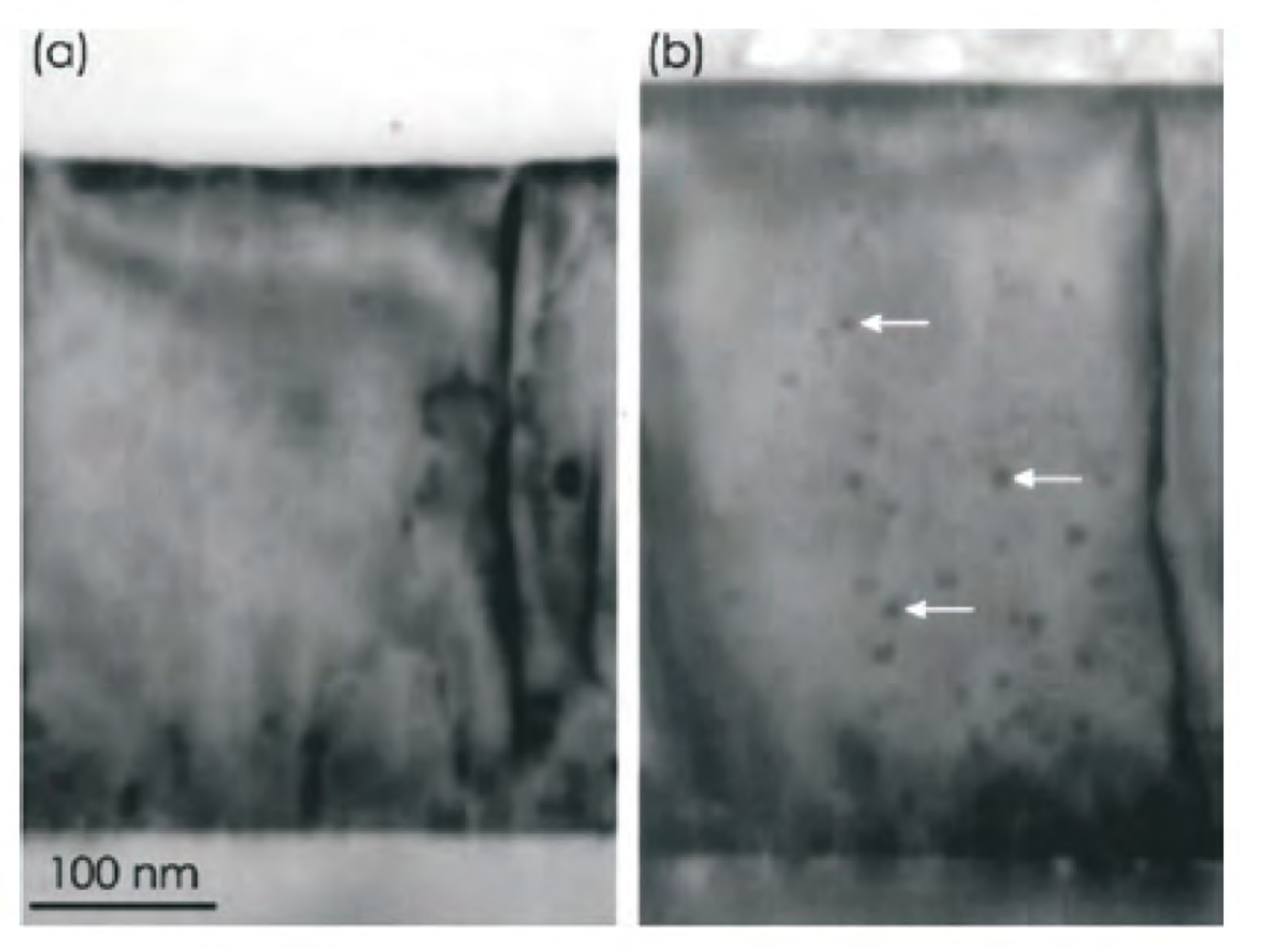}
    \caption{Bright-field TEM of (Ga,Mn)N with nominal Mn concentrations 7.6\% (a) and 13.7\%(b), respectively. The nm-scale clusters are indicated by
arrows. From \onlinecite{Dhar:2003_APL}.}
    \label{fig:Dhar1}
\end{figure}

\begin{figure}[htb]
    \centering
        \includegraphics[width=0.75\columnwidth]{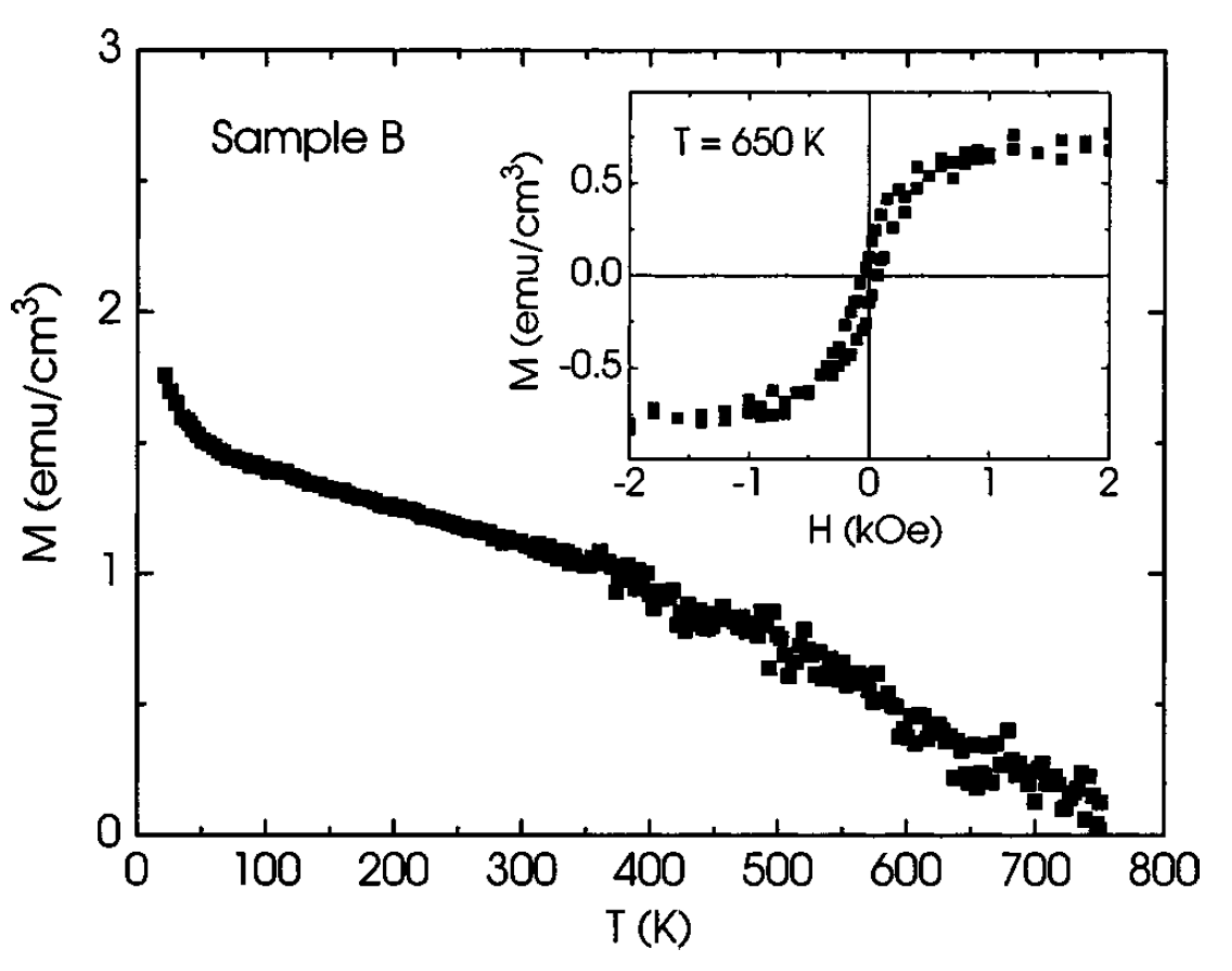}
    \caption{Magnetization at 50~Oe for a (Ga,Mn)N sample with 13.7\%  of Mn.
Inset: magnetization loop at 650~K. From \onlinecite{Dhar:2003_APL}.}
    \label{fig:Dhar2}
\end{figure}
\begin{figure}[htb]
    \centering
        \includegraphics[width=\columnwidth]{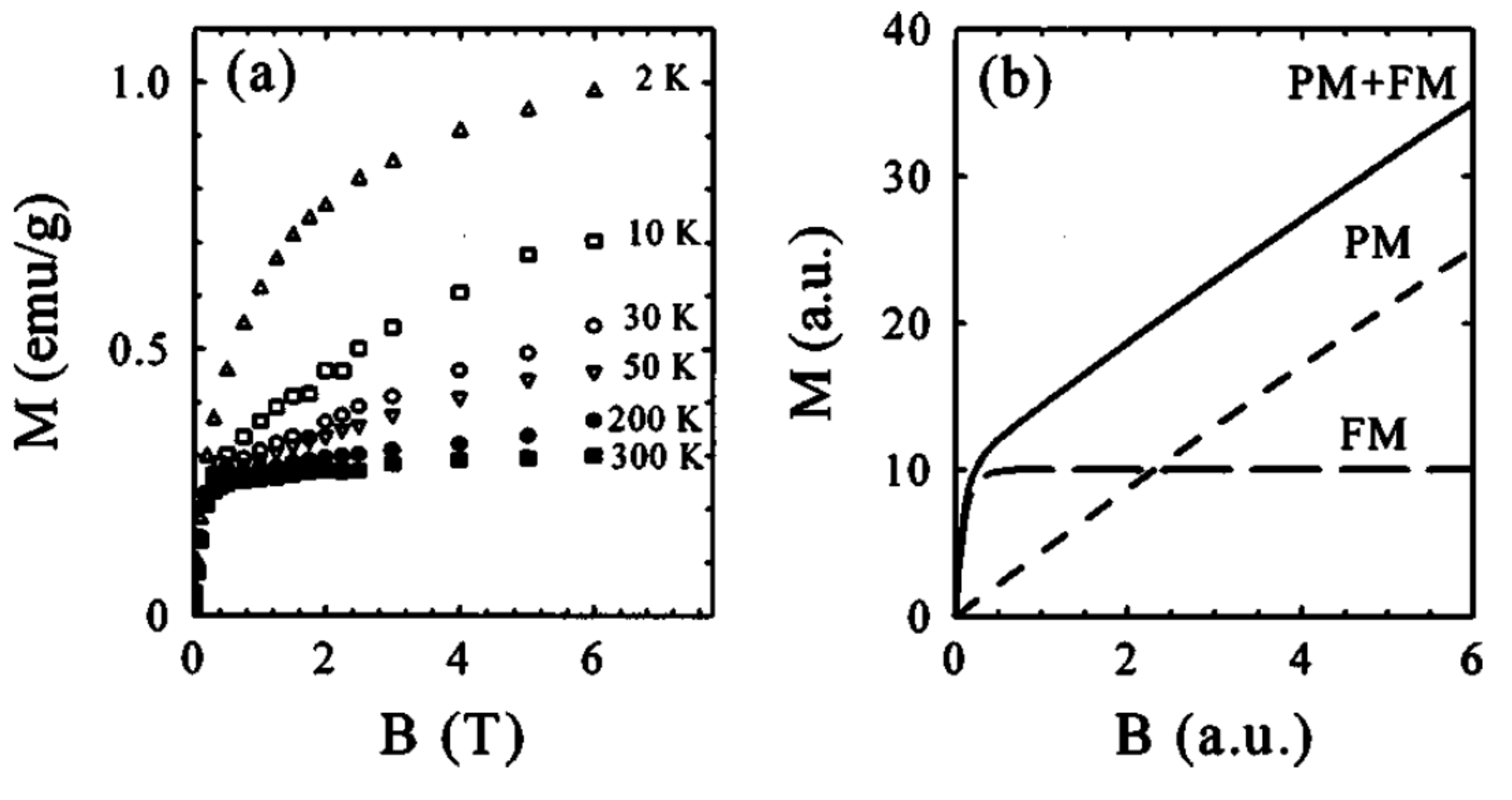}
    \caption{Evidence for various contributions to (Ga,Mn)N magnetization. (a) Magnetization as a
function of the magnetic field for several temperatures. (b) Schematic
of the total magnetization (PM + FM) decomposition into paramagnetic (PM)
and ferromagnetic (FM) contributions. From \onlinecite{Zajac:2003_JAP}.}
    \label{fig:Zajac1}
\end{figure}

Furthermore, through TEM bright-field micrographs it has been possible to recognize the presence of not better
identified secondary phases in MBE (Ga,Mn)N grown onto a GaN buffer and employing
specific growth conditions \cite{Kunert:2012_APL} or onto 4H-SiC(0001) substrates and
containing 13.7\% of Mn ions, as seen in Fig.\,\ref{fig:Dhar1}, where the comparison
between TEM micrographs from samples with respectively 7.6\% (left panel) and 13.7\%
Mn content (right panel) is given  \cite{Dhar:2003_APL}. From this work, the magnetization of
the sample with the highest Mn concentration as a function of temperature and the
magnetic field  is reported in Fig.\,\ref{fig:Dhar2}. Always worth noting is the fact
that the measured magnetization of phase-separated (Ga,Mn)N [and
generally (Ga,TM)N] consists of different components, as exemplified in
Fig.\,\ref{fig:Zajac1} and particular caution should be used to discriminating
between the various contributions, to the correct subtraction of the signal from
the substrate and to possible sources of contamination
 \cite{Bonanni:2007_PRB,Hwang:2007_JAP,Ney:2008_JMMM}.

\begin{figure}[htb]
    \centering
        \includegraphics[width=0.9\columnwidth]{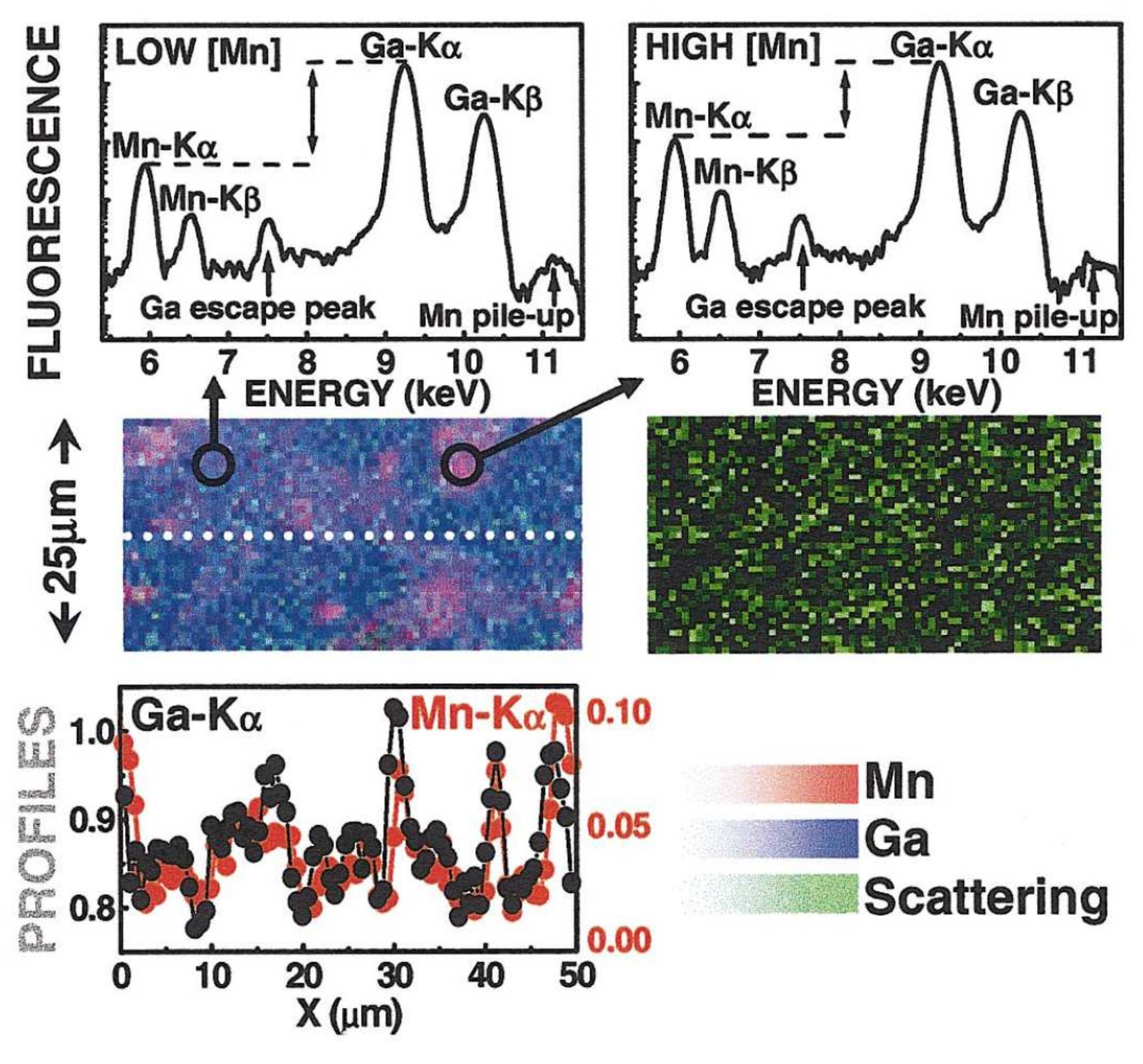}
    \caption{(Color online) Decomposition of (Ga,Mn)N with an average Mn concentration of
11\% studied by x-ray fluorescence mapping. Mn $K_{\alpha}$, Ga $K_{\alpha}$ fluorescence
line, and inelastic (Compton) scattering signal, are shown. Ga
and Mn  profiles along the dotted scan line are presented in the lower
panel indicating the formation of Mn-rich intermetallic Mn-Ga NCs. From \onlinecite{Martinez-Criado:2005_APL}.}
    \label{fig:Martinez}
\end{figure}

As mentioned, synchrotron radiation-based characterization techniques can
supply information when standard methods, such as, e.g., conventional
XRD, fail to provide the necessary sensitivity to the presence of spinodal
nanodecomposition in the investigated materials \cite{Navarro:2010_PRB}. An example
is given by the application of synchrotron radiation microprobe to the identification
and study of Mn-rich intermetallic Mn-Ga NCs in MBE (Ga,Mn)N with 11\% of Mn \cite{Martinez-Criado:2005_APL}.
A combination of fluorescence mapping with spectroscopic
techniques allows one to examine the composition of the clusters and their
crystallographic orientation, as summarized in Fig.\,\ref{fig:Martinez}.

\begin{figure}[htb]
    \centering
        \includegraphics[width=0.8\columnwidth]{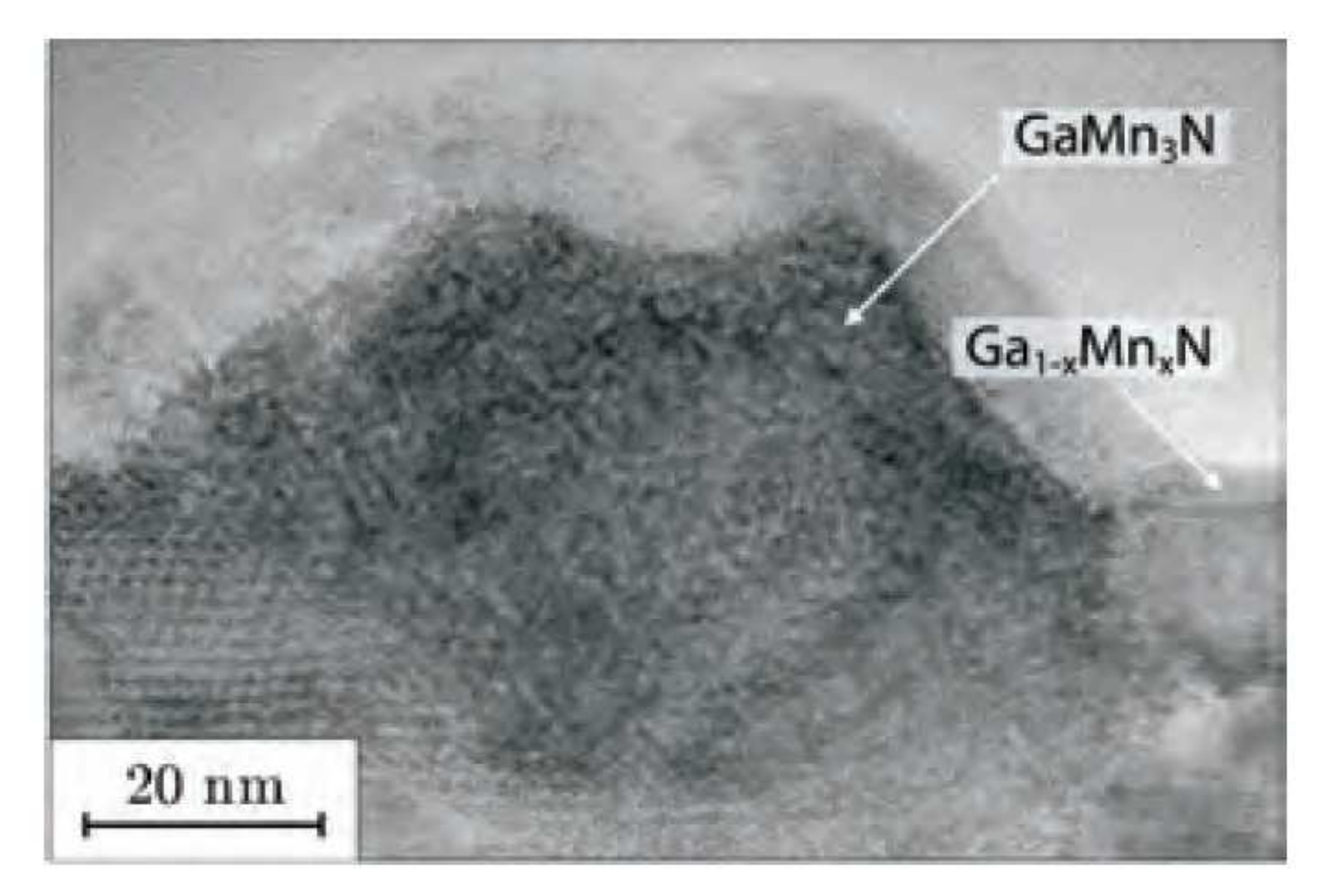}
    \caption{Cross section HRTEM of a NC in a (Ga,Mn)N layer with nominal 33\% Mn. From \onlinecite{Kocan:2006_SST}.}
    \label{fig:Kocan}
\end{figure}

It has been reported that by
employing H$_{2}$/N$_{2}$ instead of only N$_{2}$ during the plasma-assisted MBE (PAMBE) of (Ga,Mn)N
the solubility limit of Mn can be enhanced and the formation of secondary Mn-rich
phases hindered \cite{Cui:2002_APL}. Furthermore, again during a PAMBE process,
it has been shown that by switching from N-rich to Ga-rich conditions the efficiency
of Mn incorporation is diminished, while the crystalline quality is improved and
the presence of mixed domains wz-zb is suppressed \cite{Han:2007_SST}.
The amount of Mn likely to be incorporated without precipitation has been
found to depend critically in MBE on the growth temperature $T_{\mathrm{g}}$
 \cite{Kondo:2002_JCG} and above a critical value of the Mn ions supply,
GaMn$_{3}$N NCs are found to form on the samples surface \cite{Kocan:2006_SST},
as exemplified in Fig.\,\ref{fig:Kocan}.

The role of defects in the aggregation of Mn has also been taken into
account \cite{Larson:2007_PRB} and  from x-ray absorption spectroscopy
(XAS) and XMCD data complemented by $\textit{ab initio}$ calculations it has been
concluded that Mn in GaN preferentially occupies Ga sites neighboring N split
interstitial defects \cite{Keavney:2005_PRL}. The literature to date lacks a
systematic study on the possible role of threading dislocations, especially in a system
based on GaN, as preferential site for the aggregation of Mn. Energy-filtered 
TEM (EFTEM) experiments should be considered, in particular in the perspective
of identifying coherent regions (not crystallographically phase separated) with
enhanced Mn concentration.

\begin{figure}[htb]
    \centering
        \includegraphics[width=0.70\columnwidth]{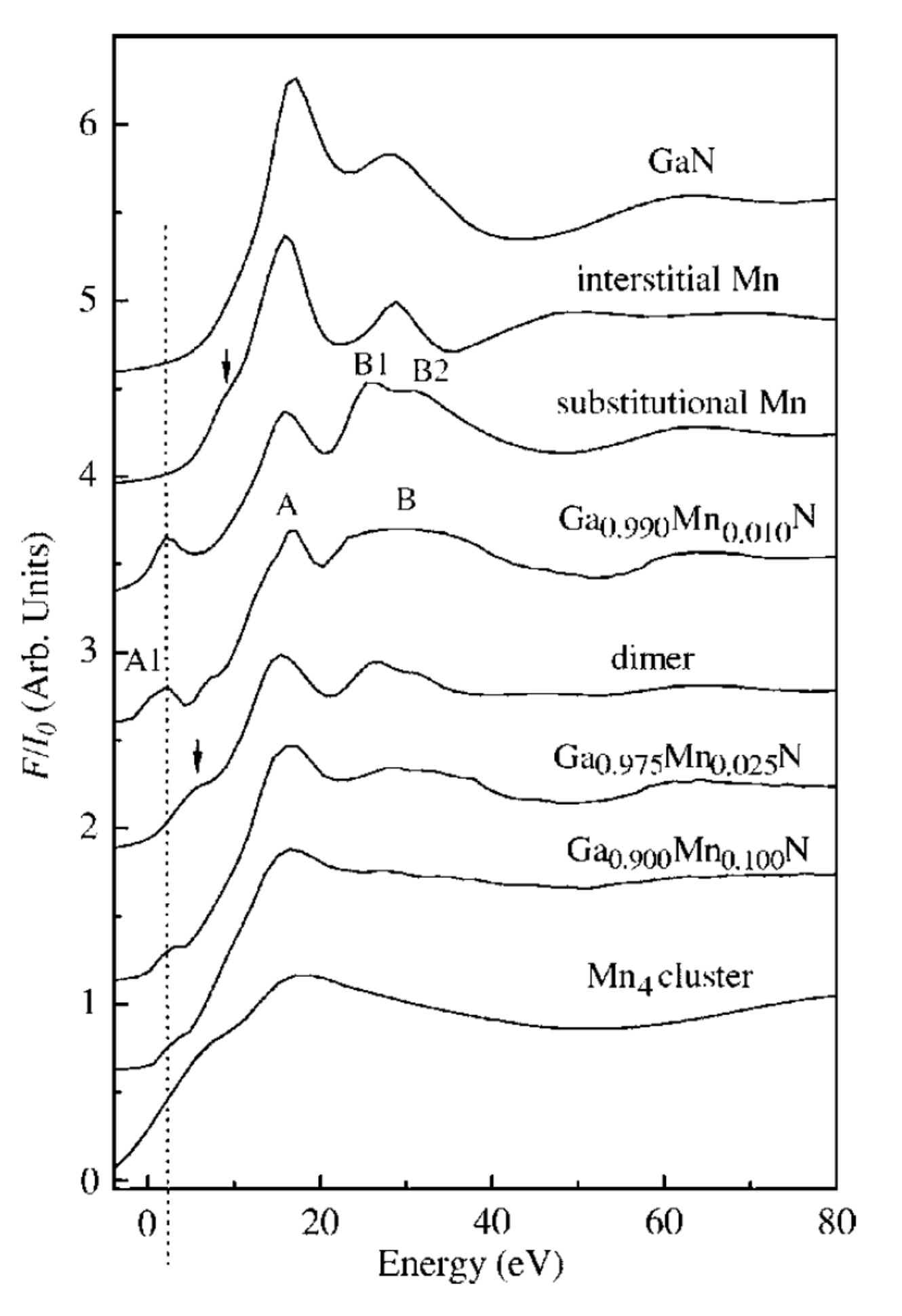}
    \caption{Comparison of experimental Mn $K$-edge XANES spectra for Ga$_{1-x}$Mn$_x$N with various Mn concentrations with computed spectra assuming various occupation sites in the GaN host crystal. From \onlinecite{Wei:2006_APL}).}
   \label{fig:Wei}
\end{figure}

The high magnetic moment of Mn and its low solubility limit into GaN, common to all TMs, 
has motivated some groups to optimize and exploit
also Mn-induced phases onto a GaN surface: e.g., magnetic
MnGa can grow with an abrupt interface and with a defined
epitaxial orientation on wz-GaN, developing
an ideal magnetic/semiconductor bilayer \cite{Lu:2006_PRL}. Furthermore,
well ordered Mn submonolayers have been deposited onto wz-GaN by evaporation
 \cite{Chinchore:2008_APL} and (Ga,Mn)N nanostructures, though with a not-investigated
structure and actual composition, and grown by
MOVPE onto GaN are reported  \cite{Gupta:2006_JCG}. A combined study by scanning
tunneling microscopy (STM) and first-principles analysis states the feasibility
of ordered Mn-induced nanostructures onto GaN(0001) at elevated temperatures,
while the growth at RT leads to disordered phases  \cite{Qi:2009_PRB}.

\subsection{Co-doping with shallow impurities and $\delta$-doping in (Ga,Mn)N}
As mentioned previously (Sec.\,\ref{sec:GaN-Mn_magnetic}), the charge state of a magnetic impurity in a semiconducting matrix,
which can be altered in GaN by codoping with electrically active impurities, is crucial
for the magnetic properties of the material. Moreover, as discussed in Sec.\,\ref{sec:pairing}, engineering of the Fermi level position
by codoping can serve to modify the lattice position (e.g., interstitial versus substitutional) and
the distribution of magnetic impurities over cation sites \cite{Dietl:2006_NM,Ye:2006_PRB}.

In addition to infrared and magnetic-resonance spectroscopy (see, \onlinecite{Graf:2003_PSSB,Wolos:2008_B,Bonanni:2011_PRB}),
x-ray absorption near-edge spectroscopy (XANES) is a widely
employed method to determine the valence state of Mn in GaN \cite{Bonanni:2011_PRB,Sato:2002_JJAP,Biquard:2003_JS,Titov:2005_PRB,Wei:2006_APL}.
As summarized in Fig.\,\ref{fig:Wei},
from the analysis of the absorption-edge prepeaks it is possible to infer the charge
state of the TM ions. Furthermore, a careful modeling of the total signal can give
information on the presence and nature of phase separation \cite{Sato:2002_JJAP,Biquard:2003_JS,Titov:2005_PRB,Wei:2006_APL}.
Another sensitive method is x-ray emission spectroscopy (XES) that was exploited to
determined the evolution of the Mn oxidation state with the Mg concentration in (Ga,Mn)N:Mg \cite{Devillers:2012_SR}.

The tuning
of the Fermi level through codoping with acceptors or donors is expected
 \cite{Dietl:2006_NM}, and in other systems already proven
 \cite{Kuroda:2007_NM,Bonanni:2008_PRL}, to affect the aggregation of TM ions
in a semiconducting host and therefore to promote the onset of spinodal nanodecomposition.
In (Ga,Mn)N the modulation of the Mn charge state has been mainly considered in
relation to its possible effect on the magnetic response. However, the reports are highly controversial: e.g.,
in some works the reduction from the neutral state Mn$^{3+}$ to Mn$^{2+}$ has been
found to decrease the FM response of (Ga,Mn)N layers \cite{Yang:2008_JPD}, while
elsewhere it has been argued that paramagnetic (Ga,Mn)N with a majority of Mn$^{3+}$
can be rendered FM via double-exchange mechanisms if Mn$^{3+}$ and Mn$^{2+}$ coexist in the same sample
 \cite{Sonoda:2007_APL}. Furthermore, codoping with acceptors (Mg) is even reported
to act not univocally, but depending on the quality of the matrix to either enhance
or quench the FM response \cite{Reed:2005_APL}. Again a hint of the crucial importance
is a characterization of the material at the nanoscale, as codoping often controls
the formation of Mn-rich NCs that can dominate the magnetic properties, especially at high temperatures.

Actually, it was demonstrated that codoping of GaN with Mn and Mg during the MOVPE growth leads
to the formation of Mn-Mg$_k$ impurity complexes, where $k$ increases up to $k = 3$ with the ratio of Mg to Mn concentration
according to the binomial distribution \cite{Devillers:2012_SR}. These complexes show appealing
optical properties, in particular a broad-band infrared emission persisting up to RT \cite{Devillers:2012_SR}.

Interestingly, MBE $p$-(Ga,Mn)N spinodally decomposed in regions of GaMn$_{3}$N is
seen to have an enhanced conductivity \cite{Kim:2003_APL_b} that could make this
phase-separated system promising for functional effects.

\begin{figure}[htb]
    \centering
        \includegraphics[width=0.80\columnwidth]{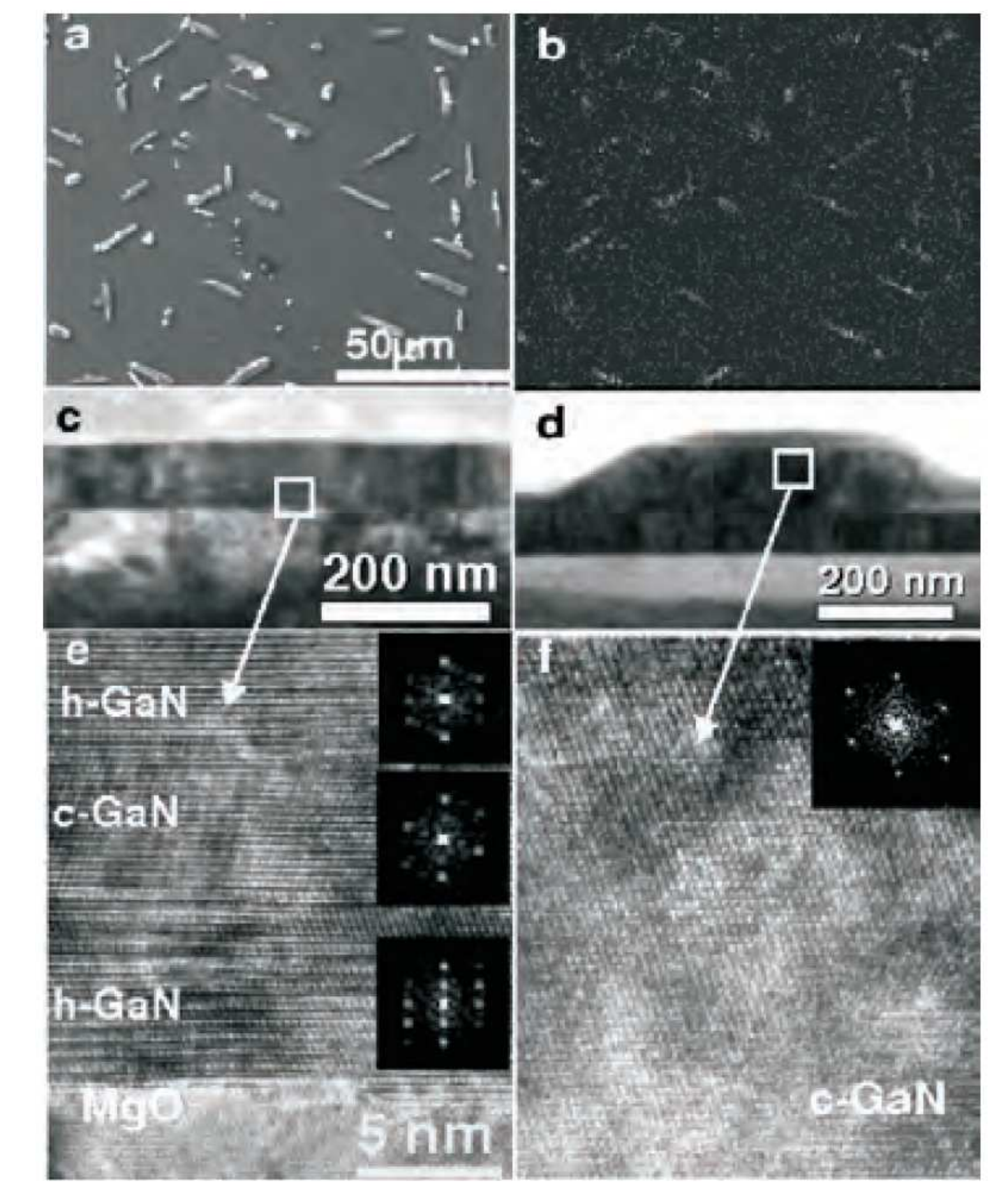}
    \caption{Multilayer Mn/GaN film on
MgO(111): top view SEM image (a) and Mn EDS map (b) showing
93\% flat film regions and 7\% Mn-rich protrusions. Bright-field TEM (c) and
HRTEM (e) images of flat film regions. Bright-field TEM (d) and HRTEM
(f) of a Mn-rich protrusion. From \onlinecite{Lazarov:2008_APL}.}
    \label{fig:Lazarov}
\end{figure}

While thermodynamically stable (Ga,Mn)N crystallizes in the wz-structure,
it has been reported that MBE layers can be locally stabilized by Mn $\delta$ doping
in the zb-phase \cite{Lazarov:2008_APL}, as evidenced in Fig.\,\ref{fig:Lazarov}.
Furthermore, the energetics of cubic and hexagonal (Ga,Mn)N arrangements has been studied
with $\textit{ab initio}$ pseudopotential calculations and indeed $\delta$ doping
has resulted likely to stabilize the zb phase \cite{Choi:2006_PRB}.

\section{\label{sec:GaN-Fe}Spinodal nanodecomposition in $\mbox{(Ga,Fe)N}$}
\subsection{Diluted (Ga,Fe)N: current status}
While extensive studies have been conducted on
(Ga,Mn)N \cite{Pearton:2003_JAP,Graf:2003_PSSB,Dietl:2004_MRS,Liu:2005_JMSME,Bonanni:2007_SST} as
promising workbench for future applications in
spintronics, until very recently,
only little was known about (Ga,Fe)N.
Formerly, and to a broad extent still now, this system is being widely considered
as semi-insulating substrate for high frequency devices, as AlGaN/GaN
high-mobility transistors \cite{Heikman:2003_JCG,Lo:2006_PRB,Muret:2007_JAP,Kashiwagi:2007_JMMM,Kubota:2009_JAP}.
In this context and generally in relation to the use of this material system in
reliable devices, careful studies of the electronic structure of (Ga,Fe)N have
been carried out and especially the knowledge of the
exact position of the Fe$^{3+/2+}$ acceptor level within the band
gap---used to predict band offsets in heterostructures
on the basis of the internal reference rule  \cite{Langer:1985_PRL}---has
been considered of great importance.  Furthermore,
the behavior of $d^{5}$ and $d^{6}$ level systems in a trigonal crystal
field of $C_{\mathrm{3V}}$ symmetry is of interest for general aspects
of group and crystal-field theory \cite{Malguth:2008_PSSB,Lo:2006_PRB}.

Furthermore, GaN doped with Fe impurities substitutional of Ga in their Fe$^{3+}$ $d^{5}$
configuration \cite{Bonanni:2007_PRB,Baur:1994_APL,Malguth:2006_PRB,Malguth:2008_PSSB}
attracted considerable attention as an ideal system for the study of
the $p$--$d$ exchange interaction in the strong coupling limit
 \cite{Pacuski:2008_PRL}. Indeed, magnetically-doped II-VI oxides and III-V nitrides, due to their
small bond length and, thus, strong $p$--$d$ hybridization, are expected to give
rise to large values of the exchange energy $N_0|\beta|$, a prediction
supported by photoemission experiments  \cite{Hwang:2005_PRB}.
Surprisingly, however, abnormally small field-induced exciton splittings in paramagnetic (Zn,Co)O
\cite{Pacuski:2006_PRB}, (Zn,Mn)O  \cite{Przezdziecka:2006_SSC,Pacuski:2011_PRB}, and (Ga,Mn)N
 \cite{Pacuski:2007_PRB,Suffczynski:2011_PRB} were reported. It was
suggested \cite{Dietl:2008_PRB} that due to the strong $p$--$d$ coupling,
the molecular and virtual crystal approximations fail in oxides and nitrides, making the apparent
exchange splitting the valence band, quantified by $N_0\beta^{\mathrm{(app)}}$, reduced in absolute value and of opposite
sign than expected, and observed in other II-VI DMSs. In this context, measurements of magnetization and
magnetoreflectivity in the free exciton region for (Ga,Fe)N
epilayers were reported \cite{Pacuski:2008_PRL,Rousset:2013_PRB}, and the obtained values
for $N_0\beta^{\mathrm{(app)}}$ supported the theoretical expectations \cite{Dietl:2008_PRB}.

Moreover, recently reported DFT
calculations supporting the notion that the spin-spin coupling in dilute (Ga,Fe)N is AF,
while it becomes ferrimagnetic when holes are introduced
into the system, have rekindled the discussion on the nature of the magnetic
interactions in TM-doped DMS  \cite{Dalpian:2009_PRB}.
Additionally, DMSs in general and (Ga,Fe)N in
particular, have become model systems to test various
implementations of the DFT to disordered
strongly correlated systems \cite{Sato:2002_SST,Mirbt:2002_JPCM,Sanyal:2003_PRB,Cui:2006_PRL}.

\subsection{Fabrication and properties of spinodally decomposed (Ga,Fe)N}
Both the above considerations and the search for high-$T_{\text{C}}$ DMSs
 \cite{Dietl:2000_S} impelled the research in various TM-doped GaN, and the
Fe-doping has been pursued through a vast palette of fabrication techniques with
variegated results. (Ga,Fe)N grown by means of radio-frequency plasma-assisted
MBE (RF-MBE) at 800$^{\circ}$C and with a Fe concentration up
to $5\times 10^{21}$ cm$^{-3}$ was shown to have superparamagnetic behavior and
HRTEM images gave hints of some inhomogeneity in the layers \cite{Kuwabara:2001_JJAP}.

The same system fabricated by ammonothermal technique \cite{Dwilinski:1998_DRM,Gosk:2003_JS} and
by a chemical transport method \cite{Gosk:2003_JS} revealed the coexistence of paramagnetic and
FM contributions (see Sec.\,\ref{sec:GaFeN_mag}), but the origin of this latter component was not
investigated.
The magnetic properties of nominally undoped and $p$-doped 
Fe-implanted GaN were reported, evidencing a FM
response characterized by magnetization hysteresis loops persisting---depending on the provided dose
\cite{Pearton:2002_JVSTA}---up to RT \cite{Abernathy:2002_JAP,Shon:2004_JAP}.

The full width at half maximum (FWHM) of the GaN(0002) XRD
rocking curves peaks acquired on
Fe implanted GaN was found to increase as a function of the Fe dose, though without
further analysis guiding to the presence of decomposition \cite{Shon:2006_APL}.
The emission channeling technique applied to Fe-implanted (Fe dose
up to $10^{19}$\,cm$^{-3}$) GaN samples confirmed the presence of a
high percentage of TM ions (up to 80\%) occupying substitutional Ga
sites of the host crystal \cite{Wahl:2001_APL}. Moreover, GaN films
doped with Fe, with concentrations up to
$~3\times 10^{19}$~cm$^{-3}$ were fabricated by
MBE at growth temperatures ($T_{\mathrm{g}}$) ranging
from 380 to $520^{\circ}$C directly on sapphire (0001) and
FM behavior with
$T_{\mathrm{C}}=100$\,K was observed only in the samples grown
at $400^{\circ}$C \cite{Akinaga:2000_APLb}.

Films of (Ga,Fe)N (Fe density
up to $6\times 10^{21}$~cm$^{-3}$) fabricated by means of MBE
at $T_{\mathrm{g}} = 500$~--~$800^{\circ}$C showed a
superparamagnetic behavior \cite{Kuwabara:2001_PE,Kuwabara:2001_JJAP}
assigned to Ga-Fe and/or Fe-N inclusions. EXAFS analysis suggests that the decrease of
$T_{\mathrm{C}}$ is caused by a structural transition from wz
to zb, and this transition was related to the origin of
FM in Fe-doped GaN films \cite{Ofuchi:2001_APL}.
The MOVPE of (Ga,Fe)N was reported, with a
focus on the actual Fe content in the layers and its effect onto the
carrier concentration \cite{Heikman:2003_JCG}.
MOVPE samples with a Fe-concentration up to 0.7\% molar were claimed to
give FM features up to RT even in the absence of holes, but without providing
information on the actual distribution of the magnetic ions in the GaN
matrix \cite{Kane:2007_PSSA,Gupta:2008_JCG}.

Parallel works \cite{Przybylinska:2006_MSEB,Bonanni:2006_PSSB} dealt with
the extensive study and characterization of this material system with the precise
objective of clarifying the correlation between the general properties---and
in particular the magnetic ones---of the system and the distribution of the magnetic
ions in the semiconducting matrix.

\subsection{Fe distribution from nanoscale characterization}
Especially the need to gain insight into the origin of the puzzling FM
signatures detected in GaN upon doping with Fe and persisting up to above RT,
has lately prompted a considerable experimental
effort in the characterization of the system at the nanoscale. As discussed in this section,
(Ga,Fe)N represents an interesting case, in which, for specific growth conditions,  NCs formed by
spinodal nanodecomposition gather in planes perpendicular to the growth direction.

\begin{figure}[htb]
    \centering
        \includegraphics[width=0.7\columnwidth]{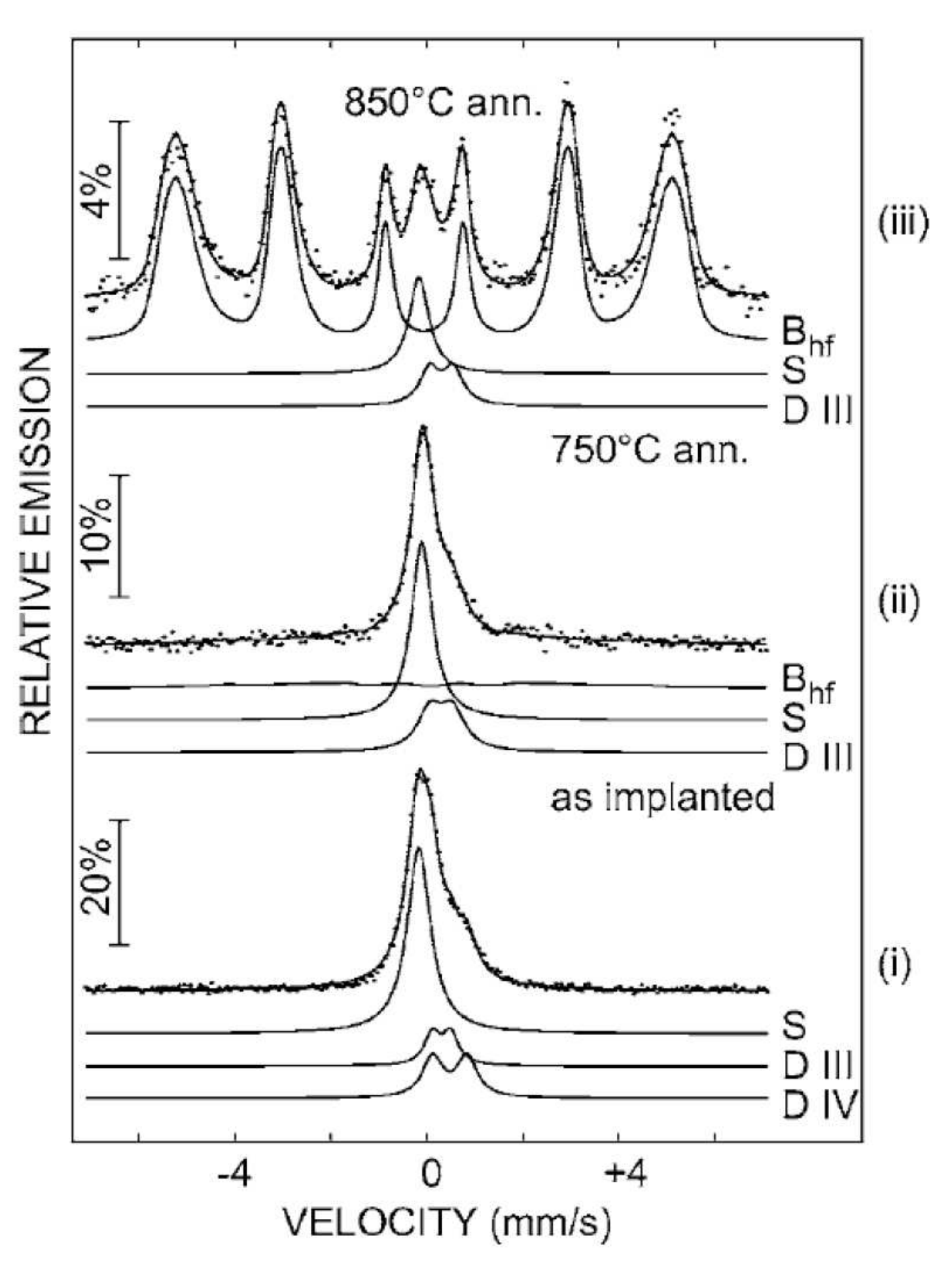}
    \caption{Room temperature conversion electron M{\"o}ssbauer
spectroscopy spectra of a $4\times 10^{16}$ cm$^{-2}$ Fe-implanted GaN sample: (i) as-implanted; (ii) annealed at 750$^{\circ}$C; (iii) annealed at 850$^{\circ}$C. From \onlinecite{Talut:2006_APL}.}
    \label{fig:Talut_06}
\end{figure}

Early steps were undertaken in the understanding of, e.g., the behavior of
Fe-implanted GaN, where a combination of {M{\"o}ssbauer spectroscopy (Fig.\,\ref{fig:Talut_06})
and HRTEM (Fig.\,\ref{fig:Talut_nanocr06}) revealed the precipitation of Fe during implantation \cite{Talut:2006_APL}.
There, $^{57}$Fe was employed as an atomic sensitive probe to investigate
the local environment of Fe by M{\"o}ssbauer spectroscopy and to study
the precipitation at very early stage. MOVPE $p$-type
Mg doped~$10^{17}$~cm$^{-3}$ wz-GaN(0001) films about 3~$\mu$m thick grown
onto sapphire (0001) substrates were
analyzed. The samples were implanted with 200~keV $^{57}$Fe ions with
fluences between $1\times 10^{16}$  and $1.6\times 10^{17}$ cm$^{-2}$
for a maximum Fe content of 1-18~at\%.

Conversion electron M{\"o}ssbauer
spectroscopy (CEMS) in constant-acceleration mode at RT was applied to investigate
the Fe lattice sites, electronic configuration, and magnetism. In samples implanted
with fluences above $8\times 10^{16}$ cm$^{-2}$, the formation of clusters was
observed already in the as-implanted state \cite{Talut:2006_APL}. As shown in Fig.\,\ref{fig:Talut_06},
the CEMS spectra of the as-implanted
state were found to consist of a singlet line and two quadrupole doublets. The singlet
could be assigned either to superparamagnetic $\alpha$-Fe or to paramagnetic
$\gamma$-Fe with an isomer shift of -0.05~mm/s. The determined values were assigned
to Fe$^{3+}$ with three $D_{\mathrm{III}}$ and four $D_{\mathrm{IV}}$ nearest N neighbors
with isomer shifts of 0.42 and 0.59~mm/s, respectively. The
magnetically isolated Fe$^{+3}$ ions were confirmed to be paramagnetic with no
FM hyperfine splitting observable.

\begin{figure}
    \includegraphics[width=0.7\columnwidth]{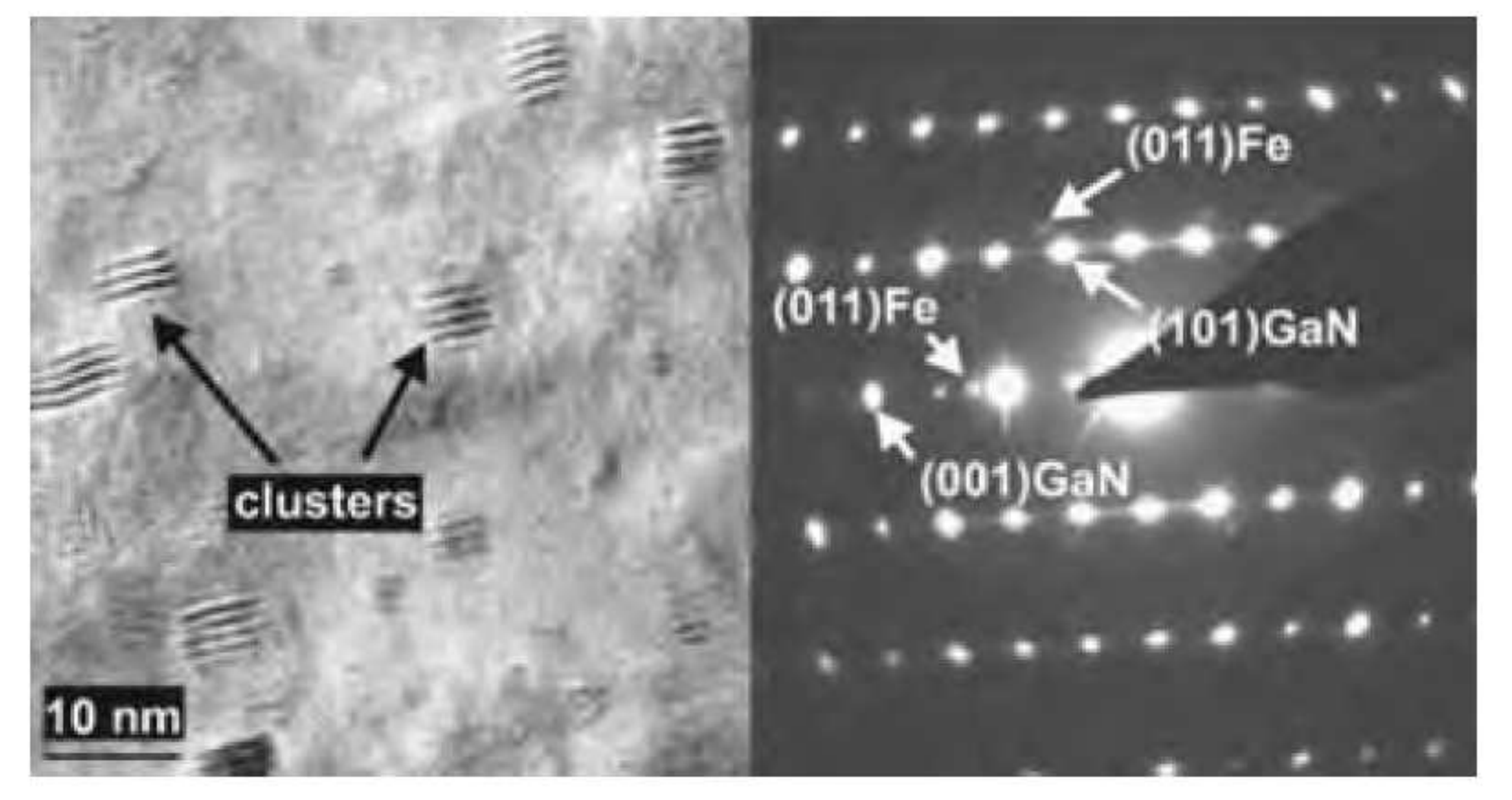}
    \caption{GaN sample implanted with $4\times 10^{16}$ cm$^{-2}$ Fe ions and annealed at 800$^{\circ}$C: left panel - HRTEM: Fe clusters in the implanted region of the sample; right panel - selected area diffraction pattern of the same region. From \onlinecite{Talut:2006_APL}.}
    \label{fig:Talut_nanocr06}
\end{figure}

The presence
of the Fe$^{+3}$ state has been taken as direct proof of its being substitutional of Ga
sites in the wz structure, where Ga is surrounded by four
N atoms and vice versa. However, only 23.6\% of the total
amount of Fe in the as-implanted samples appeared to substitute the Ga
sites. The remanent magnetic ions were regarded to be either on interstitial
sites or generating small precipitates of metallic $\gamma$- or $\alpha$-Fe. Annealing at
850$^{\circ}$C was found to trigger the formation of FM
$\alpha$-Fe, represented by a sextet with a mean magnetic hyperfine
field of 318~kOe reported in Fig.\,\ref{fig:Talut_06}.

\begin{figure}
    \includegraphics[width=0.8\columnwidth]{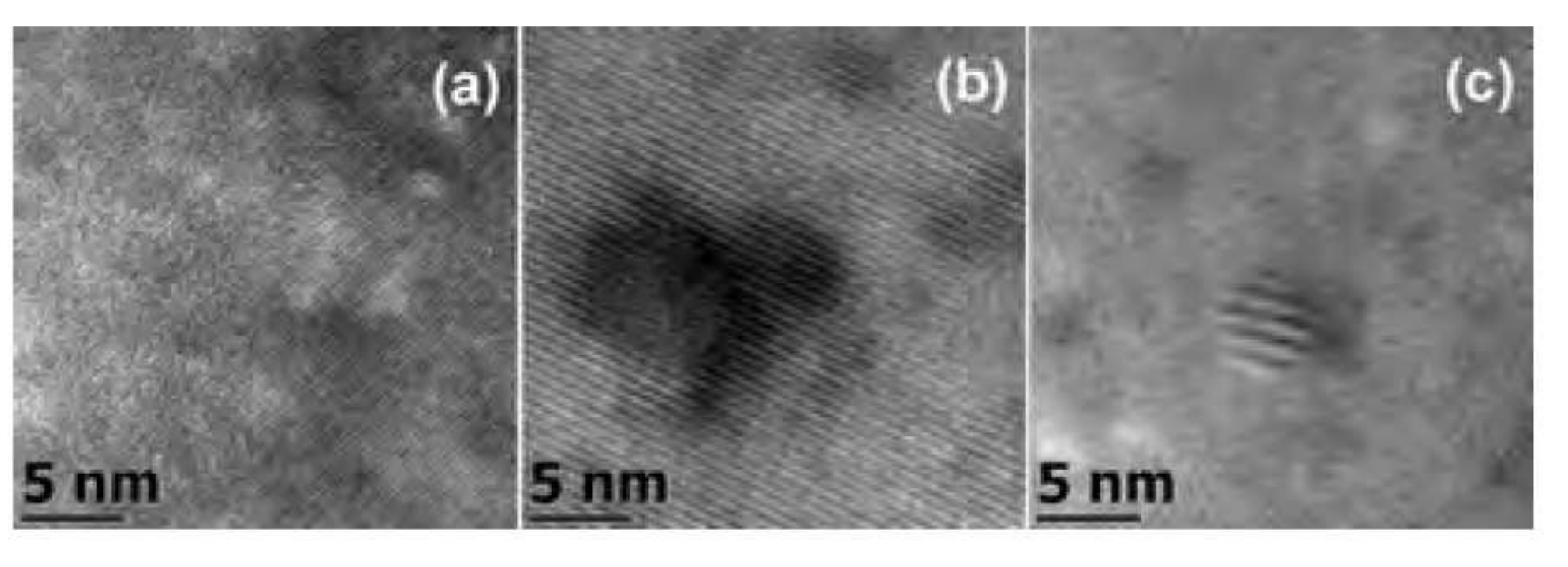}
    \caption{HRTEM of (Ga,Fe)N showing (a) dilute case; (b) coherent spinodal decomposition (chemical phase separation), and (c)   crystallographic phase separation. From \onlinecite{Bonanni:2007_PRB}.}
    \label{fig:TEM_Li}
\end{figure}

A series of works with the first systematic study of the (Ga,Fe)N system
fabricated by means of MOVPE from the impurity limit until the phase separation,
was carried out \cite{Przybylinska:2006_MSEB,Bonanni:2007_PRB,Pacuski:2008_PRL,Bonanni:2008_PRL,Rovezzi:2009_PRB,Kowalik:2012_PRB}.
It was shown that by controlling the growth parameters it is possible
to incorporate the magnetic ions in different fashions, giving rise to:
(i) a DMS with Fe substitutional of Ga, in the charge state Fe$^{+3}$ and
responsible for the paramagnetic response of the samples; (ii) a system with spinodal
decomposition (chemical phase separation) in Fe-rich regions structurally coherent with the matrix;
and (iii) a crystallographic phase-separated material consisting of Fe$_{x}$N NCs embedded
in the GaN host and likely to account for the FM signatures, as discussed in Sec.\,\ref{sec:GaFeN_mag}.

The different arrangements mentioned are evidenced by the HRTEM images
in Fig.\,\ref{fig:TEM_Li}, where (a) dilute, (b) chemically separated, and
(c) crystallographically separated (Ga,Fe)N are, respectively, reported.
Energy dispersive x-ray spectroscopy (EDS) measurements could confirm the enhanced
concentration of Fe both in the (b) coherent regions  and in the (c) incoherent
precipitates \cite{Bonanni:2007_PRB}.

\begin{figure}
    \includegraphics[width=0.8\columnwidth]{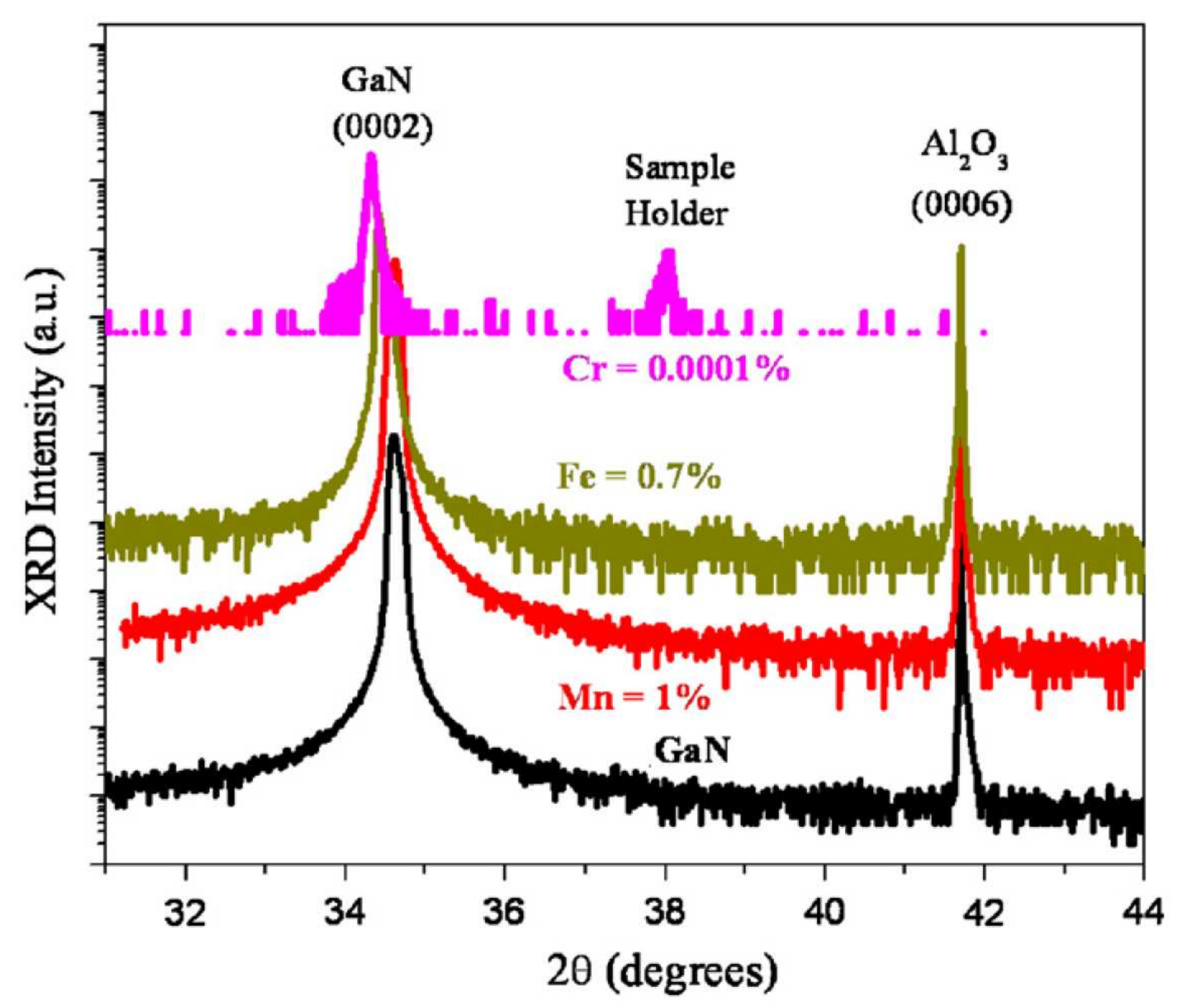}
    \caption{(Color online) HRXRD for different MOVPE grown (Ga,TM)N. From \onlinecite{Gupta:2008_JCG}.}
    \label{fig:Gupta_XRD}
\end{figure}

While often laboratory XRD does not evidence
any phase separation in (Ga,Fe)N samples giving a FM response---an example is
given in Fig.\,\ref{fig:Gupta_XRD}, where MOVPE (Ga,Fe)N with 0.7\% Fe ions does not
show diffraction peaks from secondary phases---SXRD measurements on MOVPE (Ga,Fe)N with more than 0.4\% Fe, reveal the presence
of diffraction peaks identified as the (002) and (111) of the phase
$\varepsilon$-Fe$_3$N---as reported in Fig.\,\ref{fig:SXRD_GaFeN}---a compound known
to be FM with $T_{\mathrm{C}}=575$\,K.

The solubility limit in (Ga,Fe)N was found to be around 0.4\% of
the Fe ions under optimized MOVPE growth conditions \cite{Bonanni:2007_PRB}. As
discussed in Sec.\,\ref{sec:pairing}, an order of magnitude lower solubility limit
of Fe comparing to Mn (see Sec.\,\ref{sec:GaN-Mn}) under similar
growth conditions is to be linked to a different sign of the chemical forces
between TM adatoms on the (0001)GaN surface during the epitaxy: according to {\em ab initio}
studies the pairing interaction that is repulsive for surface cation Mn dimers becomes
attractive in the case of Fe ions \cite{Gonzalez:2011_PRB}. Actually, a reduced magnitude of FM response
with increasing the growth rate \cite{Bonanni:2008_PRL} and lowering growth temperature
\cite{Navarro:2010_PRB} constitutes an experimental hint about the importance of aggregation at
the growth surface in the formation of Fe-rich NCs in (Ga,Fe)N.

\begin{figure}
    \includegraphics[width=0.9\columnwidth]{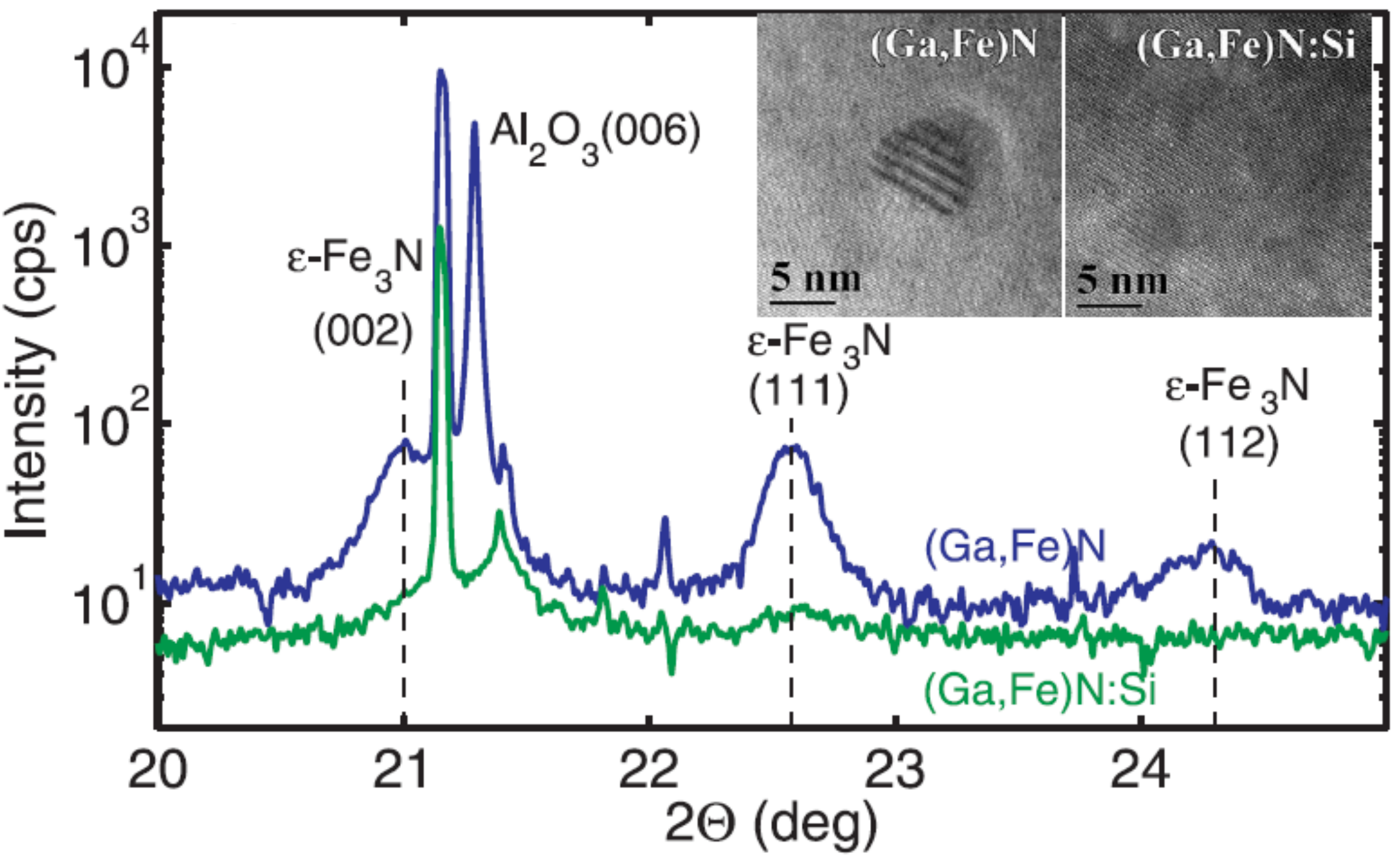}
    \caption{(Color online) SXRD for MOVPE grown (Ga,Fe)N with Fe density above the solubility limit. Secondary phases identified as $\varepsilon$-Fe$_3$N are evidenced in accord with HRTEM data (inset). The effect of codoping by Si (see, Sec.\,\ref{sec:codoping})
    is also shown.  Adapted from \onlinecite{Bonanni:2008_PRL}.}
    \label{fig:SXRD_GaFeN}
\end{figure}

An interesting aspect of spinodal nanodecomposition in (Ga,Fe)N was discovered by HRTEM \cite{Navarro:2011_PRB} and confirmed
by x-ray photoemission electron microscopy (XPEEM) and XAS \cite{Kowalik:2012_PRB}.
It was found that Fe-rich NCs tend to accumulate in a plane adjacent to the film surface.
This observation allows one to develop a method of controlling  the NCs position \cite{Navarro:2012_APL}.
As shown in Fig.\,\ref{fig:TEM_plane}, it was demonstrated that for an employed growth mode the NCs gathered in a plane
at which the Cp$_2$2Fe source flow was interrupted, i.e., at the interface between the (Ga,Fe)N layer and the GaN cap.  This
means that Fe-rich regions move together with the growth front as long as Fe is supplied, reemphasizing the notion
that the aggregation of TM ions occurs at the growth surface.

The command
over the location of the NC array is a major step toward applications of decomposed systems. Furthermore, a highly nonrandom
distribution of NCs over the film volume indicates that the visualization of spinodal nanodecomposition can be quite challenging, as
NCs may reside outside the probed region.

\begin{figure}
    \includegraphics[width=0.9\columnwidth]{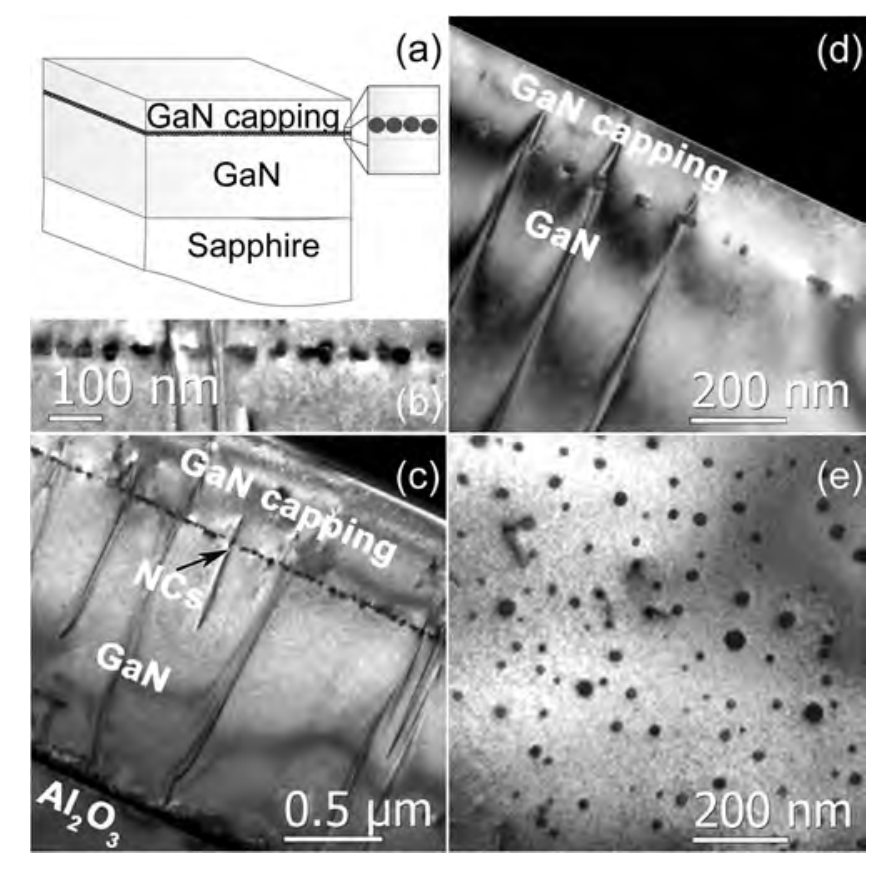}
    \caption{Control over the spatial distribution of Fe-rich
    nanocrystals (NCs) in (Ga,Fe)N. (a) Schematic layout of the structure; (b) cross-sectional TEM: magnification
of the region containing the array of NCs for the sample reported in (c);
(c) and (d) cross section TEM images of the samples, showing the spatial
distribution of the NCs into a planar array perpendicular to the
growth direction and located at the interface to the cap layer, 500\,nm and 150\,nm below the sample surface,
respectively; (e) plane-view TEM image of the sample in (c), giving the in-plane
distribution of NCs. From \onlinecite{Navarro:2012_APL}).}
    \label{fig:TEM_plane}
\end{figure}

\subsection{Phase diagram of the spinodal decomposition in (Ga,Fe)N}
\label{sec:phase_diagram}
It is known that an increase of $T_{\mathrm{g}}$ promotes the aggregation of
the TM ions incorporated in the semiconductor host and therefore brings the system
far from the dilute state. Moreover, various Fe$_{x}$N phases with specific magnetic
and structural properties are expected to be stable up to different temperatures.
The MOVPE (Ga,Fe)N material system was studied as a function of
$T_{\mathrm{g}}$ in the range between 800 and 950$^{\circ}$C and for samples
with a total concentration of Fe ions in the range $1-4\times 10^{20}$ cm$^{-3}$
 \cite{Navarro:2010_PRB}. In that work,
SXRD, EXAFS and XANES, combined with HRTEM and SQUID magnetometry  permitted
one to detect and to identify particular Fe$_{x}$N phases in samples fabricated
at different $T_{\mathrm{g}}$, as reported in Fig.\,\ref{fig:phasediagram},
and to establish  establish a correlation between
the existence of the specific phases and the magnetic response of the system.
It was found that already a 5\% variation in the growth temperature is
critical for the aggregations of new Fe$_{x}$N species and it could be confirmed
that an increase in the growth temperature promotes the onset of spinodal
decomposition, resulting in an enhanced density of Fe-rich NCs
in the matrix and in a consequent increase of the FM response of the system.
\begin{figure}
    \centering
        \includegraphics[width=0.8\columnwidth]{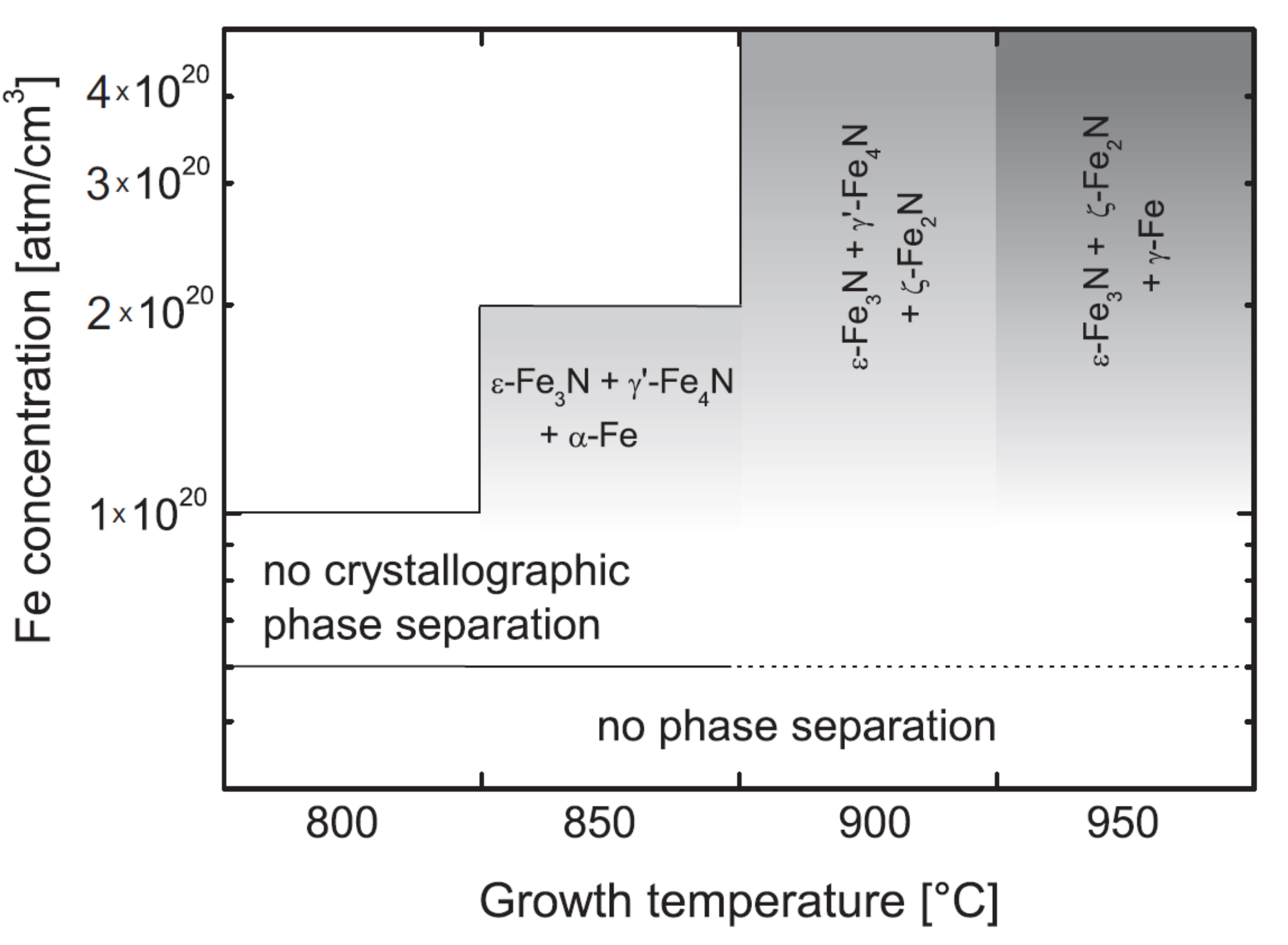}
    \caption{A phase diagram of MOVPE (Ga,Fe)N as a function of the growth temperature.
    From \onlinecite{Navarro:2010_PRB}.}
    \label{fig:phasediagram}
\end{figure}

In Fig.\,\ref{fig:SXRD_T} SXRD spectra for (Ga,Fe)N samples grown
at different temperatures, are reported. For layers fabricated at 800$^{\circ}$C
there is no evidence of secondary phases and only diffraction peaks originating
from the sapphire substrate and from the GaN matrix are revealed, in agreement
with HRTEM measurements showing no crystallographic phase separation.
Moving to a $T_{\mathrm{g}}$ of $850^{\circ}$C different diffraction peaks belonging
to secondary phases become evident, giving proof that at this $T_{\mathrm{g}}$ and
at the given growth parameters the system undergoes spinodal decomposition and
is phase separated. It was reported \cite{Bonanni:2007_PRB} that when growing
(Ga,Fe)N at this temperature, one dominant Fe-rich phase is formed, namely
wz $\varepsilon$-Fe$_3$N, for which two main diffraction peaks are identified in SXRD,
corresponding to the (002) and the (111) reflexes, respectively. A closer analysis
of the (111)-related feature and a fit with two Gaussian curves centered at
$35.2^{\circ}$ and $35.4^{\circ}$, gives evidence of the presence of the (110)
reflex from cubic metallic $\alpha$-Fe. Moreover, the broad feature appearing
around 38$^{\circ}$ is associated with the (200) reflex of fcc
 $\gamma$'-Fe$_4$N that crystallizes in an inverse perovskite
structure \cite{Jack:1952_AC}.

\begin{figure}
    \includegraphics[width=0.8\columnwidth]{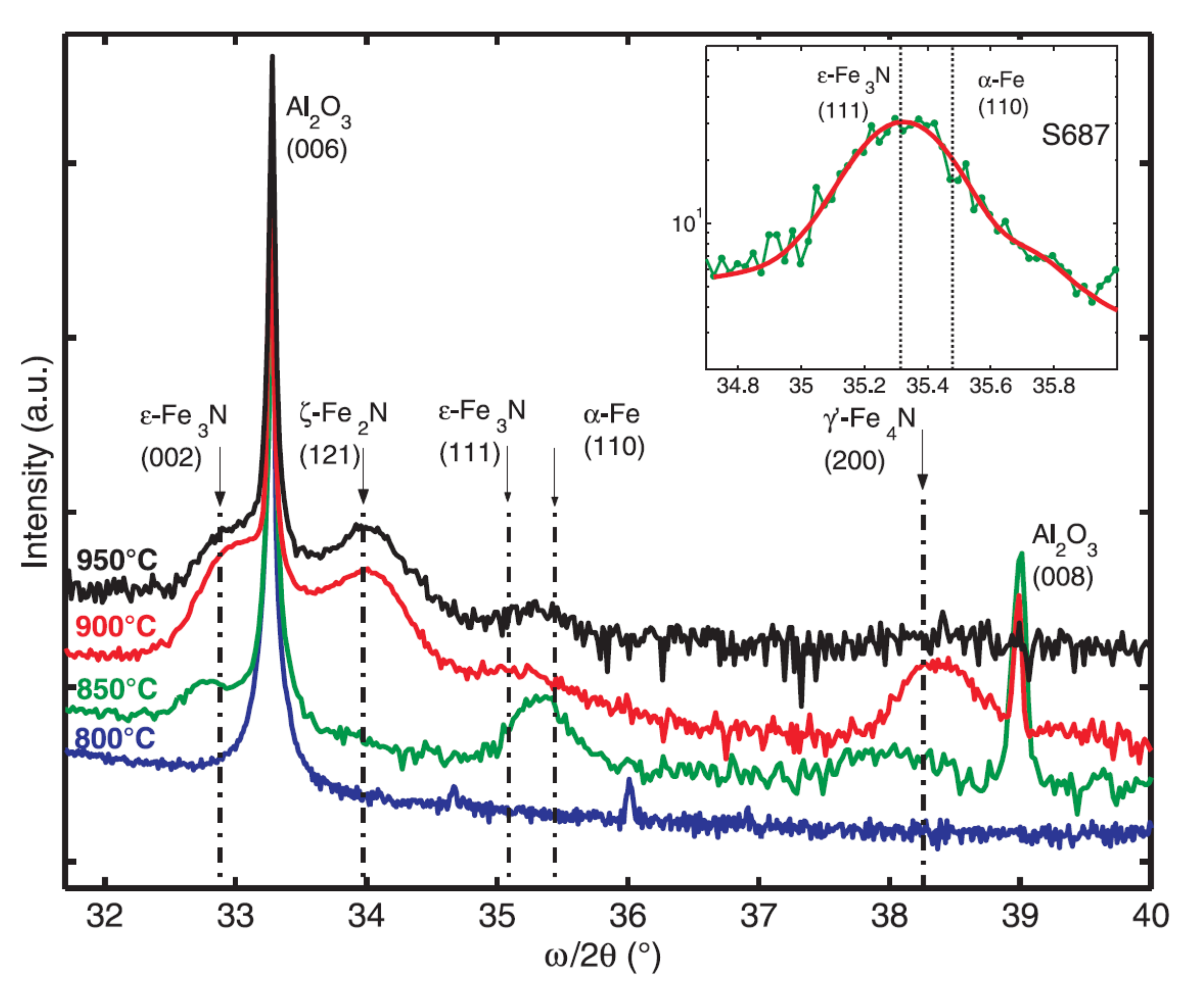}
    \caption{Color online) SXRD spectra for (Ga,Fe)N layers deposited at different growth temperatures. Inset: peak at 35.3$^{\circ}$ deconvoluted into two components assigned to diffraction maxima (111) of $\varepsilon$-Fe$_3$N  and (110) of $\alpha$-Fe [experiment (dotted line) and fit (smooth line)]. From \onlinecite{Navarro:2010_PRB}.}
    \label{fig:SXRD_T}
\end{figure}

As the growth temperature is increased to $900^{\circ}$C there is no contribution
left from the (110) $\alpha$-Fe phase, and the signal from the (111) of
$\varepsilon$-Fe$_3$N is significantly quenched, indicating the reduction in either
size or density of the specific phase. Furthermore, an intense peak is seen at
34$^{\circ}$, corresponding to the (121) contribution from orthorhombic $\zeta$-Fe$_2$N.
This phase crystallizes in the $\alpha$-PbO$_2$-like structure, where the Fe atoms
show a slightly distorted hexagonal close packing (hcp), also found
for $\varepsilon$-Fe$_3$N \cite{Jacobs:1995_JAC}.
At a growth temperature of $950^{\circ}$C the diffraction peak of (200)
$\gamma$'-Fe$_4$N recedes, indicating the decomposition of this fcc
phase at temperatures above $900^{\circ}$C, in agreement with the phase diagram
for freestanding Fe$_x$N \cite{Jacobs:1995_JAC}, reporting cubic $\gamma$'-Fe$_4$N
as stable at low temperatures. Only the (002) $\varepsilon$-Fe$_3$N- and the (121)
$\zeta$-Fe$_2$N-related diffraction peaks are preserved with a constant intensity
and position with increasing temperature, suggesting that at high $T_{\mathrm{g}}$
these two phases and their corresponding orientations are noticeably stable.

Following a procedure based on the Williamson-Hall formula method
 \cite{Williamson:1953_AM,Lechner:2009_APL}, the approximate average NCs
size is obtained from the FWHM of the diffraction peaks in the radial
($\omega/2\theta$) scans. The FWHM of the (002) $\varepsilon$-Fe$_3$N, of the (200)
$\gamma$'-Fe$_4$N, and of the (121) $\zeta$-Fe$_2$N diffraction peaks are comparable
for samples grown at different temperatures, indicating that the average size of
the corresponding NCs is also constant, as summarized in
Fig.\,\ref{fig:nanosize}.

\begin{figure}[htb]
    \centering
        \includegraphics[width=0.9\columnwidth]{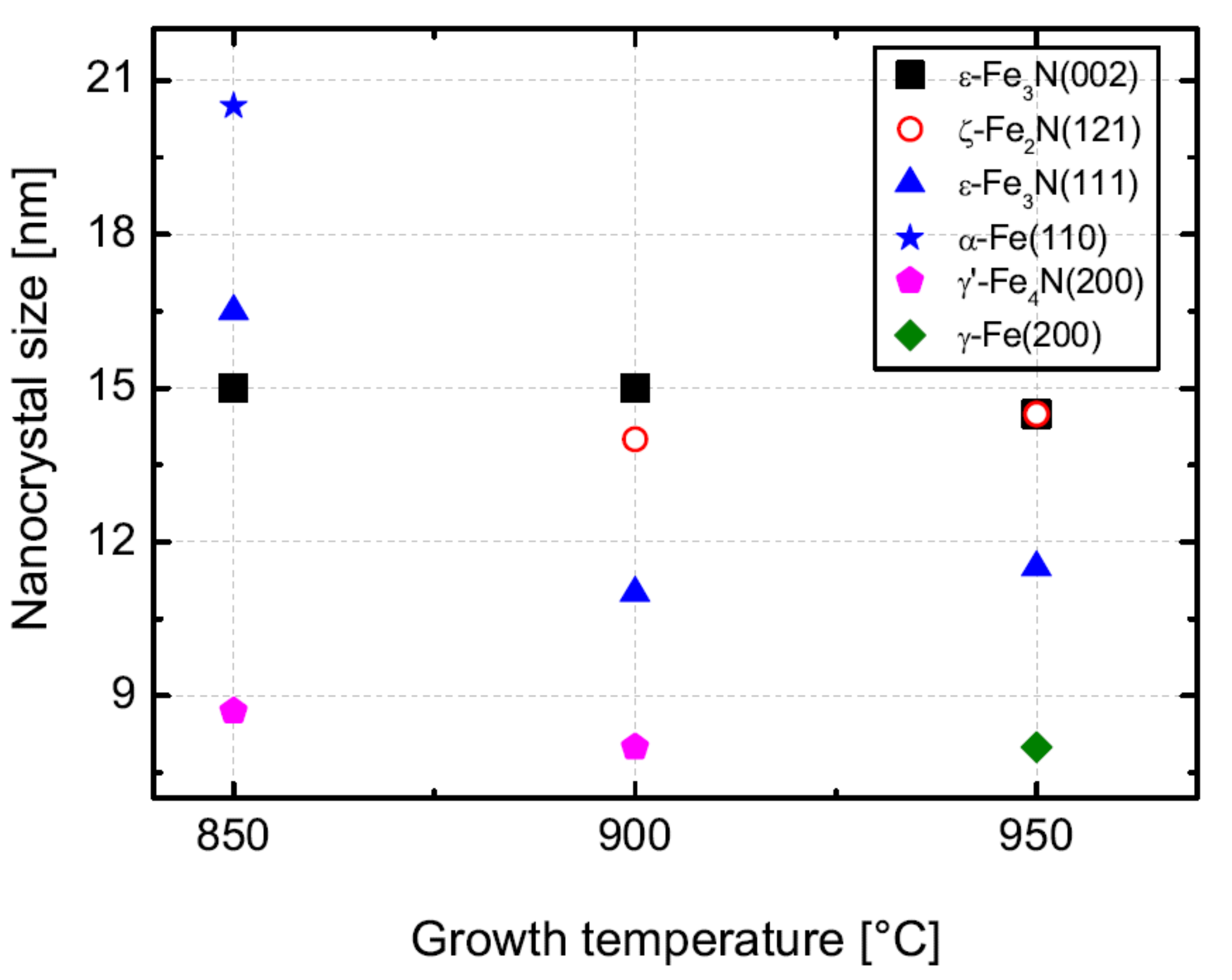}
    \caption{(Color online) Average size versus growth temperature $T_{\mathrm{g}}$ of nanocrystals in the different Fe$_x$N phases, as determined from SXRD. From \onlinecite{Navarro:2010_PRB}.}
    \label{fig:nanosize}
\end{figure}

By summing up the SXRD, EXAFS, and HRTEM findings, a phase diagram of the Fe-rich
phases formed in (Ga,Fe)N as a function of the growth temperature has been
constructed and reported in Fig.\,\ref{fig:phasediagram}, showing the dominant
phases for each temperature interval.
Moreover, according to the Fe versus N phase diagram the orthorhombic
phase ($\zeta$-Fe$_2$N) contains a higher percentage of nitrogen \cite{Jack:1952_AC}
compared to the hexagonal one ($\varepsilon$-Fe$_3$N), and this suggests
that the higher the growth temperature, the more nitrogen is introduced into
the system. Furthermore, similar to the case of (Ga,Mn)As (\ref{sec:GaAs}) and
(Ga,Mn)N (\ref{sec:GaN-Mn}), NCs in (Ga,Fe)N can be built from compounds containing
Ga. The presence of planar arrays of Ga$_x$Fe$_{4-x}$N NCs was evidenced in a recent
work \cite{Navarro:2012_APL}.

\subsection{Co-doping with shallow impurities in (Ga,Fe)N}
\label{sec:codoping}
Remarkably, HRTEM, SXRD, and SQUID data reveal that
the aggregation of Fe ions in a GaN host, and therefore the onset of spinodal
decomposition, can be affected by codoping with shallow impurities, Si donors and Mg acceptors
 \cite{Bonanni:2008_PRL,Navarro:2011_PRB}.

In Fig.\,\ref{fig:TEM_PRL} HRTEM images data for the two
relevant spinodally decomposed initial regimes, namely (i) (Ga,Fe)N with embedded
Fe-rich NCs [Fig.\,\ref{fig:TEM_PRL}(a)] evidenced by moir\'e fringes
contrast and (ii) (Ga,Fe)N showing coherent chemical separation
[Fig.\,\ref{fig:TEM_PRL}(c)] generating mass contrast show the reduced aggregation
of the Fe-rich regions as a consequence of codoping with Si donors
[Fig.\,\ref{fig:TEM_PRL}(b),(d)].

\begin{figure}
\includegraphics[width=0.7\columnwidth,clip]{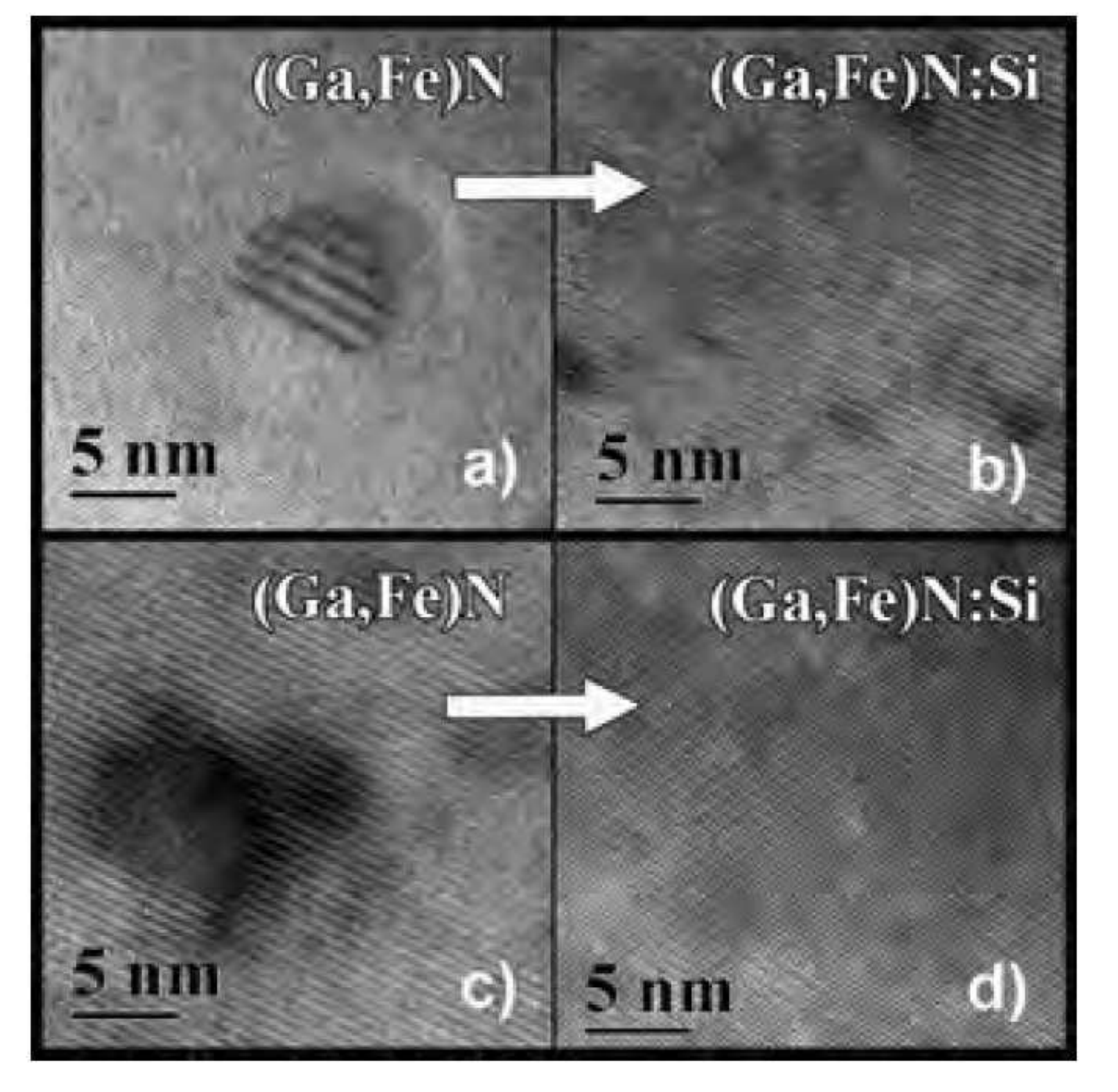}
\caption{HRTEM with mass contrast (b,c) of (Ga,Fe)N
revealing the presence of Fe$_3$N precipitates (a), spinodal decomposition
(c), and the effect of codoping by Si (b,d) preventing the formation of
the Fe-rich regions. Adapted from  \onlinecite{Bonanni:2008_PRL}.}
\label{fig:TEM_PRL}
\end{figure}

This effect is further confirmed by the SXRD results given already in
Fig.\,\ref{fig:SXRD_GaFeN}, where the introduction of shallow donor impurities is
found to efficiently hamper the precipitate aggregation, so that the SXRD
diffraction peaks corresponding to $\varepsilon$-Fe$_3$N are suppressed
in the case of the codoped samples \cite{Bonanni:2008_PRL}.
The quenching of the FM contribution in the codoped layers is further
validated by the reduced number of average Fe ions
involved in the FM response in codoped (Ga,Fe)N samples.


In order to explain these key findings, it was noted, as discussed theoretically in Sec.\,\ref{sec:pairing}, that except for
Mn in II-VI compounds \cite{Kuroda:2007_NM}, owing to the presence of the open $d$
shells in the vicinity of the Fermi level, the nearest-neighbor pair of TM cations in
semiconductors shows a large binding energy that promotes the magnetic ions
aggregation.
However, as also discussed in Sec.\,\ref{sec:pairing} from a theoretical perspective,
if carriers introduced by codoping can be trapped by these ions,
the pair binding energy will be altered, usually reduced by the corresponding Coulomb
repulsion between charged TM impurities \cite{Dietl:2006_NM,Ye:2006_PRB}.

The effect of shallow Si donors on the Fe aggregation and consequently
on the onset of spinodal decomposition in (Ga,Fe)N is quite clear, since
the presence of the midgap electron trap, i.e., the Fe$^{+3}$/Fe$^{+2}$
state, is well established in GaN  \cite{Malguth:2008_PSSB}.
It was found that the Fe K-edge
probed by the XAS shifts under Si doping from
a position expected for the Fe$^{3+}$ oxidation state toward that specific
to the Fe$^{+2}$ configuration  \cite{Rovezzi:2009_PRB}, as evidenced by the XANES
spectra in Fig.\,\ref{fig:EXAFS_Si}.

\begin{figure}
    \includegraphics[width=0.75\columnwidth]{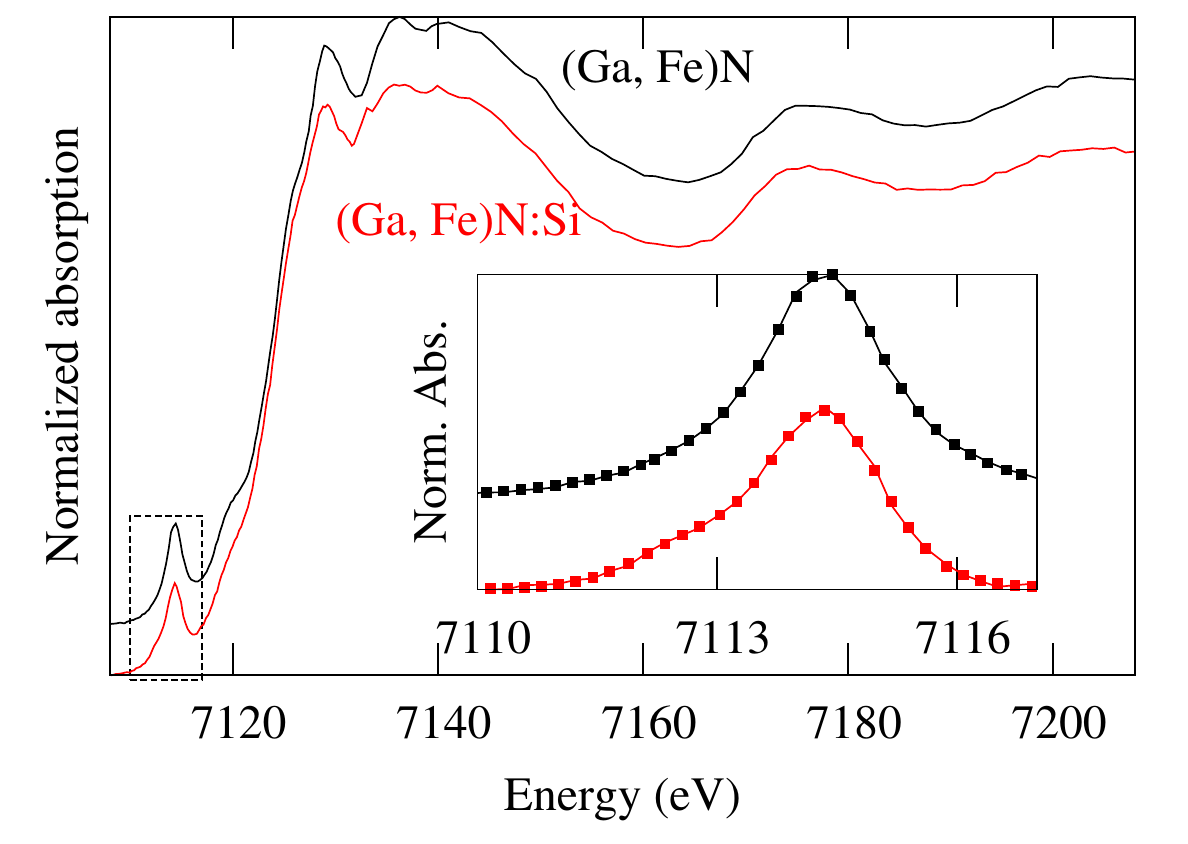}
    \caption{(Color online) XANES spectra of (Ga,Fe)N and (Ga,Fe)N:Si. The region marked by the dashed box is expanded in the inset. Here the experimental data (lines) after a background subtraction are compared to the fitted values (squares). In addition to the peak at $7114.3\pm 0.1$~eV assigned to the Fe$^{3+}$  charge state, a shoulder at $7112.7\pm 0.1$~eV is visible in the Si-doped sample pointing to the reduction of a part of the Fe ions to the Fe$^{2+}$ charge state by Si doping. From \onlinecite{Bonanni:2008_PRL}.}
    \label{fig:EXAFS_Si}
\end{figure}

In contrast, the role of additional
acceptors (Mg) is more complex in (Ga,Fe)N. The level Fe$^{+3}$/Fe$^{+4}$ is expected
to reside rather in the valence band \cite{Malguth:2008_PSSB}, but it was
suggested that the potential introduced by a TM ion in GaN (and ZnO)
is deep enough to bind a hole  \cite{Dietl:2008_PRB}, shifting
the Fe$^{+3}$/Fe$^{+4}$ up to the GaN band gap \cite{Pacuski:2008_PRL}.
If this is the case, Mg codoping could also hamper Fe aggregation.

It was found experimentally that the influence of Mg depends crucially on the
way it was introduced to the system: uniform Mg codoping reduces the probability of incorporation of Fe,
diminishing in this way the density of NCs in the samples \cite{Navarro:2011_PRB}. In contrast, codoping
of Mg in the $\delta$ fashion promotes the NC formation, an effect discussed
theoretically in terms of pairing energies of clusters containing different
numbers of Fe and Mg cations in GaN \cite{Navarro:2011_PRB}. The influence of Mg codoping
on magnetic properties is discussed below (Sec.\,\ref{sec:GaFeN_mag}).

Based on these results, the previously observed effect of codoping
on ferromagnetism in (Ga,Mn)N, and assigned to the dependence of the double-exchange 
mechanisms of the spin-spin coupling on the position of the Fermi level
with respect to the center of the $d$ band  \cite{Reed:2005_APL}, has to be reconsidered.

\subsection{Magnetic properties of (Ga,Fe)N}
\label{sec:GaFeN_mag}

The comprehensive nanocharacterization of (Ga,Fe)N allows to put on a more firm basis
the origin of
a complex magnetic response reported repeatedly for this system. Two examples
are given in Figs.\,\ref{fig:VanVleck} and \ref{fig:three_components}, in which various contributions
to magnetization are clearly seen. The richness of the magnetic response correlates
with the multiphase character of (Ga,Fe)N with Fe concentrations beyond the
solubility limit, as discussed in Secs.\,\ref{sec:GaN-Fe}C-E.

\begin{figure}
    \includegraphics[width=0.9\columnwidth]{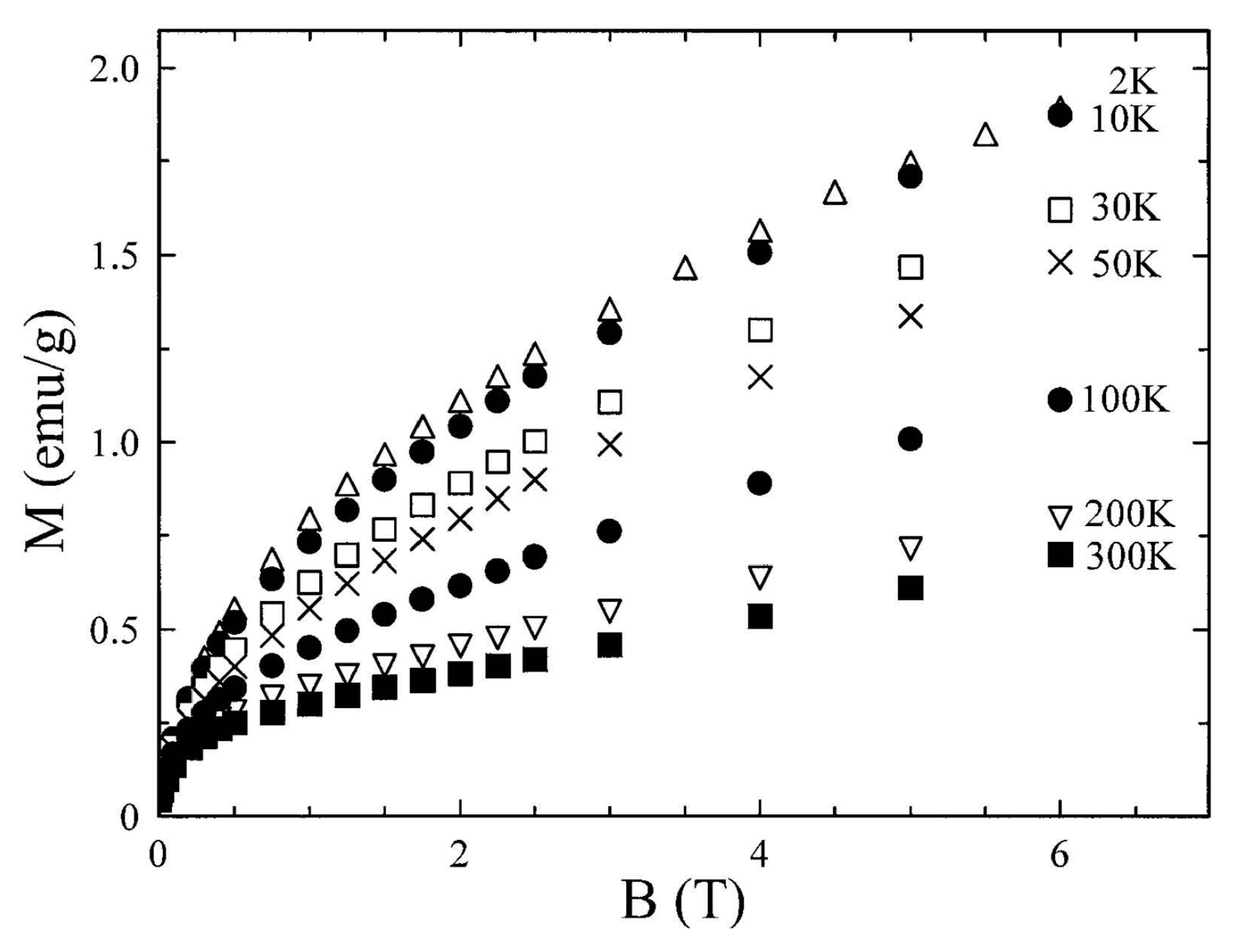}
    \caption{ Magnetization of (Ga,Fe)N bulk crystals obtained by a chemical transport method showing a nonsaturating linear in field component assigned to a Van Vleck-type paramagnetism of Fe$^{2+}$ ions, together with a high-$T_{\text{C}}$ ferromagnetic contribution found in weak fields. From \onlinecite{Gosk:2003_JS}.}
    \label{fig:VanVleck}
\end{figure}

\begin{figure}[htb]
    \centering
        \includegraphics[width=0.7\columnwidth, angle =-90]{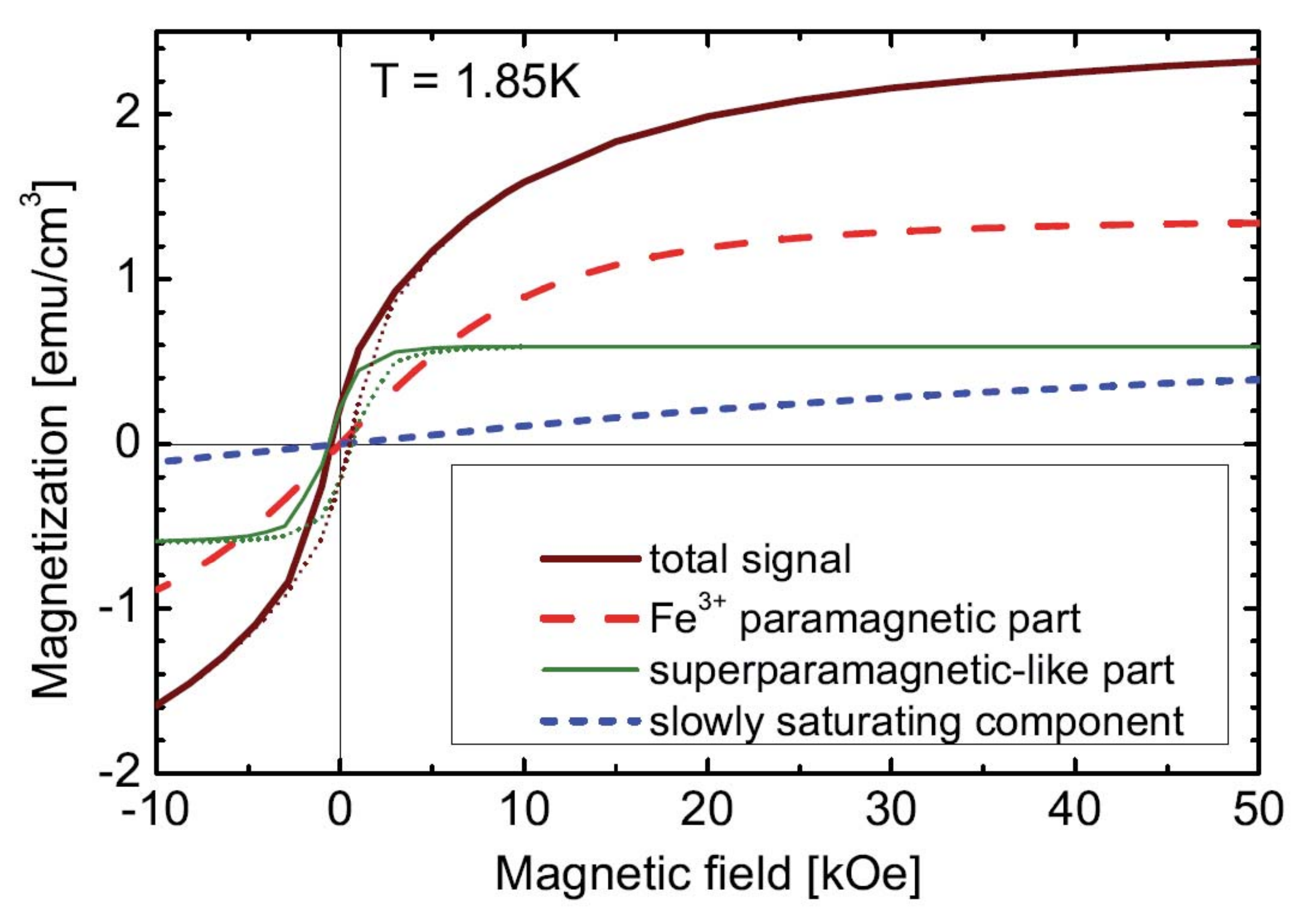}
    \caption{(Color online) Crystallographically phase-separated (Ga,Fe)N - contributions to the total magnetization as a function of the magnetic field : (i) paramagnetic from noninteracting Fe$^{3+}$, (ii) high-$T_{\mathrm{C}}$ superparamagneticlike or superferromagneticlike from the FM NCs and (iii) slowly saturating component, presumably coming from AF NCs. From \onlinecite{Navarro:2010_PRB}.}
    \label{fig:three_components}
\end{figure}

In general, according to studies of magnetization \cite{Gosk:2003_JS,Bonanni:2007_PRB,Navarro:2010_PRB,Navarro:2011_PRB},
electron paramagnetic resonance (EPR) \cite{Bonanni:2007_PRB,Malguth:2006_PRB}, ferromagnetic resonance (FMR) \cite{Grois:2014_NT},
infrared spectroscopy \cite{Malguth:2006_PRB,Malguth:2006_PRBb}, magnetooptics \cite{Pacuski:2008_PRL,Rousset:2013_PRB}, and
XMCD \cite{Kowalik:2012_PRB} Fe dopants
appear in distinct magnetic phases in (Ga,Fe)N,
whose relative importance depends on growth conditions, Fe concentration, and codoping by donors or acceptors.

{\em Brillouin paramagnetism of Ga-substitutional Fe$^{3+}$ ions}:
Because of a nonzero but relatively low solubility limit, a paramagnetic contribution from diluted Fe$^{3+}$ ions is always
present (provided that the density of compensating donors is sufficiently small), as confirmed by a quantitative interpretation of EPR spectra
in terms of weakly interacting localized spins $S = 5/2$ occupying Ga-substitutional sites \cite{Bonanni:2007_PRB,Malguth:2006_PRB}.
These spins give rise to the Brillouin-like dependence of magnetization on the magnetic field and
temperature visible in magnetooptical \cite{Pacuski:2008_PRL,Rousset:2013_PRB}
and magnetization measurements \cite{Bonanni:2007_PRB,Pacuski:2008_PRL,Bonanni:2008_PRL,Navarro:2010_PRB}.

{\em Van Vleck paramagnetism due to Fe$^{2+}$ ions}:
In addition to the Brillouin-like term described previously, a term linear in the magnetic field was found to contribute
to (Ga,Fe)N magnetization, and assigned to the Van Vleck paramagnetism of Fe$^{2+}$ ions that can be present
in (Ga,Fe)N due to residual or purposely introduced donors \cite{Bonanni:2007_PRB,Gosk:2003_JS,Malguth:2006_PRBb}.

\begin{figure}
    \includegraphics[width=0.9\columnwidth]{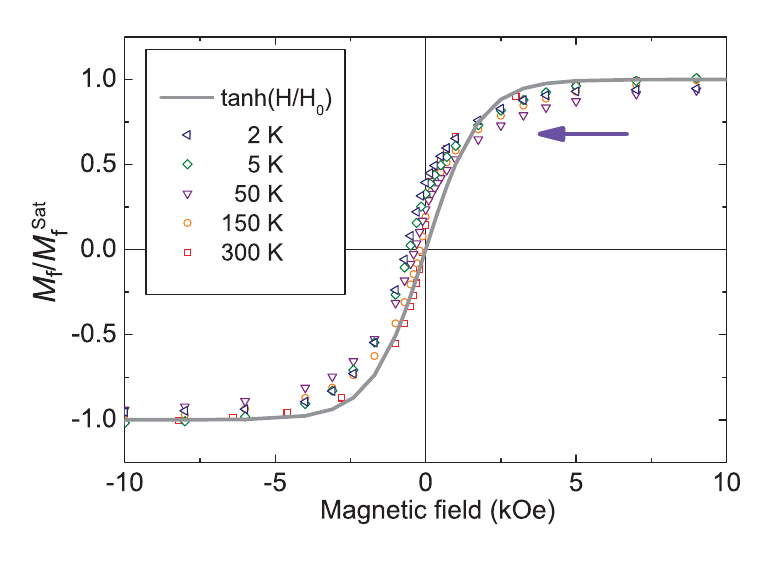}
    \caption{(Color online) Normalized magnetization at various
temperatures for (Ga,Fe)N codoped by Mg in $\delta$-like fashion with
embedded Fe=rich nanocrystals. Points: experimental data; solid line: fit a
Langevin-type function with $H_0 = 800$ Oe; arrow: sweep direction
of the magnetic field. From \onlinecite{Navarro:2011_PRB}.}
    \label{fig:loops}
\end{figure}

{\em High-$T_{\text{C}}$ ferromagnetism}:
A FM-like contribution to magnetization of (Ga,Fe)N, dominating at high temperatures,  shows characteristics specific to
the whole family of high-$T_{\text{C}}$ DMSs and dilute magnetic oxides \cite{Coey:2010_NJP,Sawicki:2013_PRB}. As presented in
Fig.\,\ref{fig:loops}, for any temperature and orientation of the magnetic field magnetic hysteresis loops $M(H,T)$
are leaning and narrow, so that the magnitude of spontaneous magnetization $M_s(T)$
is much smaller than the saturation
magnetization $M_{\text{Sat}}$. Remarkably, $M_s(T)$
persists up to above RT and the values of normalized magnetization are approximately described
by a temperature-independent Langevin function $M(H)/M_{\text{Sat}} = \tanh(H/H_0)$, where
in the case under consideration $H_0 \simeq 800$ Oe.

This behavior is in contrast to the one of  (Ga,Mn)As, which is characterized by a squarelike shape and
and strong temperature dependence of hystereses.
At the same time, if interpreted in terms of superparamagnetism it indicates that, for some NCs, the
blocking temperature $T_{\rm b}$ is higher than RT. This conclusion was confirmed
via zero-field cooled (ZFC) and field cooled (FC)
SQUID magnetometry measurements at a low magnetic field of 50~Oe \cite{Bonanni:2007_PRB},
as shown in Fig.\,\ref{fig:FC_ZFC}. However, the high magnitudes of  $T_{\rm b}$ were surprising
in view of the relatively small values of NC diameters, shown in
Fig.\,\ref{fig:nanosize}.

\begin{figure}
    \includegraphics[width=0.8\columnwidth]{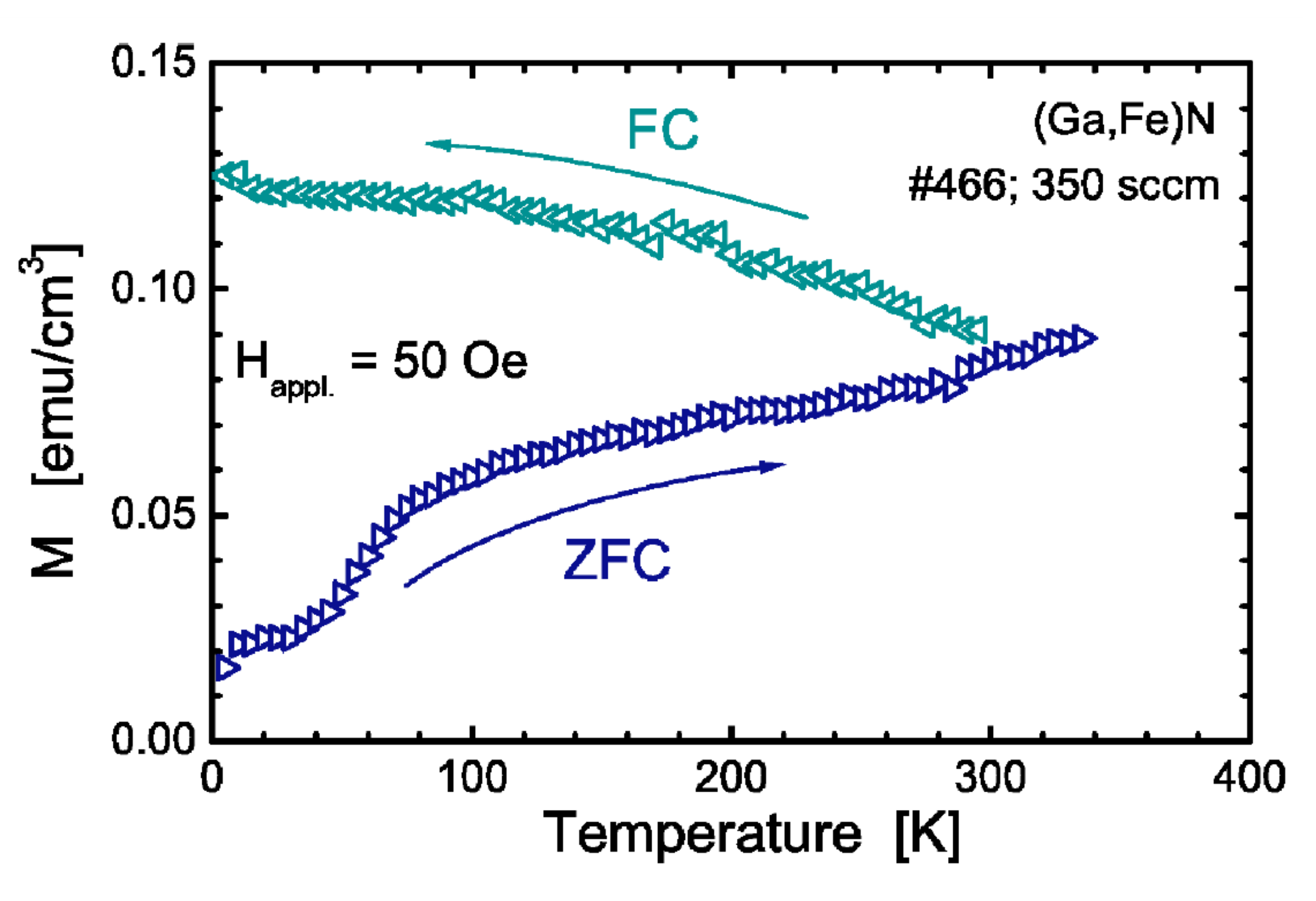}
    \caption{(Color online) ZFC-FC magnetization at
$H_{\mbox{\tiny{\emph{appl.}}}}= 50$~Oe for a (Ga,Fe)N sample with crystallographic phase separation. From \onlinecite{Bonanni:2007_PRB}.}
    \label{fig:FC_ZFC}
\end{figure}

It was suggested that dipole-dipole interactions
between densely packed magnetic constituents may lead to
a magnetization that can be parametrized by a
temperature-independent Langevin-type function \cite{Coey:2010_NJP},
a behavior refer to as superferromagnetism \cite{Sawicki:2013_PRB}.  According to
TEM \cite{Navarro:2010_PRB} and XPEEM \cite{Kowalik:2012_PRB}, NCs tend to aggregate in planes perpendicular
to the growth direction in the case of (Ga,Fe)N,
which increases the density of NCs and thus supporting the superferromagnetism scenario \cite{Navarro:2011_PRB}.

This interpretation requires also that the
$T_{\mathrm{C}}$ of individual NCs is well above RT.
According to the results
presented in Fig.\,\ref{fig:Arrot}, the value of $M_{\text{Sat}}$ determined from Arrott plots as a function
of temperature for the same type of samples, pointed to a $T_{\mathrm{C}} = 575$~K . This value is
in accord with the identification of
the dominant secondary phases via SXRD and TEM as $\varepsilon$-Fe$_3$N
in the samples under consideration \cite{Bonanni:2007_PRB}. According to the phase diagram discussed
in Sec.\,\ref{sec:phase_diagram}, the abundance of ferromagnetic NCs ($\varepsilon$-Fe$_3$N, $\gamma$'-Fe$_4$N,
$\alpha$'-Fe, and their derivatives) increases with lowering the epitaxy temperature.

\begin{figure}
    \includegraphics[width=0.9\columnwidth]{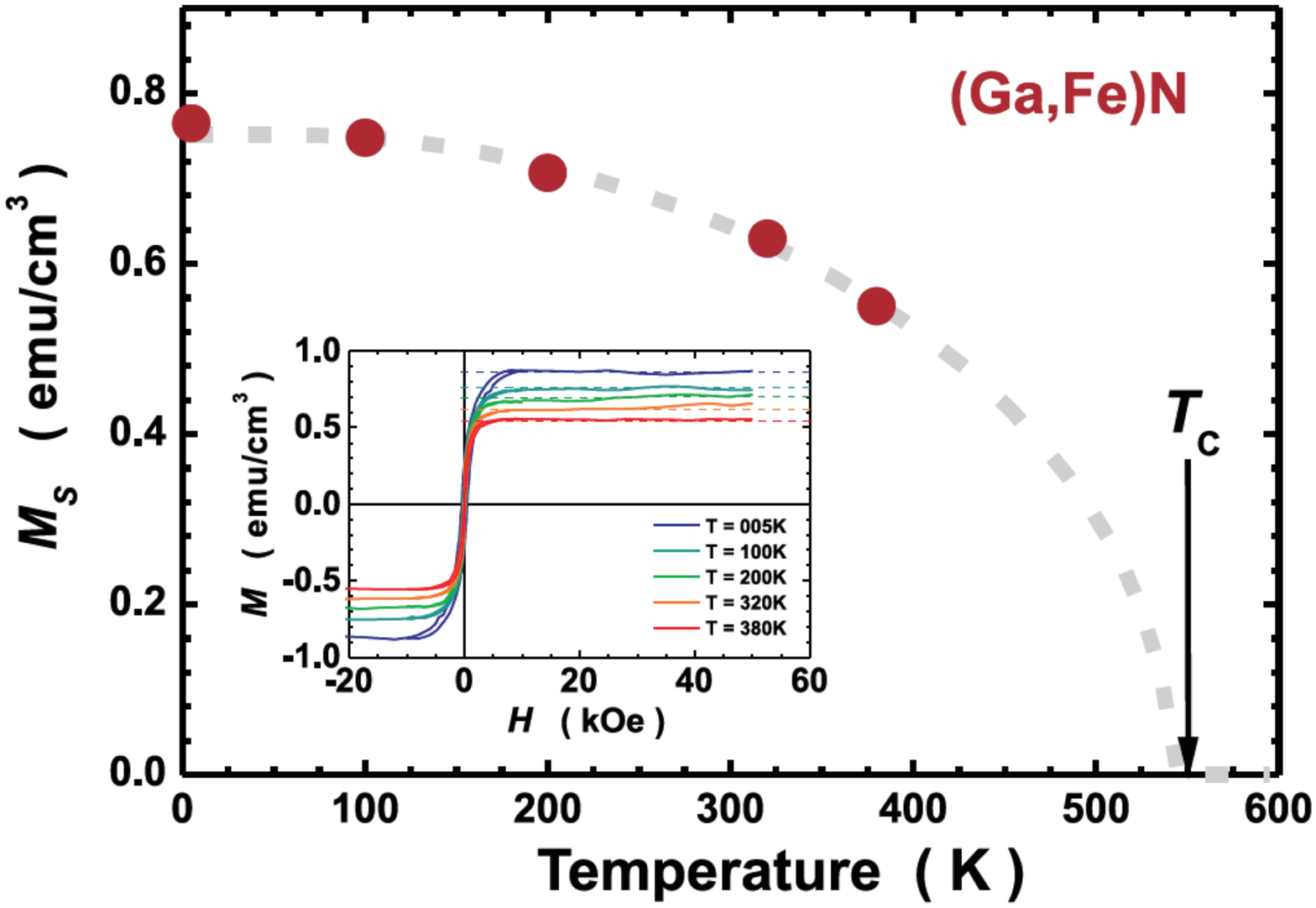}
    \caption{(Color online) Saturation magnetization determined from the Arrott plots as a function of temperature for spinodally decomposed (Ga,Fe)N: the evaluated $T_{\mathrm{C}}$ of 575~K corresponds to the Curie temperature of  $\varepsilon$-Fe$_3$N. Adapted from \onlinecite{Bonanni:2007_PRB}.}
    \label{fig:Arrot}
\end{figure}

Recently, FMR was detected at RT in spinodally decomposed
(Ga,Fe)N  in which Ga$_x$Fe$_{4-x}$N NCs formed a planar array \cite{Grois:2014_NT}.
The magnetic anisotropy was found to be primarily uniaxial with the hard axis normal to the NCs plane
and to have a comparably weak in-plane hexagonal symmetry.

{\em Antiferromagnetic contribution:}
According to the phase diagram discussed
in Sec.\,\ref{sec:phase_diagram}, an increase in  the epitaxy temperature results in the formation
of NCs with a higher degree of nitridation, for which AF coupling can prevail. In particular,
$\gamma\prime$-GaFe$_3$N exhibits a weak AF coupling with a Curie-Weiss temperature of
-20\,K \cite{Houben:2009_CM}, whereas orthorhombic $\zeta$-Fe$_2$N  is AF below 9~K \cite{Hinomura:1996_NC}. Furthermore, it is
also possible that wz-FeN NCs formed by chemical phase separation exhibit
AF properties as, according to theoretical predictions presented in Sec.\,\ref{sec:pairing},
the coupling between Fe cation pairs is AF in GaN.
These considerations led to the conclusion that a weakly saturating component of magnetization visible in Fig.\,\ref{fig:three_components},
rather than resulting from a Van Vleck paramagnetism could originate from antiferromagnetically ordered
Fe-rich NCs \cite{Navarro:2010_PRB}.

The presence of various Fe-rich phases with peculiar magnetic properties has
been proven, but now it is mandatory to explore the routes that can lead to
a single-phase system, especially in the perspective of exploiting hybrid AF/FM systems
for, e.g., AF spintronics \cite{Shick:2010_PRB}.

\section{\label{sec:Ge-Mn}Spinodal nanodecomposition in $\mbox{(Ge,Mn}$}

\subsection{Mn dilution in germanium}

Almost all the theoretical predictions on the magnetic properties of magnetic
semiconductors rely on the ideal dilution of magnetic TM
atoms substituting the semiconducting host. \textit{d} levels of Mn atoms
in a tetrahedral environment are split into two energy levels: \textit{e$_{g}$}
(twice degenerated) and \textit{t$_{2g}$} (3 times degenerated) levels.
In germanium, \textit{t$_{2g}$} orbitals strongly hybridize with \textit{p}
orbitals in the germanium valence band forming bonding and antibonding states.
The resulting magnetic moment per Mn atom is 3$\mu_{B}$ and two holes
(when activated) are created in the germanium valence band. \textit{Ab-initio}
calculations have further shown that exchange coupling between Mn atoms mediated
by holes oscillates following the Rudermann-Kittel-Kasuya-Yoshida (RKKY)
magnetic interaction \cite{Zhao:2003_PRL,Continenza:2006_PRB}. Among all TM atoms,
Mn is the only one to provide both ferromagnetism and high
localized spin moments in germanium. The Zener model complemented with mean
field approximation further predicts $T_{\text{C}}$ as high as 80\,K in
Ge$_{0.975}$Mn$_{0.025}$ with a hole density of 3.5$\times$10$^{20}$
cm$^{-3}$ \cite{Dietl:2000_S}. However exchange coupling was shown to be
highly anisotropic, i.e., to depend on the crystal orientation between
two Mn atoms \cite{Continenza:2006_PRB}. Hence, for high Mn concentrations
(up to a few percent), the relative position of Mn atoms in the germanium
crystal lattice should be known in detail to understand magnetic properties.
Finally, from  band structure calculations, Mn dilution in germanium is more
favorable than in silicon to induce half-metallic character (100\% spin
polarization at the Fermi level) \cite{Stroppa:2003_PRB}. To conclude about theoretical
works, (Ge,Mn) might be a very promising candidate as a DMS in future spintronic devices
compatible with mainstream silicon technology.

Surprisingly, only few works have been published on ferromagnetism in TM-doped
germanium until 2002. The evidence of FM order in an epitaxial layer of
Mn-doped germanium was first reported in 2002 by Park \cite{Park:2002_S}.
Ge$_{1-x}$Mn$_{x}$ ($0.6<x<3.5$\%) films were grown by LT-MBE on Ge(001) and GaAs(001) substrates. They exhibit
$p$-type doping with hole densities up to 10$^{19}$--10$^{20}$ cm$^{-3}$ and
$T_{\text{C}}$ increases linearly with Mn concentration from
25 up to 116\,K. Moreover they could demonstrate the interplay between band
carriers and Mn spins by measuring AHE as shown in Fig.\,\ref{jm_fig1}(a). They could
further modulate AHE by the application of a gate voltage as low as 0.5 V.
However the magnetic moment per Mn atom (1.4-1.9)$\mu_{B}$) was much less than
the expected value for Mn substituting Ge (3 $\mu_{B}$), and they
reported the presence of small unidentified precipitates. Mn dilution in germanium
is thus questionable.

\subsection{From Mn dilution to phase separation}

Considering the very low solubility of Mn in germanium
(10$^{-6}$\%) \cite{Woodbury:1955_PR}, out-of-equilibrium growth techniques are
required to dope germanium films with a few percent of Mn. Indeed, according to the
binary phase diagram \cite{Massalski:1990_B} and DFT calculations \cite{Arras:2011_PRB}, the stable (Ge,Mn) alloy with the lowest
Mn content is Ge$_{8}$Mn$_{11}$ and contains 57.9 at.\% of Mn. Since the 1980s,
seven stable (Ge,Mn) alloys have been synthesized:
$\epsilon$-GeMn$_{3.4}$ \cite{Ohoyama:1961_JPSJ_a},
$\epsilon_{1}$-GeMn$_{3.4}$, $\varsigma$-Ge$_{2}$Mn$_{5}$ \cite{Ohba:1987_AC},
$\kappa$-Ge$_{3}$Mn$_{7}$, $\eta$-Ge$_{3}$Mn$_{5}$ \cite{Ohoyama:1961_JPSJ_b},
GeMn$_{2}$ \cite{Ellner:1980_JAC} and Ge$_{8}$Mn$_{11}$ \cite{Ohba:1984_AC}.
Although some of them are FM, they all exhibit a metallic character
that makes them poor candidates for spin injection in nonmagnetic semiconductors
due to conductivity mismatch \cite{Fert:2001_PRB}. Other (Ge,Mn) alloys could be
prepared by melting Ge and Mn under very high pressure ($\simeq$4-6 GPa) by Takizawa.
Metastable alloys such as Ge$_{5}$Mn$_{3}$ \cite{Takizawa:1987_JSSC}, GeMn \cite{Takizawa:1988_JSSC},
Ge$_{4}$Mn \cite{Takizawa:1990_JSSC} or GeMn$_{3}$ \cite{Takizawa:2002_JPCM} could be obtained.
All these (Ge,Mn) alloys are Ge rich and thus can exhibit a semiconducting
character required for direct spin injection in semiconductors. We note in this context that
pressure associated with the strain imposed by the Ge host can stabilize Ge rich Ge$_{n}$Mn$_m$
nanocrystals ($n > m$) embedded in the Ge matrix.  As discussed in Sec.\,\ref{structure},
LT-MBE can serve to obtain such nanocomposites via spinodal nanodecomposition.

The very low solubility of Mn in Ge was further theoretically demonstrated
using \textit{ab initio} calculations. For instance, in order to increase
Mn concentration, mixing substitutional and interstitial Mn in germanium lowers
the free energy of the system \cite{Continenza:2007_JMMM,Arras:2012_PRB}.
At last, except in the work by Zeng \cite{Zeng:2008_PRL}, most groups have
experimentally observed inhomogeneous Mn-doped germanium films. Inhomogeneities
can be either secondary phase precipitates such as Ge$_{3}$Mn$_{5}$ clusters or
Mn-rich nanostructures due to spinodal decomposition. Indeed
spinodal decomposition leads to the formation of Mn-rich nanometer sized areas
[either a metastable (Ge,Mn) alloy or Ge lattice with high Mn content] surrounded
with an almost pure germanium matrix. In the next section, we thoroughly
review the results obtained on the (Ge,Mn) material.

\subsubsection{Review of experimental results}
\label{sec:MCD_GeMn}
As already mentioned, in order to prevent the formation of stable metallic (Ge,Mn) phases,
out-of-equilibrium growth techniques are required. In the following,
we summarize Mn implantation in germanium and MBE growth of thin (Ge,Mn) films.
\onlinecite{Ottaviano:2006_JAP} first performed Mn implantation in germanium (up to 4\%) at
240 and $270^{\circ}$C substrate temperatures.
Before and after annealing (to improve the crystalline quality) Mn-rich precipitates
are observed by TEM. These precipitates are
amorphous before annealing and Ge$_{3}$Mn$_{5}$ clusters after annealing.
In addition, x-ray absorption measurements showed that Mn-rich precipitates
only form in the deeper part and substitutional Mn are detected close to
the film surface \cite{Ottaviano:2007_APL}. Ferromagnetic behavior was further observed
by magnetooptical Kerr effect measurements up to 270\,K (before annealing)
and 255\,K (after annealing). However the magnetic signal clearly arises from
many different magnetic phases including Ge$_{3}$Mn$_{5}$.
\onlinecite{Passacantando:2006_PRB} found similar results using different implantation doses in germanium wafers. At low implantation dose, Mn-rich precipitates are amorphous whereas they form Ge$_3$Mn$_5$ clusters at higher implantation doses as shown in Fig.\,\ref{jm_fig1}(b).
The second technique widely used to grow Mn-doped Ge films is low-temperature MBE. Following the first
results published by  \onlinecite{Park:2002_S}, many groups have attempted to dilute
large amounts of Mn in Ge in order to raise $T_{\text{C}}$ up to RT.  \onlinecite{D'Orazio:2004_JMMM} investigated the magnetic and electrical properties of
thin epitaxial Ge$_{1-x}$Mn$_{x}$ films ($0.027 <x< 0.044$) exhibiting ferromagnetism
up to 250\,K. \onlinecite{Pinto:2005_PRB} and \onlinecite{Morresi:2006_MSEB} interpreted experimental
observations in the framework of magnetic polaron
percolation. However further structural characterizations
showed that (Ge,Mn) epilayers contained Ge$_{3}$Mn$_{5}$ clusters
surrounded with germanium containing almost 1.5\% of substitutional paramagnetic
Mn atoms. Similar results were also reported by others \cite{Padova:2007_SS,Padova:2008_PRB,Bihler:2006_APL}.
Finally, using high-resolution TEM and energy dispersive x-ray spectroscopy, \onlinecite{Sugahara:2005_JJAP}
showed the phase separation between Mn-rich amorphous clusters and a pure Ge matrix in thin epitaxial (Ge,Mn) films [see Fig.\,\ref{jm_fig1}(c)].

\begin{figure}[h!]
\begin{center}
\includegraphics[width=\columnwidth]{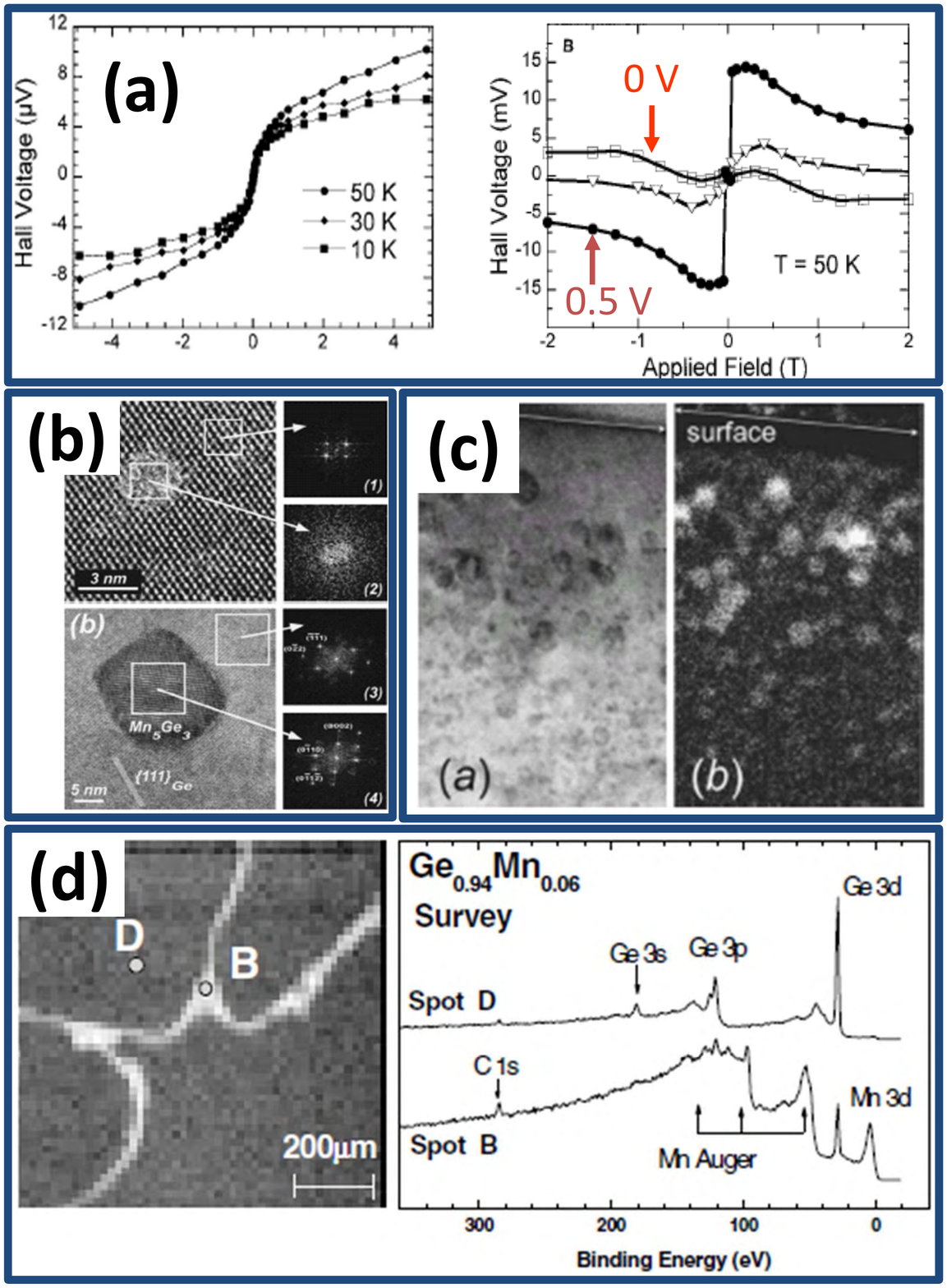}
\end{center}
\caption{(color online) Characterization of (Ge,Mn) samples obtained by various methods. (a) Anomalous Hall effect measured at 10, 30 and 50\,K in a Mn$_{0.023}$Ge$_{0.937}$ film grown by LT-MBE (left) and voltage control of AHE (right). From \onlinecite{Park:2002_S}. (b) HRTEM images of a Ge(100) wafer implanted at a dose of 1$\times$10$^{16}$ (up) and 4$\times$10$^{16}$ (down) Mn$^{+}$ ions/cm$^{2}$. At low dose, Mn-rich areas are amorphous whereas at high dose, they form Ge$_3$Mn$_5$ clusters. From \onlinecite{Passacantando:2006_PRB}. (c) TEM image of a GeMn film grown by LT-MBE showing amorphous Ge$_{1-x}$Mn$_x$ clusters (left) and corresponding energy dispersive x-ray spectroscopy image (right). Bright areas are Mn-rich. From \onlinecite{Sugahara:2005_JJAP}. (d) Scanning photoelectron microscopy image of a Ge$_{0.94}$Mn$_{0.06}$ single crystal (left) and the survey spectra obtained at a bright spot (B) and a dark spot (D) respectively (right). From \onlinecite{Kang:2005_PRL}.}
\label{jm_fig1}
\end{figure}

The second metallic (Ge,Mn) phase that usually forms during MBE growth is
Ge$_{8}$Mn$_{11}$. This phase exhibits two magnetic transitions at 150\,K
(AF/FM) and $T_{\text{C}} =285$\,K (FM/paramagnetic).
Ge$_{8}$Mn$_{11}$ precipitates were observed by  \onlinecite{Park:2001_APL} in thin Ge$_{1-x}$Mn$_{x}$
films ($0 <x< 0.12$) epitaxially grown between 250 and
$300^{\circ}$C. Using different methods to grow bulk single crystals,
the same Ge$_{8}$Mn$_{11}$ phase could be detected by \onlinecite{Cho:2002_PRB,Kang:2005_PRL} and
 \onlinecite{Biegger:2007_JAP} performing chemical analysis at the micrometer scale, as shown in Fig.\,\ref{jm_fig1}(d).
More recently, by investigating Mn-doped (up to 4\%) Ge films epitaxially grown
on Ge(001) using various TEM techniques,  \onlinecite{Wang:2008_APL} could find the coexistence of
Ge$_{8}$Mn$_{11}$ and Ge$_{2}$Mn$_{5}$ clusters. Moreover,
by well adjusting growth parameters, these crystalline precipitates could be
replaced by undefined Mn-rich nanostructures.

Since 2005, many groups reported the absence of any secondary phase
precipitation in (Ge,Mn) films epitaxially grown at low temperature ($ < 100^{\circ}$C).
However, due to the very low solubility of Mn in Ge, the formation of Mn-rich
nanostructures seems unavoidable mostly as a result of spinodal nanodecomposition.
These nanostructures are so hard to detect that only highly sensitive techniques
such as TEM with nanoscale chemical analysis, 3DAP or XRD/XAS using
synchrotron radiation can be used. From the low-temperature MBE growth of
Ge$_{0.94}$Mn$_{0.06}$ films, \onlinecite{Sugahara:2005_JJAP} found elongated Mn-rich amorphous
precipitates surrounded with pure germanium. These precipitates
exhibit ferromagnetism up to 100\,K as given by MCD
measurements. Their diameter is close to 5\,nm and the Mn content is between 10
and 20\%. Using almost the same growth conditions, \onlinecite{Ahlers:2006_PRB,Bougeard:2006_PRL,Bougeard:2009_NL} observed the formation
of Mn-rich cubic clusters coherently strained on the Ge matrix in addition to the
precipitation of Ge$_{3}$Mn$_{5}$
clusters. Assuming zero Mn content
in the Ge matrix, the maximum Mn content in the clusters is close to 15\%. Their
diameter is slightly less than 5 nm. Moreover they exhibit superparamagnetic behavior
and a $T_{\text{C}}$ value close to 200\,K. Still using low-temperature MBE growth and
according to magnetic measurements only,  \onlinecite{Jaeger:2006_PRB} concluded that Ge$_{1-x}$Mn$_{x}$
films ($x=0.04$ and 0.2) contain two cluster populations (undefined Mn-rich
precipitates and Ge$_{3}$Mn$_{5}$ clusters) along with diluted Mn atoms in the
germanium matrix. The overall population of Mn-rich precipitates
behaves as a spin glass. The freezing temperatures are 12 and 15\,K in
Ge$_{0.96}$Mn$_{0.04}$ and Ge$_{0.8}$Mn$_{0.2}$ films, respectively. Finally \onlinecite{Li:2007_PRB} also
found elongated Mn-rich precipitates but coherently strained on the surrounding
Ge matrix. In as-grown Ge$_{0.95}$Mn$_{0.05}$ samples, these
precipitates show ferromagnetism at low temperature: remanence vanishes above 12\,K.
Post-growth annealing at $200^{\circ}$C leads to a rather substantial improvement in magnetic
and electrical properties by converting interstitial Mn into substitutional Mn.
They exhibit remanence up to 125\,K,  the magnetic moment per Mn atom reaches
$1.5\mu_{B}$ instead of $1.0\mu_{B}$ in as-grown samples. Moreover, annealing
triggers strong positive MR and AHE in (Ge,Mn) films.

To summarize, the key issue of (Ge,Mn) material is Mn-dilution: the very low
solubility of Mn in Ge always results in Mn segregation and in the formation
of Mn-rich precipitates. Low-temperature growth techniques favor the formation
of metastable (Ge,Mn) phases. As a consequence, slight differences in growth
parameters can result in much different magnetic and electrical properties.
In the following sections, we show how to control spinodal nanodecomposition
in (Ge,Mn). The structure, magnetic, and electrical properties of (Ge,Mn) films
with spinodal nanodecomposition are reviewed.

\subsection{\label{structure}Growth and structure of thin (Ge,Mn) films with spinodal decomposition}

\subsubsection{Sample preparation}

Growth was performed using solid sources MBE
by codepositing Ge and Mn evaporated from standard Knudsen effusion
cells \cite{Jamet:2006_NM,Devillers:2006_PSSC,Devillers:2007_PSSA,Devillers:2007_PRB}. The deposition
rate was kept constant and quite low ($\simeq 0.2$\, {\AA}s$^{-1}$). Germanium
substrates were epiready Ge(001) wafers with a residual $n$-type doping and
resistivity of $10^{15}$\,cm$^{-3}$ and 5\,$\Omega$cm, respectively. After thermal
desorption of the surface oxide, a 40\,nm-thick Ge buffer layer was grown at
250$^{\circ}$C, resulting in a $2 \times 1$ surface reconstruction as observed
by RHEED (see Fig.\,\ref{jm_fig2}). Next,
80\,nm-thick Ge$_{1-x}$Mn$_{x}$ films were subsequently grown  at low substrate
temperature (between $T_{g}=80^{\circ}$C and $T_{g} =200^{\circ}$C).
Mn content has been determined by x-ray fluorescence measurements performed
on thick samples ($\simeq 1$\,$\mu$m thick) and complementary Rutherford
back scattering (RBS) on thin Ge$_{1-x}$Mn$_{x}$ films grown on silicon.
Mn concentrations range from 1 to 11\% Mn.

\begin{figure}[h!]
\begin{center}
\includegraphics[width=\columnwidth]{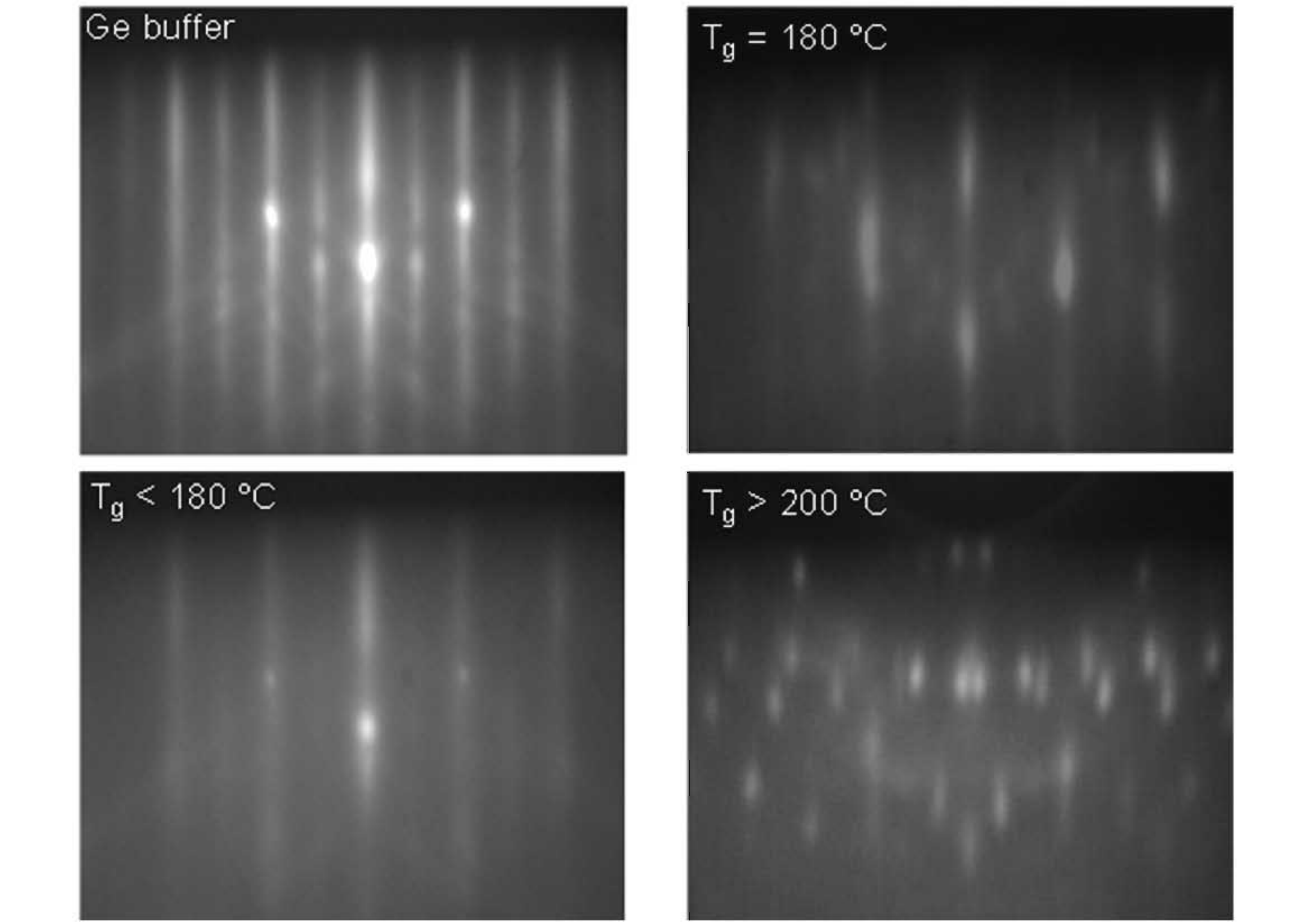}
\end{center}
\caption{RHEED patterns recorded for different growth temperatures of Ge$_{1-x}$Mn$_{x}$. Adapted from \onlinecite{Devillers:2007_PRB}.}
\label{jm_fig2}
\end{figure}

For Ge$_{1-x}$Mn$_{x}$ films grown at substrate temperatures below 180$^{\circ}$C,
after the first monolayer (ML) deposition, the  $2 \times 1$ surface reconstruction
almost totally disappears. After depositing a few MLs (corresponding almost to 5\,nm),
a slightly diffuse $1 \times 1$ streaky RHEED pattern and a very weak $2 \times 1$
reconstruction (Fig.\,\ref{jm_fig2}) indicate the predominantly 2D growth of a
single crystalline film exhibiting the same lattice parameter as the Ge buffer layer.
Increasing the layer thickness leads to an amplification of the surface roughness
which is expected for the low-temperature epitaxial growth of
germanium \cite{Chason:1989_JVSTB,Venkatasubramanian:1993_JVSTB}. For growth temperatures above
180$^{\circ}$C additional spots appear in the RHEED pattern during the
Ge$_{1-x}$Mn$_{x}$ growth (Fig.\,\ref{jm_fig2}). These spots correspond to the formation
of very small Ge$_{3}$Mn$_{5}$ crystallites.

\subsubsection{Morphology of (Ge,Mn) films}

From TEM images, vertical elongated nanostructures, i.e.,
nanocolumns were observed as shown in Fig.\,\ref{jm_fig3}(a). These observations are in very good agreement with the theoretical predictions of Sec.~\ref{sec:theory}. The formation of the konbu phase as a consequence of 2D spinodal decomposition in (Ge,Mn) is clearly demonstrated here as well as in (Zn,Cr)Te in Sec.~\ref{sec:ZnTe}. Similar elongated nanostructures were also obtained in (Ge,Mn) by several groups in comparable growth conditions [Fig.\,\ref{jm_fig3}(b-e)]. Nanocolumns span the entire Ge$_{1-x}$Mn$_{x}$
film thickness. Whatever the growth temperature and Mn concentration (except 0.1\%),
Ge$_{1-x}$Mn$_{x}$ films always exhibit the presence of nanocolumns with
their axis along the growth direction [001]. Depending on growth conditions,
the average columns diameter and density range between 2 and 7\,nm  and between
10\,000  and 40\,000\,$\mu$m$^{-2}$, respectively.

\begin{figure}[h!]
\begin{center}
\includegraphics[width=\columnwidth]{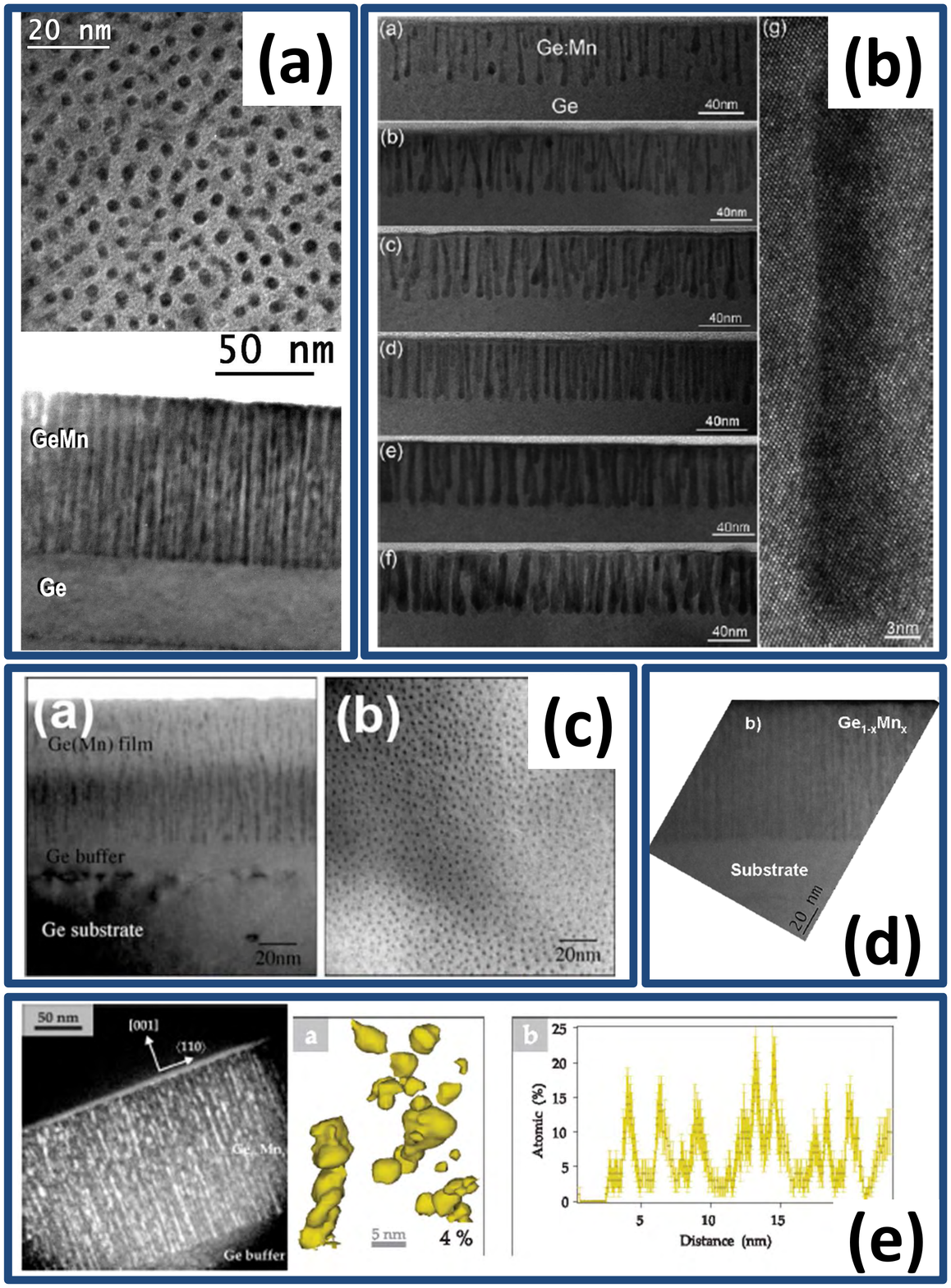}
\end{center}
\caption{(Color online) Characterization of Ge$_{1-x}$Mn$_x$ thin films grown by LT-MBE. (a) TEM plane view (up) and cross section (down) images of Ge$_{0.9}$Mn$_{0.1}$ containing Mn-rich nanocolumns. From \onlinecite{Devillers:2007_PRB}. (b) TEM cross section images of films with $x= 2.5\%, 4, 4.5, 5.5, 7$, and 12\% from top to bottom. The morphology of Mn-rich nanocolumns evolves from tadpolelike to cylinders when increasing Mn content. From \onlinecite{Wang:2010_JAC}. (c) and (d) TEM images of Ge$_{0.95}$Mn$_{0.05}$  [from \onlinecite{Li:2007_PRB}] and Ge$_{0.94}$Mn$_{0.06}$ [from \onlinecite{Le:2011_JPCS}], respectively, showing the presence of nanocolumns. (e) Vertical self-assembly of roughly spherical Mn-rich clusters: TEM cross section image in dark field with $x=7.3$\% (left) and  atom probe tomography data for $x=2$\% (right). From \onlinecite{Bougeard:2009_NL}.}
\label{jm_fig3}
\end{figure}

Further evidence of 2D spinodal nanodecomposition was provided by analyzing the periodic structures of the nanocolumns \cite{Hai:2011_JAP,Yada:2011_JAP}. Figure \ref{jm_fig_tanaka}(a) presents a plan view TEM image showing that the nanocolumns are uniform in diameter ($\sim 3$\,nm), and form either rectangular or triangular lattice structures, as indicted by thin dashed lines in Fig.\,\ref{jm_fig_tanaka}(a). According to the Fourier transform of the TEM image, depicted in Fig.\,\ref{jm_fig_tanaka}(b), the average distance between the nanocolumns is $\sim 9$\,nm.

\begin{figure}[h!]
\begin{center}
\includegraphics[width=\columnwidth]{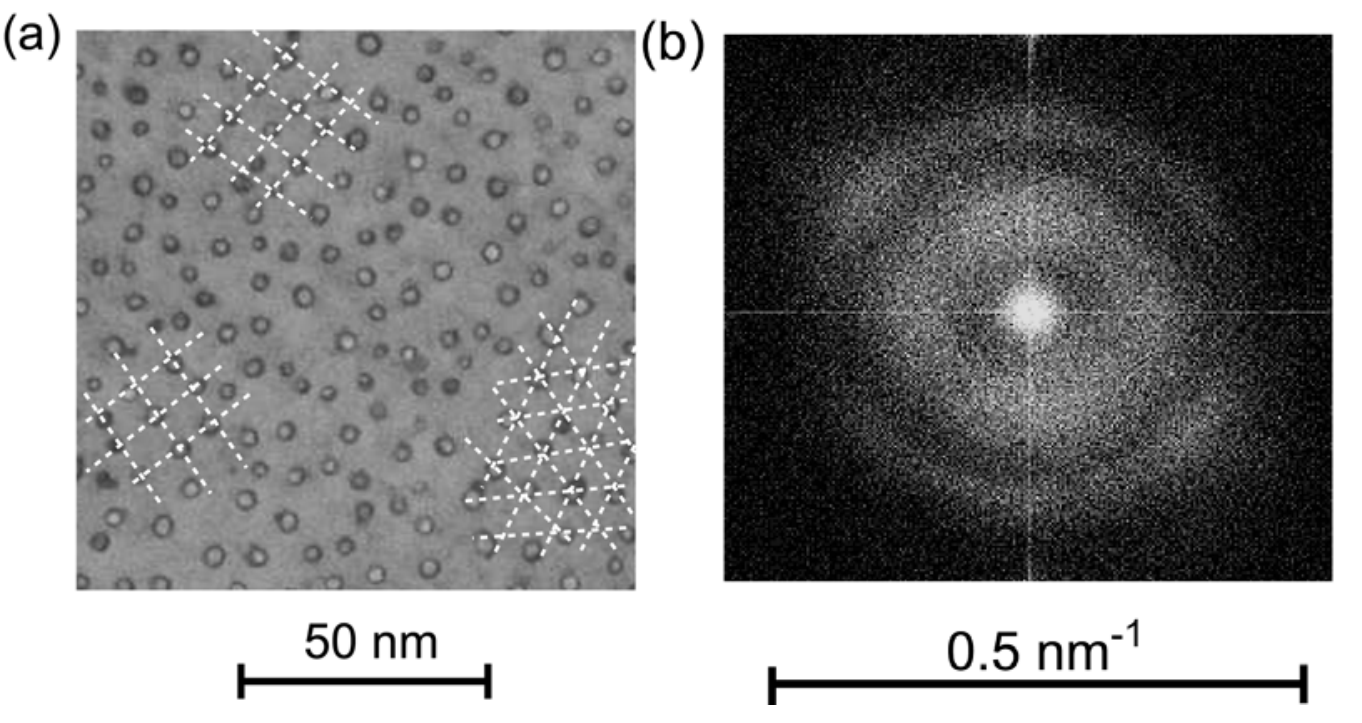}
\end{center}
\caption{(a) Plan view TEM image of Mn-rich nanocolumns embedded in a Ge matrix fabricated by MBE of a 30 nm-thick Ge$_{0.94}$Mn$_{0.06}$ film onto a Ge(001) substrate at 100$^{\circ}$C with a growth rate of 150 nm/h.  (b) Fourier transform of the contrast of a $0.36\times0.36$\,$\mu$m$^2$ TEM image. Ring structures up to second harmonic are clearly seen, indicating strong 2D spinodal decomposition. Adapted from \onlinecite{Hai:2011_JAP}.}
\label{jm_fig_tanaka}
\end{figure}

The morphology of Ge$_{1-x}$Mn$_{x}$ films was further investigated using
grazing incidence small angle x-ray scattering (GISAXS) from synchrotron radiation
and atomic force microscopy (AFM), as depicted in Fig.\,\ref{jm_fig4}.

\begin{figure}[h!]
\begin{center}
\includegraphics[width=\columnwidth]{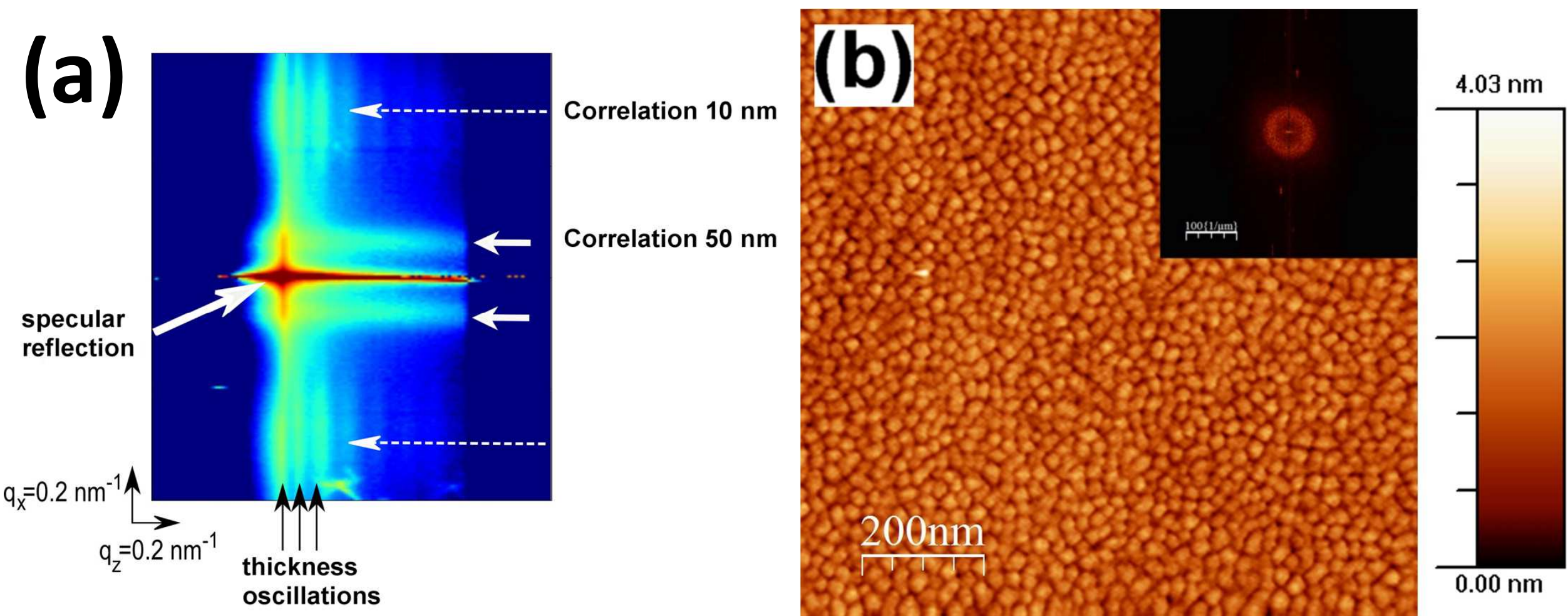}
\end{center}
\caption{(Color online) Characterization of a Ge$_{0.93}$Mn$_{0.07}$ film grown at
150$^{\circ}$C. (a) GISAXS spectrum; the nanocolumns
diameter is 5\,nm and their density 8100 $\mu$m$^{-2}$. Adapted from \onlinecite{Tardif:2010a_PRB}.
(b) Corresponding AFM image of the film surface. The maximum roughness amplitude is
of the order of 1 nm and corresponds to the formation of small islands at the film
surface. The fast Fourier transform (inset) of this picture exhibits
the presence of a correlation ring between surface islands of the order of 30\,nm.
From \onlinecite{Devillers:2008_PhD}.}
\label{jm_fig4}
\end{figure}

As shown in Fig.\,\ref{jm_fig4}(a), in addition to the specular reflection peak in the GISAXS spectrum,
two pairs of correlation peaks were observed. The first one at large $q_{x}$
value corresponds to a correlation length of $\simeq 10$ nm which is the average
distance between nanocolumns. The oscillations along $q_{z}$ are thickness
oscillations and stand for the finite thickness of the Ge$_{1-x}$Mn$_{x}$ film.
It demonstrates that nanocolumns span the entire film thickness. Furthermore
there is a sizable contrast of electronic density between nanocolumns
and the surrounding matrix. At low $q_{x}$ value, a second pair of correlation peaks
corresponds to a distance of almost 50\,nm. Moreover these peaks are much more
elongated along $q_{z}$ which is characteristic of a surface effect. Indeed,
correlations are no more related to the presence of nanocolumns but to the surface
roughness, as shown in Fig.\,\ref{jm_fig4}(b). In order to investigate the columns chemical
composition, electron energy loss spectroscopy (EELS) has been performed in cross section and plane view. The corresponding EFTEM images close to the Mn $L_{2,3}$ edge are displayed in
Fig.\,\ref{jm_fig5} along with the corresponding Mn chemical profiles.

\begin{figure}[h!]
\begin{center}
\includegraphics[width=\columnwidth]{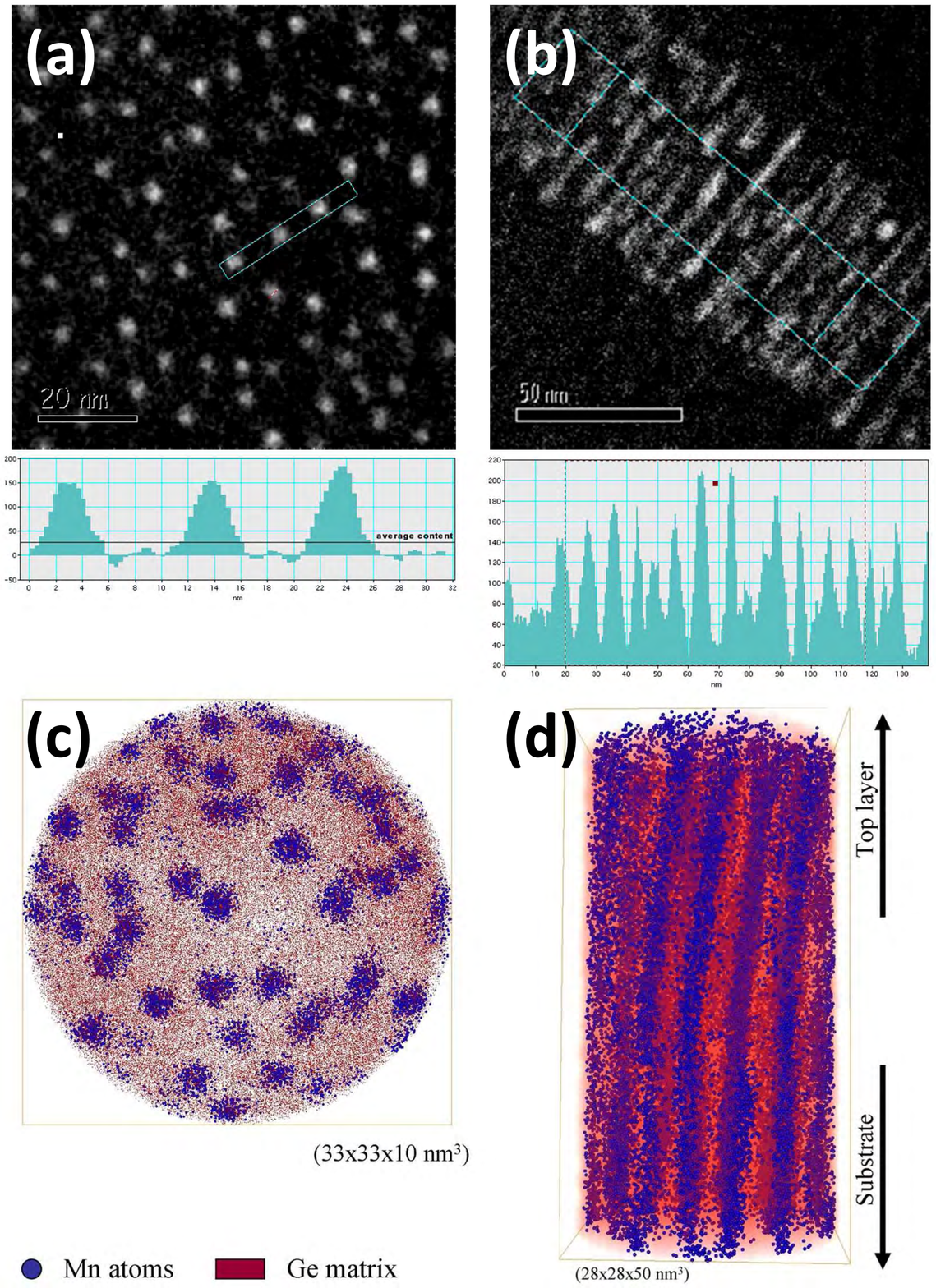}
\end{center}
\caption{(Color online) Characterization of a
Ge$_{0.94}$Mn$_{0.06}$ film grown at 130$^{\circ}$C. (a), (b) Cross sectional and plan view EFTEM images.  (c), (d)  3DAP reconstructed 3D images.  From \onlinecite{Mouton:2012_JAP}. Below the EFTEM images are the corresponding Mn chemical profiles (bright areas are Mn-rich). From \onlinecite{Jamet:2006_NM}.}
\label{jm_fig5}
\end{figure}

From EFTEM images, one can conclude that nanocolumns are Mn-rich and surrounded
with an almost pure Ge matrix. The Mn signal in the Ge matrix is indeed below
the resolution limit of EELS spectroscopy estimated to be around 1\%. The most appropriate technique to estimate with more accuracy the Mn content in the Ge matrix is 3DAP. It leads to an average Mn concentration inside the Ge matrix below 0.05\% (\cite{Mouton:2012_JAP}).
As a consequence, the Mn concentration in the nanocolumns shown in
Fig.\,\ref{jm_fig5}
is close to 30\%. Hence the composition of nanocolumns is close to
Ge$_{2}$Mn. Depending on the growth conditions, nanocolumns with Mn
concentrations between 5 and 50\% could be grown, all of them exhibiting FM
properties \cite{Devillers:2007_PRB}.

\subsubsection{Lateral and vertical control of nanocolumns}

In order to achieve lateral control of nanocolumns (diameter and density),
the growth temperature and Mn concentration were varied to study their respective influence. For this purpose, two series
of Ge$_{1-x}$Mn$_{x}$ films were grown at two different growth temperatures:
$T_{g}=100$ and $T_{g}=150^{\circ}$C. In each case, the Mn
concentration $x$ was varied between 1 and 11\%. All the samples were
then observed by TEM in plane view: they all exhibit Mn segregation in Mn-rich
nanocolumns surrounded with an almost pure Ge matrix. The results are shown
in Fig.\,\ref{jm_fig6}.

\begin{figure}[h!]
\begin{center}
\includegraphics[width=\columnwidth]{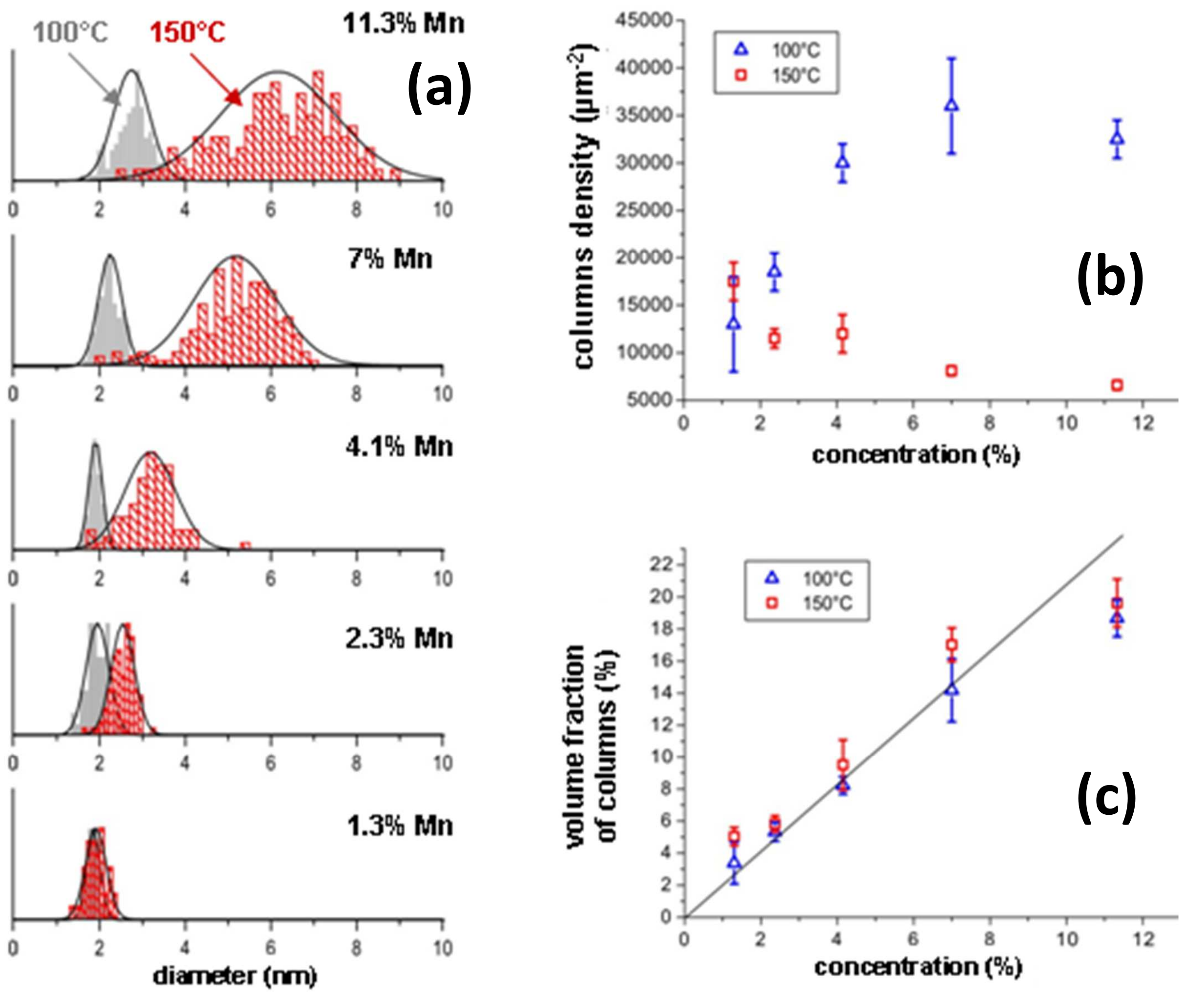}
\end{center}
\caption{(Color online) Lateral control of nanocolumns. (a) Numbers of (Ge,Mn) nanocolums with a given radius (size distribution) extracted from plane view TEM images for samples with various average Mn concentrations $x$ (1.3\%, 2.3\%, 4.1\%, 7\%, and 11.3\%)  for two growth temperatures 100$^\circ$C (filled bars) and 150$^\circ$C (dashed bars). (b) Density (in $\mu$m$^{-2}$) and (c) volume fraction of nanocolumns as a function of $x$ for these two growth temperatures (open triangles and open squares, respectively). From \onlinecite{Devillers:2007_PRB}.}
\label{jm_fig6}
\end{figure}

For samples grown at $100^\circ$C with Mn concentrations below 5\% the nanocolumns
mean diameter is $1.8\pm 0.2$\,nm [see Fig.\,\ref{jm_fig6}(a)]. For higher Mn concentrations, it only slightly
increases up to 2.8\,nm at 11.3\%. The evolution of columns' density as a function
of the Mn concentration is reported in Fig.\,\ref{jm_fig6}(b). As seen, by increasing Mn concentration
up to 7\% a significant increase of the columns density from 13\,000 to
35\,000 $\mu$m$^{-2}$ was observed. Then the density seems to reach a plateau corresponding
to almost 40\,000 $\mu$m$^{-2}$. Hence, increasing Mn concentration in samples
grown at $100^{\circ}$C allows one to control the columns' density while keeping their
diameter constant. Increasing the Mn content in the samples grown at 150$^\circ$C
from 1.3 to 11.3\% leads to a slight decrease of the columns density as shown
in Fig.\,\ref{jm_fig6}(b). However, according to Fig.\,\ref{jm_fig6}(a), their
average diameter increases significantly and size
distributions become much broader. For the highest Mn concentration (11.3\%)
very small columns with a diameter of 2.5\,nm coexist with
very large columns with a diameter of 9\,nm. Therefore increasing Mn concentration
in samples grown at $150^{\circ}$C allows to control the columns' diameter while
keeping their density constant around 10\,000\,$\mu$m$^{-2}$. To summarize, by
combining two growth parameters (temperature and Mn concentration), it is possible to
independently control the diameter and density of nanocolumns. This ability is of
great importance from a fundamental point of view to characterize nanocolumns
but also for potential applications in spintronic devices. In Fig.\,\ref{jm_fig6}(c),
the volume fraction occupied by the columns in the film is reported as a function
of Mn concentration showing a clear linear dependence for Mn contents up to almost
7\%. The nonlinear behavior above 7\% can indicate that the mechanism of Mn
incorporation is different in this concentration range, leading to an increase
of the Mn concentration in the columns and/or in the matrix.

\begin{figure}[h!]
\begin{center}
\includegraphics[width=\columnwidth]{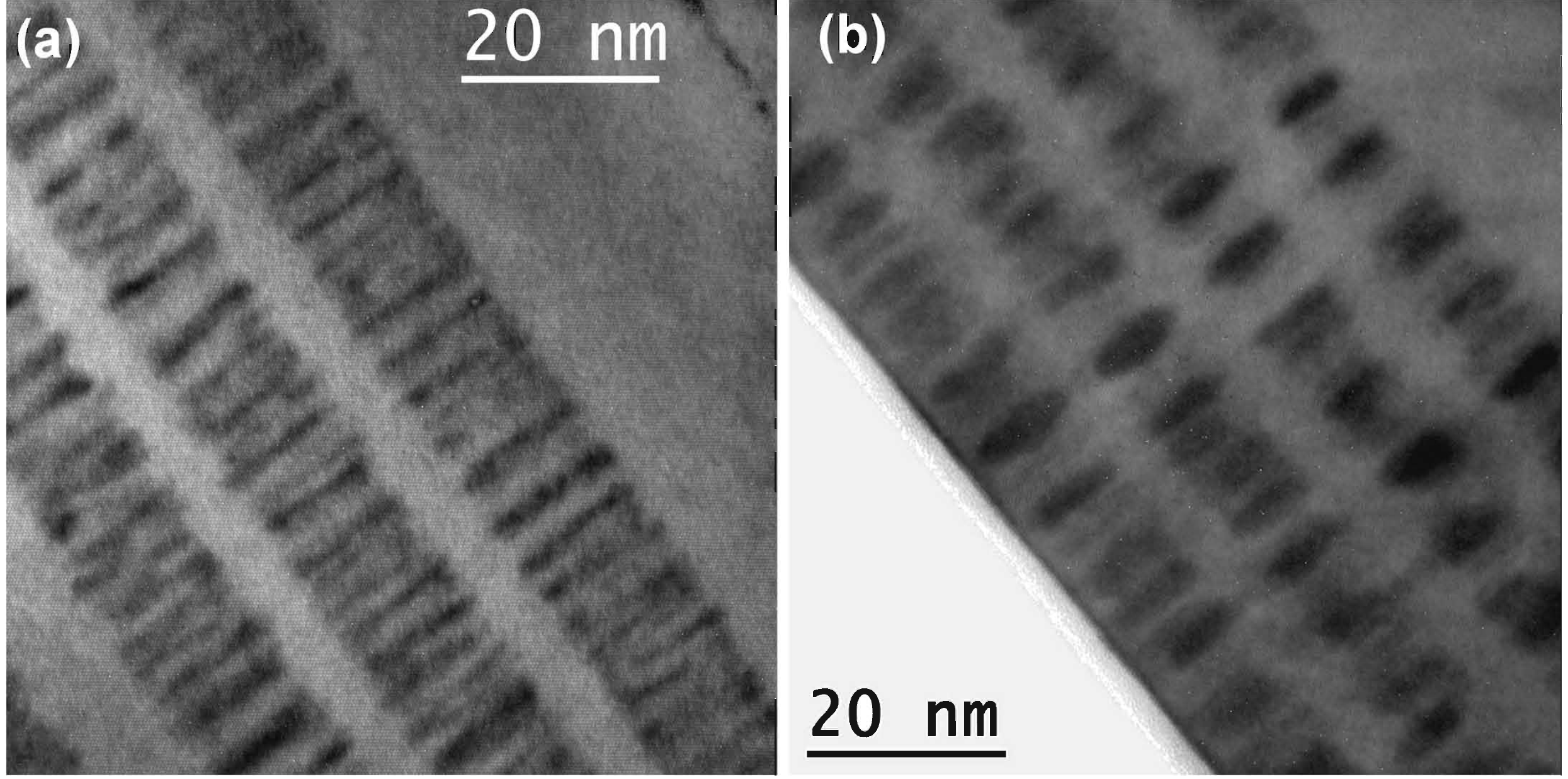}
\end{center}
\caption{(Color online) HRTEM images of (GeMn/Ge)$_{4}$ superlattices
observed along the
[110] crystal axis. From \onlinecite{Devillers:2008_PhD}. In (a), no vertical
correlation is observed whereas in (b)
the vertical positions are clearly correlated and columns well aligned.
In both cases, the germanium spacer is 5 nm thick.}
\label{jm_fig7}
\end{figure}

Considering the vertical geometry of nanocolumns and the Mn concentration contrast
between the columns and the Ge matrix, vertical transport
will mainly take place through nanocolumns. Since nanocolumns are FM
(see Sec.\,\ref{magnetic}), they could be used as conduction channels for spin
injection in nonmagnetic semiconductors such as Si and Ge. The vertical control of
nanocolumns is thus an important issue from a fundamental point of view as well as
for potential applications in spintronics. For this purpose, (GeMn/Ge)
superlattices have been grown using different growth conditions. The (GeMn/Ge)$_{4}$ superlattice
in Fig.\,\ref{jm_fig7}(a) was grown in \textit{standard} conditions (0.2\,\AA/s, 6\% of Mn,
$T_{g}=100^{\circ}$C) leading to the formation of small nanocolumns with
no vertical self-organization. The superlattice in Fig.\,\ref{jm_fig7}(b) was grown at higher
temperature ($140^{\circ}$C) using a lower growth rate (0.13 \AA/s) in order to
favor the surface diffusion of Mn. Moreover the Mn content was increased up to
10\% to grow larger columns. Assuming that the lattice parameters within
the columns and the Ge matrix are different, larger columns lead to an enhanced
residual strain field which favors the nucleation of nanocolumns on top of
each other as in quantum dot superlattices \cite{Tersoff:1996_PRL}. By this method,
the vertical self-organization of nanocolumns could be achieved \cite{Wang:2011_NRL}.
Since the interface between nanocolumns and Ge spacer are quite abrupt, 
the diffusion of Mn along the growth direction is limited and then
restricted to the top surface layer as discussed next.

\subsubsection{Two-dimensional spinodal decomposition}

In order to understand the growth mechanism of nanocolumns,
the first growth stage of (Ge,Mn) films on Ge(001) were investigated. 
For this purpose, STM images have been recorded after depositing 0.44\,\AA\ of Mn
on Ge(001) at $80^{\circ}$C [Fig.\,\ref{jm_fig8}(a)].

\begin{figure}[h!]
\begin{center}
\includegraphics[width=\columnwidth]{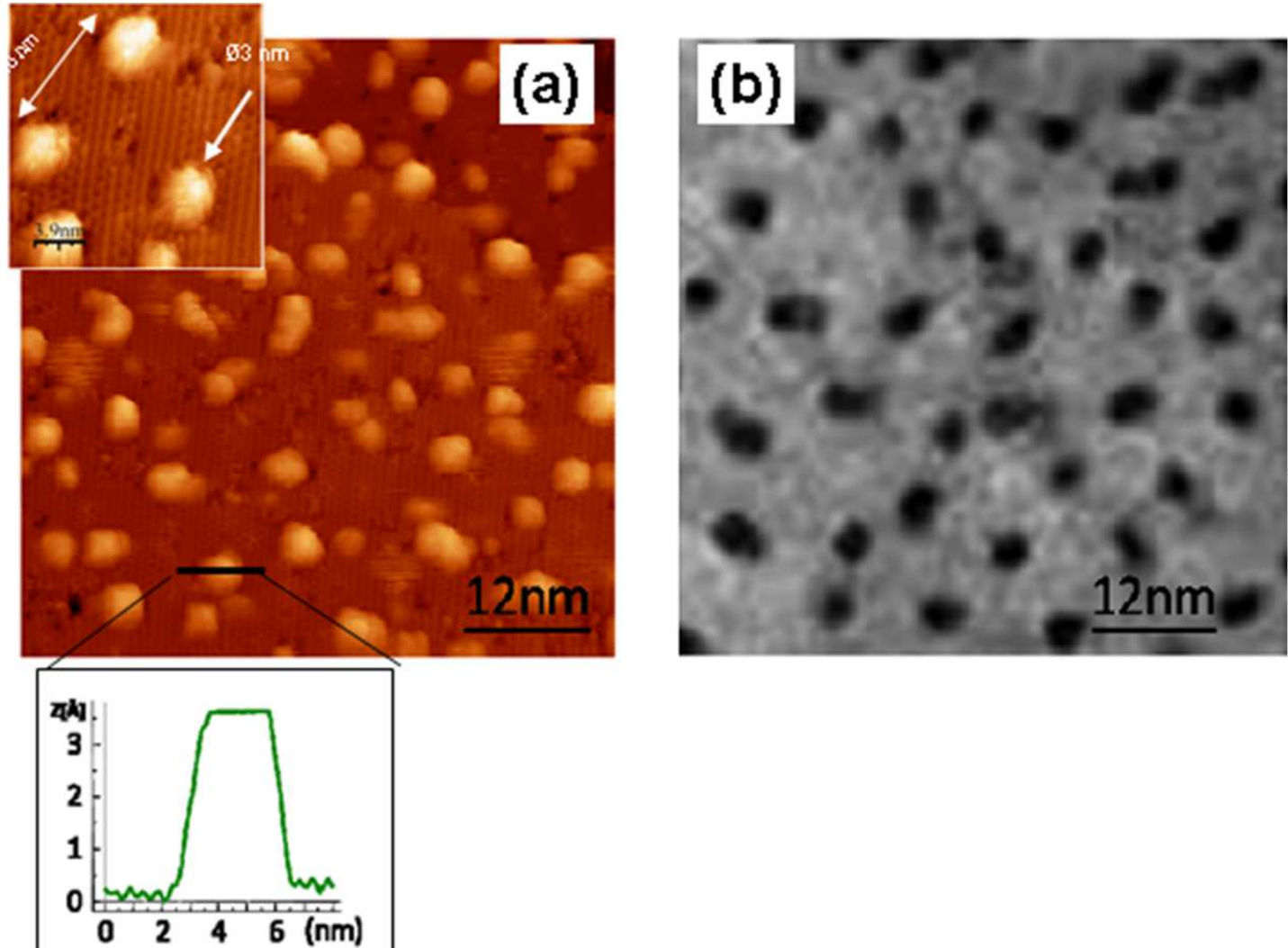}
\end{center}
\caption{(Color online) Formation of nanocolums from self-assembled nanoislands.
 (a) STM image of a germanium surface after depositing 0.44\,\AA\ of Mn
at 80$^{\circ}$C (from \onlinecite{Lethanh:2006_priv}). The inset shows the diameter
of self-assembled islands
($\simeq 3$ nm) and the distance between nearest-neighbors ($\simeq 10$ nm).
The reported $z$-profile gives an estimation of their height: 3.5\,{\AA}. (b) Plan
view TEM image of a 80\,nm thick Ge$_{0.94}$Mn$_{0.06}$ film grown at $130^{\circ}$C.
Adapted from \onlinecite{Jamet:2006_NM}.}
\label{jm_fig8}
\end{figure}

The images show the formation of self-assembled nanoislands. Their size and density are
further comparable to the size and density of nanocolumns in as-grown (Ge,Mn) films
[Fig.\,\ref{jm_fig8}(b)]. Therefore, nanoislands are most probably Mn-rich nuclei
at the origin of the formation of nanocolumns. Accordingly, this observation
implies that the columns size and density are fully determined at the first growth
stage. This growth mode leading to the formation of the konbu phase is summarized in Sec.~\ref{sec:theory}.
It results from 2D spinodal decomposition due to Mn-Mn chemical pair
attraction followed by layer-by-layer growth \cite{Fukushima:2006_JJAP,Zheng:2010_APL}.
The expected scenario for the growth of Mn-rich nanocolumns is illustrated
in Fig.\,\ref{jm_fig9}.

\begin{figure}[h!]
\begin{center}
\includegraphics[width=\columnwidth]{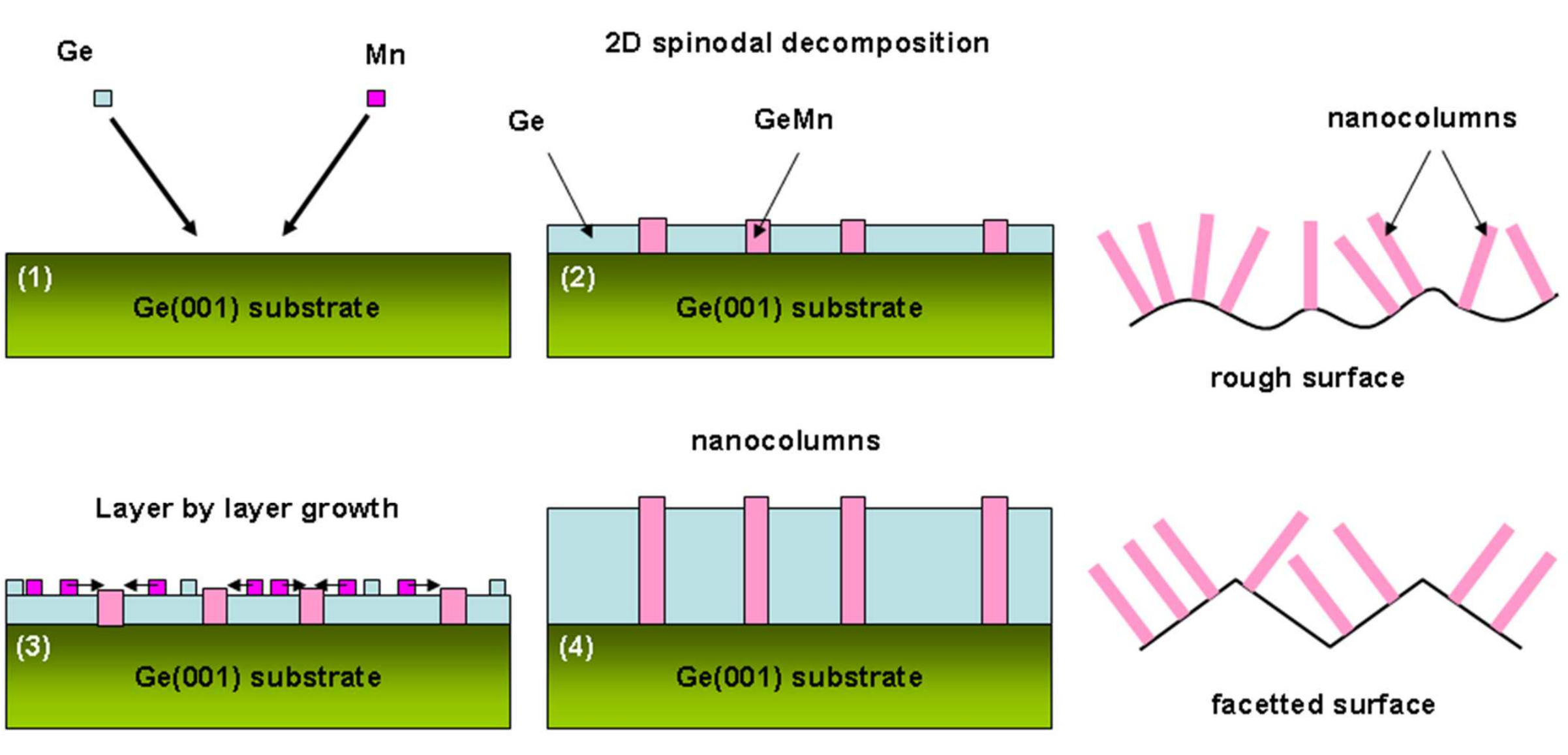}
\end{center}
\caption{(Color online) Nanocolumn growth and geometry. Left: four-steps scenario for the growth of Mn-rich nanocolumns
in germanium.
(1) Ge and Mn atoms are codeposited by LT-MBE on the germanium surface.
(2) Because of strong Mn-Mn pair attraction, a two-dimensional spinodal decomposition
takes place at the first growth stage setting the size and density of nanocolumns.
(3) Layer-by-layer growth with surface diffusion of Mn atoms. Because of Mn-Mn pair
attraction, Mn atoms aggregate on top of Mn-rich areas. (4) Mn-rich nanocolumns
perpendicular to the initial surface are obtained. Right: expected columns geometries
for a rough and facetted initial surface. From \onlinecite{Jamet:2010_unp}.}
\label{jm_fig9}
\end{figure}
As a consequence, nanocolumns will always grow perpendicular to the initial
surface. Therefore, they will exactly follow the initial surface roughness or
facets (Fig.\,\ref{jm_fig9}). In order to support this conclusion, thin (Ge,Mn)
films were grown in \textit{standard} growth conditions on a rough GaAs(001) surface as well as
on a facetted Ge surface \cite{Yu:2010_PRB,Yu:2011_JAP}.

\begin{figure}[h!]
\begin{center}
\includegraphics[width=\columnwidth]{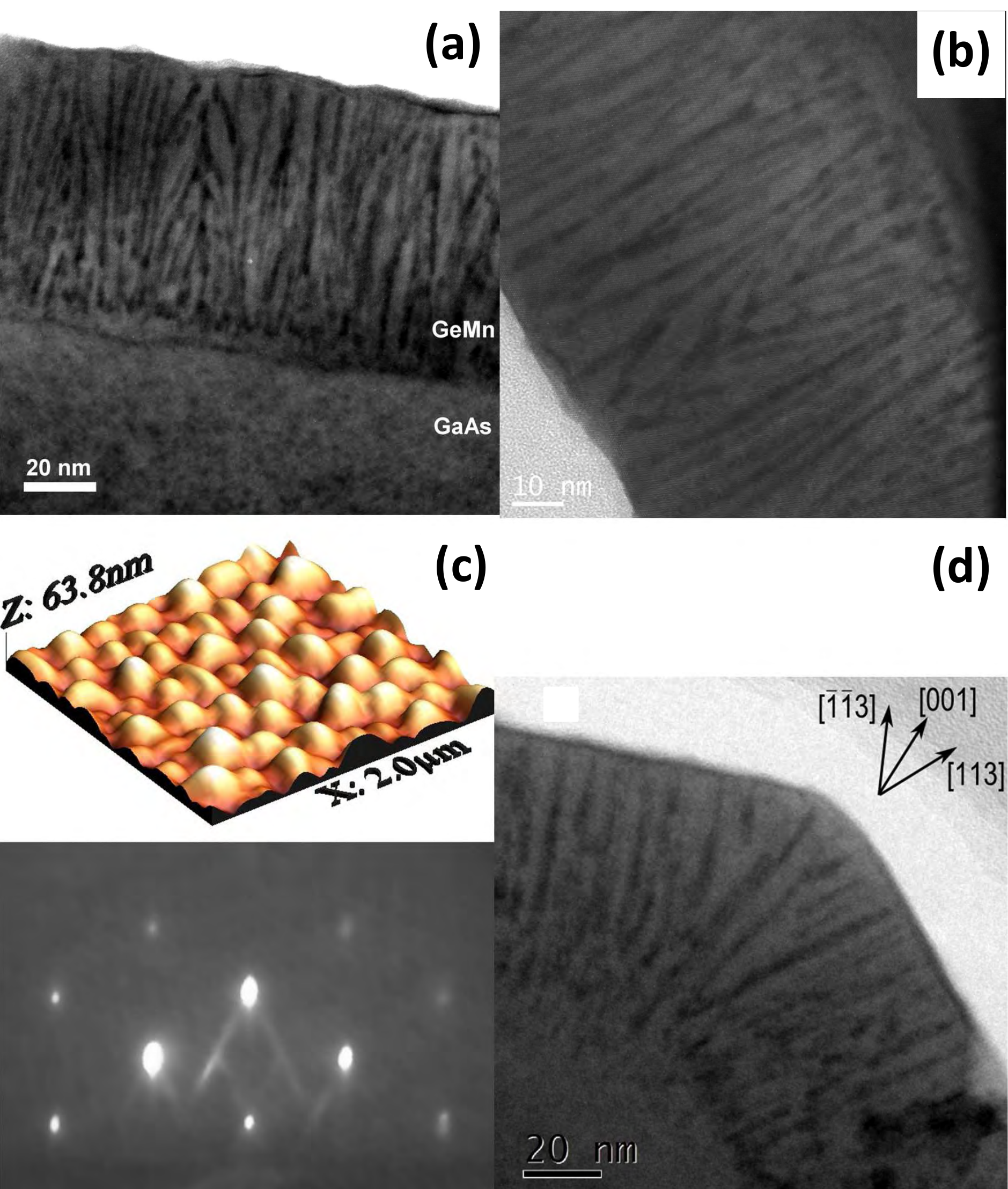}
\end{center}
\caption{(Color online) (a) Low and (b) HRTEM cross section images of a 80\,nm-thick
Ge$_{0.9}$Mn$_{0.1}$ film grown at $100^{\circ}$C on GaAs(001). Before the growth,
the native oxide layer was first thermally removed from the GaAs(001) substrate
at $600^{\circ}$C leading to a rough Ga-rich surface. (c) AFM image and RHEED
pattern of a facetted Ge surface obtained after anisotropic chemical etching
revealing \{113\} facets. (d) TEM cross section image of a 80\,nm-thick
Ge$_{0.94}$Mn$_{0.06}$ film grown at $100^{\circ}$C on the facetted Ge surface. From \onlinecite{Yu:2010_PRB}.}
\label{jm_fig10}
\end{figure}

The resulting (Ge,Mn) films were then observed by TEM in cross section
(Fig.\,\ref{jm_fig10}).
On GaAs(001), Mn-rich nanocolumns are bent following the initial surface
roughness. On the facetted Ge surface, nanocolumns have their axis perpendicular
to the \{113\} facets  \cite{Yu:2010_PRB}. Hence 2D spinodal decomposition and further
layer-by-layer growth are most probably the mechanisms responsible for
the formation of Mn-rich nanocolumns.

\subsubsection{Crystal structure of nanocolumns}

The crystalline structure of (Ge,Mn) films was investigated using high-resolution
TEM observations as well as XRD and XAS. High resolution TEM
observations in cross section show that Ge$_{1-x}$Mn$_{x}$ films grown below
180$^{\circ}$C are single crystalline in epitaxial relationship with the substrate
[Fig.\,\ref{jm_fig11}(a)]. Furthermore nanocolumns can hardly be distinguished. The interface between
the Ge buffer layer and the Ge$_{1-x}$Mn$_{x}$ film is flat and no defect propagates
from the interface into the film. Diffraction scans performed in $\theta/2\theta$
mode were acquired on a high-resolution diffractometer using synchrotron radiation.
The scans reveal only the (004) Bragg peak of the germanium
crystal, confirming the good epitaxial relationship between the layer and
the substrate, and the absence of secondary phases to a good accuracy proven by a high
ratio of the main peak intensity to noise, of the order of 10$^7$ [inset in Fig.\,\ref{jm_fig11}(a)].

\begin{figure}[h!]
\begin{center}
\includegraphics[width=\columnwidth]{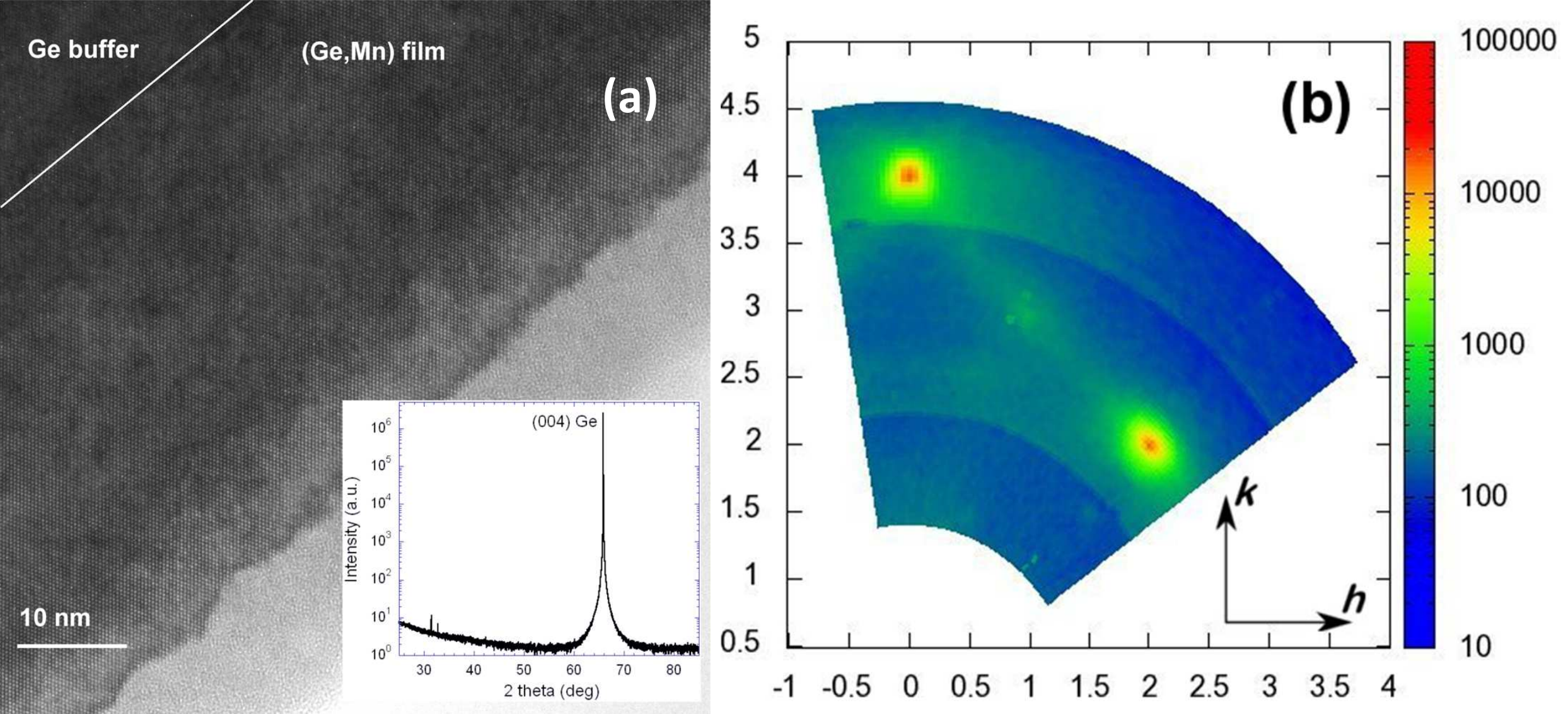}
\end{center}
\caption{(Color online) (a) HRTEM image of a 80\,nm thick Ge$_{0.94}$Mn$_{0.06}$
sample grown at 130$^{\circ}$C. Inset: XRD showing
the single (004) Bragg peak of pure Ge. (b) $(h,k,0)$ map of the reciprocal space
performed on a Ge$_{0.94}$Mn$_{0.06}$ film grown at 150$^{\circ}$C. Adapted from \onlinecite{Tardif:2010a_PRB}.}
\label{jm_fig11}
\end{figure}

In order to get further insight into the crystalline structure of nanocolumns,
grazing incidence SXRD was studied.
The reciprocal space of a Ge$_{0.94}$Mn$_{0.06}$ film grown at
150$^{\circ}$C [Fig.\,\ref{jm_fig11}(b)] was mapped. In addition to the (400) and (220) Bragg peaks of
germanium, two additional features were found: a broad (130) peak (which is
forbidden in the diamond lattice) and a diffuse line connecting (400) and (220)
Bragg peaks. At this stage, none of them could be related to the crystalline structure of
nanocolumns \cite{Tardif:2010a_PRB}. (220) and (400) Bragg peaks further exhibit correlation rings related to anisotropic elastic strain in the germanium matrix between nanocolumns \cite{Tardif:2010a_PRB}. Since the internal
crystalline structure of nanocolumns could not be resolved using XRD, three conclusions
can be drawn: (i) nanocolumns exhibit the same diamond lattice as germanium
with high Mn content, (ii) they are made of a metastable (Ge,Mn) phase showing
a diffraction pattern similar to that of pure germanium and (iii) they are amorphous.
In the following, the decisive role of the growth temperature on the columns crystalline structure is discussed.

\begin{figure}[h!]
\begin{center}
\includegraphics[width=\columnwidth]{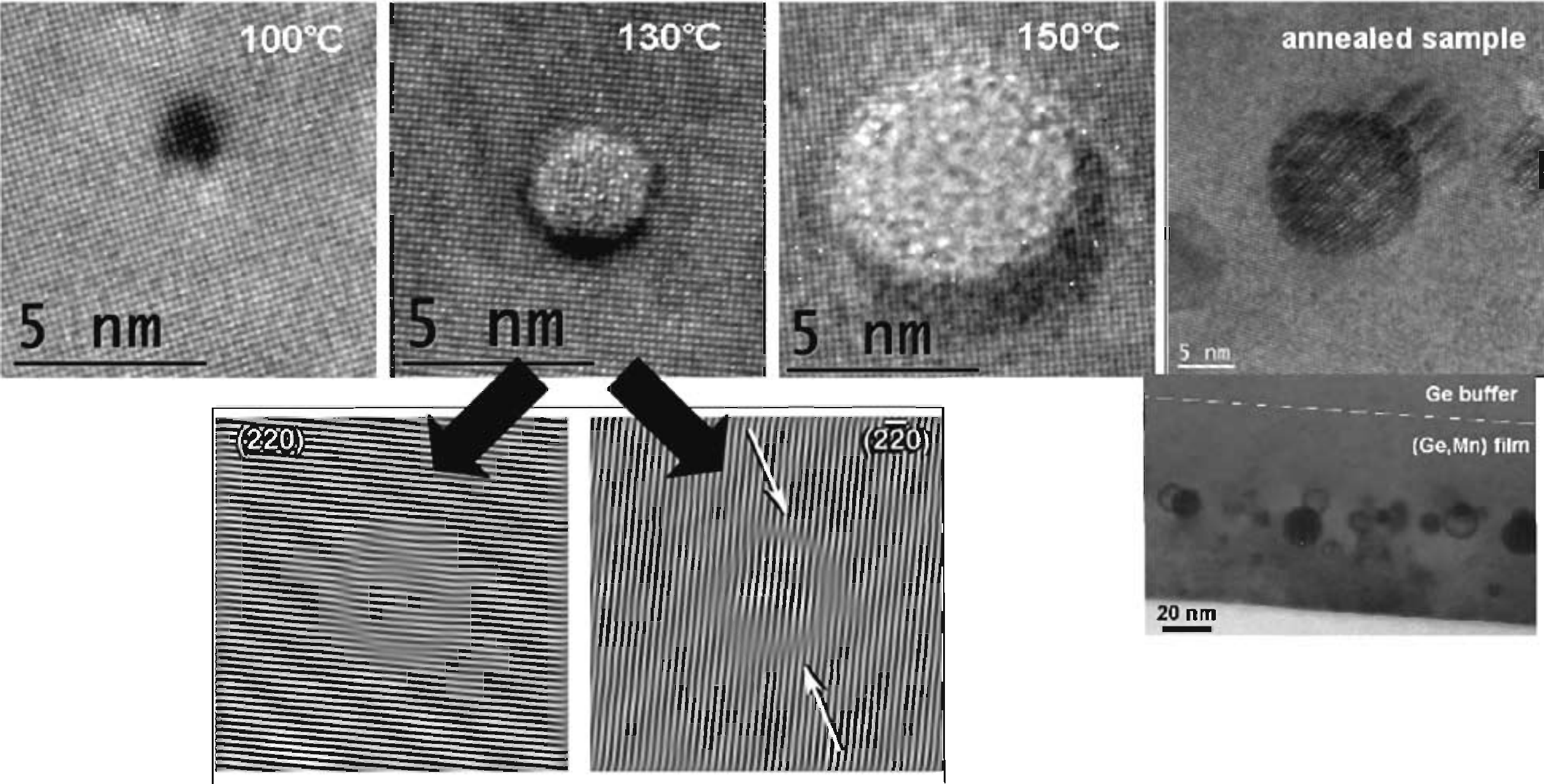}
\end{center}
\caption{(Color online) Crystalline structure of nanocolumns as a function of growth temperature.
From left to right: Ge$_{0.99}$Mn$_{0.01}$ film grown at $100^{\circ}$C;
Ge$_{0.94}$Mn$_{0.06}$ film grown at $130^{\circ}$C; Ge$_{0.887}$Mn$_{0.113}$
film grown at $150^{\circ}$C; Ge$_{0.94}$Mn$_{0.06}$ film grown at $130^{\circ}$C
and annealed at $650^{\circ}$C for 15 min. Adapted from \onlinecite{Jamet:2006_NM} and \onlinecite{Devillers:2007_PRB}.}
\label{jm_fig12}
\end{figure}

High-resolution TEM observations in plan view clearly demonstrate an evolution
of the columns structure with growth temperature (Fig.\,\ref{jm_fig12}).
For $T_{g}=100^{\circ}$C, small nanocolumns are fully strained on the surrounding
Ge matrix and clearly exhibit a perfect cubic crystal. For $T_{g}=130^{\circ}$C,
crystalline columns were still observed. However, performing (220) and (2$\bar{2}$0)
Bragg filtering of diffraction spots, a pair of dislocations
at the interface between the column and the Ge matrix is clearly visible. This leads to a lattice
expansion of almost 4\% in the column. Nanocolumns are thus only partially
strained. For $T_{g}=150^{\circ}$C, nanocolumns are amorphous. Finally,
for samples grown at $T_{{g}}> 180^{\circ}$C or annealed at high temperature,
nanocolumns have collapsed into stable Ge$_{3}$Mn$_{5}$ clusters.
For annealed samples, these clusters are further well buried inside
the germanium film away from the surface.


Since magnetic properties highly depend on the relative position of Mn atoms
in the nanocolumns, the local chemical environment of Mn atoms was probed by
EXAFS measurements \cite{Rovezzi:2008_APL}. By fitting EXAFS oscillations, the chemical environment of Mn atoms in the nanocolumns was found very similar to that of a Mn atom in an elementary tetrahedron of the Ge$_{3}$Mn$_{5}$ unit cell. This elementary tetrahedron corresponds to the building block of nanocolumns. However, at this stage, the structure of crystalline columns grown at low temperature still remains unknown. Nevertheless, a possible phase which exhibits a crystalline structure and composition compatible with the experimental findings was proposed by  \onlinecite{Arras:2012_PRB}. This phase is called the $\alpha$-phase and corresponds to a new class of materials based on the insertion of Mn atoms into a simple cubic crystal of Ge. \onlinecite{Arras:2012_PRB} showed that the incorporation of this phase (with a Mn content ranging from 10 to 50\%) in the form of nanocolumns into the pure Ge matrix is energetically favorable.

\subsection{\label{magnetic}Magnetic properties}

In this section, the magnetic properties of four different
Ge$_{1-x}$Mn$_{x}$ films grown at different temperatures are investigated: crystalline columns
with and without partial relaxation through the formation of a pair of dislocations,
amorphous nanocolumns and Ge$_{3}$Mn$_{5}$ clusters. For this purpose, highly sensitive SQUID magnetometry
as well as EPR was thoroughly used.

\subsubsection{\label{columns1}Crystalline (Ge,Mn) nanocolumns / Ge(001)}

Magnetic measurements on diluted Ge$_{0.999}$Mn$_{0.001}$ films
grown at 100$^{\circ}$C are first reported. Only a paramagnetic signal at low temperature
by SQUID and the six isotropic hyperfine absorption lines characteristic of
paramagnetic Mn atoms in EPR spectra (not shown) could be measured. This result is in good agreement
with TEM observations and EXAFS measurements. Indeed the absence of any contrast
in TEM images rules out the presence of nanocolumns and EXAFS oscillations could be
fitted with a single substitutional environment.

\begin{figure}[h!]
\begin{center}
\includegraphics[width=\columnwidth]{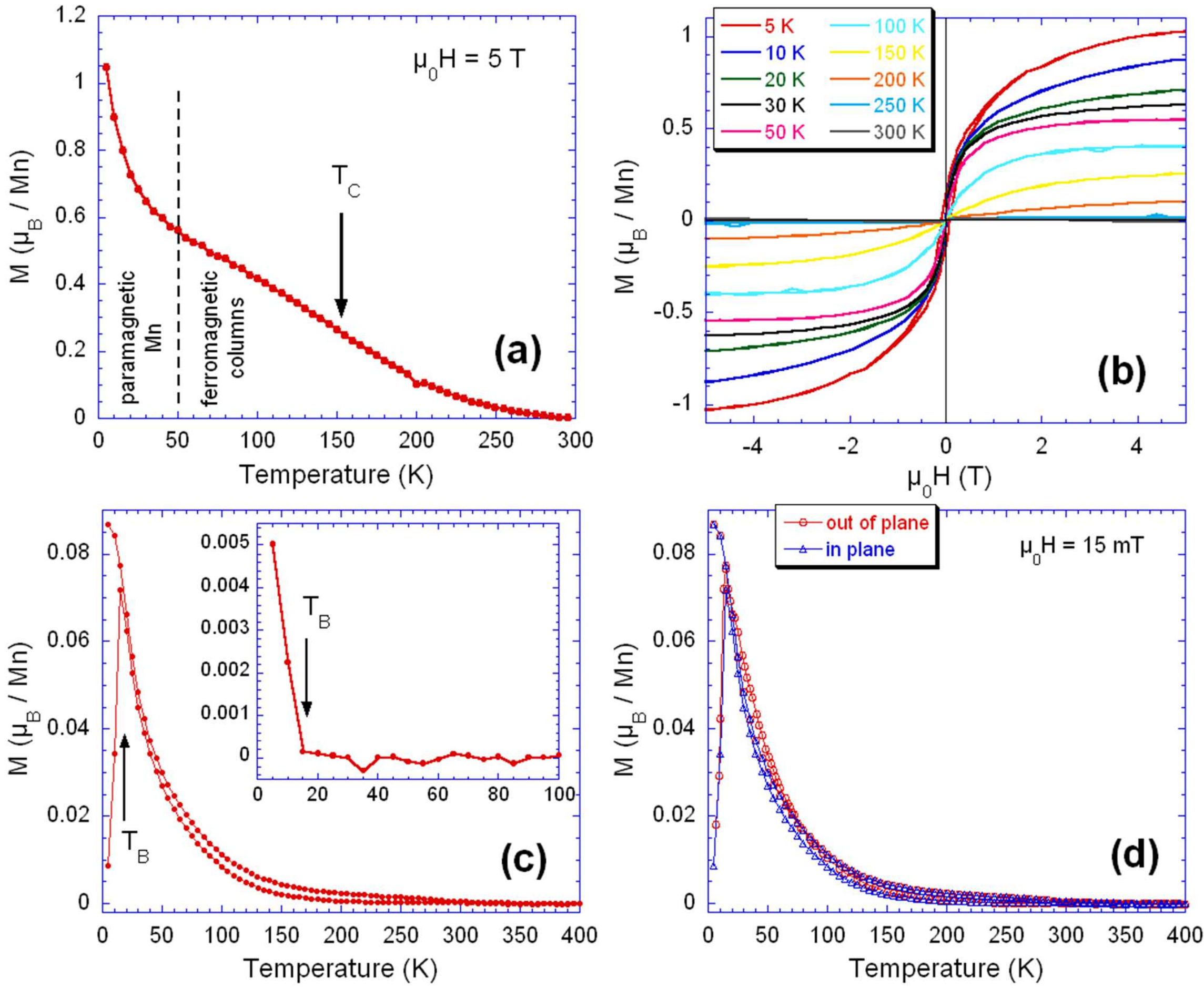}
\end{center}
\caption{(Color online) Magnetic characterization of a Ge$_{0.9}$Mn$_{0.1}$ film grown at $100^{\circ}$C.
(a) Temperature dependence of the saturation magnetization (in $\mu_{B}$/Mn). The applied field is 5\,T.
(b) Magnetization curves for different temperatures and (c) ZFC-FC measurements, pointing
to $T_{\text{C}} \approx 150$\,K of nanocolumns.
The in-plane applied field is 0.015\,T. Inset: Magnetic remanence after maximum
field cooling under 5\,T. (d) ZFC measurements performed with the field parallel and
perpendicular to the film plane. From \onlinecite{Jain:2010_APL}.}
\label{jm_fig13}
\end{figure}

For more concentrated films grown at $T_{\rm{g}} =100^{\circ}$C, small nanocolumns
spanning the whole film thickness are observed by TEM. Moreover they are well
crystalline and in perfect epitaxial relationship with the Ge matrix
(see Fig.\,\ref{jm_fig12}).
Assuming that the Mn concentration in the Ge matrix is below 0.05\%,
the resulting Mn content in the columns is close to 50\%. SQUID measurements
performed on a Ge$_{0.9}$Mn$_{0.1}$ film are reported in Fig.\,\ref{jm_fig13}.
Two different contributions in Fig.\,\ref{jm_fig13}(a) can be identified: a paramagnetic signal from isolated Mn atoms and the FM nanocolumns.
It has been shown recently that the paramagnetic Mn atoms are located within a thin shell around nanocolumns \cite{Dalmas:2015_PRB}. The $T_{\text{C}}$ magnitude of the columns is
of the order of 150\,K, according to Figs.\,\ref{jm_fig13}(b,c)  \cite{Jain:2010_APL}.
In Fig.\,\ref{jm_fig13}(b), magnetization curves are reversible above 15\,K,
also confirmed by ZFC-FC curves and the dependence of remanence with temperature
[Fig.\,\ref{jm_fig13}(c)]. These nanocolumns are superparamagnetic with a blocking temperature of
$15 \pm 5$\,K. Moreover the narrow shape of the ZFC peak is related to the narrow size
distribution of nanocolumns grown in this temperature range (see Fig.\,\ref{jm_fig6}).
To
determine the magnetic anisotropy of these columns, ZFC-FC
 measurements were performed with the field parallel and perpendicular
to the film plane. The susceptibility perpendicular to the plane seems slightly
higher than that in the plane but a strong diamagnetic signal from the substrate
and paramagnetic signal from diluted Mn atoms makes it difficult to quantify
precisely this anisotropy.

\begin{figure}[h!]
\begin{center}
\includegraphics[width=\columnwidth]{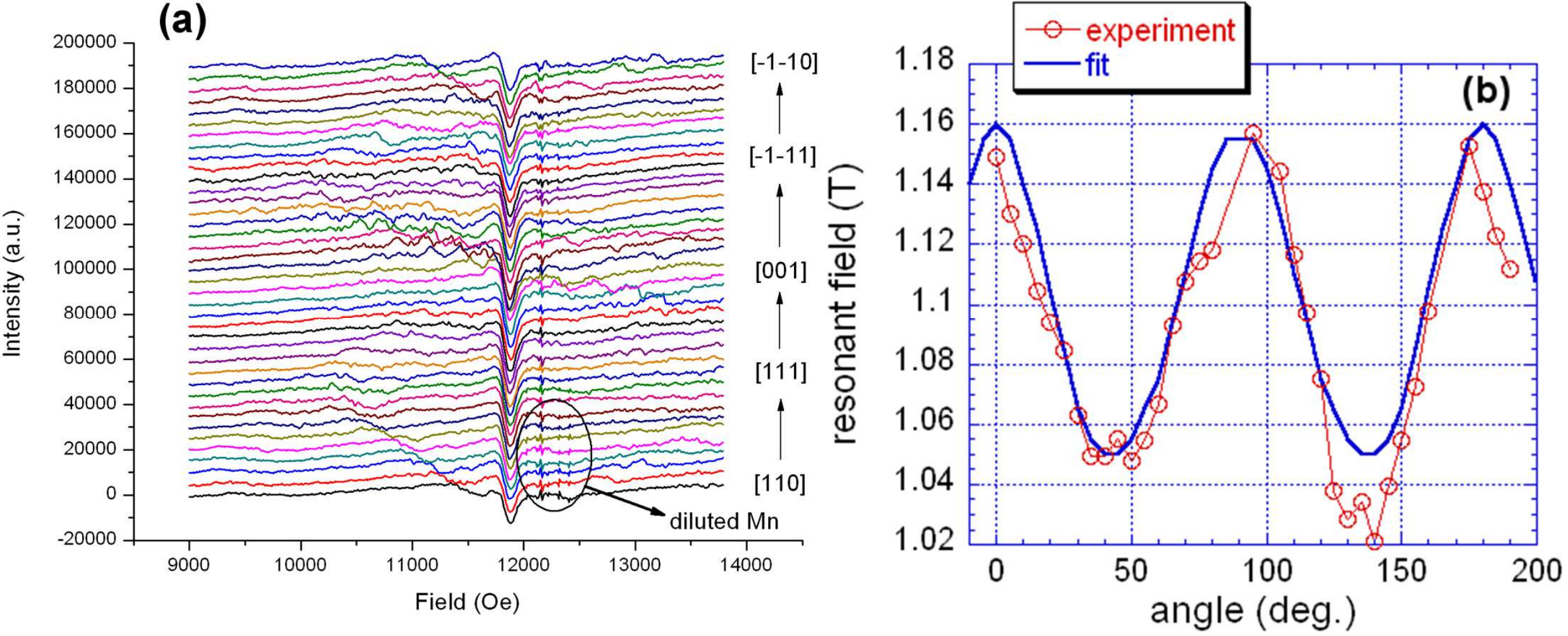}
\end{center}
\caption{(Color online) (a) EPR absorption spectra of a Ge$_{0.9}$Mn$_{0.1}$ film grown at
$100^{\circ}$C as a function of the static field orientation at 5\,K. The angle
is defined between the [110] direction and the applied magnetic field.
The working frequency is 34\,GHz ($Q$-band setup). Two different
contributions can be observed: a FM broad peak of weak anisotropy and six hyperfine
peaks from diluted paramagnetic Mn atoms. (b) Angular dependence of the resonance field.
From \onlinecite{Jain:2010_APL}.}
\label{jm_fig14}
\end{figure}

EPR technique is hence used to differentiate the FM signal of nanocolumns
from the paramagnetic diluted Mn atoms and estimate magnetic anisotropy \cite{Jain:2010_APL}.
The EPR spectra are observed as a function of the angle of applied magnetic field.
The field is applied out-of-plane from the direction [110] to direction
[$\bar{1}\bar{1}0$] passing through the [001] direction. In Fig.\,\ref{jm_fig14} the EPR
spectra of a Ge$_{0.9}$Mn$_{0.1}$ film with crystalline nanocolumns exhibit
the hyperfine Mn lines corresponding to isolated paramagnetic Mn atoms and also a weak
FM line attributed to nanocolumns. The angular dependence of
the resonance field in Fig.\,\ref{jm_fig14}(b) is fitted using the Smit-Beljers
formalism \cite{Smit:1955_PRR}. For this purpose, it is assumed that nanocolumns are
noninteracting and single domain at 5\,K. The magnetic anisotropy energy of
a single nanocolumn is written as: $F=K_{2}m_{z}^{2}+K_{4}(m_{x}^{2}m_{y}^{2}
+m_{x}^{2}m_{z}^{2}+m_{y}^{2}m_{z}^{2})$ where $x$, $y$, and $z$ axes are the [100],
[010], and [001] crystal axes respectively. $m_{x}$, $m_{y}$, and $m_{z}$ are
the coordinates of a unit vector along the column magnetization direction 
and $K_{2}$ and $K_{4}$ are the second and fourth order magnetic anisotropy constants.
Anisotropy fields are thus defined as  $\mu_{0}H_{a2}=2 K_{2}/M_{S}$ and
$\mu_{0}H_{a4}=2 K_{4}/M_{S}$. From EPR data, the anisotropy fields can be deduced from EPR data:
$\mu_{0}H_{a2}\simeq - 0.09$\,T for the second order and $\mu_{0}H_{a4}\simeq-0.11$\,T
for the fourth order with $\gamma/\gamma_{e}\simeq 1.07$ (where $\gamma_{e}$ is
the gyromagnetic ratio of a free electron). Hence, the nanocolumns exhibit
a perpendicular uniaxial anisotropy and a cubic anisotropy with the easy axis along
[111]. The presence of cubic anisotropy supports the crystallinity of nanocolumns.
Also, cubic anisotropy and uniaxial anisotropy constants are of the same order of
magnitude making the two crystal axes [$110$] and [$001$] close in energy. It thus
explains the small difference between the ZFC curves for the field parallel and
perpendicular to the film plane in the SQUID measurements of Fig.\,\ref{jm_fig13}(d).
Considering
the saturation magnetization of these small nanocolumns $M_{S}=140 \pm 20$ kA/m,
the uniaxial anisotropy constant can be calculated as $K_{2}\simeq-0.63\times 10 ^{4}$
J/m$^{3}$ which corresponds to shape anisotropy
$\mu_{0}M_{S}^{2}/4\simeq 0.62\times 10^{4}$ J/m$^{3}$ assuming that
nanocolumns are infinite cylinders \cite{Aharoni:1996_B}.

\begin{figure}[h!]
\begin{center}
\includegraphics[width=\columnwidth]{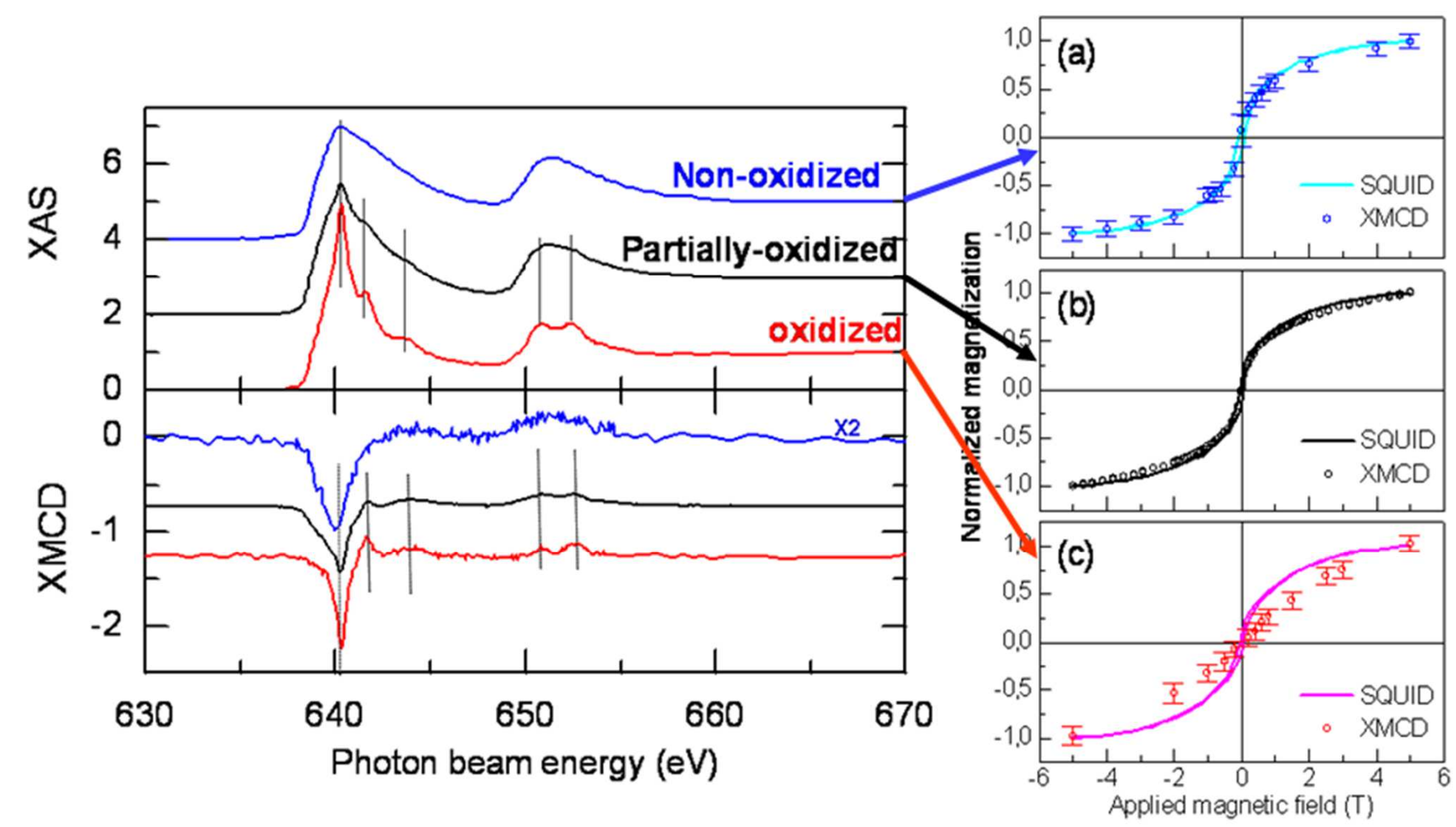}
\end{center}
\caption{(Color online) Isotropic XAS spectra at the Mn $L_{2,3}$ edge and
corresponding XMCD signal recorded at 5\,K under 5\,T. Data were acquired in the total
electron yield mode. Measurements were performed on a Ge$_{0.9}$Mn$_{0.1}$ film grown
at $100^{\circ}$C exhibiting different surface states: nonoxidized, partially
oxidized, and oxidized. The spectra are shifted vertically for clarity.
The XMCD signal recorded at the Mn $L_{3}$ edge is then compared with the SQUID
signal at 5\,K. Both XMCD and SQUID signals are normalized to their maximum value.
Adapted from \onlinecite{Tardif:2010b_APL}.} \label{jm_fig15}
\end{figure}
Finally in order to verify that the magnetic signal detected by SQUID magnetometry
really comes from Mn-rich nanocolumns, XMCD measurements were carried out at the Mn $L_{2,3}$ edge
\cite{Gambardella:2005_PRB,Ahlers:2009_APL}. In Fig.\,\ref{jm_fig15},
when the Ge$_{0.9}$Mn$_{0.1}$ film is properly
capped with a Si/Ge layer to prevent surface oxidation \cite{Tardif:2010b_APL}, the XMCD
signal fits the SQUID measurements. Moreover the magnetic moment per Mn atom
deduced by sum rules \cite{Thole:1992_PRL} from XMCD data also corresponds well to
the magnetic moment measured by SQUID: $\simeq 1\mu_{B}$/Mn.
Electrical properties have been intensively investigated in such (Ge,Mn) films grown on semi-insulating GaAs substrates \cite{Yu:2010_PRB}. A negative giant MR signal has been detected at low temperature due to spin scattering on FM nanocolumns. In addition, two positive MR contributions were observed: the first one arises from ordinary or Lorentz MR and the second one was attributed to geometrically enhanced MR as a consequence of the conductivity contrast between the nanocolumns and the matrix \cite{Solin:2000_S}. Then AHE was demonstrated at low temperature but the interpretation of magnetotransport in such an inhomogeneous system remains a challenging task \cite{Yu:2011_JAP}.

\subsubsection{\label{columns2}Crystalline (Ge,Mn) nanocolumns with partial relaxation / Ge(001)}

Ge$_{0.94}$Mn$_{0.06}$ films grown in a narrow temperature range around
130$^{\circ}$C exhibit partial lattice relaxation (see Fig.\,\ref{jm_fig12}) and
ferromagnetism above RT up to 400\,K as shown
in Fig.\,\ref{jm_fig16} \cite{Jamet:2006_NM}.
This high-$T_{\text{C}}$ FM behavior was also observed in nanowires and
quantum dots \cite{Kazakova:2005_PRB,Meulen:2009_NL,Cho:2008_CM,Xiu:2010_NM}.

\begin{figure}[h!]
\begin{center}
\includegraphics[width=\columnwidth]{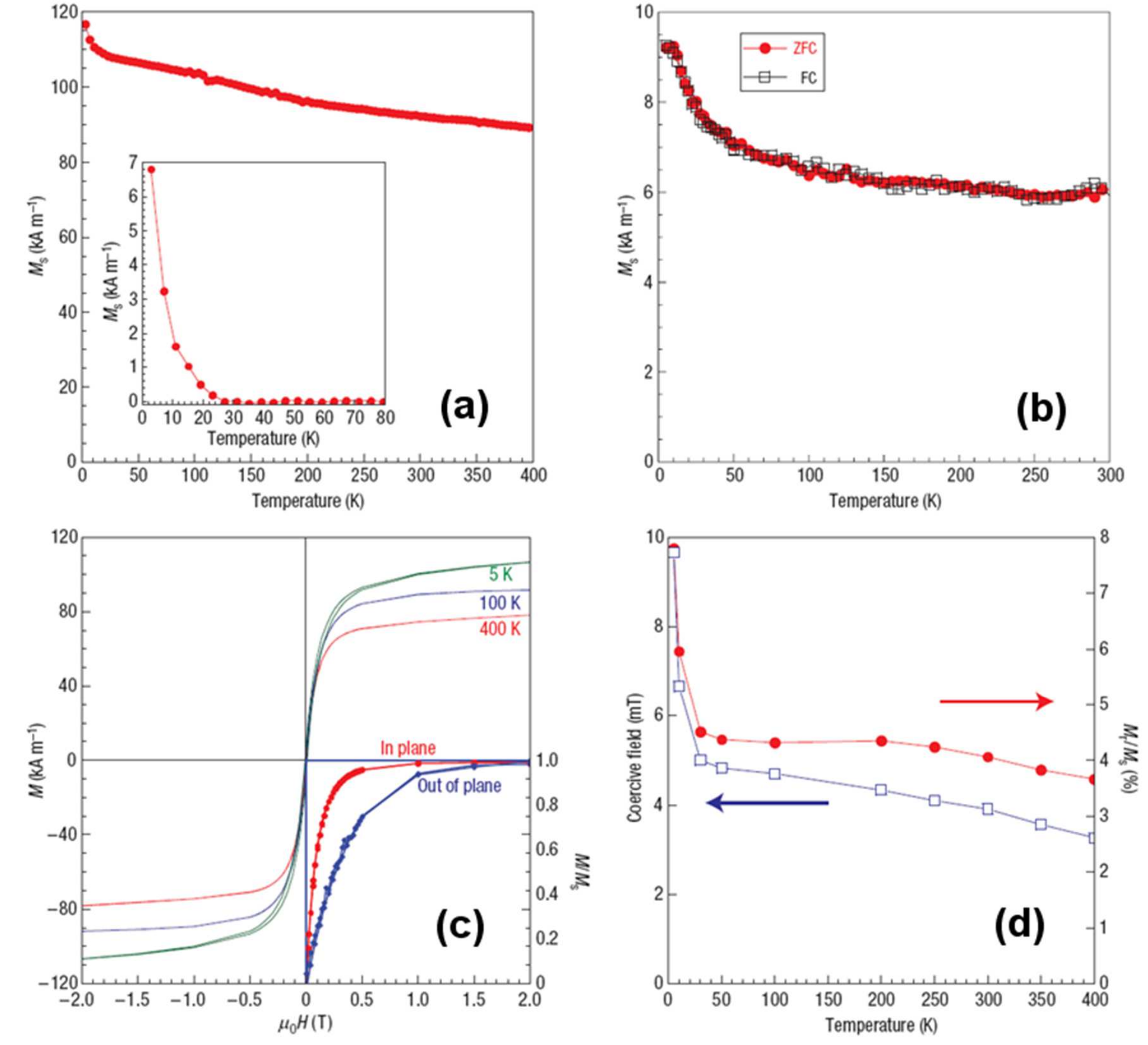}
\end{center}
\caption{(Color online) Magnetic characterization of Ge$_{0.94}$Mn$_{0.06}$ films grown around
$130^{\circ}$C. (a) Temperature dependence of the saturation magnetization measured at 2\,T.
The inset shows the extrapolated matrix signal at low temperature after subtracting
the nanocolumns magnetic signal. (b) ZFC-FC measurements performed at 0.01\,T.
Both curves are superimposed. (c) Magnetization loops at 5, 100, and 400\,K, after
subtracting the diamagnetic contribution from the substrate. The inset demonstrates
the easier saturation in-plane at 250\,K. (d) Coercive field ($\mu_{0}H_{C}$) and
remanent magnetization ($M_{r}/M_{S}$) versus temperature. $\mu_{0}H_{C}$ and
$M_{r}/M_{S}$ are given with a precision of the order of 10\%. From \onlinecite{Jamet:2006_NM}.}
\label{jm_fig16}
\end{figure}

Figure~\ref{jm_fig16}(a)
shows the temperature dependence of the magnetization at 2\,T, measured
by SQUID magnetometry. The magnetic moment per Mn atom is 4.7\,$\mu_{B}$ at 3\,K.
This value is close to 5$\mu_{B}$ expected for isolated Mn$^{2+}$ ions according
to Hund's rule. The large magnetization at high temperature is consistent with
a FM phase with $T_{\text{C}}> 400$\,K. ZFC-FC curves superimpose from 3 to
300\,K [Fig.\,\ref{jm_fig16}(b)],
thus ruling out the presence of superparamagnetic particles except
if their blocking temperature exceeds 300\,K. Considering the Mn distribution
in the GeMn films, this high $T_{\text{C}}$ FM phase can be attributed to
the nanocolumns. In the low-temperature range, the magnetization increases
when decreasing the temperature [inset in Fig.\,\ref{jm_fig16}(a)]. The corresponding saturation
magnetization is small ($\simeq 9$\,kA/m), and the additional susceptibility is
described by a Curie-Weiss temperature between 10 and 15\,K. Similar behavior was
reported for a strongly diluted Ge$_{1-x}$Mn$_{x}$ layer. Assuming that isolated Mn atoms
are present in the film, all magnetically active with a magnetic moment of
3$\mu_{B}$ \cite{Schulthess:2001_JAP}, the Mn concentration
in the nanocolumns reaches 33.7\% (their composition is close to Ge$_{2}$Mn).
The magnetic moment per Mn atom is close to 4.9$\mu_{B}$ from the saturation
magnetization $\simeq$680\,kA/m. Magnetization loops [Fig.\,\ref{jm_fig16}(c)] all exhibit
a pronounced S shape, a low remanent magnetization and a low coercive field
[Fig.\,\ref{jm_fig16}(d)]: in-plane and out-of-plane directions act as hard magnetic axes.
The saturation is easier with the field applied in-plane [inset in Fig.\,\ref{jm_fig16}(c)].

Nanocolumns are very close to each other and a strong magnetostatic or carrier
mediated coupling is expected. Energy calculations using the Fourier
formalism \cite{Beleggia:2004_JMMM} were performed for an infinite square lattice of nanocolumns
defined by the average experimental parameters (spacing 10\,nm, diameter 3\,nm and
height 80\,nm), taking into account the long-range dipolar interaction and
dipolar self-energy of the nanocolumns. Three configurations have been tested:
saturated out-of-plane ($\uparrow\uparrow$), saturated in-plane
($\rightarrow\rightarrow$) and antiparallel out-of-plane with $M = 0
(\uparrow\downarrow)$. Results are given in units of $\mu_{0}M_{S}^{2}V/2$
where $V$ is the volume of the nanocolumns and $M_{S}$ their magnetization
at saturation. The smallest energy, $E_{\uparrow\downarrow}\simeq 0.01$,
corresponds to the antiparallel out-of-plane configuration: dipolar interactions
favor an antiparallel arrangement of first neighbor nanocolumns. This finding is
in good agreement with the low remanent magnetizations, low coercive fields and
S shaped in-plane and out-of-plane magnetization curves measured by SQUID.
In real samples, the remanent magnetization is not zero because nanocolumns are
randomly distributed, leading to many frustrated magnetic configurations.
The model however fails to predict the right saturated configuration: we
calculate $E_{\rightarrow\rightarrow}=0.46$ and $E_{\uparrow\uparrow} =0.08$,
while experimentally the magnetization is easier to saturate in-plane
($E_{\rightarrow\rightarrow}<E_{\uparrow\uparrow})$.
Thus an additional in-plane magneto-crystalline anisotropy has to be considered, competing with shape
anisotropy, which favors the in-plane saturated state. This behavior has previously been
observed in arrays of electrodeposited cobalt nanowires \cite{Darques:2004_JPC}.
In (Ge,Mn) nanocolumns, this magnetocrystalline anisotropy originates
from the broken in-plane cubic symmetry (columns are fully strained in only
one direction) leading to a natural uniaxial anisotropy combined with large
magnetoelastic effects (columns are 4\% contracted). This feature also explains
the very high (at least higher than 400\,K) blocking temperature of nanocolumns.

As previously discussed in III-V and II-VI FM
semiconductors \cite{Dietl:2001_PRB,Kossacki:2004_PE}, strain can modify the valence band
dispersion and induce a large magnetic anisotropy. Carrier mediated FM
coupling between the nanocolumns would also favor an in-plane saturated state.
Note that in calculating the self-energy that the column aspect
ratio can be overestimated since the columns are not perfect cylinders.
Finally, these (Ge,Mn) films grown on Ge exhibit large positive MR attributed 
to geometrically enhanced MR \cite{Solin:2000_S} and more importantly AHE 
that might be attributed to the spin polarization of holes \cite{Jamet:2006_NM}. 
That makes this system a promising candidate for spintronics applications.

\subsubsection{\label{columns3}Amorphous (Ge,Mn) nanocolumns / Ge(001) }

For Ge$_{1-x}$Mn$_{x}$ films grown at higher temperatures ($T>150^{\circ}$C) and
high Mn concentration ($\geq 10$\%), amorphous nanocolumns are observed in
TEM (see Fig.\,\ref{jm_fig12}).
As for small crystalline nanocolumns grown at 100$^{\circ}$C
(see Sec.\,\ref{columns1}), the Mn concentration in the columns is close to 50\%.
These amorphous nanolumns have $T_{\text{C}}$ close to 170~K.
The contribution from isolated paramagnetic Mn atoms is still observed at low temperature,
as shown in Fig.\,\ref{jm_fig17}.

\begin{figure}[h!]
\begin{center}
\includegraphics[width=\columnwidth]{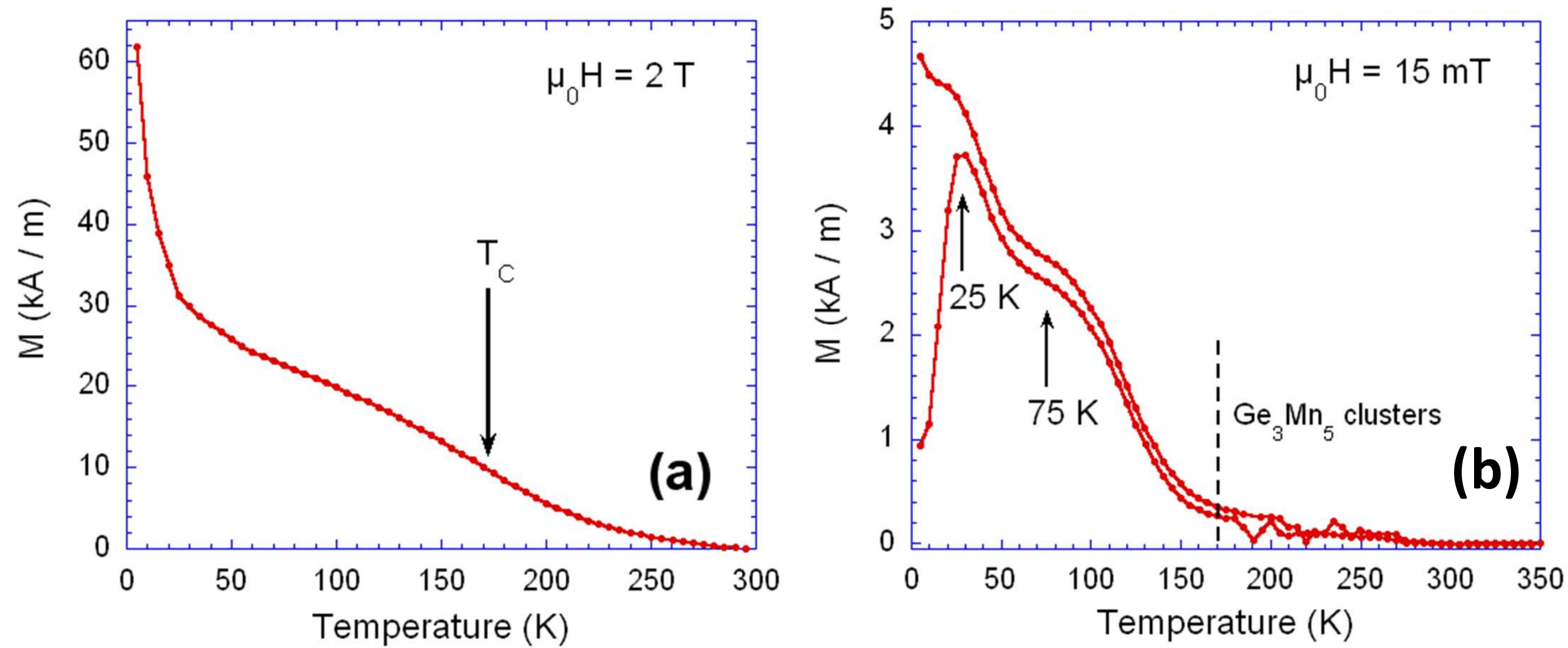}
\end{center}
\caption{(Color online) Magnetic characterization of a Ge$_{0.9}$Mn$_{0.1}$ film grown above 150$^{\circ}$C
with amorphous nanocolumns.
(a) Temperature dependence of the saturation magnetization in
the magnetic field of 2\,T
applied in the film plane along the [110] direction. (b) ZFC-FC measurements
performed at 0.015 T. The ZFC curve exhibits two peaks with blocking temperatures
around 25 and 75\,K, and points to $T_{\text{C}} \simeq 170$\,K of
nanocolumns. Nonzero magnetization at higher temperatures is due to
Ge$_{3}$Mn$_{5}$ NCs. From \onlinecite{Jain:2010_APL}.}
\label{jm_fig17}
\end{figure}

In ZFC-FC curves [Fig.\,\ref{jm_fig17}(b)] two peaks are present at around 25\,K and 75\,K in the ZFC
curve as well as a weak contribution from Ge$_{3}$Mn$_{5}$ clusters. Those peaks are
in agreement with the broad size distribution observed by TEM (see Fig.\,\ref{jm_fig6}).

\begin{figure}[h!]
\begin{center}
\includegraphics[width=\columnwidth]{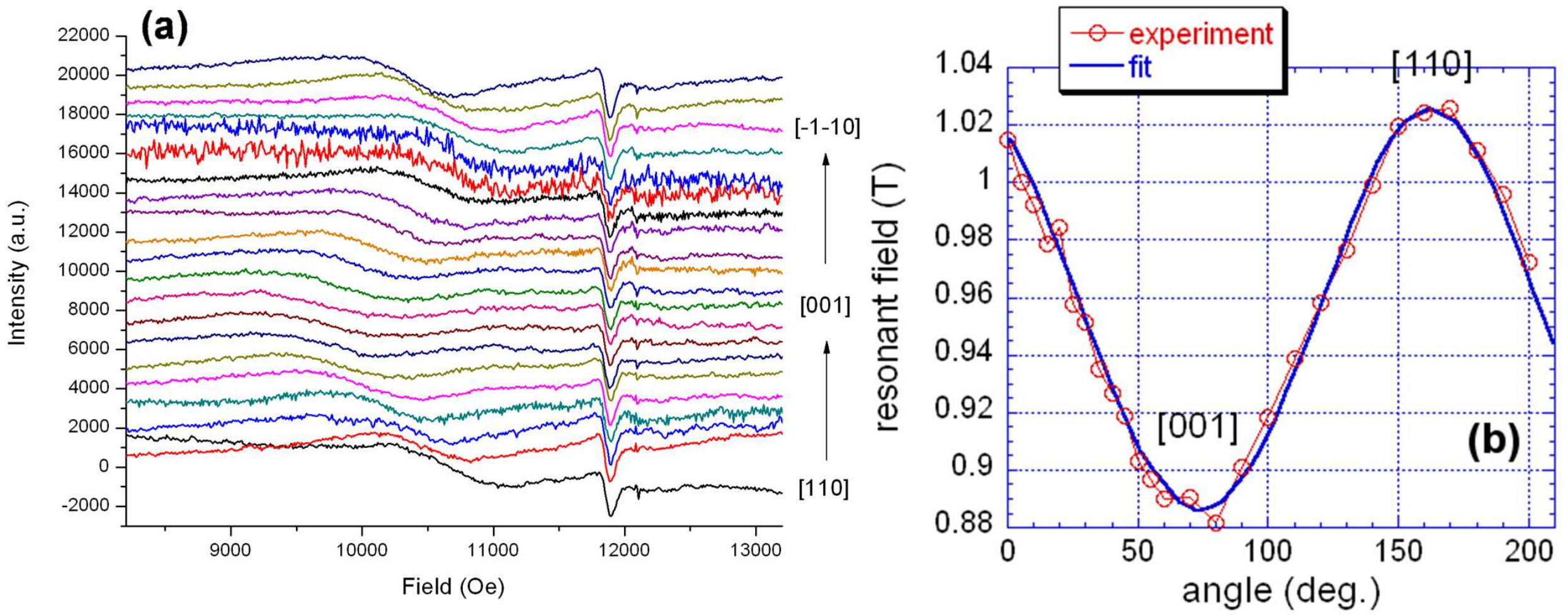}
\end{center}
\caption{(a) EPR absorption spectra of a Ge$_{0.9}$Mn$_{0.1}$ film grown
at 150$^{\circ}$C as a function of the static field orientation at 5 K.
The angle is defined between the [110] direction and the applied magnetic field.
The working frequency is 34 GHz (Q-band set-up). A single FM
broad peak of weak anisotropy is observed. (b) Angular dependence of the resonance field.
From \onlinecite{Jain:2010_APL}.}
\label{jm_fig18}
\end{figure}

The magnetic anisotropy of these columns is studied by EPR \cite{Jain:2010_APL}.
The data are shown
in Fig.\,\ref{jm_fig18}
and the angular dependence of the resonance field is fitted using the
Smit-Beljers formalism \cite{Smit:1955_PRR} assuming that the nanocolumns are noninteracting
and single domain. A single perpendicular uniaxial anisotropy is found;
the corresponding anisotropy field is  $\mu_{0}H_{a2}\simeq - 0.09$\,T and
$\gamma/\gamma_{e}\simeq1.24$. From the saturation magnetization of amorphous
nanocolumns $M _{S} =220 \pm 20$\,kA/m, the anisotropy energy density is calculated
to be $K _{2}\simeq -1.1 \times 10 ^{4}$ J/m$^{3}$. This value corresponds to
that of shape anisotropy $\mu_{0}M_{S}^{2}/4 =  1.5\times 10^4$\,J/m$^3$ of
nanocolumns assuming their aspect ratio is high \cite{Aharoni:1996_B}. As expected
for amorphous nanocolumns, no cubic magneto-crystalline contribution is found.
In the case of uniaxial magnetic anisotropy, the blocking
temperature of nanocolumns can be easily estimated using the classical N\'eel-Brown formula \cite{Jamet:2001_PRL}:
$KV=25k_{B}T_{B}$, where $K$ is the anisotropy constant, $V$ the columns volume,
$k_{B}$ the Boltzmann constant and $T_{B}$ the blocking temperature. From the average
diameter and length of nanocolumns (6\,nm and 80\,nm respectively) given by TEM
observations, one can estimate their volume and the blocking temperature:
$T_{B}\simeq 72$\,K which corresponds to the second broad peak in the ZFC curve.
The first peak with a maximum close to 25\,K can correspond either to shorter magnetic
nanocolumns with an average length of $\simeq(25/72)\times80$\,nm = 28\,nm or to 
the manifestation of a disordered magnetic state within the thin shell around nanocolumns and containing substitutional Mn atoms \cite{Dalmas:2015_PRB}.
In the first assumption it means that increasing the growth temperature partly activates Mn diffusion
along the growth direction, which leads to the formation of elongated clusters
rather than continuous nanocolumns spanning the whole film thickness.

\subsubsection{\label{cluster}Ge$_{3}$Mn$_{5}$ clusters / Ge(001)}

The growth of Ge$_{1-x}$Mn$_{x}$ films at high temperature ($>$180$^{\circ}$C)
on Ge(001) or annealing Ge$_{1-x}$Mn$_{x}$ films grown at low temperature above
600$^{\circ}$C leads to the formation of randomly distributed spherical
Ge$_{3}$Mn$_{5}$ clusters. The RHEED pattern of annealed samples is very similar
to that of Ge as in HRTEM the clusters are observed away from the surface (Fig.\,\ref{jm_fig12}).


The structure of Ge$_{3}$Mn$_{5}$ clusters was investigated using grazing incidence SXRD. The mean cluster diameter was estimated to be $10.6 \pm 1$\,nm in agreement with TEM observations. Moreover 97\% of the clusters have their $c$-axis perpendicular to the film plane and 3\% in plane. Finally, by searching for all possible Ge$_{3}$Mn$_{5}$ reflections in the plane parallel to the sample surface, a slight in-plane distortion of the hexagonal lattice was found \cite{Jain:2011_JAP}. Magnetic properties were studied using SQUID
magnetometry and EPR. Temperature-dependent magnetization measurements show the presence of two magnetic phases:
the Ge$_{3}$Mn$_{5}$ with $T_{\text{C}}$ of $300 \pm 5$\,K and the paramagnetic
contribution of isolated Mn atoms.
Hysteresis curves at 5\,K show that most
of Ge$_{3}$Mn$_{5}$ clusters exhibit perpendicular magnetic anisotropy.
Since the clusters are spherical, perpendicular anisotropy arises from
magnetocrystalline anisotropy. This result is in good agreement with XRD data. Indeed, bulk Ge$_{3}$Mn$_{5}$ crystal is hexagonal
with uniaxial magnetic anisotropy along the $c$-axis. From hysteresis loops recorded with the field perpendicular to the film plane, the fraction of magnetic signal from Ge$_{3}$Mn$_{5}$
clusters is estimated to be 70\%. The remaining signal likely comes from amorphous Mn-rich precipitates with very low coercive field. From the anisotropy field $\mu_{0}H_{a}\simeq 0.6$\,T and using the bulk $M _{S}$ value
(1100 kA/m) \cite{Forsyth:1990_JPCM}, the anisotropy constant can be calculated:
$K=\mu_{0}H_{a} M_{S}/2 \simeq 3.3 \times 10 ^{5}$ J/m$^{3}$ which is less than
the reported bulk value $4.2 \times 10^{5}$ J/m$^{3}$ \cite{Tawara:1963_JPSJ}. This feature
is explained by the in-plane distortion of the Ge$_{3}$Mn$_{5}$ crystal observed by XRD which is due to the epitaxy on the germanium matrix. It introduces
an additional in-plane magnetoelastic component to magnetic anisotropy. Finally the same anisotropy field and perpendicular anisotropy were obtained from EPR measurements in good agreement with SQUID data \cite{Jain:2011_JAP}.

\section{\label{sec:Ge-Fe}Spinodal nanodecomposition in $\mbox{(Ge,Fe)}$}

\subsection{Introduction}

Apart from Mn  other TM atoms such as Cr, Co or Fe have also been incorporated into Ge films to obtain FM properties \cite{Kioseoglou:2004_APL,Tsui:2004_PRB,Kim:2005_JAP}.
In particular, FM order was predicted in Fe-doped germanium films \cite{Weng:2005_PRB,Zhou:2004_JMMM} and experimentally demonstrated in Fe-implanted films \cite{Venugopal:2002_JAP}, Fe-doped germanium nanowires \cite{Cho:2008_ChemMater}
or quantum dots \cite{Xiu:2010_JACS} as well as in Fe-doped bulk Ge single crystals \cite{Choi:2003_JAP}.

In order to enhance the solubility limit of Fe,  LT-MBE   was used to grow Ge$_{1-x}$Fe$_{x}$ films
with $x$ up to 17.5\% \cite{Shuto:2006_JAP}.
In this section, we describe the growth, characterization, and magnetic properties of
such Ge$_{1-x}$Fe$_{x}$ thin films, in particular, the relation between growth or annealing
temperatures, defect formation, Fe distribution, and resulting magnetism \cite{Shuto:2006_JAP,Shuto:2006_PSSC,Shuto:2007_APL,Wakabayashi:2014_JAP,Wakabayashi:2014_PRB}.

\subsection{MBE growth}
Ge$_{1-x}$Fe$_{x}$ thin films were grown on Ge (001) substrates
by LT-MBE. After a LT
Ge buffer layer was grown at the substrate temperature $T_{\rm g}$ of
100$^{\circ}$C, a 16-nm-thick Ge$_{1-x}$Fe$_{x}$ film was grown at $T_{\rm g} = 100, 200, 300,$ or $400^{\circ}$C. The Fe content ($x$) was varied from 2.0 to
24.0\%. {\em In situ} RHEED was used
to monitor the crystallinity and surface morphology of the Ge buffer layer
and the Ge$_{1-x}$Fe$_{x}$ film during the epitaxy.
Although the diffraction pattern of
the LT Ge buffer layer revealed intense and sharp $2\times 2$ streaks,
the pattern was quickly changed at the initial stage
of Ge$_{1-x}$Fe$_{x}$ epitaxy.
The Ge$_{1-x}$Fe$_{x}$ films grown at $T_{\rm g} = 100$ and $200^{\circ}$C
showed 2D growth mode and exhibited diamond-type crystal structure.

When the Ge$_{1-x}$Fe$_{x}$ film was grown at
$T_{\rm g} = 300$ (or $400^{\circ}$C), a spotty pattern was clearly observed
in the RHEED image,
indicating a 3D growth mode or surface roughening. Thus,
the growth mode of the Ge$_{1-x}$Fe$_{x}$ film was changed from the 2D
mode to the 3D mode between $T_{\rm g} = 200$ and
$300^{\circ}$C. This growth mode change at around $T_{\rm g} = 300^{\circ}$C
is caused by the precipitates of Fe-Ge FM compounds
in the film (crystallographic phase separation), as discussed next.


\subsection{Structural and chemical characterization}
Figure~\ref{tanaka-gefe_fig2}(a) shows a HRTEM image of a Ge$_{1-x}$Fe$_{x}$ film ($x = 9.5$\%)
grown at $T_{\rm g} = 200^{\circ}$C, projected along the exact direction
of Ge[110]. The image indicates that the Ge$_{1-x}$Fe$_{x}$ layer was
homogeneously and epitaxially grown on the Ge buffer layer without any
threading dislocations, and that the surface was atomically flat with
a height of roughness less than 1\,nm.  These features were consistent
with the RHEED observations, as mentioned earlier. However, when the lattice
image was projected along the angle slightly tilted from the Ge[110]
direction, clusterlike dark contrast regions appeared in the epitaxially
grown Ge$_{1-x}$Fe$_{x}$ layer, as shown in Fig.\,\ref{tanaka-gefe_fig2}(b).
It is well recognized in TEM observations that the contrast due to the
difference in chemical compositions is enhanced when the projection direction
is slightly tilted from the exact zone axis.  Using spatially resolved
transmission electron diffraction (TED) and EDS, the crystal structure and chemical composition of
the dark region [denoted as *1 in Fig.\,\ref{tanaka-gefe_fig2}(b)] and
surrounding bright region [denoted as *2 in Fig.\,\ref{tanaka-gefe_fig2}(b)]
were investigated. The TED image of point *1 (the right inset) showed
the diffraction pattern of the diamond structure projected from its [110]
direction with weak extra spots caused by stacking fault defects.
On the contrary, point *2 (the left inset) in the bright region exhibited
the diffraction pattern of the diamond structure without any extra spots.
This means that no other crystal structures except the diamond crystal
structure were formed both in the dark and in the bright regions, although
the dark region included small stacking fault defects.

The local Fe composition
at point *1 measured by EDS was $\sim 12$\%, which is higher than that
($\sim 4$\%) at point *2.  Therefore, the contrast in the HRTEM image shown
in Fig.\,\ref{tanaka-gefe_fig2}(b) was due to the nonuniform distribution
of the Fe composition.   Similar structural features were also observed
for a Ge$_{1-x}$Fe$_{x}$ film with $x = 6.0$\%,
in which dark and bright contrast regions appeared in the TEM image caused
by nonuniform Fe distribution.  Small stacking fault defects were also detected
in the dark contrast regions.  The local Fe composition in the dark and
bright contrast regions of this film was about 8.0\% and 1.5\%, respectively.
Thus, both the Fe compositions in dark and bright contrast regions increased
with increasing $x$.

Thus, RHEED, HRTEM, TED, and EDS studies lead to the conclusion that
Ge$_{1-x}$Fe$_{x}$ films grown at 200$^{\circ}$C on Ge(001) maintain the diamond
crystal structure without any intermetallic Fe-Ge compounds or Fe inclusions.
Similarly, the diamond structure is preserved
under annealing (Ge,Fe) below 600$^{\circ}$C \cite{Wakabayashi:2014_PRB}.
However, the films grown at 200$^{\circ}$C or annealed  exhibit a nonuniform distribution of Fe
composition $x$, i.e., they undergo spinodal nanodecomposition in the form
of the chemical phase separation.

It is worth noting that since EDS provides chemical information
averaged over the specimen width, the amplitude of actual $x$ fluctuations can
be even larger than implied by the values of $x$ determined for points *1 and *2.

\begin{figure}
\includegraphics[angle=0,width=0.8\columnwidth]{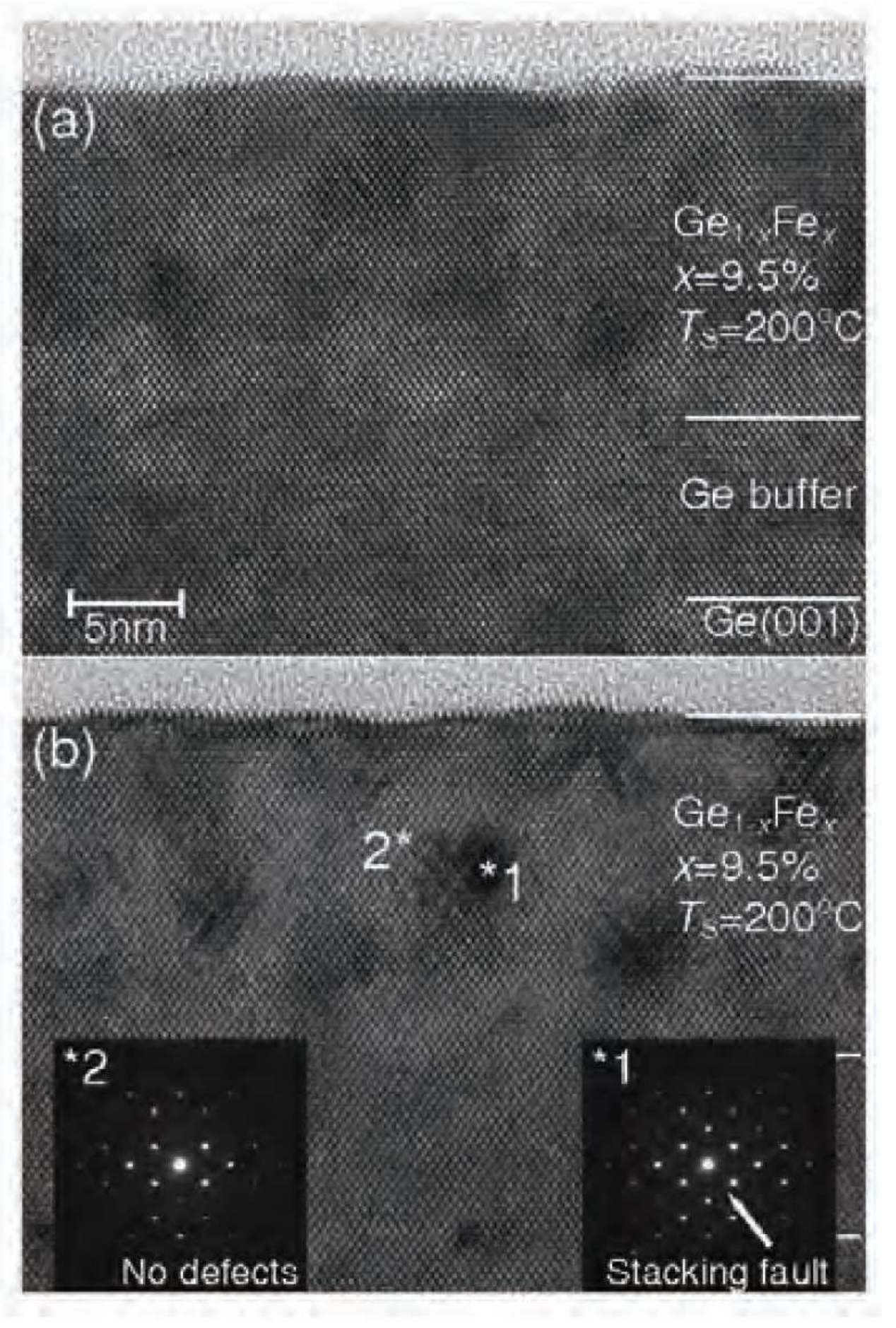}
\caption{\label{tanaka-gefe_fig2} HRTEM lattice images projected along
(a) the exact direction of the Ge[110] axis and (b) the direction slightly
tilted from the Ge[110] axis for Ge$_{1-x}$Fe$_{x}$ thin films with $x = 9.5$\%
grown at $T_{\rm g} = 200^{\circ}$C.
Insets: Diffraction  images at the dark region (*1) and at the surrounding bright region (*2).
From \onlinecite{Shuto:2007_APL}.}
\end{figure}

\subsection{Magneto-optical characterization}
\label{sec:MCD_GeFe}
\begin{figure}
\includegraphics[angle=0,width=0.8\columnwidth]{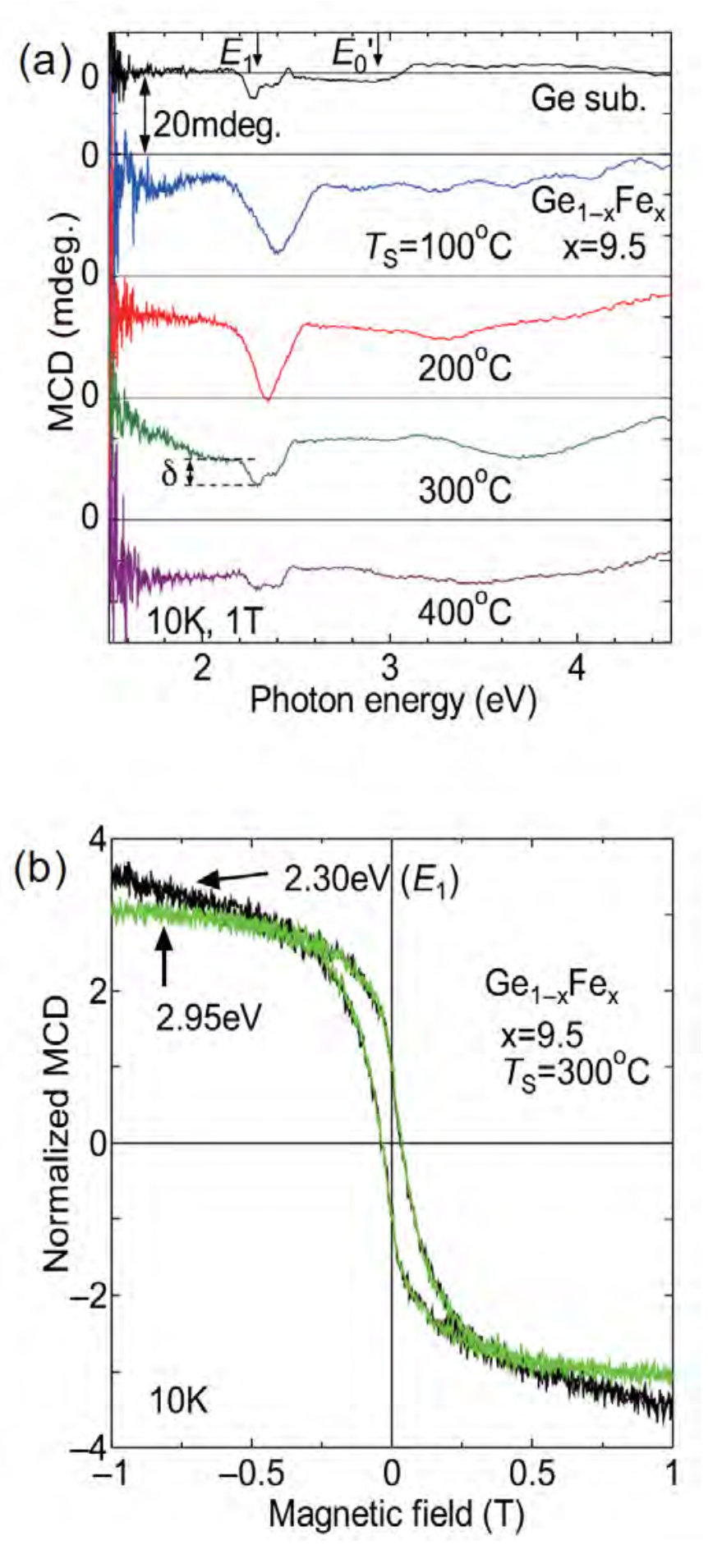}
\caption{\label{tanaka-gefe_fig3} (a) Growth temperature dependence of the MCD spectra
of Ge$_{1-x}$Fe$_{x}$ ($x=9.5$\%) films grown at $T_{\rm g} = 100 - 400^{\circ}$C,
measured at 10\,K with magnetic field of 1\,T perpendicular to the sample plane with
reflection configuration. (b) Magnetic-field dependence of MCD intensities at 2.30\,eV
($E_{1}$) and at other energy point (2.95\,eV) for the Ge$_{1-x}$Fe$_{x}$ ($x=9.5$\%)
films grown at $T_{\rm g} = 300^{\circ}$C, where the data were normalized by their
MCD intensities at the magnetic field of 0\,T. Adapted from \onlinecite{Shuto:2006_PSSC}.}
\end{figure}

Figure \ref{tanaka-gefe_fig3}(a) shows  MCD spectra taken at 10\,K
in reflection configuration for
 Ge$_{1-x}$Fe$_{x}$ ($x = 9.5$\%) films grown at
$T_{\rm g} = 100, 200, 300$, and $400^{\circ}$C,  with magnetic field perpendicular
to the sample plane.  As see, when the Ge$_{1-x}$Fe$_{x}$ film was grown
at $T_{\rm g} = 100$ (or $200^{\circ}$C), its MCD spectrum showed an $E_1$
peak whose intensity was significantly enhanced compared with that of bulk
Ge [uppermost curve in Fig.\,\ref{tanaka-gefe_fig3}(a)], implying the presence
of $s,p$--$d$ exchange splitting of bands. An offsetlike broad signal was also
observed in the whole energy range examined here. The magnetic-field
dependence of the MCD intensity at the critical point $E_1$ exhibited
a FM hysteresis loop, and the hysteresis shape at any other
photon energies in the offsetlike signal was identical with that at E1,
indicating the Ge$_{1-x}$Fe$_{x}$ film grown at $T_{\rm g} = 100 - 200^{\circ}$C
was magnetically homogeneous without Fe-Ge FM precipitates.
Note that the $E_{0}$ of the sample was not resolved
due to light interference caused by adsorbed moisture on
the sample surface \cite{Shuto:2006_JAP}.

When the Ge$_{1-x}$Fe$_{x}$
film was grown at $T_{\rm g} = 300$ (or $400^{\circ}$C), the MCD intensity
at the critical point E$_{1}$ was not enhanced at all, but the MCD spectrum
of bulk Ge was overlapped with an offsetlike MCD signal: The magnitude of the
E$_{1}$ peak [denoted as $\delta$ shown in Fig.\,\ref{tanaka-gefe_fig3}(a)]
that was measured from the bottom of the offsetlike signal level was equal
to that of bulk Ge. The MCD hysteresis loop measured at E$_{1}$ (2.30\,eV)
was not identical with that measured at other energy points (for example, 2.95\,eV)
as shown in Fig.\,\ref{tanaka-gefe_fig3}(b), indicating crystallographic phase separation.
The MCD hysteresis loop at E$_{1}$ can be divided into two components, i.e.,
the paramagnetic component of the host Ge matrix and the FM
component of FM precipitates.

\begin{figure}
\includegraphics[angle=0,width=0.8\columnwidth]{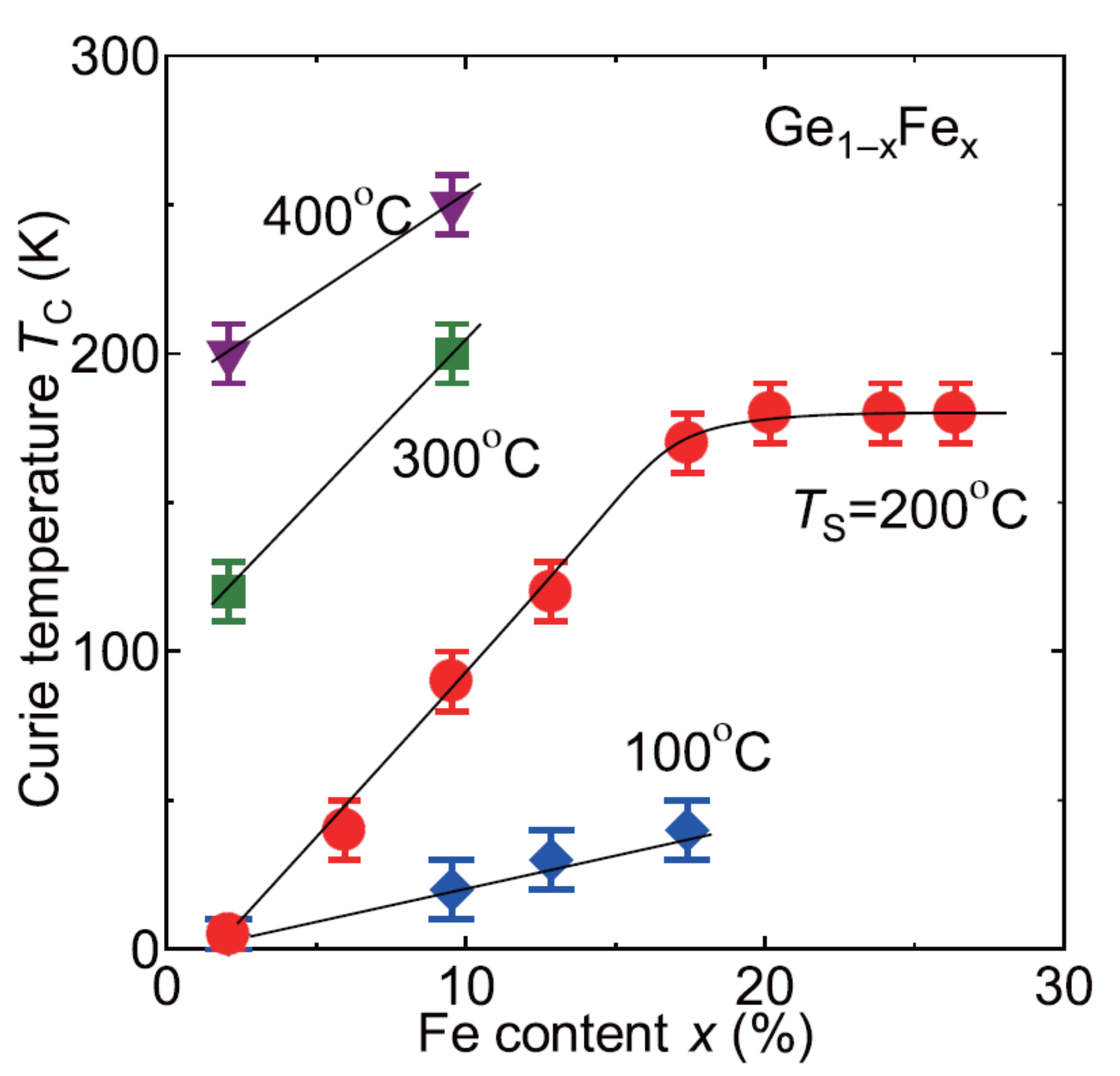}
\caption{\label{tanaka-gefe_fig4} Curie temperature $T_{\rm C}$ of Ge$_{1-x}$Fe$_{x}$
films as a function of Fe content $x$. Ge$_{1-x}$Fe$_{x}$ films were grown at
$T_{\rm g} = 100, 200, 300,$ and $400^{\circ}$C. From \onlinecite{Shuto:2006_PSSC}.}
\end{figure}

Figure \ref{tanaka-gefe_fig4} shows $T_{\rm C}$ values of
Ge$_{1-x}$Fe$_{x}$ films as a function of Fe content $x$, where $T_{\rm C}$
was evaluated from the Arrott plots using temperature-dependent MCD hysteresis
loops at the $E_{1}$ critical point.

For the Ge$_{1-x}$Fe$_{x}$ films grown at
$T_{\rm g} = 200^{\circ}$C, $T_{\rm C}$ linearly increased with increasing $x$
in the range between $x = 0$ and 17.5\%. Above $x = 17.5$\%, $T_{\rm C}$ was
saturated at around 170 - 180\,K  \cite{Shuto:2007_APL}.
$T_{\rm C}$ of Ge$_{1-x}$Fe$_{x}$
films grown at $T_{\rm g} = 100^{\circ}$C also linearly increased with
increasing $x$, but $T_{\rm C}$ was significantly reduced in comparison
with that of the samples grown at $T_{\rm g} = 200^{\circ}$C.

When
Ge$_{1-x}$Fe$_{x}$ films were grown at $T_{\rm g} = 300$ and $400^{\circ}$C,
their $T_{\rm C}$ values were higher than those of the samples grown at
$T_{\rm g} = 100$ and $200^{\circ}$C. For example, $T_{\rm C}$ values were 120
and 200\,K in the cases of $T_{\rm g} = 300$ and $400^{\circ}$C, respectively,
even when $x$ was only 2.0\%. On the contrary, $T_{\rm C}$ values were below 10\,K
in the cases of $T_{\rm g} = 100$ and $200^{\circ}$C with $x = 2.0$\%.
This behavior of $T_{\rm C}$ at $T_{\rm g} = 300$ and $400^{\circ}$C can be
attributed to the precipitates of FM compounds described previously.

The apparent values of $T_\text{C}^{(app)}$ not only increase with the growth temperature but can be enhanced
by annealing \cite{Wakabayashi:2014_PRB}. In either case, the magnitude of $T_\text{C}^{(app)}$ correlates with the nonuniformity of the Fe distribution and the stacking fault density \cite{Wakabayashi:2014_PRB,Wakabayashi:2014_JAP}.  Meanwhile, channeling Rutherford backscattering and particle-induced x-ray emission measurements revealed that about 15\% of the Fe atoms reside in the tetrahedral interstitial sites and that the substitutional Fe concentration, in agreement with the decomposition scenario, is barely correlated with the magnitude of $T_\text{C}^{(app)}$ \cite{Wakabayashi:2014_JAP}. At the same time, all the (Ge,Fe) films show a weak spin-glass-like behavior in a low-temperature region (below $\sim 26$\,K), which is insensitive to the annealing temperature. However,  the ferromagnetism associated with the nonuniform Fe distribution  dominates magnetic properties of the system.

In summary, studies of Ge$_{1-x}$Fe$_{x}$ as a function of
epitaxy temperature and TM content clearly show an evolution of the alloy character
from a dilute magnetic semiconductor to a decomposed system, first involving
chemical and then crystallographic phase separation. This evolution is accompanied by
changes in spectral dependences of MCD that is initially enhanced in the region
of optical transitions specific to the host semiconductor but as nanodecomposition progresses
this magnetization-dependent enhancement starts to extend over a wide spectral region
indirectly pointing to a metallic character of Fe-rich regions. The apparent
$T_{\rm C}$ values determined from the MCD hysteresis loops show a clear correlation with the degree
of nanocomposition, i.e., with the upper limit of local TM concentrations as well as with the appearance of structural defects.

\section{\label{sec:ZnTe}Spinodal nanodecomposition in $\mbox{(Zn,Cr)Te}$}
\subsection{Introduction}
Among various DMSs, Cr-doped II-VI compounds have been studied for a long time, attracting attention due to their peculiar magnetic and magnetooptical properties. Excitonic magnetospectroscopy indicated that the $p$--$d$ exchange interaction is FM \cite{Mac:1996_PRB}, in contrast to the AF interaction in Mn-doped II-VI compounds. According to the Schrieffer-Wolff transformation,  the FM $p$--$d$ interaction results from the position of the Cr 3$d$ donor level {\em above} the top of the valence band  \cite{Kacman:2001_SST,Mizokawa:1997_PRB}.

Concerning the spin-dependent coupling between Cr spins in (II,Cr)VI, it was theoretically predicted that the superexchange interaction becomes FM for $d^4$ ions in the tetrahedral crystal field \cite{Blinowski:1996_PRB}. Later, a FM ordering of Cr spins in II-VI compounds was found by DFT first-principle calculations \cite{Sato:2002_SST}. Hence, the theoretically expected ferromagnetism can be understood in the picture of double-exchange interaction between the Cr $3d$ electrons in the localized level within the band gap \cite{Sato:2002_SST,Fukushima:2004_JJAP} or as FM superexchange interaction of highly localized Cr spins \cite{Blinowski:1996_PRB}. In either case, the FM interaction between Cr spins is short-range, extending over few nearest-neighbors.

Experimentally, the FM behavior was observed in (Zn,Cr)Te \cite{Saito:2002_PRB} and (Zn,Cr)Se \cite{Karczewski:2003_JSNM} with Cr contents reaching a few percent, and later ferromagnetism at RT was achieved in Zn$_{1-x}$Cr$_x$Te with a high Cr content $x = 0.2$ \cite{Saito:2003_PRL}. The FM transition temperature $T_{\text{C}}$ was reported to increase almost linearly with Cr content $x$, reaching 300\,K at $x = 0.2$, in agreement with the {\em ab initio} prediction \cite{Fukushima:2004_JJAP}. In addition, the intrinsic nature of ferromagnetism appeared to be confirmed by the similar magnetic-field dependences of magnetization and MCD \cite{Saito:2003_PRL}. However, according to \onlinecite{Karczewski:2003_JSNM}, the ferromagnetism in Zn$_{1-x}$Cr$_x$Se resulted from the precipitation of a FM compound, as $T_{\text{C}}$ was independent of $x$.

It was also found that the FM properties of (Zn,Cr)Te were significantly affected by the codoping of donor or acceptor impurities; the ferromagnetism was suppressed by codoping with acceptor impurity nitrogen (N) \cite{Ozaki:2005_APL} while it was enhanced due to codoping with iodine (I) donor impurities  \cite{Ozaki:2006_PRL}. In particular, for a fixed Cr content $x = 0.05$, the FM transition temperature $T_{\text{C}}$ increased up to 300\,K with the codoping of iodine at a concentration of the order of $10^{18}$\,cm$^{-3}$, from $T_{\text{C}}\sim 30$\,K in a layer without codoping (referred to as undoped sample). However, it was difficult to understand the realization of  high-temperature ferromagnetism at a low Cr content, $x \simeq 0.05$, which is below the percolation limit of the zb structure, if we assume a short-range interaction between Cr spins. Moreover, the combined analysis of the nanoscale chemical probing using TEM and the magnetization measurements with SQUID revealed that there is a systematic correlation between FM properties and the heterogeneity of the Cr distribution \cite{Kuroda:2007_NM}. In I-doped (Zn,Cr)Te crystals exhibiting ferromagnetism at RT, it was found that the Cr distribution is strongly inhomogeneous, i.e., nanoscale regions are formed, in which the concentration of Cr cations is high. According to the current understanding, in these Cr-rich volumes, Cr spins are ordered ferromagnetically due to the short distance between them. These FM clusters give rise to superparamagnetism controlled by the magnetic anisotropy of individual Cr-rich NCs and interactions among them.

The electronic structure of the Cr ions and host bands was investigated in Zn$_{1-x}$Cr$_x$Te ($x = 0.03$ and 0.15)
using XMCD and photoemission spectroscopy \cite{Kobayashi:2008_NJP}.
It was concluded that neither double-exchange mechanism nor  carrier-induced ferromagnetism are important, and
the inhomogeneous distribution of Cr atoms dominantly influences the FM
properties of Zn$_{1-x}$Cr$_x$xTe.

In the case of (Zn,Cr)Te, similar to other DMSs discussed, both chemical and crystallographic phase separations can be induced depending on Cr content and growth conditions \cite{Kuroda:2007_NM,Kobayashi:2012_PB}.
The pairing energy of Cr ions, which is the driving force of the phase separation (see, Secs.\,\ref{sec:pairing}), is expected to vary with the Cr charge state \cite{Dietl:2006_NM}. Since the Cr 3$d$ electrons form deep donor and acceptor levels in the band gap of ZnTe \cite{Godlewski:1980_JPC}, the studies of (Zn,Cr)Te demonstrated that the shift of the Fermi level produced by a change in the concentration of electrically active defects or shallow impurities alters the Cr charge state, which affects the Coulomb interaction between Cr ions and hence their aggregation \cite{Kuroda:2007_NM}.

In this section, we review recent experimental studies on spinodal nanodecomposition in (Zn,Cr)Te grown by MBE \cite{Kuroda:2007_NM,Ishikawa:2009_P,Nishio:2009_MRSP,Kobayashi:2012_PB} as well as associated magnetooptical \cite{Kuroda:2007_NM,Ozaki:2006_PRL} and magnetotransport properties \cite{Kuroda:2007_NM}. In the first part (Sec.\,\ref{sec:ZnTe_chemical}) we describe the properties of (Zn,Cr)Te films grown in the [001] crystallographic direction under the standard condition of MBE growth for the host binary compound ZnTe -- at a substrate temperature of $T_{\mathrm{g}} = 300^{\circ}$C and at a fixed growth rate of $\sim 1$\,{\AA}/sec. We also discuss the effect of codoping by I donors and N acceptors as well as varying the stoichiometry by changing the ratio of Zn and Te fluxes. The influence of the growth rate and temperature on the structural and magnetic properties of (Zn,Cr)Te:I is presented in Sec.\,\ref{sec:ZnTe_crystallographic}. The accumulated data demonstrate the possibility of controlling the Cr aggregation by growth conditions and codoping with shallow impurities, corroborating the theoretical considerations on the pairing energy presented in Sec.\,\ref{sec:pairing}, as well as provide experimental illustration of the dairiseki and konbu phases discussed theoretically in Secs.\,\ref{sec:dairiseki} and \ref{sec:konbu}. This control over the size and shape of Cr-rich regions substantiates the prospects of spinodal nanotechnology discussed in Sec.\,\ref{sec:prospects}.

\subsection{Chemical phase separation and its control by the Fermi level position}
\label{sec:ZnTe_chemical}
\subsubsection{Visualization of Cr distribution}
Figure \ref{kuroda_fig1} shows the EDS mapping images of the local Cr composition in I-doped, undoped, and N-doped Zn$_{1-x}$Cr$_x$Te films with relatively low and high values of the average Cr content $x \simeq 0.05$ and 0.2.
In the case of the low Cr composition $x \simeq 0.05$, the Cr mapping images exhibit clear differences in the uniformity of Cr distribution between particular samples \cite{Kuroda:2007_NM}: uniform distributions within the spatial resolution of 2--3\,nm in the undoped films grown under Te-rich flux [Fig.\,\ref{kuroda_fig1}(a)] and in the N-doped film [Fig.\,\ref{kuroda_fig1}(d)] versus nonuniform distributions in the undoped film grown under Zn-rich flux [Fig.\,\ref{kuroda_fig1}(b)] and in the I-doped film [Fig.\,\ref{kuroda_fig1}(c)]. In the case of nonuniform distributions, the typical length scale of the compositional fluctuation is estimated to be 20--50\,nm. According to the spot analysis of the EDS spectra, a typical value of the local Cr composition inside the Cr-rich regions is $x \sim 0.1$.

The TEM observation of the series of the films with the Cr composition $x \simeq 0.05$ revealed that the structural properties are not much affected by the codoping of iodine or nitrogen, or the growth under different Zn/Te flux ratios. As a typical example of the series of the films, the result of the I-doped film is shown in Fig.\,\ref{kuroda_fig2}. The TEM image in Fig.\,\ref{kuroda_fig2}(a) exhibits mostly zb crystalline structure, but there appear many stacking faults along the \{111\} planes. It was suggested that strain associated with a preferential formation of TM cation dimers along the [$\bar{1}$10] direction accounts for the appearance of such faults during the growth \cite{Birowska:2012_PRL}. In addition, the electron diffraction from regions close to the bottom [Fig.\,\ref{kuroda_fig2}(b)] and the top [Fig.\,\ref{kuroda_fig2}(c)] of the Zn$_{1-x}$Cr$_x$Te layer exhibits different images, indicating the deterioration of crystallinity as the growth proceeds. In the diffraction image close to the bottom [Fig.\,\ref{kuroda_fig2}(b)], there appear additional weak spots at one-third positions between the fundamental spots of a hexagonal arrangement of the zb structure, corresponding to a triplet periodicity in the stacking sequence along the \{111\} planes in the observed stacking faults. On the other hand, the diffraction image close to the film top [Fig.\,\ref{kuroda_fig2}(c)] shows hexagonal patterns revealing the presence of zb nanocrystals with different crystallographic orientations at the end of the growth process. However, any apparent precipitates of other crystal structures were detected neither in the lattice image nor in the diffraction pattern.

\begin{figure}[h!]
\begin{center}
\includegraphics[width=1.0\columnwidth]{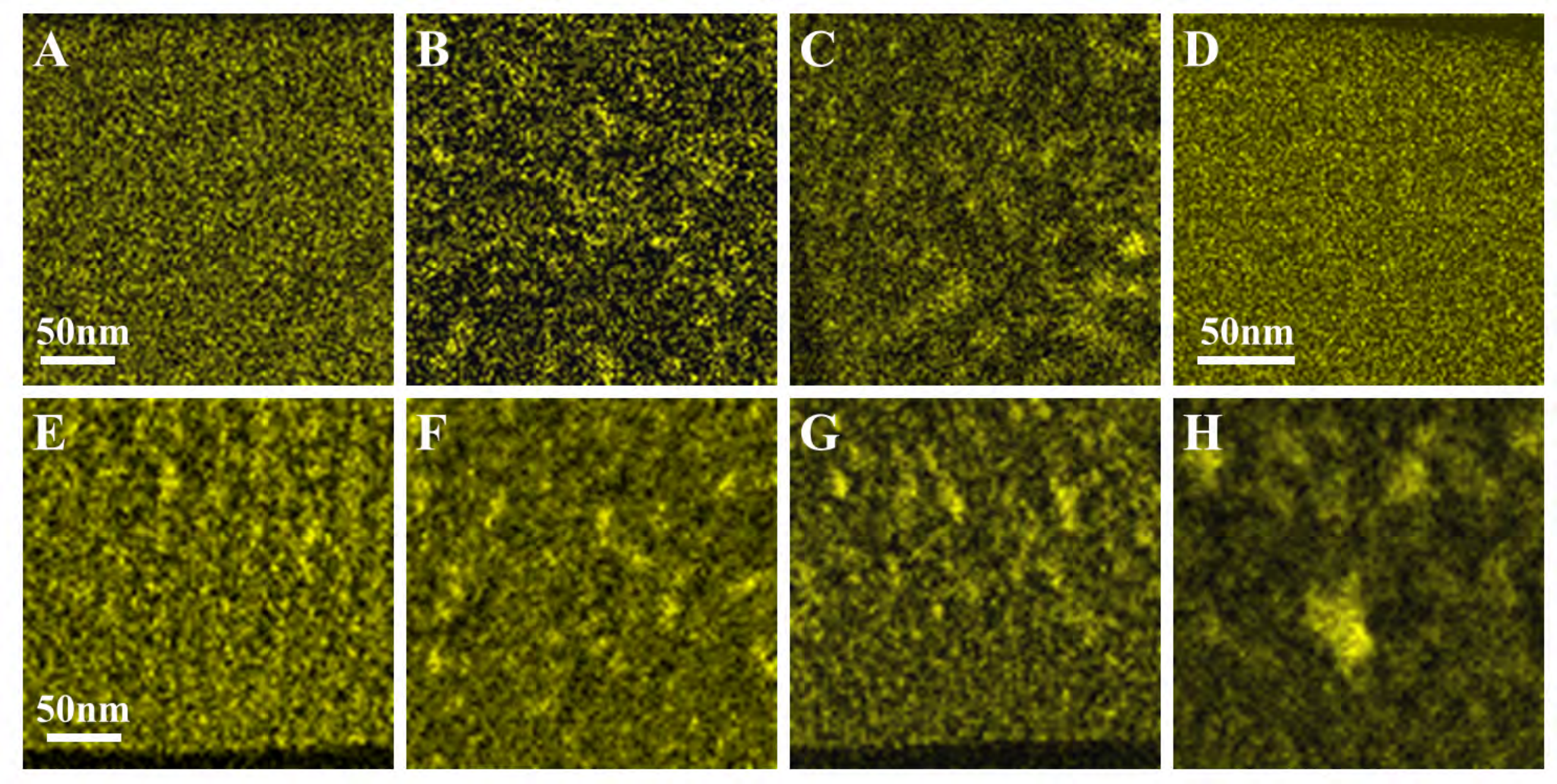}
\end{center}
\caption{(Color online) Mapping images of the emission intensity of the EDS Cr K$_{\alpha}$  line for cross-sectional pieces of Zn$_{1-x}$Cr$_x$Te films with average Cr contents (a)-(d) $x \simeq 0.05$  and (e)-(h) $x \simeq 0.2$. (a), (e): undoped films grown under Te-rich flux; (b), (f): undoped films grown under Zn-rich flux; (c), (g), (h): I-doped films with iodine concentrations of (c), (g) $\sim 2\times 10^{18}$\,cm$^{-3}$ and
(h) $\sim 1\times10^{19}$\,cm$^{-3}$; (d): N-doped film with a nitrogen concentration of $\sim 2\times 10^{20}$\,cm$^{-3}$. All these films were grown in the [001] crystallographic direction using GaAs(001) substrates under the standard condition of the MBE growth for ZnTe (the substrate temperature $T_g = 300^{\circ}$C and the growth rate around 1{\AA}/sec). Adapted from \onlinecite{Kuroda:2007_NM} and \onlinecite{Ishikawa:2009_P}.}
\label{kuroda_fig1}
\end{figure}

\begin{figure}[h!]
\begin{center}
\includegraphics[width=0.9\columnwidth]{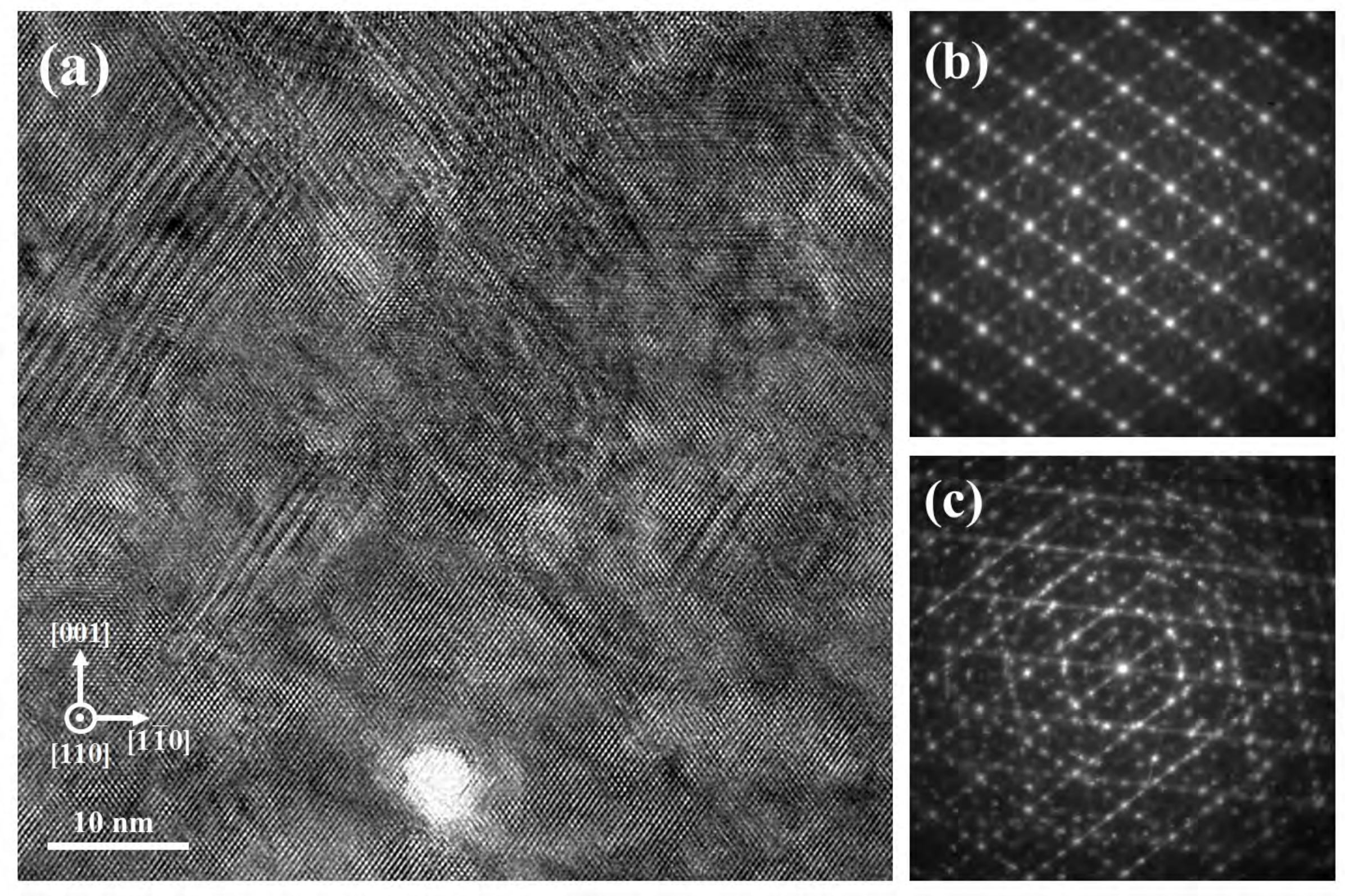}
\end{center}
\caption{TEM and electron diffraction images of the I-doped Zn$_{0.95}$Cr$_{0.05}$Te film [sample (a) in Fig.\,\ref{kuroda_fig1}). (a) Cross-sectional lattice image. (b) Electron diffraction image from a region close to the interface with the ZnTe buffer layer  and  (c) from a region close to the surface. From \onlinecite{Kuroda:2007_NM}.}
\label{kuroda_fig2}
\end{figure}

For the high Cr composition of $x \sim 0.2$, the Cr distribution is always non-uniform, irrelevant of the codoping or the Zn/Te flux ratio, but the degree of inhomogeneity changes depending on the codoping or the Zn/Te flux ratio \cite{Ishikawa:2009_P}. As shown in Fig.\,\ref{kuroda_fig1}(e)-(h), the inhomogeneity of Cr distribution in the undoped film grown under Te-rich flux is not significant, whereas the Cr distribution in the I-doped films and the undoped film grown under Zn-rich flux is strongly inhomogeneous. In particular, the size of Cr-rich regions becomes larger in the I-doped film with a higher iodine concentration of $ \sim 1\times10^{19}$\,cm$^{-3}$ [Fig.\,\ref{kuroda_fig1}(h)]. A typical length scale of the Cr-rich regions is estimated to be 30--70\,nm. According to the EDS spot analysis, a maximum value of the local Cr composition in the Cr-rich regions is $x = 0.4$--0.5. It is concluded from these results that the inhomogeneity of Cr distribution is enhanced either by codoping with iodine or by the growth under the Zn-rich condition for both low and high Cr contents.

\subsubsection{Superparamagnetic behavior due to the $\mbox{Cr}$ aggregation}

The measurements of the temperature and magnetic-field dependences of magnetization $M(T,H)$ using SQUID revealed that the magnetic properties also depend significantly on the Zn/Te flux ratio or codoping with donor or acceptor impurities \cite{Kuroda:2007_NM,Ishikawa:2009_P}. In Fig.\,\ref{kuroda_fig3}, the temperature dependence of magnetization is compared for undoped films grown under Te-rich flux and I-doped films. As seen, superparamagnetic features, such as the irreversibility between magnetizations determined under FC and ZFC conditions, and a cusp in the $M(T)$ dependence taken in the ZFC process, are more pronounced in the case of the I-doped films than for the undoped films grown under Te-rich flux. The data in Fig.\,\ref{kuroda_fig3}(a) indicate that superparamagnetic features appear at low temperatures also for the Zn$_{0.95}$Cr$_{0.05}$Te film for which the Cr distribution is uniform according to Fig.\,\ref{kuroda_fig1}(a). This means that Cr aggregates with diameters below spatial resolution of EDS (2--3\,nm) are present in such samples. The magnitude of blocking temperature $T_{\mathrm{b}}$, i.e., a temperature value corresponding to the maximum in the $M(T)$ dependence taken in the ZFC process, is much higher for the I-doped films than in the case of undoped films grown in the Te-rich flux. The paramagnetic Curie-Weiss temperature  $\theta_P$ obtained from the linear fitting of the temperature dependence of the inverse magnetic susceptibility $\chi(T)$ is also higher in the case of the I-doped films compared to the undoped films grown under Te-rich flux.

The magnitudes of three characteristic temperatures, $T_{\mathrm{b}}$, $\theta_P$, and the apparent Curie temperature $T_{\text{C}}^{\text{(app)}}$ deduced from the Arrott plot analysis of the $M(H)$ dependence, are plotted in Fig.\,\ref{kuroda_fig4} for samples with the average Cr contents
$x \simeq 0.05$ and 0.2 as a function of the iodine concentration and the Te/Zn flux ratio. In the case of $x \simeq 0.05$, these temperatures  are closely correlated with the uniformity of the Cr distribution:  $T_{\text{C}}^{\text{(app)}}$, $T_{\mathrm{b}}$, and  $\theta_P$ assume maximum values at the iodine concentration  $\sim 2\times10^{18}$\,cm$^{-3}$ for the I-doped films and Te/Zn flux ratio 0.7 (Zn-rich flux) for the undoped films, which correspond to the most pronounced inhomogeneity of the Cr distribution. On the other hand, in the case of $x \sim 0.2$,  $T_{\text{C}}^{\text{(app)}}$ and  $\theta_P$ are already high even in the undoped film grown under Te-rich flux and do not increase much for the growth under Zn-rich flux or the codoping of iodine. In contrast, $T_{\mathrm{b}}$ shows a sizable change depending on the flux ratio or the I doping; in the undoped films, $T_{\mathrm{b}}$ increases with the decrease of the Te/Zn flux ratio, and in the I-doped films, $T_{\mathrm{b}}$ increases with the increase of iodine concentration at first and reaches a maximum at $ \sim 5\times10^{18}$\,cm$^{-3}$, and then decreases gradually with a further increase of iodine content. These findings indicate that the magnitude of $T_{\mathrm{b}}$ shows a close correlation with the degree of inhomogeneity in the Cr distribution at high Cr compositions.

\begin{figure}[h!]
\begin{center}
\includegraphics[width=0.9\columnwidth]{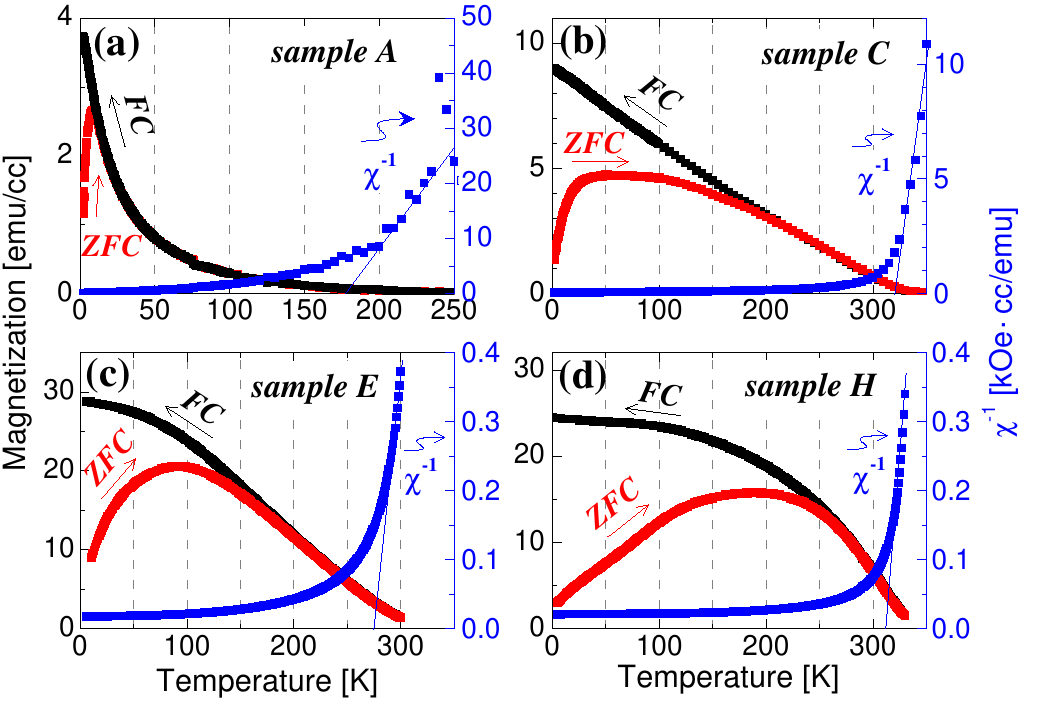}
\end{center}
\caption{(Color online) Temperature dependence of magnetization and the inverse magnetic susceptibility of Zn$_{1-x}$Cr$_x$Te films: (a) undoped Zn$_{0.95}$Cr$_{0.05}$Te film grown under Te-rich flux (sample A in Fig.\ref{kuroda_fig1}); (b) I-doped Zn$_{0.95}$Cr$_{0.05}$Te film (sample C); (c) undoped Zn$_{0.81}$Cr$_{0.19}$Te film grown under Te-rich flux (sample E); (d) I-doped Zn$_{0.81}$Cr$_{0.19}$Te film (sample H). Adapted from \onlinecite{Kuroda:2007_NM}.}
\label{kuroda_fig3}
\end{figure}

\begin{figure}[h!]
\begin{center}
\includegraphics[width=1.0\columnwidth]{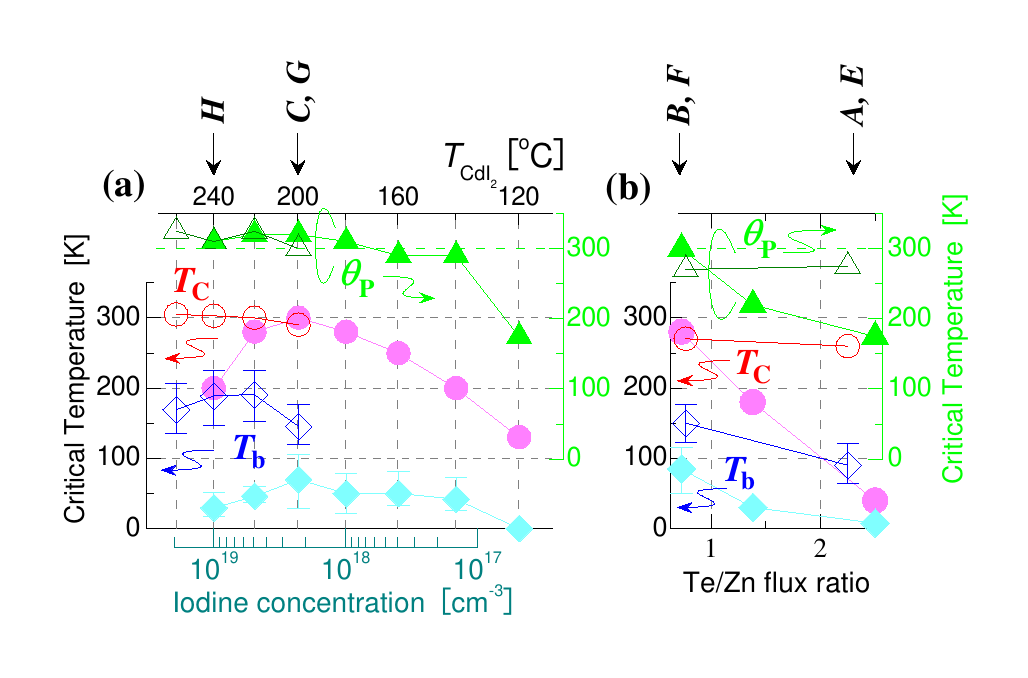}
\end{center}
\caption{(Color online) Three characteristic temperatures for I-doped and undoped Zn$_{1-x}$Cr$_x$Te films with $x \simeq 0.05$ and $x \sim 0.2$: the apparent Curie temperature  $T_{\text{C}}^{\text{(app)}}$ (circles), the paramagnetic Curie-Weiss temperature  $\theta_P$ (triangles) and the blocking temperature $T_{\mathrm{b}}$ (diamonds) are plotted as a function of (a) the iodine concentration  for I-doped films and (b) the Te/Zn flux ratio for undoped films. For clarity, the vertical scale for $\theta_P$ (right) is shifted from that for  $T_{\text{C}}^{\text{(app)}}$ and $T_{\mathrm{b}}$ (left). The data for $x \simeq 0.05$ and 0.2 are represented by full and open symbols, respectively. The labels A-H above the figure serve to identify the films shown in Fig.\,\ref{kuroda_fig1}. From \onlinecite{Ishikawa:2009_P}.}
\label{kuroda_fig4}
\end{figure}

These experimental observations demonstrate that the magnetic properties of (Zn,Cr)Te films are closely correlated with the uniformity of Cr distribution. In particular, non-uniform Cr distributions give rise to higher values of characteristic temperatures introduced above. Since the FM exchange interaction between Cr spins is considered to be of short-range \cite{Bergqvist:2004_PRL}, the long-range FM order cannot be expected for average Cr compositions below the percolation limit for the nearest-neighbor coupling in the fcc lattice, $x < 0.20$, if we assume a random distribution of Cr ions. On the other hand, if the alloy is phase-separated into regions with low and high Cr contents, Cr spins inside the Cr-rich regions can order ferromagnetically, so that the crystal containing these FM nanoclusters is expected to exhibit superparamagnetic features. Below the blocking temperature $T_{\mathrm{b}}$, whose magnitude is given by the product of the mean cluster volume $V$ and the density of the magnetic anisotropy energy $K$, FM-like properties are observed in superparamagnetic systems, including magnetization hysteresis and remanence. A correlation between the values of $T_{\mathrm{b}}$ and the inhomogeneity of Cr distribution confirms this interpretation. In addition, rather broad peaks in the $M(T)$ curves obtained under the ZFC condition reflect a dispersion in values of blocking temperatures. The apparent Curie temperature  $T_{\text{C}}^{\text{(app)}}$, corresponding to temperature at which hysteretic behaviors due to the magnetic anisotropy of the clusters disappear entirely, is determined by an upper bound of the $T_{\mathrm{b}}$ distribution. On the other hand, the paramagnetic Curie-Weiss temperature  $\theta_P$, deduced from the fitting of the linear dependence in the $\chi^{-1}(T)$ curves in the high-temperature range, is determined by interactions between Cr spins inside the Cr-rich NCs and, therefore, is virtually independent of their volume $V$. This is consistent with the observation that $\theta_P$ is less dependent on the uniformity of the Cr distribution than $T_{\text{C}}^{\text{(app)}}$ and $T_{\mathrm{b}}$.

Altogether, the data imply that zb Cr-rich (Cr,Zn)Te with the ZnTe lattice parameter is characterized by $T_{\text{C}} \simeq 300$\,K. A ferromagnetic ground state was theoretically predicted for such zb-CrTe \cite{Zhao:2005_PRB}.

\subsubsection{Mechanism of $\mbox{Cr}$ aggregation}

Similarly to the most combinations of magnetic elements and host semiconductors, the solubility of Cr in ZnTe is low and the incorporation of Cr beyond a certain limit results in the crystallographic or chemical phase separation. The driving force of the phase separation is an attractive interaction between magnetic cations. The key to understanding the observation that the uniformity of Cr distribution varies with the Zn/Te flux ratio or the codoping of donor or acceptor impurities is the dependence of the attractive interaction on the Fermi level position within the band gap \cite{Dietl:2006_NM,Ye:2006_PRB}. In the case of intrinsic ZnTe, Cr assumes the 2+ charge state. However, this charge state can be changed by trapping a hole or an electron since the substitutional Cr forms the deep donor Cr$^{2+/3+}$ and acceptor Cr$^{1+/2+}$ levels within the band gap of the host ZnTe \cite{Godlewski:1980_JPC}. Therefore, the codoping with shallow impurities changes the Cr valence and, thus, the Cr-Cr interaction. In particular, an additional Coulomb repulsion between those Cr ions that have trapped a carrier hinders the Cr aggregation, leading to a uniform distribution of Cr cations. This is in contrast to a non-uniform distribution driven by the attractive interaction between isoelectronic Cr$^{2+}$ ions, associated with the contribution of TM 3$d$ states to bonding, as discussed in Sec.\,\ref{sec:pairing}.

In order to understand the observed phenomena based on this model, one should teke into account that ZnTe crystals, due to the native formation of Zn vacancies, have a tendency to become $p$-type even without intentional doping \cite{Baron:1998_JAP}. In (Zn,Cr)Te films grown under the Te-rich condition, the Zn vacancies are formed, so that a part of the Cr ions assumes the Cr$^{3+}$ charge state. The growth under the Zn-rich condition or  codoping with iodine donors restore the isoelectronic Cr$^{2+}$ configuration, as the Zn vacancies are suppressed by a surplus supply of Zn or compensated by the I donors. In the I-doped films, the degree of inhomogeneity and the value of the blocking temperature $T_{\mathrm{b}}$ (and also  $T_{\text{C}}^{\text{(app)}}$ for $x \simeq 0.05$) attain a maximum when the iodine and Zn vacancy concentrations become equal, so that the charge state of all Cr ions is 2+. In contrast, codoping with N increases the concentration of Cr$^{3+}$ ions (as observed by XMCD \cite{Yamazaki:2011_JPCM}), which diminishes attractive forces between Cr cations and results in their uncorrelated distribution.

\subsubsection{Magnetooptical and magnetotransport properties}

Magnetooptical properties were investigated for crystals exhibiting a high-temperature ferromagnetism assigned, as explained previously, to a nonuniform Cr distribution. Figure \ref{kuroda_fig5} shows the result of the MCD measurement on an I-doped Zn$_{1-x}$Cr$_x$Te film with a relatively low Cr content of $x = 0.07$ and $T_{\text{C}}^{\text{(app)}} \simeq 300$\,K \cite{Ozaki:2006_PRL}. As shown in Fig.\,\ref{kuroda_fig5}(a), the MCD spectrum exhibits a broad band below the band-gap energy 2.38\,eV of ZnTe, in contrast to a sharp peak in the spectrum of ZnTe.  According to Fig.\,\ref{kuroda_fig5}(b), the MCD intensity at a fixed photon energy shows virtually identical magnetic-field dependence as magnetization measured by SQUID. These features of MCD were also observed in undoped films $x = 0.07$ grown under Te-rich flux \cite{Kuroda:2005_STAM}, in which smaller scale inhomogeneities in the Cr distribution lead to lower value of $T_{\text{C}}^{\text{(app)}}$, of the order of 100\,K.

\begin{figure}[h!]
\begin{center}
\includegraphics[width=0.9\columnwidth]{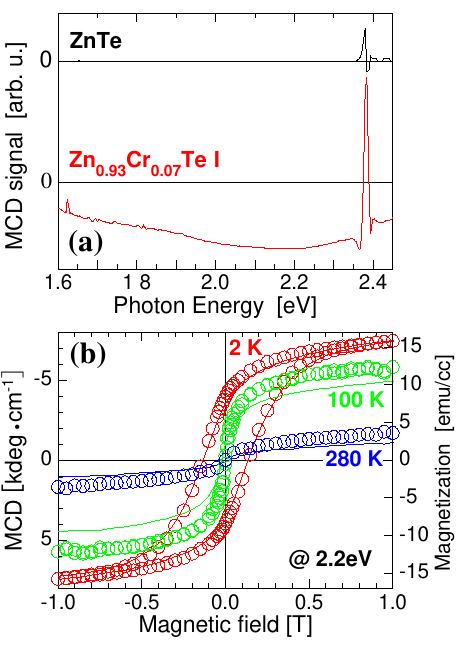}
\end{center}
\caption{(Color online) Results of MCD studies on an I-doped Zn$_{0.93}$Cr$_{0.07}$Te film. The measurement was performed in the transmission mode in the magnetic field perpendicular to the film plane (Faraday configuration). (a) The MCD spectra at 2\,K and 1\,T, with the reference data for a ZnTe film. (b) Magnetic-field dependence of the MCD intensities at a photon energy of 2.2\,eV (circles), together with magnetization values measured by SQUID (lines). From \onlinecite{Ozaki:2006_PRL}.}
\label{kuroda_fig5}
\end{figure}

The broadband in the MCD spectrum could be attributed to optical transitions from the valence band to the Cr 3$d$ level in the middle of the band gap, with a significant broadening of the absorption energy due to the lattice relaxation \cite{Godlewski:1980_JPC} or may originate from the presence of Cr-rich regions that affect MCD in a wide spectral region via magnetization-dependent boundary conditions for light propagation (the Kerr effect). The consistency between magnetooptical and magnetization data, observed also in the case of decomposed (Ga,Mn)As, (Ge,Mn),  and (Ge,Fe), as discussed in Secs.\,\ref{sec:MCD_GaAs}, \ref{sec:MCD_GeMn},  and \ref{sec:MCD_GeFe}, respectively, suggests that in these systems magnetooptical response is dominated by regions with high TM concentrations, which account for the FM properties.

In addition, the anomalous Hall effect was observed in an I-doped Zn$_{1-x}$Cr$_x$Te film (\onlinecite{Kuroda:2007_NM}, Supplementary Information). The observation of the magnetooptical and magnetotransport properties peculiar to DMSs even in phase-separated (Zn,Cr)Te layers suggests a possibility of new functionalities in hybrid structure consisting of TM-rich nanocrystals embedded in a semiconductor matrix, as discussed in Sec.\,\ref{sec:prospects}.

\subsection{Crystallographic phase separation and konbu phase}
\label{sec:ZnTe_crystallographic}
\subsubsection{Structural nanocharacterization}

Structural, compositional, and magnetic properties were investigated for a series of I-doped Zn$_{1-x}$Cr$_x$Te films varying substrate temperature during the growth, growth rate, and crystallographic orientation of the substrate \cite{Nishio:2009_MRSP}. It was found that the substrate temperature is a critical factor controlling the phase separation. For the lower average Cr content $x \simeq 0.05$, the structural properties of I-doped Zn$_{1-x}$Cr$_x$Te films change with the substrate temperature $T_{\mathrm{g}}$, while the uniformity of Cr distribution is not much different. Figure \ref{kuroda_fig6} shows the results of TEM and EDS studies of I-doped Zn$_{1-x}$Cr$_x$Te ($x \simeq 0.05$) films grown in the [001] direction at various values of $T_{\mathrm{g}}$. As shown in the TEM and diffraction images (left column), the crystallinity exhibits a marked variation with $T_{\mathrm{g}}$; at an intermediate temperature of $T_{\mathrm{g}} = 270^{\circ}$C, the crystal consists dominantly of the zb structure, but the stacking faults along the \{111\} planes are observed in the lattice image and additional spots corresponding to the triplet periodicity of the stacking faults appear in the diffraction image. The crystallinity is deteriorated by decreasing $T_{\mathrm{g}}$ down to $240^{\circ}$C, exhibiting polycrystalline features in addition to the stacking faults, while it looks much improved by increasing $T_{\mathrm{g}}$ up to $360^{\circ}$C, the TEM and diffraction images exhibiting almost perfect zb structure without the stacking faults. On the other hand, as shown in the Cr mapping images (right column), the Cr distribution is inhomogeneous in all the films, but the degree of inhomogeneity seems to be slightly reduced with the increase of $T_{\mathrm{g}}$. The results of XAS measurements at Cr $K$-edge for the same series of I-doped Zn$_{1-x}$Cr$_x$Te ($x \simeq 0.05$) films suggested a change in the local crystallographic structure depending on the substrate temperature \cite{Ofuchi:2009_JPCS}, which is correlated with the results of the TEM observation.

\begin{figure}[h!]
\begin{center}
\includegraphics[width=0.9\columnwidth]{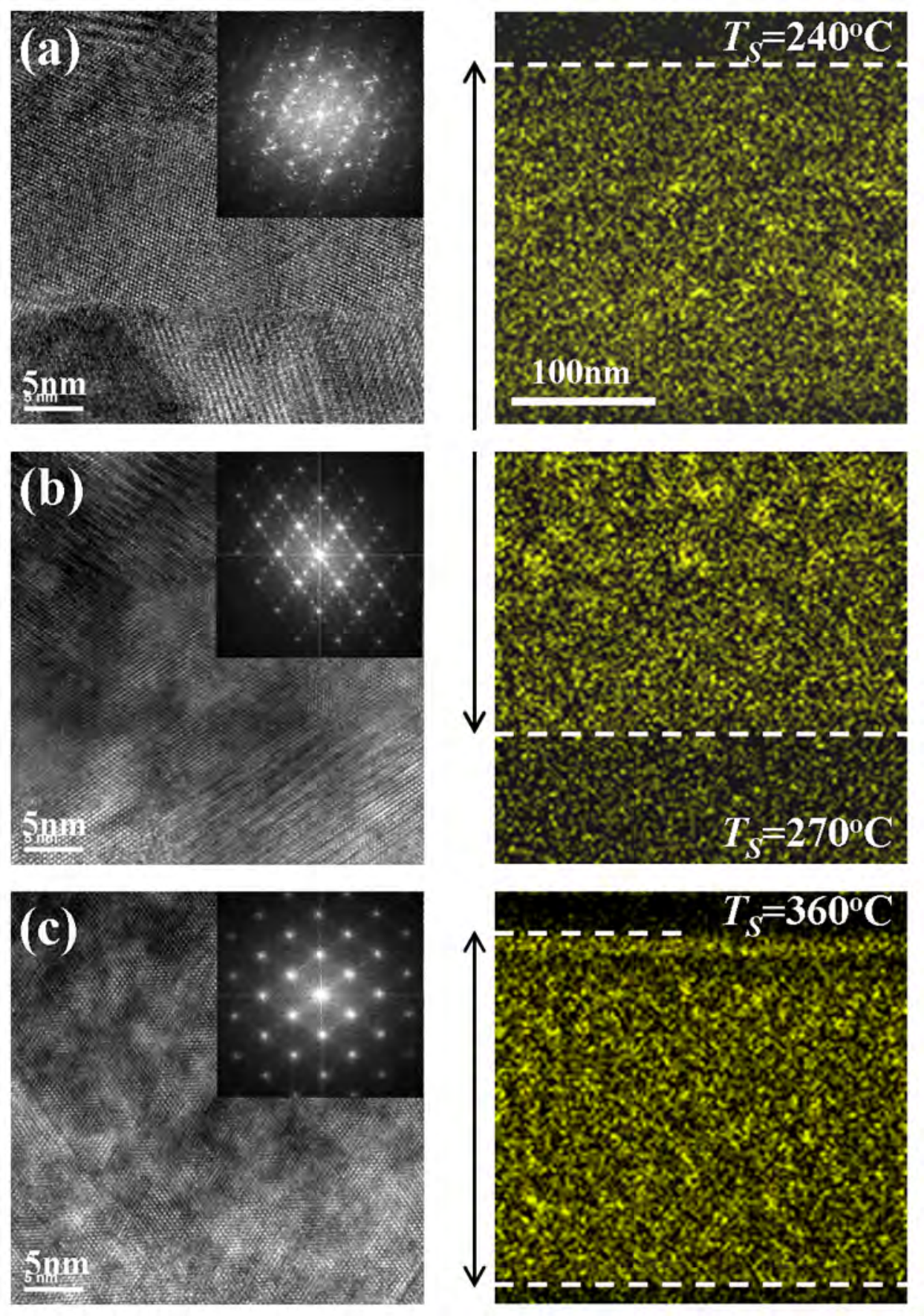}
\end{center}
\caption{(Color online) TEM and diffraction images (left) and EDS mapping images of Cr (right) for cross-sectional pieces of I-doped Zn$_{1-x}$Cr$_x$Te films ($x \simeq 0.05$) grown in the [001] direction at various substrate temperatures. (a) $T_{\mathrm{g}} = 240^{\circ}$C; (b) $T_{\mathrm{g}} = 270^{\circ}$C; (c) $T_{\mathrm{g}} = 360^{\circ}$C. In the EDS mapping images, the range of the Zn$_{1-x}$Cr$_x$Te:I layer and the boundary with the ZnTe buffer layer are indicated by arrows and dashed lines, respectively. From \onlinecite{Nishio:2009_MRSP}.}
\label{kuroda_fig6}
\end{figure}

A natural explanation of these results would be that lattice defects appearing at low  $T_{\mathrm{g}}$ lower kinetic barriers for Cr diffusion, so that the Cr aggregation is efficient even at low $T_{\mathrm{g}}$.

In contrast, for the higher average Cr content $x\sim 0.2$, the substrate temperature significantly affects the form of the phase separation \cite{Nishio:2009_MRSP}. Figure \ref{kuroda_fig7} presents Cr mapping images of the films grown onto either (001) or (111) substrates at intermediate and high substrate temperatures,  $T_{\mathrm{g}} = 300$ and $360^{\circ}$C. As shown in Fig.\,\ref{kuroda_fig7}(a), the  Cr-rich regions form isolated NCs at $T_{\mathrm{g}} = 300^{\circ}$C, whereas according to Figs.\,\ref{kuroda_fig7}(b,c), in the case of films grown at $T_{\mathrm{g}} = 360^{\circ}$C, Cr-rich regions look like continuous nanocolumns aligned approximately along the $\langle111\rangle$ direction of the host crystal. As estimated by the EDS spot analysis, a maximum value of the local Cr composition is $x = $0.4--0.5, in the case of  Cr-rich both NCs and nanocolumns.

\begin{figure}[h!]
\begin{center}
\includegraphics[width=0.9\columnwidth]{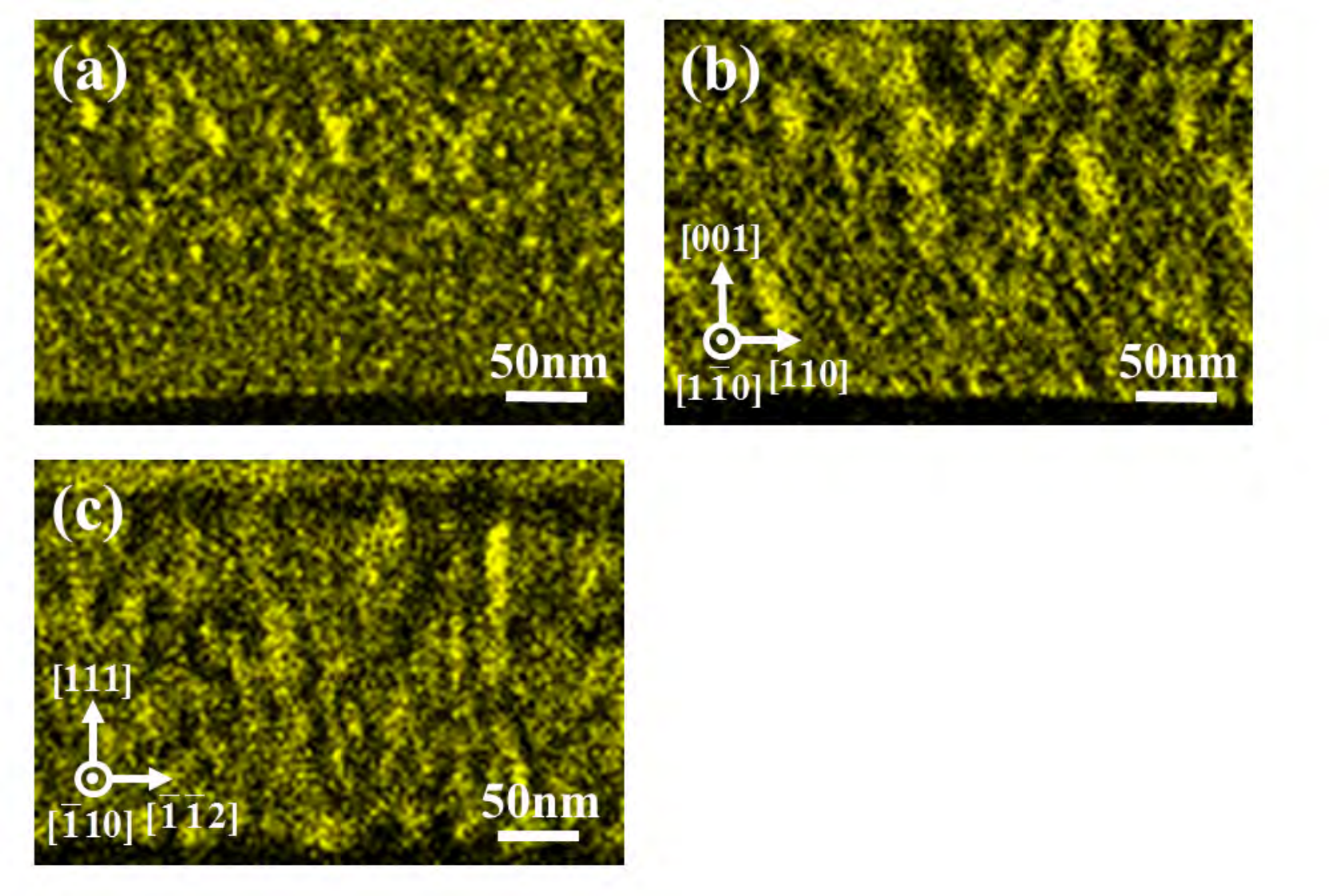}
\end{center}
\caption{(Color online) Cross-sectional EDS mapping images of Cr for Zn$_{1-x}$Cr$_x$Te:I films ($x \simeq 0.25$) grown (a) at a substrate temperature $T_{\mathrm{g}} = 300^{\circ}$C in the [001] direction, (b) at $T_{\mathrm{g}} = 360^{\circ}$C in the [001] direction, and (c) at $T_{\mathrm{g}} = 360^{\circ}$C in the [111] direction, imposed by crystallographic orientations of GaAs substrates. Cr-rich regions are formed as isolated NCs in (a), while they are elongated and form continuous nanocolumns in (b), (c). From \onlinecite{Nishio:2009_MRSP}.}
\label{kuroda_fig7}
\end{figure}

Figure \ref{kuroda_fig8} shows TEM and EELS images allowing to shed some light on the crystal structure of Cr-rich regions in an I-doped Zn$_{1-x}$Cr$_x$Te films ($x = 0.25$) grown at $T_{\mathrm{g}} = 360^{\circ}$C \cite{Kobayashi:2012_PB}. In the TEM image in Fig.\,\ref{kuroda_fig8}(a), moir\'e fringes appear in many regions, which suggests that these regions are composed of a mixed phase with different crystal structures, pointing to the presence of crystallographic phase separation. By comparing the TEM image [Fig.\,\ref{kuroda_fig8}(a)] with the EELS image [Fig.\,\ref{kuroda_fig8}(b)] of the same area, it is confirmed that the moir\'e regions in the TEM image correspond to the regions with a high value of Cr content. In a magnified lattice image in Fig.\,\ref{kuroda_fig8}(c), the moir\'e region is boarded by the two \{111\}$_{zb}$ planes, and there appears another region exhibiting a different structure, identified as a hexagonal structure from the arrangement of diffraction spots in the fast Fourier transform image shown in Fig.\,\ref{kuroda_fig8}(d). The lattice parameters deduced from the spacing between the spots are $c = 6.32$\,{\AA} and $a = 4.18$\,{\AA}, close to the values reported for bulk CrTe in the NiAs structure \cite{Dijkstra:1989_JPC,Ohta:1993_JPC}. This suggests that the precipitates consist of NiAs-type CrTe or of nonstoichiometric Cr$_{1-\delta}$Te in a hexagonal structure. According to the TEM and diffraction images in Fig.\,\ref{kuroda_fig8}, the $c$-axis of elongated hexagonal NCs is nearly parallel to the \{111\}$_{\text{zb}}$ planes of the host. In the case of $T_{\mathrm{g}} = 360^{\circ}$C, a 3D analysis using atom probe tomography revealed that the Cr-aggregated regions were formed as thin plates with the base plane nearly parallel to the \{111\}$_{\text{zb}}$ planes \cite{Kodzuka:2012_PhD}.

\begin{figure}[h!]
\begin{center}
\includegraphics[width=0.9\columnwidth]{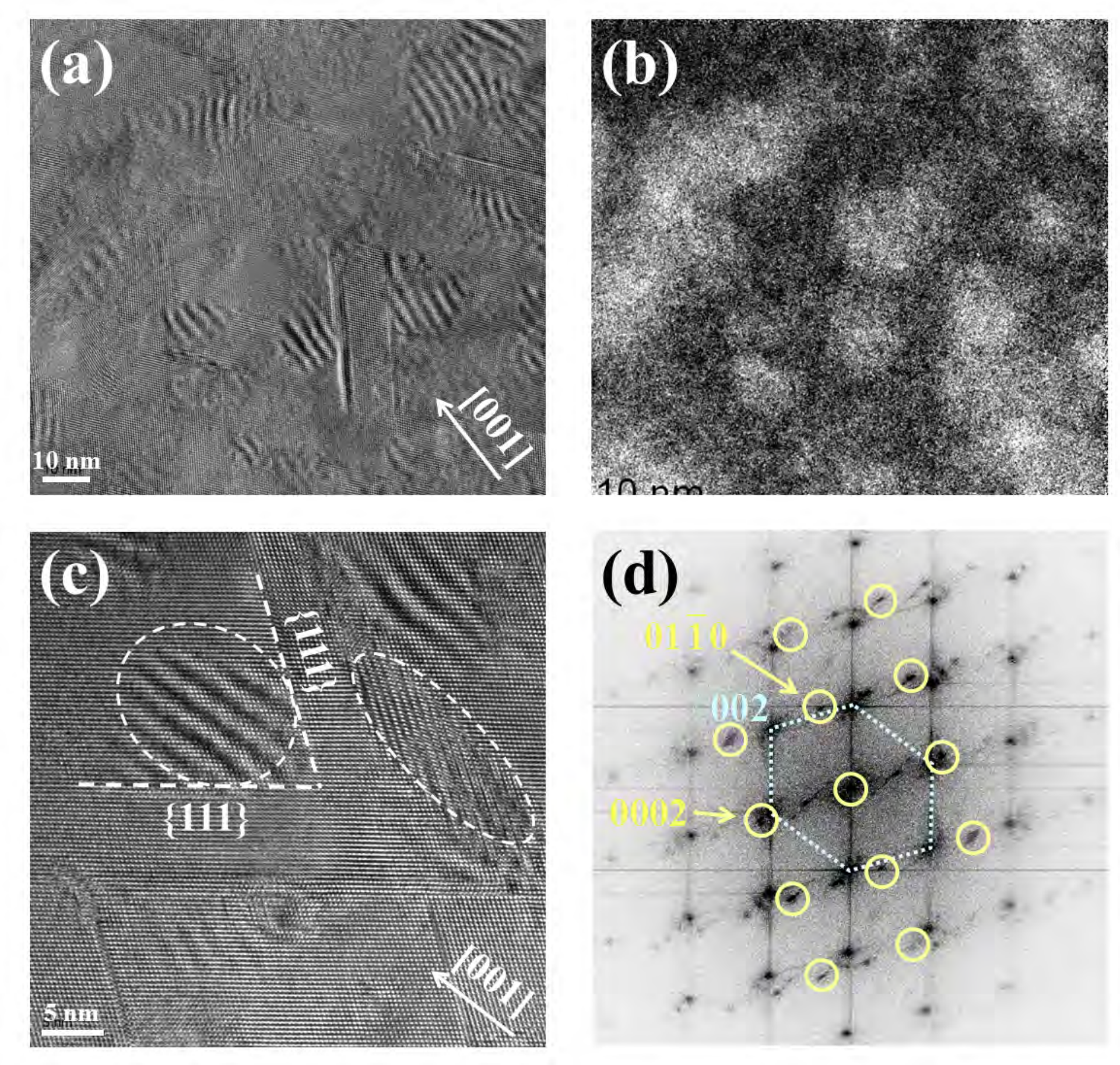}
\end{center}
\caption{(Color online) TEM and EELS images of the I-doped Zn$_{0.75}$Cr$_{0.25}$Te film grown in the [001] direction at $T_{\mathrm{g}}\ = 360^{\circ}$C [the same film as shown in Fig.\,\ref{kuroda_fig7}(b)] giving evidence for crystallographic phase separations.  (a) TEM image. (b) Cr mapping image of the same area obtained by EELS. (c) Lattice image in a magnified scale. A moir\'e region boarded by the two \{111\}$_{\text{zb}}$ planes and a region of a different crystallographic structure are observed (encircled by broken lines). (d) FFT image determined from the data in (c). The set of diffraction spots of a hexagonal arrangement (connected by dotted line) originates from the zb structure and that of a rectangular arrangement (denoted by circles) originates from the hexagonal structure whose $c$-plane is almost parallel to the \{111\}$_{\text{zb}}$ planes. From \onlinecite{Kobayashi:2012_PB}.}
\label{kuroda_fig8}
\end{figure}

These observations of the crystallographic phase separation support theoretical computations indicating that zb-CrTe with the lattice parameter of ZnTe is unstable against the formation of CrTe in the NiAs crystal structure \cite{Zhao:2005_PRB}.

In the conventional  $\omega$--2$\theta$  XRD scan, which probes XRD from the film plane, only maxima corresponding to the zb host crystal were detected for the (001) Zn$_{1-x}$Cr$_x$Te:I films with $x \sim 0.2$. However, hexagonal Cr$_{1-\delta}$Te precipitates that have principal diffraction planes inclined against the film plane, can be detected in the  $\omega$ scan, as shown in Fig.\,\ref{kuroda_fig9}.  This measurement was performed with the incident and reflected x rays in the $(1\bar{1}0)$ plane of the host zb structure. By locking $2\theta$  to the values corresponding to the ($\bar{1}$101) and ($\bar{1}$102) planes of the hexagonal Cr$_{1-\delta}$Te, the diffractions from these planes were detected in an almost symmetric way with respect to $\omega$. The values of $\omega$  of the diffraction peaks are close to the ones expected from the crystallographic relation of these NCs to the host crystal revealed by the TEM observation. On the other hand, in the same  $\omega$-scan measurement in another configuration with the incident and reflected x rays in the (110)$_{\text{zb}}$ plane, the intensity of the diffraction from the hexagonal nanocrystals was much reduced \cite{Kobayashi:2013_P}. This result suggests that the hexagonal Cr$_{1-\delta}$Te tends to stack preferentially on the $(111)_{\text{zb}}$ plane [or Zn-terminated (111)$A$ plane] rather than $(\bar{1}\bar{1}\bar{1})_{\text{zb}}$ plane [or Te-terminated (111)$B$ plane]. This preference could be understood from the arrangement of atoms on the interface between the hexagonal Cr$_{1-\delta}$Te and the host zb crystals.

\begin{figure}[h!]
\begin{center}
\includegraphics[width=0.9\columnwidth]{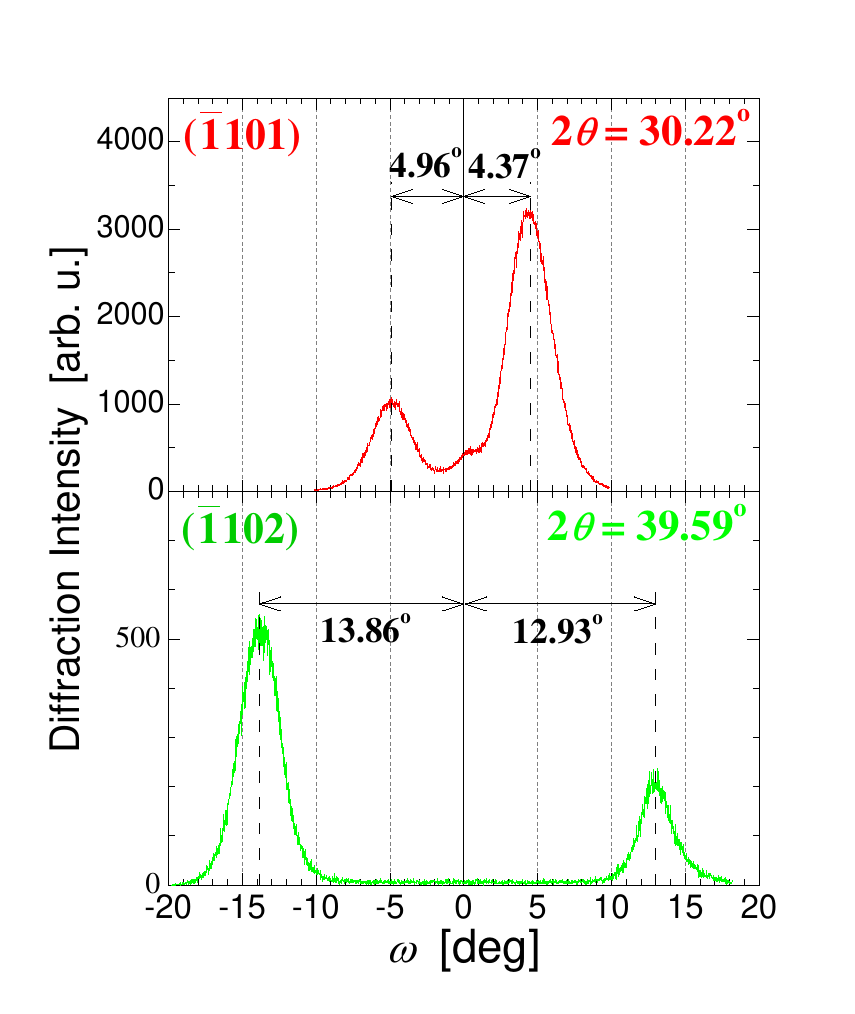}
\end{center}
\caption{(Color online)XRD  $\omega$-scan profiles of an I-doped Zn$_{0.73}$Cr$_{0.27}$Te film grown in the [001] orientation at $T_{\mathrm{g}} = 390^{\circ}$C. The diffraction profiles from the plane ($2\theta  = 30.22^{\circ}$) (upper panel) and the plane ($2\theta   = 39.59^{\circ}$) (lower panel) are plotted, respectively. This measurement was performed in the configuration with the incident and reflected x-ray in the (1$\bar{1}$0) plane of the host zb structure.  From \onlinecite{Kobayashi:2012_PB}.}
\label{kuroda_fig9}
\end{figure}

The intensity of the diffraction from the hexagonal Cr$_{1-\delta}$Te gives us an estimate of the amount of the precipitates formed in the crystal. The  $\omega$-scan measurement on the series of Zn$_{1-x}$Cr$_x$Te:I films grown at different substrate temperatures revealed how the amount of the precipitates depends on the growth temperature. As seen in Fig.\,\ref{kuroda_fig10}, in which the diffraction intensity from the ($\bar{1}$101) plane of the hexagonal Cr$_{1-\delta}$Te is plotted as a function of the substrate temperature $T_{\mathrm{g}}$, the hexagonal precipitates are formed in a larger quantity at a higher $T_{\mathrm{g}}$.

\begin{figure}[h!]
\begin{center}
\includegraphics[width=0.9\columnwidth]{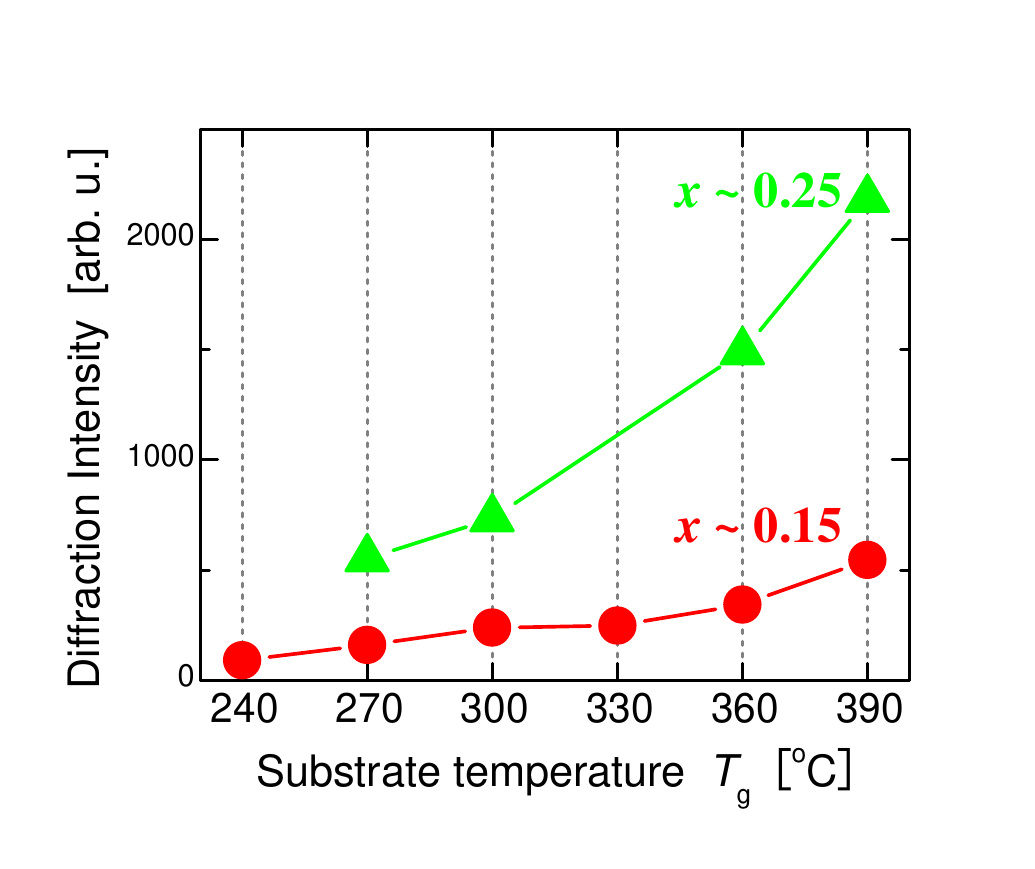}
\end{center}
\caption{(Color online) The diffraction intensity from the ($\bar{1}$101) plane of the hexagonal Cr$_{1-\delta}$Te nanocrystals plotted as a function of the substrate temperature $T_{\mathrm{g}}$ for the two series of Zn$_{1-x}$Cr$_x$Te:I films with $x \simeq 0.15$ and 0.25. From Kuroda {\em et al.}, unpublished.}
\label{kuroda_fig10}
\end{figure}

As described previously, it was revealed that the substrate temperature during the growth significantly affects the Cr aggregation; the Cr-aggregated regions are formed in almost spherical NCs at a low $T_{\mathrm{g}}$ while they are formed in thin plates at higher $T_{\mathrm{g}}$. In addition, the amount of the hexagonal precipitates increases with the increase of $T_{\mathrm{g}}$. The observed variation in the shape of Cr-aggregated regions with the substrate temperature reflects a difference in the dimensionality of the spinodal decomposition, which results in the formation of the dairiseki or konbu phase, as discussed in Secs.\,\ref{sec:dairiseki} and \ref{sec:konbu}. According to a theoretical consideration \cite{Fukushima:2006_JJAP}, the spinodal decomposition in the layer-by-layer growth mode results in the formation of one-dimensional columnar regions with high content of the magnetic element. With an enhanced migration of impinging atoms on the growing surface at a higher substrate temperature, Cr atoms tend to aggregate in such places on the top surface where Cr-aggregated areas are already formed in the layer just below the growing surface. As a result, the Cr-rich regions form continuous nanocolumns, instead of isolated clusters appearing in the case of slow surface migration. Once Cr-aggregated regions are formed inside the crystal, they can transform at sufficiently high $T_{\mathrm{g}}$ into a stable Cr$_{1-\delta}$Te hexagonal compound that assumes crystal orientation insuring the best match of the atom positions with the host, i.e., minimization of the interfacial energy.

\subsubsection{Magnetic properties}

Three characteristic temperatures  $T_{\text{C}}^{\text{(app)}}$,  $\theta_P$, and $T_{\mathrm{b}}$ are presented in Fig.\,\ref{kuroda_fig11} as a function of $T_{\mathrm{g}}$ for the two series of I-doped Zn$_{1-x}$Cr$_x$Te films with the low and high average Cr contents of $x \simeq 0.05$ (full symbols) and 0.25 (empty symbols). In the case of $x \simeq 0.05$, the film grown at the lowest $T_{\mathrm{g}}$ of $240^{\circ}$C did not exhibit ferromagnetism even at 2\,K, presumably due to the deterioration of crystallinity. In contrast, the films obtained at $T_{\mathrm{g}}$ from 270 to 390$^{\circ}$C show high-temperature ferromagnetism, consistent with the presence of Cr-rich regions.
The observed values $T_{\text{C}}^{\text{(app)}} \simeq 300$\,K and  $\theta_P\simeq 330$\,K for films with  $x \simeq 0.25$ are consistent with the fact that Cr$_{1-\delta}$Te in bulk form exhibits ferromagnetism with $T_{\text{C}}$ = 325--360\,K for a relatively a small amount of Cr deficiency   \cite{Shimada:1996_PRB},  though $T_{\text{C}}$ decreases when the amount of Cr deficiency increases further on \cite{Ohta:1993_JPC}. Furthermore, in agreement with TEM data pointing to smaller volumes of Cr-rich regions in the case of samples with lower $x$, the magnitudes of $T_{\mathrm{b}}$ are significantly lower for $x \simeq 0.05$ compared to the values found for $x \simeq 0.25$. However, rather surprisingly, $T_{\text{C}}^{\text{(app)}}$ and  $\theta_P$ tend to decrease from 300 down to 250\,K with $T_{\mathrm{g}}$ for samples with $x \simeq 0.05$.
This result points to a complex interplay between the degree of host crystallinity, TM aggregation, and resulting $T_{\text{C}}$ of embedded NCs.

\begin{figure}[h!]
\begin{center}
\includegraphics[width=0.9\columnwidth]{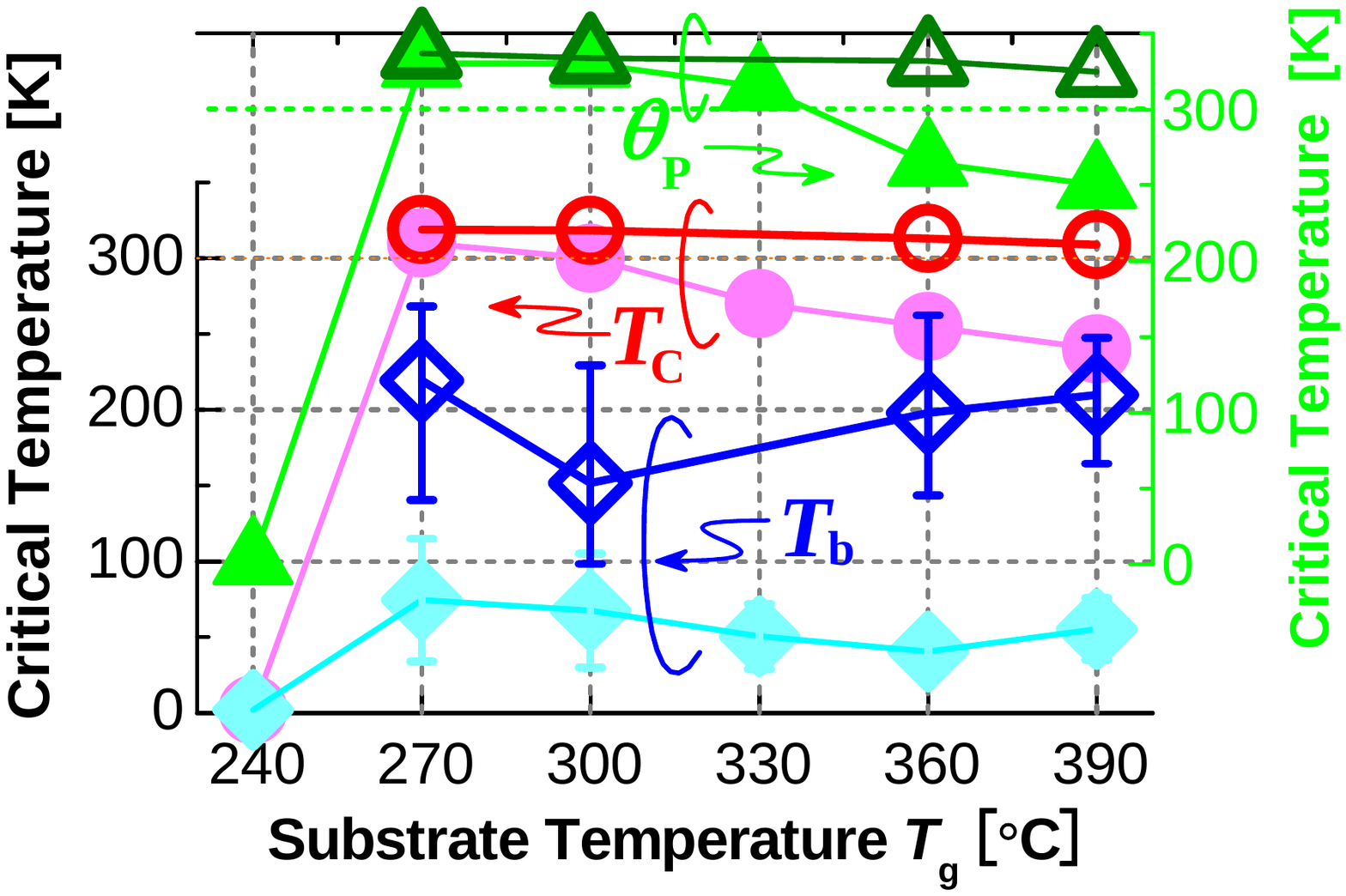}
\end{center}
\caption{(Color online) Plot of three characteristic temperatures  $T_{\text{C}}^{\text{(app)}}$ (circles),  $\theta_P$ (triangles), and $T_{\mathrm{b}}$ (diamonds) as a function of the substrate temperature $T_{\mathrm{g}}$ for I-doped Zn$_{1-x}$Cr$_x$Te films with the low ($x  \simeq 0.05$) and high ($x \simeq 0.25$) average Cr contents (full and empty symbols, respectively). From Kuroda {\em et al.}, unpublished.}
\label{kuroda_fig11}
\end{figure}

It can be expected that the strain and shape of NCs affect their magnetic anisotropy.
Figure \ref{kuroda_fig12} presents magnetization loops $M(H)$ for magnetic fields perpendicular and parallel to the film plane for the I-doped (111) Zn$_{1-x}$Cr$_x$Te film ($x = 0.22$), in which Cr-rich NCs show the elongated shape with long axes almost perpendicular to the film plane \cite{Nishio:2009_MRSP}. As seen,  hysteretic behaviors are more pronounced (i.e., the coercive force is larger) for the perpendicular magnetic field. This suggests that the easy direction is parallel to the long axis of the NCs, the expected result for the shape-dependent direction of the demagnetization field and weak in-plane crystalline magnetic anisotropy.

\begin{figure}[h!]
\begin{center}
\includegraphics[width=0.9\columnwidth]{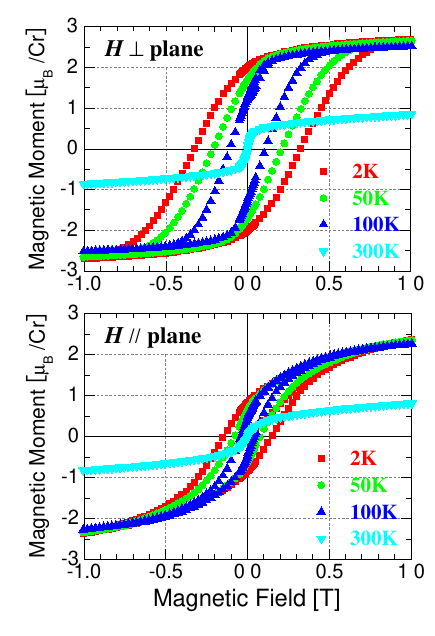}
\end{center}
\caption{(Color online) Magnetization loops in Zn$_{0.78}$Cr$_{0.22}$Te:I film grown in the [111] direction at $T_{\mathrm{g}} = 360^{\circ}$C [the same film as shown in Fig.\,\ref{kuroda_fig7}(c)]. The upper (lower) panel shows the results obtained for magnetic fields perpendicular (parallel) to the film plane. The coercive field $H_c$ for the perpendicular configuration ($\mu_0H_c = 0.33$\,T at 2\,K) is much larger than for the parallel configuration ($\mu_0H_c = 0.14$\,T at 2\,K).  From \onlinecite{Nishio:2009_MRSP}.}
\label{kuroda_fig12}
\end{figure}

\subsection{Summary}
The comprehensive studies of (Zn,Cr)Te described in this section show that the Cr distribution is affected significantly by codoping with donor or acceptor impurities or by the growth under different Zn/Te flux ratios. This behavior originates from the influence of codoping and growth conditions upon the position of the Fermi level in the band gap, which controls the Cr charge state: from Cr$^{+1}$ for high I donor concentrations to Cr$^{3+}$ at high density of N acceptor impurities or Zn vacancies. These findings corroborate the influence of Coulomb interactions upon the chemical forces between TM ions. The magnetic properties are closely correlated with the heterogeneity of the Cr distribution. In particular, the Cr-rich regions give rise to the appearance of high-temperature ferromagnetism. The systematic investigations on (Zn,Cr)Te films grown under various MBE conditions revealed that the growth temperature plays a crucial role in the phase separation: the crystallographic structure and the shape of the Cr-rich regions change with the substrate temperature during the growth. The ability to fabricate in a self-organized fashion hybrid systems consisting of FM nanoclusters embedded in the semiconductor matrix, as realized in (Zn,Cr)Te and other DMSs, opens doors for studies of various functionalities of these novel nanocomposites.

\section{Prospects of spinodal nanotechnology \label{sec:prospects}}

It is believed that the application of embedded metallic NCs will revolutionize the performance of various commercial devices, such as flash memories. Similarly colloidal or embedded semiconducting NCs, i.e., semiconductor quantum dots and nanowires, are extensively studied as perspective media for lightning, low current semiconductor lasers, solar cells, single-photon emitters and detectors, quantum processors and memories.

In view of the results presented, the incorporation of TM impurities into semiconductors opens the door to fabrication of dense arrays of TM-rich NCs coherently embedded into a semiconductor matrix. Depending on growth and processing conditions as well as on the combination of host semiconductor, TM impurity, and codoping the self-assembled NCs can be metallic or semiconducting, form nanodots or nanocolumns, exhibit FM, ferrimagnetic, or AF spin ordering persisting usually to above RT. Thus, spinodal nanotechnology not only allows one to explore the feasibility of device fabrication with a new and versatile bottom-up method but also considerably enlarges the spectrum of possible functionalities offered by nanoscale heterogeneous systems \cite{Katayama-Yoshida:2007_PSSA,Dietl:2008_JAP}.

As an example, Fig.\,\ref{fig:HKY} presents results of Monte Carlo simulations in which the Cr flux was altered during the 2D growth of CrTe nanocolumns (the {\em konbu} phase) embedded in ZnTe \cite{Fukushima:2007_PSSC}. This example suggests that dense arrays of various nanodevices can be fabricated by selecting appropriate growth protocols. For instance, by reducing the flux for time corresponding to the growth of a few monolayers, the formation of magnetic tunnel junctions is predicted for nanocolumns of a FM metal \cite{Fukushima:2007_PSSC}. Such junctions can serve for low-power high-density magnetic storage, including spin-torque magnetic random access memories and, if sufficiently high TMR is found, for the field programmable logic, i.e., TMR-based  connecting/disconnecting switches \cite{Reiss:2006_APL}, and even all-magnetic logic, characterized by low-power consumption and radiation hardness. Furthermore, nanocolumns might form racetracks for domain-wall based 3D memories \cite{Thomas:2007_S}.

\begin{figure}[tb]
\includegraphics[angle=0,width=0.45\textwidth]{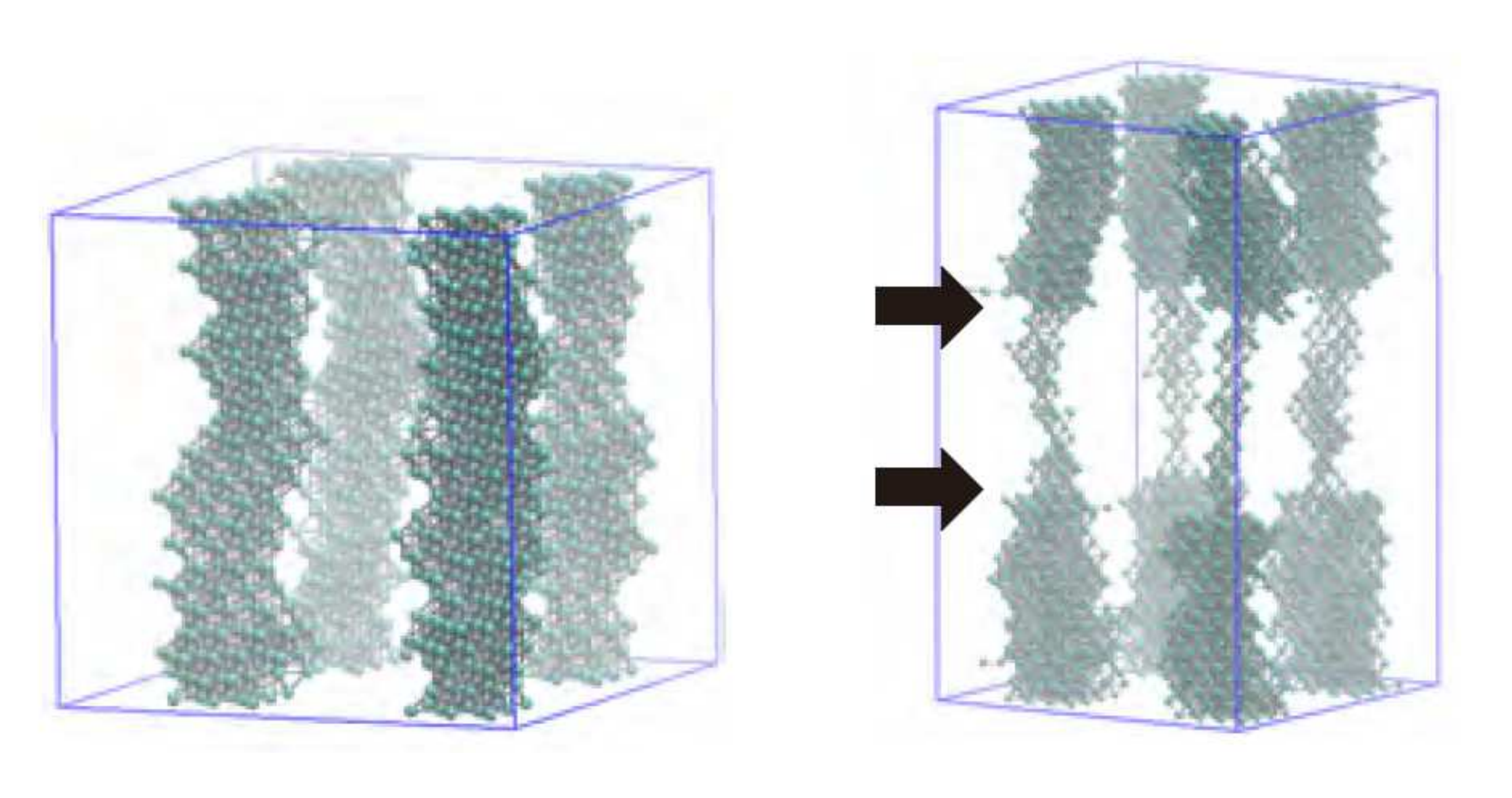}
\caption{(Color online) Demonstration of the nanocolumn shape control
by Monte Carlo simulations of (Zn,Cr)Te epitaxy (the konbu phase).
Positions of Cr cations are shown.
(a) Nanoscale seeding by zinc-blende CrTe. (b) During the deposition
(between two arrows),
the Cr concentration is reduced to control the shape of the metallic and
FM CrTe nanocolumns embedded in ZnTe.
Adapted from \onlinecite{Fukushima:2007_PSSC}.}
\label{fig:HKY}
\end{figure}

Another kind of possible spintronic applications makes use of the coupling between FM leads and carriers in a semiconductor matrix. In particular, in the proposal of a scalable processor \cite{Dery:2007_N}, the magnetization-dependent interaction between neighbor FM contacts is mediated by spin currents injected from the FM contact to the semiconductor. Furthermore, in the pioneering design of quantum processers involving the couplings between single spins on quantum dots, nanomagnets serve to introduce differences in spin resonance frequencies of particular dots, which makes it possible to address individual spins \cite{Loss:1998_PRA}. Actually, top-down methods were already employed to pattern nanostructures with small magnets whose stray fields controlled spin currents \cite{Wrobel:2004_PRL} and spin resonance frequencies \cite{Pioro-Ladriere:2008_NP} at the nanoscale.

As discussed in Secs.\,\ref{sec:GaAs}, \ref{sec:Ge-Mn}, \ref{sec:Ge-Fe}, and \ref{sec:ZnTe}, a number of decomposed alloys exhibits spin-related magnetotransport and magnetooptical phenomena typically persisting up to $T_{\text{C}}$ of the relevant NCs. In particular, AHE was observed in (Ge,Mn) \cite{Jamet:2006_NM} and (Zn,Cr)Te \cite{Kuroda:2007_NM}. A sizable magnitude of AHE together with heterogeneity-induced mixing between diagonal and off-diagonal conductivity tensor components resulted in a large positive MR of (Ge,Mn) \cite{Jamet:2006_NM}. A related effect was also observed in granular films with MnSb \cite{Akinaga:2000_APLc} and MnAs \cite{Yokoyama:2006_JAP,Kudrin:2014_PRB} NCs. Furthermore, TMR was found in a device containing MnAs and GaAs:MnAs electrodes \cite{Hai:2006_APL,Hai:2008_PRB}. All these findings could possibly be exploited in magnetic-field sensors with characteristics  (e.g., impedance) that could be tailored over a wide range.

At the same time, as discussed in Secs.\,\ref{sec:GaAs}, \ref{sec:Ge-Mn}, \ref{sec:Ge-Fe}, and \ref{sec:ZnTe}, MCD was reported for decomposed (Ga,Mn)As, (Ge,Mn), (Ge,Fe), and (Zn,Cr)Te as well as for other systems such as (Ga,Mn)P \cite{Monette:2010_JAP}. A combination of strong and spectrally broad MCD specific to FM metals and weak losses characterizing the semiconductor hosts suggests possible applications of decomposed semiconductor alloys as optical isolators \cite{Amemiya:2006_APL} as well as 3D tunable photonic crystals and spatial light modulators for advanced photonic applications \cite{Park:2002_JJAP}.

Metallic nanocolumns and nanodots could also serve as high quality nanocontacts in nanoelectronics and optoelectronics as well as metallic elements in nanoplasmonics. Dense arrays of nanocolumns are also attractive for thermoelectric applications (power generators and coolers) as a specific form of density of states in 1D systems is expected to result in a significant enhancement of the Seebeck and Peltier effects \cite{Vu:2011_APE,Shinya:2014_JJAP,Vu:2014_JJAP,Vu:2014a_JJAP,Vu:2015_JJAP}. Furthermore, the growth process presented in Fig.\,\ref{fig:HKY} can be adapted to fabricate devices for all-metal nanoelectronics based, for instance, on single-electron transistors. Actually, the Coulomb blockade was demonstrated for a single MnAs NC, tunnel-coupled to MnAs electrodes patterned lithographically \cite{Hai:2010_NN}. A long spin relaxation time, of the order of 10 $\mu$s, was evaluated from these data.

If, in contrast, nanocolumns were semiconducting, appropriate band-gap engineering could improve efficiency of photovoltaic solar cells by leading to spatial separation of photoelectrons and photoholes in such all-semiconductor superstructures \cite{Oshitani:2011_APE,Tani:2010_APE,Tani:2011_APE,Tani:2012_JJAP,Tani:2012_PB,Tani:2012_JNCS}. As one more example of possible functionalities worth mentioning is the case of catalysts for automotive-emissions control. Here,  by using decomposed alloy with spatially separated NCs containing the relevant metal, e.g., Pt, the destructive process of metal agglomeration could be much reduced \cite{Kizaki:2008_APE,Kizaki:2013_CPL,Hamada:2011_JACS,Katz:2011_JACS}.

It worth adding that crystallographic phase separation has been extensively studied in III-V semiconductors doped with rare earths \cite{Buehl:2010_PRB,Clinger:2012_JAP,Kawasaki:2013_NL}. In these systems,  self-assembled rock-salt ErAs, TbAs, or ErSb nanocrystals, in the form of either nanodots or nanocolumns embedded in zinc-blende GaAs,  (In,Ga)As, or GaSb hosts, respectively were found  depending on epitaxy conditions. The suitability of these nanocomposites for thermoelectric, plasmonic, and THz  applications has already been addressed experimentally \cite{Clinger:2012_JAP,Lu:2014_NL,Salas:2015_APL}.

\section{Summary and outlook\label{sec:summary}}

We reviewed the recent progress in the understanding of high-temperature ferromagnetism in a range of magnetically doped semiconductors, including primarily (Ga,Mn)As, (Ga,Mn)N, (Ga,Fe)N, (Ge,Mn), (Ge,Fe), and (Zn,Cr)Te.  As we have emphasized, the abundance of contradicting views on the mechanisms accounting for surprisingly large magnitudes of Curie temperature $T_{\text{C}}$ resulted from intertwined theoretical and experimental challenges requiring the application of cutting-edge computational and materials nanocharacterization methods that have mostly become available only recently. In this way, semiconductors doped by transition metals (TMs) have emerged as outstanding systems to test our understanding of unanticipated relationships among (i) growth conditions, codoping, and processing; (ii) alloy nanostructure; and (iii) pertinent macroscopic properties.

The key ingredient of the materials families described in this review is spinodal nanodecomposition leading to the formation of TM-rich nanocrystals (NCs) either commensurate with the TM-depleted host (chemical phase separation) or precipitating in another crystallographic and/or chemical form (crystallographic phase separation). Whether magnetic ions are distributed randomly or form aggregates is determined by the competition between attractive forces among TM cations (revealed by {\em ab initio} computations) and kinetic barriers for TM diffusion at the growth surface or in the bulk. Accordingly, the alloy decomposition is more efficient at high growth temperatures and slow growth rates, and  depends also on the Fermi level position (codoping) that changes TMs valence and their diffusion coefficients. Remarkably, according to both theoretical simulations and experimental results the TM-rich regions (condensed magnetic semiconductors - CMSs) assume the form of either nanodots (the {\em dairiseki} phase) or nanocolumns (the {\em konbu} phase). Furthermore, the nanodots can be distributed randomly or accumulate in a plane adjacent to the interface or surface. Similarly, nanocolumns can extend along the growth direction or assume another spatial orientation.

A rich spectrum of forms assumed by spinodal nanodecomposition, such as chemical phase separation or aggregation of precipitates in one plane, have elucidated the reason why uncovering the presence of a nonrandom distribution of magnetic ions was so challenging in DMS research and, in particular, why the application of standard in-house structure characterization techniques (e.g., x-ray diffraction) was often misleading. The comprehensive element-specific nanocharacterization investigations described have demonstrated the existence of a tight correlation between the presence of TM-rich NCs and high-$T_{\text{C}}$ ferromagnetism. Within this scenario the puzzling $T_{\text{C}}$ independence of the average TM concentration $x$ has been explained. However, it has been found that the reverse is not true, i.e., variations of $T_{\text{C}}$ with $x$ do not prove a uniform distribution of TM ions. Furthermore, according to the evidence, spontaneous magnetization of decomposed systems is usually smaller than expected from the TM concentration, as typically a part of TM ions is distributed randomly giving rise to a paramagnetic, and not FM  or superparamagnetic response. It also happens that NCs formed by spinodal decomposition are weak ferromagnets or even antiferromagnets [see, e.g., the data for(Ga,Fe)N in Sec.\,\ref{sec:GaN-Fe}] or that coupling between them results rather in a superferromagnetic behavior than in a superparamagnetic phase \cite{Sawicki:2013_PRB}.

It was often argued that the presence of the magnetic circular dichroism (MCD) points to a random distribution of TM ions over cation sites \cite{Ando:2006_S}. Actually, because of boundary conditions for electromagnetic waves, the dielectric function of heterogeneous media contains contributions from all constituents. Hence, TM-rich NCs not only give a specific contribution to the magnetic response but, in general, can also affect, in a magnetization-dependent fashion, magnetooptical properties, as observed for (Ga,Mn)As, (Ge,Mn), (Ge,Fe), and (Zn,Cr)Te (Secs.\,\ref{sec:GaAs}, \ref{sec:Ge-Mn}, \ref{sec:Ge-Fe}, and \ref{sec:ZnTe}, respectively). In particular, metallic NCs rather than enhancing MCD only at host critical points of the Brillouin zone produce a large MCD signal over a wide spectral range. However, no enhancement of MCD has been reported for (Ga,Mn)N and (Ga,Fe)N (Secs.\,\ref{sec:GaN-Mn} and \ref{sec:GaN-Fe}, respectively).  Presumably, in the case of (Ga,Mn)N,  Mn-rich NCs are wide band-gap insulators, whereas in (Ga,Fe)N the NCs aggregate in a narrow plane parallel to the film surface, which reduces their coupling to light.

A similar question arises to what extent magnetotransport studies could tell whether the TM distribution is uniform or one deals rather with spinodal nanodecomposition. A weak negative contribution to the Hall signal, resembling the anomalous Hall effect (AHE), can originate from stray fields generated by magnetic NCs.  More often, spin-dependent coupling between host charge current and magnetic NCs results in a sizable AHE, as observed in (Ge,Mn) and (Zn,Cr)Te, and mentioned in Secs.\,\ref{sec:Ge-Mn} and \ref{sec:ZnTe}. Particularly challenging is the question of magnetoresistance (MR). Here positive MR, specific to semiconductor-metallic nanocomposites \cite{Solin:2000_S}, may appear, particularly if the magnitude of AHE is significant, the effect being discussed for (Ge,Mn) in Sec.\,\ref{sec:Ge-Mn}. Moreover, as in other nonmagnetic and magnetic semiconductors, one expects a range of MRs associated with quantum localization phenomena \cite{Dietl:2008_JPSJ}. In particular, strong spin disorder scattering near $T_{\text{C}}$ accounts for colossal negative MR. However, a weak negative MR away from $T_{\text{C}}$, according to straightforward and parameter-free theoretical modeling, originates rather from the influence of magnetic flux upon interference of scattered carrier de Broglie waves (weak localization MR) than from spin-disorder scattering. This is in contrast with magnetically doped {\em metals} in which effects of spin-disorder scattering dominate, as competing scattering mechanisms and associated localization effects are typically weak.

Despite the recent progress, there is a number of challenging and open questions ahead. As an example we note that an important theoretical issue is the determination of kinetic barriers for TM diffusion at the growth surface and in the bulk, crucial parameters to simulate quantitatively the NC assembly during epitaxy, codoping, or postgrowth annealing. Furthermore, since CMSs assume a form imposed by the matrix, their chemical composition and associated properties are by no means obvious. In particular, it is hardly known whether open $d$ shells remain localized in CMSs (as in the parent DMS) or rather a Mott-Hubbard transition occurs, leading to itinerant magnetism  specific to certain end compounds such as MnAs and CrTe. It was actually suggested that Mn-rich zinc-blende (zb) NCs in decomposed (Ga,Mn)As contain only about 20\% of Mn and retain the properties of (Ga,Mn)As in which spins are localized \cite{Lawniczak-Jablonska:2011_PSS}. A related question concerns the crystalline magnetic anisotropy of particular combinations of CMSs and hosts. This anisotropy, together with shape anisotropy and dipolar or exchange coupling between NCs, accounts for macroscopic magnetic properties, including the apparent magnitude of $T_{\text{C}}$ and the character of magnetic hystereses. A theoretical and experimental evaluation of this anisotropy awaits for future studies.

Furthermore, FM-like features brought about by uncompensated spins at the surface of antiferromagnetic (AF) NCs, as discussed theoretically  \cite{Dietl:2007_PRB}, have not yet been put into evidence experimentally in the materials considered here. We also note that the FM proximity effect or the exchange bias in the case of AF CMSs can lead to spin polarization of a semiconductor surrounding a given NC. Such induced polarization will persist up to the spin ordering temperature of the CMS and, according to the RKKY theory, will extend over a distance of the order of the inverse Fermi vector in the presence of band carriers and perhaps over two or three bond lengths in their absence. This effect was examined in Fe/(Ga,Mn)As heterostructures \cite{Maccherozzi:2008_PRL} but not yet in decomposed alloys, in which it can lead to an erroneous conclusion that the host is intrinsically FM up to high temperatures.

While theoretical and experimental results described in this review are concerned with a limited set of compounds, there is a common ground to expect that the developed methodology, the observed phenomena, and the conclusions drawn from theoretical and experimental results apply to a much broader class of magnetically doped materials and also to many other alloys. We are quite certain that with further development and with  wide-spread use of powerful nanocharacterization tools, the family of decomposed alloys will steadily grow. Future work will also show the role of structural defects, residual impurities (e.g. H) or contaminates (e.g. Fe-rich nanoparticles) in the appearance of high-$T_{\text{C}}$ ferromagnetism in semiconductors and oxides.

It is clear that the understanding of the origin of high-$T_{\text{C}}$ ferromagnetism and detailed knowledge on the TM distribution are preconditions for a meaningful design of devices that could exploit the properties of magnetically doped semiconductors and insulators. As discussed in Sec.\,\ref{sec:prospects}, the demonstrated and predicted functionalities of decomposed magnetic alloys might be of interest not only for spintronics but also for electronics, photonics, plasmonics, photovoltaics, thermoelectrics, and catalysis. In general terms, the future studies of these systems will contribute to address the timely question of to what extent and when bottom-up technologies will start to be competitive with the top-down approaches dominating today.

\section*{List of abbreviations}
\begin{longtable}{l l}
1D, 2D, 3D& One-, two-, three-dimensional \\
AF& Antiferromagnetic \\
AP& Atom probe \\
AFM& Atomic force microscopy \\
AHE& Anomalous Hall effect \\
CEMS & Conversion electron M{\"o}ssbauer spectroscopy \\
CMS &Condensed magnetic semiconductor \\
CPA &Coherent potential approximation \\
DFT &Density functional theory \\
DMS &Dilute magnetic semiconductors \\
DFS& Dilute ferromagnetic semiconductor \\
EDS &Energy dispersive x-ray spectroscopy \\
     &(the same as EDX) \\
EELS &Electron energy loss spectroscopy \\
EPR &Electron paramagnetic resonance \\
EF& Energy-filtered \\
EXAFS &Extended x-ray Absorption Fine Structure\\
FC &Field cooled \\
fcc &face-centered cubic \\
FM &Ferromagnetic \\
FMR &Ferromagnetic resonance \\
FR & Ferrimagnetic\\
GGA &Generalized gradient approximations \\
GISAXS&Grazing incidence small angle x-ray scattering \\
GMR &Giant magnetoresistance \\
HR &High resolution \\
LDA &Local density approximation \\
LSDA &Local spin density approximation \\
LT &Low temperature  \\
MBE &Molecular beam epitaxy  \\
MOVPE &Metal organic vapor phase deposition \\
    &(the same as MOCVD)  \\
MCD& Magnetic circular dichroism \\
MR &Magnetoresistance  \\
NC &Nanocrystal \\
RHEED &Reflection high-energy electron diffraction\\
RPA & Random phase approximation \\
RKKY &Ruderman-Kittel-Kasuya-Yosida \\
SEM &Scanning electron microscopy \\
SIMS &Secondary ions mass spectroscopy  \\
SQUID &Superconducting quantum interference device  \\
STM &Scanning tunneling microscopy \\
SXRD &Synchrotron x-ray diffraction  \\
TEM &Transmission electron microscopy  \\
TM &Transition metal \\
TMR &Tunneling magnetoresistance  \\
wz &wurtzite \\
XANES  &X-ray absorption near edge structure\\
XAS &X-ray absorption spectroscopy \\
XES &X-ray emission spectroscopy \\
XMCD &X-ray magnetic circular dichroism \\
XPEEM &X-ray photoemission electron microscopy \\
XRD &X-ray diffraction \\
ZFC &Zero-field cooled \\
zb &zinc-blende
\end{longtable}

\begin{acknowledgments}
The work of T.\,D. and A.\,B.
was supported by the European Research Council through the FunDMS Advanced Grant (No.\,227690)
within the "Ideas" 7th Framework Programme of the EC,
by Wroclaw Research Centre EIT+ within the project
"The Application of Nanotechnology in Advanced Materials" - NanoMat
(P2IG.01.01.02-02-002/08) co-financed by the European Regional Development
Fund (operational Programme Innovative Economy, 1.1.2),
and also (T.\,D.) by National
Center of Science in Poland (Decision No.\,2011/02/A/ST3/00125).
A.\,B. was supported by the Austrian Fonds
zur {F\"{o}rderung} der wissenschaftlichen Forschung --
FWF (P18942, P20065, P22471, P22477, and P26830) 
and by the NATO Science for Peace Programme (Project 984735).
K.\,S. acknowledges Ministry of Education, Culture, Sports, Science and Technology (MEXT)  
KAKENHI (Grant numbers, 22740256, 22104012 and 26286074), the GCOE program 
"Core Research and Engineering of Advanced Materials-Interdisciplinary Education Center for Materials Science"
by MEXT.  K.\,S. also acknowledges the financial support from Strategic Japanese-German Cooperative Program "Computational design and evaluation of spintronics materials" and PREST from Japan Science and Technology Agency (JST).
 M.\,J. and A.\,Ba. was supported by the French Agence Nationale pour
la Recherche (ANR project GeMO) and the Nanoscience Foundation of Grenoble
(RTRA project IMAGE). M.\,T. and P.\,N.\,H. acknowledge the financial support
from the Grant-in-Aid for  Specially Promoted Research (No. 23000010) and Grant-in-Aid for Young Scientists (A)
(No. 24686040) by JSPS, the Project for Developing Innovation Systems of MEXT,
the FIRST program, the Global COE program (C04), and the Advanced Technology Institute (ATI) foundation.
The study of S.\,K. was partially
supported by Grant-in-Aids for Scientific Research from Ministry of Education,
Science, Sports and Culture of the Japanese Government.
H.\,K.-Y. acknowledges the financial
support provided by the Japan Society for the
Promotion of Science (JSPS) 'Core-to-Core' Program "Computational Nano-Materials
Design on Green Energy", and Advanced Low Carbon Technology Research and Development
Program (ALCA) of Japan Science and Technology
Agency (JST) "Spinodal Nanotechnology
for Super-High Efficiency Energy Conversion". H.\,K.-Y. and M.\,T. also acknowledge
the stimulated discussion in the meeting of the Cooperative Research
Project of the Research Institute of Electrical Communication, Tohoku University.
H.\,K.-Y. thanks the Future Research Initiative Group Support Project on Computational 
Nano-Materials Design: New Strategic Materials, Osaka University.
\end{acknowledgments}





%

\end{document}